\documentclass[11pt]{article}
\usepackage{amsmath,amsfonts,amsthm,amssymb}
\usepackage[hmargin={1.0in, 1.0in}, vmargin={1.0in, 1.0in}]{geometry}
\usepackage{setspace}
\usepackage[utf8]{inputenc}
\usepackage{lastpage}
\usepackage{extramarks}
\usepackage{chngpage}
\usepackage{soul,color}
\usepackage{graphicx,float,wrapfig}
\usepackage{layout}
\usepackage{fullpage}
\usepackage{natbib}
\usepackage{accents}
\usepackage[titletoc]{appendix}
\usepackage{multirow}
\usepackage{xcolor}
\usepackage{array}

\usepackage{hyperref}[]
\hypersetup{
	colorlinks=true,
	linkcolor=black,
	filecolor=magenta,      
	urlcolor=cyan,
	citecolor=black,
}
\usepackage{cleveref}
\newcommand{\ignore}[1]{}{}

\newcommand{\colorlinks}[2]{{\hypersetup{allcolors=#1}#2}}
\newcommand\thickbar[1]{\accentset{\rule{.4em}{1pt}}{#1}}

\def\bfx{\mathbf x}
\def\bfy{\mathbf y}

\def\bfX{\mathbf X}
\def\bfY{\mathbf Y}

\def\bs{{\textbf{s}}}
\def\bd{{\textbf{d}}}

\DeclareMathOperator*{\argmin}{arg\,min}
\DeclareMathOperator*{\argmax}{arg\,max}

\makeatletter
\def\thm@space@setup{\thm@preskip=1pt
\thm@postskip=1pt}
\makeatother

\newtheorem{theorem}{Theorem}
\newtheorem{assumption}{Assumption}

\newtheorem{proposition}{Proposition}

\newtheorem*{definition*}{Definition}
\theoremstyle{definition}
\newtheorem{remark}{Remark}

\newtheorem{simulation}{Simulation}

\newcommand{\blind}{1}

\usepackage{titlesec}
\titlespacing*\section{0pt}{2pt plus 2pt minus 2pt}{0pt plus 2pt minus 2pt}
\titlespacing*\subsection{0pt}{2pt plus 2pt minus 2pt}{0pt plus 2pt minus 2pt}
\titlespacing*\subsubsection{0pt}{0pt plus 2pt minus 2pt}{0pt plus 2pt minus 2pt}

\usepackage{xr}
\makeatletter
\newcommand*{\addFileDependency}[1]{
  \typeout{(#1)}
  \@addtofilelist{#1}
  \IfFileExists{#1}{}{\typeout{No file #1.}}
}
\makeatother

\begin{document}
\setlength{\abovedisplayskip}{2pt}
\setlength{\belowdisplayskip}{2pt}
\setlength{\abovedisplayshortskip}{1pt}
\setlength{\belowdisplayshortskip}{1pt}
\setlength\intextsep{4pt}

\if1\blind
{
	\title{A Composite Likelihood-based Approach for Change-point Detection in Spatio-temporal Processes}	
	\author{
		Zifeng Zhao$^{1}$, Ting Fung Ma$^{2}$, Wai Leong Ng$^{3}$, Chun Yip Yau$^{4}$\\
		University of Notre Dame$^{1}$, University of South Carolina$^{2}$,\\
		Hang Seng University of Hong Kong$^{3}$ and Chinese University of Hong Kong$^{4}$
		}
	\date{}	
	\maketitle
} \fi

\if0\blind
{
	\title{A Composite Likelihood-based Approach for Change-point Detection in Spatio-temporal Processes}
	\author{}
	\date{}
	\maketitle
	\vspace{-2cm}
} \fi

\vspace{-8mm}

\begin{abstract}
    This paper develops a unified and computationally efficient method for change-point estimation \textcolor{black}{along the time dimension} in a non-stationary spatio-temporal process. By modeling a non-stationary spatio-temporal process as a piecewise stationary spatio-temporal process, we consider simultaneous estimation of the number and locations of change-points, and model parameters in each segment. A composite likelihood-based criterion is developed for change-point and parameter estimation. Under the framework of increasing domain asymptotics, theoretical results including consistency and distribution of the estimators are derived under mild conditions. In contrast to classical results in fixed dimensional time series that the localization error of change-point estimator is $O_{p}(1)$, exact recovery of true change-points is possible in the spatio-temporal setting. More surprisingly, the consistency of change-point estimation can be achieved without any penalty term in the criterion function.	In addition, we further establish consistency of the change-point estimator under the infill asymptotics framework where the time domain is increasing while the spatial sampling domain is fixed. A computationally efficient pruned dynamic programming algorithm is developed for the challenging criterion optimization problem. Extensive simulation studies and an application to the U.S. precipitation data are provided to demonstrate the effectiveness and practicality of the proposed method.
\end{abstract}

\noindent\textit{Keywords}: 
Dynamic programming; increasing domain asymptotics; infill asymptotics; multiple change-points; pairwise likelihood; asymptotic distribution.

\section{Introduction}\label{sec:intro}
With the advances in data technology, large datasets observed sequentially over long periods are becoming increasingly available. For analyzing such data, stationary models are often inadequate. Instead of developing complicated models for describing the non-stationary behavior, it is often more intuitive and effective to incorporate a change-point model which segments the data into stationary pieces. \textcolor{black}{This makes change-point analysis increasingly popular in recent decades across many applications, such as climate science~\colorlinks{black}{\citep{killick2010detection, Lu2010, Kelly2014}}, finance~\colorlinks{black}{\citep{ang2012regime,FRYZLEWICZ2014}}, genetics~\colorlinks{black}{\citep{shen2012change,Fearnhead2018}}, and signal processing~\colorlinks{black}{\citep{wang2004change,harle2016bayesian}}.}

Change-point estimation has been extensively studied for time series data, see e.g.\ \cite{Davis2006, Aue2009, Shao2010a, Matteson2014, Preuss2015, Jiang2021}. Recently, change-point analysis has also gained popularity in high-dimensional statistics \citep{Cho2015, Wang2018, Wang2021, chen2022inference} and functional data analysis \citep{Berkes2009, Aston2012, Aue2018}. However, \textcolor{black}{change-point estimation} for spatio-temporal processes remains largely unexplored with only a handful of works in the literature. Moreover, existing methods are subject to various limitations. For example, Bayesian methods such as \cite{Majumdar2005} and \cite{Altieri2015} usually require very specific model structures without theoretical guarantees and are computationally intensive. \cite{Gromenko2017} focuses on at most one change-point in the mean function of a spatio-temporal process. Furthermore, all existing literature require the separability of space-time covariance. The major difficulty of change-point analysis for spatio-temporal processes stems from the theoretical and computational challenges of spatio-temporal modeling under the increase in dimensions of both space and time with the presence of unknown change-points.

In this paper, we propose a likelihood-based procedure for multiple change-point estimation \textcolor{black}{along the time dimension} in a spatio-temporal process with spatial locations at a possibly irregular grid. \textcolor{black}{We take this approach since the parametric model, such as regression based mean functions and Mat\'ern class based space-time covariance functions, is one of the main workhorses in the spatio-temporal literature.} Our procedure adopts pairwise likelihood to alleviate computational difficulty of full likelihood for spatio-temporal data while maintaining statistical efficiency. The proposed method is a general approach that can handle a wide range of spatio-temporal models including both separable and non-separable space-time covariance. Furthermore, it works under model misspecification and can detect changes beyond first and second moments. In spatial statistics, there are mainly two asymptotic framework, namely the increasing domain asymptotics and the spatial infill asymptotics. The choice of asymptotic frameworks plays an important role in establishing theoretical properties of the statistical methodology. Both frameworks are common in practice, see \cite{stein1999interpolation}, \cite{zhang2005towards} and \cite{bevilacqua2020unifying} for more discussions. Thus, for completeness, we establish the theoretical results of the proposed change-point estimation procedure under both the increasing domain and infill asymptotics in the spatial dimension.

The contribution of this paper is two-fold. First, in terms of statistical theory, we show that in contrast to the commonly used likelihood-based change-point estimation approach for multivariate time series \citep[e.g.][]{Davis2006, Ma2016}, the pairwise likelihood for spatio-temporal data induces an edge effect around each change-point. This edge effect is non-ignorable under the spatio-temporal setting as the spatial dimension grows and can cause inconsistency of change-point estimation. \textcolor{black}{To tackle this problem, we carefully design a compensating mechanism which modifies the pairwise likelihood by introducing a marginal likelihood term to correct the edge effect.}

Interestingly, unlike traditional criterion functions such as BIC and minimum description length (MDL) which involve a penalty term, the consistency of the change-point estimation can be achieved solely by the modified pairwise likelihood. Moreover, in contrast to classical results in fixed dimensional time series that the asymptotic error of change-point estimation is $O_{p}(1)$, we show that exact recovery of true change-points is possible in the spatio-temporal setting under the increasing domain asymptotics. To further achieve consistent model selection for each segment and enhance finite sample performance, an MDL-based criterion function is developed. We prove that, even under possible model misspecification, the number and locations of change-points can be consistently estimated under mild conditions. The asymptotic distributions of the estimated change-points and the spatio-temporal model parameters in each stationary segment are also derived. In addition, we further establish the consistency of the number and locations of the change-point estimator under infill asymptotics where the time domain is increasing while the spatial sampling domain is fixed. To the best of our knowledge, this paper is the first to study change-point estimation in spatio-temporal data under the spatial infill setting.

Second, in terms of statistical computing, we develop a computationally efficient algorithm for the optimization of the criterion function. Computational feasibility is a major challenge in change-point estimation since it involves optimization over a large number of change-point configurations. Popular optimization methods, such as binary segmentation~\citep{Vostrikova1981} and genetic algorithm \citep{Davis2006}, are fast but only provide approximate solutions. On the other hand, dynamic programming~\citep{Jackson2005} provides exact solutions but incurs quadratic computational cost. In this paper, we adapt the pruned exact linear time (PELT) algorithm in \cite{Killick2012} (originally designed for univariate time series) to the spatio-temporal setting. The new algorithm is computationally efficient and proves to provide an asymptotically exact solution.

\textcolor{black}{We remark that change-point estimation based on pairwise likelihood with an MDL penalty has previously been studied in \colorlinks{black}{\cite{Ma2016}} under the multivariate time series setting. However, as evident from our discussion above, there are notable differences between the two works. First, our work focuses on the more complex spatio-temporal setting with a growing spatial dimension. Importantly, we show the pairwise likelihood procedure in \colorlinks{black}{\cite{Ma2016}} leads to inconsistent change-point estimation due to an edge effect under our setting (see details later), and we instead design a new criterion function with both pairwise and marginal likelihood to achieve consistency. Second, we operate under both the increasing domain and spatial infill asymptotics, and discover new phenomena for change-point estimation such as exact recovery and consistency without penalty, which requires substantially different technical arguments than the ones in \colorlinks{black}{\cite{Ma2016}}.
}

The rest of the paper is organized as follows. Section \ref{sec:background} provides the background and derivation of the composite likelihood based criterion for change-point estimation. The main results under increasing domain asymptotics including estimation consistency and asymptotic distribution of the estimators are presented in Section \ref{sec:main}. Numerical experiments and an application to the U.S. precipitation data are given in Section \ref{sec:num}. Section \ref{sec:conclusion} concludes. All technical proofs, detailed results under the infill asymptotics, and additional numerical studies can be found in the supplement.

\section{Background}\label{sec:background}
\subsection{Settings and notations}\label{subsec:setting}
On a set of spatial locations $\mathcal{S}$ with cardinality $S=\text{Card}(\mathcal{S})$, 
consider a spatio-temporal process
\begin{align*}
\bfY=\{y_{t,\bs}: t\in [1,T],~\bs\in \mathcal{S}\} = \{\bfy_t: 1\leq t\leq T, t\in\mathbb{N}^+\}   \,,
\end{align*}
where $\bfy_t=\{y_{t,\bs}: \bs \in\mathcal{S} \}$ denotes the observations of all spatial locations at time $t$, \textcolor{black}{and given any two positive integers $t_1\leq t_2$, we denote $[t_1,t_2]$ as the set $\{t_1,t_1+1,\cdots,t_2\}.$} There are in total $S\cdot T$ observations. We focus on $\mathcal{S}\subset \mathbb{R}^2$, while our result can be easily generalized to $\mathbb{R}^d$ with $d\geq 3$.

\textbf{\textcolor{black}{Data generating process}}: We assume that $\bfY$ can be partitioned into $m_o+1$ stationary segments \textcolor{black}{along the time dimension}\footnote{Note that potentially there can also be structural breaks across the space $\mathcal{S}$, however, since there is no natural order for $\mathcal{S}$, the search space is much larger, so here we do not consider this case.}. \textcolor{black}{In other words, there are $m_o$ unknown change-points $0<\tau_1^o<\cdots<\tau_{m_o}^o<T$ in the spatio-temporal process. Our asymptotic results hold given that as $S,T$ increase, the normalized change-point $\tau_j^o/T$ converges to a limit $\lambda_j^o$ for all $j$, where $0<\lambda_1^o<\cdots<\lambda_{m_o}^o <1$. However, for clarity of presentation, in the rest of the paper, we simply set $\tau_j^o=[T\lambda_j^o]$ for $j=1,\cdots,m_o$, where for a number $z\in \mathbb R_{\geq 0}$, we denote $[z]$ as the closest integer to $z$.} For notational convenience, we further define $\tau_0^o=0$ and $\tau_{m_o+1}^o=T$, and $\lambda_0^o=0$ and $\lambda_{m_o+1}^o=1.$

Let \textcolor{black}{$T_j^o=\tau_j^o-\tau_{j-1}^o$} be the length of the $j$th stationary segment for $j=1,\ldots,m_o+1$. For convenience, we can \textcolor{black}{re-index} the $j$th stationary segment of $\bfY$ as a stationary process \textcolor{black}{$\bfX_j^o=\{\dot\bfx_t^{(j)}: t\in[1,T_j^o]\}$ with $\dot\bfx_t^{(j)}=\{\dot x_{t,\bs}^{(j)}: \bs\in\mathcal{S}\}$}, such that
\begin{align} \label{xy}
\dot\bfx_{t-\tau_{j-1}^o}^{(j)}=\bfy_t\,, \quad t=\tau_{j-1}^o+1,\ldots,\tau_j^o\,.
\end{align}
The observed spatio-temporal process $\bfY$ can then be written as
\begin{align*}
\bfY=(\dot\bfx_1^{(1)},\ldots,\dot\bfx_{T_1^o}^{(1)},\dot\bfx_1^{(2)},\ldots,\dot\bfx_{T_2^o}^{(2)},\ldots,\dot\bfx_1^{(m_o+1)},\ldots,\dot\bfx_{T_{m_o+1}^o}^{(m_o+1)})\,.
\end{align*}

As in \cite{Davis2008} and \cite{Aue2009}, we first assume for simplicity that the data across different segments $\bfX_j^o$ ($j=1,\ldots,m_o+1$) are independent. This assumption is commonly found in spatio-temporal change-point literature~\citep[e.g.][]{Altieri2015, Gromenko2017}. We refer to Remark \ref{dep.seg} and \Cref{sec:relaxIS} of the supplement for discussions on its relaxations.

\textcolor{black}{\textbf{Model and parameterization}}: We adopt a parametric approach and model each stationary segment by a member of a pre-specified finite class of models, $\mathcal M$. Each element in $\mathcal M$ is a model indexed by an integer-valued vector $\xi$ that represents the model order. In other words, $\xi$ determines the form of the parametric model. Given $\xi$, the model can be fully specified by a $d(\xi)$-dimensional parameter $\theta(\xi)$ in a compact parameter space  $\Theta(\xi) \subset \mathbb{R}^{d(\xi)}$. We refer to $\xi$ as the model order and $\theta(\xi)$ as the model parameter.

For the $j$th stationary segment $\bfX_j^o$, we assume there exists a pseudo-true parametric model in $\mathcal{M}$, indexed by a model order vector $\xi_j^o$ of dimension $c_j^o$, that provides the best fit for the data~(see \Cref{ass.model.id} later for the technical definition). Given $\xi_j^o$, the exact model for $\bfX_j^o$ is fully specified by a model parameter $\theta_j^o=\theta_j(\xi_j^o)$ of dimension $d_j^o=d_j(\xi_j^o)$. \textcolor{black}{Importantly, note that we do not require $\mathcal{M}$ to cover the true data generating process of $\bfX_j^o$.} This modeling framework is flexible. In particular, on each stationary segment, it allows a general spatio-temporal model such that
\begin{align}
y_{t,\bs}=\mu_{t,\bs}+\varepsilon_{t,\bs}.
\label{GeneralModel}
\end{align}
Here, $\mu_{t,\bs}$ is the mean process which can take various regression forms such as $z_{t,\bs}^\top\beta$ with $z_{t,\bs}$ being the covariate associated with $(t,\bs)$, and $\varepsilon_{t,\bs}$ is the error process whose space-time covariance can take various parametric forms of separable or non-separable spatio-temporal dependence. 

\textcolor{black}{\textbf{Notation}: Denote $\psi_j^o=(\xi_j^o,\theta_j^o)$ as the true model parameter set for the $j$th stationary segment $\bfX_j^o=\{\dot\bfx_1^{(j)},\ldots,\dot\bfx_{T_j^o}^{(j)}\}$ and denote $\Psi^o = \{\psi_1^o,\ldots,\psi_{m_o+1}^o\}$. Furthermore, denote $\Lambda^o = \{\lambda_1^o,\ldots,\lambda_{m_o}^o\}$ as the set of true (normalized) change-points. To avoid confusion, in the following, we use $\Lambda=(\lambda_1,\cdots,\lambda_m)$ to denote a generic set of $m$ change-points. Furthermore, we denote $\bfX_1,\bfX_2,\cdots,\bfX_{m+1}$ as a generic partition of $\bfY$ imposed by $\Lambda$, where $\bfX_j=\{\bfx_1^{(j)},\ldots,\bfx_{T_j}^{(j)}\}$ is of length $T_j=\tau_{j}-\tau_{j-1}$ with $\tau_j=[T\lambda_j].$ Note that $\bfX_1,\bfX_2,\cdots,\bfX_{m+1}$ depend on $\Lambda$ implicitly, which is suppressed for notational simplicity. In particular, we have that $\bfX_j=\bfX_j^o$ for all $j=1,2,\cdots,m_o+1$ if $\Lambda=\Lambda^o$. We use $\psi_j=(\xi_j,\theta_j)$ to denote a generic model parameter set for the $j$th segment $\bfX_j$, where $\xi_j$ is of dimension $c_j$ and $\theta_j$ is of dimension $d_j.$ Denote $\Psi=\{\psi_1,\cdots,\psi_{m+1}\}$.}

\textcolor{black}{\textbf{An illustrative example}: To build more intuition, we conclude this section with a concrete example where the model class $\mathcal{M}$ consists of Gaussian space-time AR~(STAR) models. On a stationary segment, an STAR model of order $q$ takes the form
\begin{align*}
y_{t,\bs}=\mu+\sum_{i=1}^q \rho_iy_{t-i,\bs}+\varepsilon_{t,\bs},
\end{align*}
where $\varepsilon_t=\{\varepsilon_{t,\bs}, \bs\in \mathcal{S}\}$ is a Gaussian process that can take $K$ possible parametric forms of spatial dependence, such as exponential or Mat\'ern covariance. Thus, the model order can be specified by $\xi=(i_1,i_2)$, where $i_1\in\{1,\cdots,p\}$ indicates the temporal autoregressive order and $i_2\in\{1,\cdots,K\}$ indicates the parametric form of the spatial dependence. Given $\xi$, the model parameter $\theta(\xi)$ consists of the temporal AR coefficients $(\mu,\rho_1,\cdots,\rho_q)$ and the spatial covariance parameters of $\varepsilon_t$.}

\subsection{Composite likelihood and pairwise likelihood}\label{subsec:CLPL}
Although (full) likelihood based methods generally achieve high statistical efficiency, 
when the likelihood function involves high-dimensional inverse covariance matrices or
integrals, computations become infeasible. 
To overcome this limitation, \cite{Lindsay1988} considers the composite likelihood, which is a weighted product of likelihoods for some subsets of the data. By specifying the subsets, different classes of composite likelihood are obtained. One popular class is the pairwise likelihood~(PL), which is the product of the bivariate densities of all possible pairs of observations,
\begin{equation} \label{PL.weight.form}
\mathcal{L}_P(\theta; \{x_i\}_{i=1}^n)=\prod_{i,j }\mathcal{L}(\theta;x_{i},x_{j})^{w_{i,j}}\,,
\end{equation}
where $w_{i,j}$ are the weights. Composite likelihood often enjoys computational efficiency while statistical efficiency is retained; see \cite{Lindsay1988,Varin2011}.
Owing to its flexibility and attractive
asymptotic properties, composite likelihood has been widely used in genetics \citep{Larribe2011}, longitudinal data \citep{Bartolucci2016}, time series \citep{Davis2011}, and spatio-temporal statistics \citep{Bevilacqua2012, HuserDavison2014}.

\textcolor{black}{Given a generic segment $\bfX_j=\{\bfx_1^{(j)},\ldots,\bfx_{T_j}^{(j)}\}$} and a model parameter $\psi_j$, the classical pairwise likelihood is defined as the product of the bivariate densities $f(\cdot,\cdot;\psi_j)$ of each distinct pair $(x_{t,\bs}^{(j)},x_{t',\bs'}^{(j)})$ in $\bfX_j$. However, in many spatio-temporal processes, the dependence between observations diminishes quickly as the time lag or spatial distance increases. Therefore, it suffices to consider pairs in $\bfX_j$ that are up to a small time lag $k$ and a small spatial distance $d$ apart. 

In particular, define $\mathcal{N}\equiv\mathcal{N}(\bs)=\{\bs'|\bs'\in\mathcal{S}, \bs' \neq \bs, \text{dist}(\bs,\bs')\leq d\}$ as a distance-based neighborhood of location $\bs$, where the distance can be Euclidean. Note that $\mathcal{N}$ depends on $d$ implicitly.
{\color{black} Given $(k,\mathcal{N})$, for a time lag $0< i\leq k$, define the index set
\begin{align*}
    P_{i,\mathcal{N}}^{(j)}= \bigcup_{t=1}^{T_j-i} \big\{(t,i,\bs_1,\bs_2): \bs_1\in\mathcal{S}, \bs_2 \in \{\bs_1\}\cup\mathcal{N}(\bs_1)\big\},
\end{align*}
which collects all pairs of observations $(x_{t,\bs_1}^{(j)},x_{t+i,\bs_2}^{(j)})$ in $\bfX_j$ that are exactly $i$ time units apart and at most $d$ spatial distance away. For the time lag $i=0$, we further define $P_{0,\mathcal{N}}^{(j)}=\bigcup_{t=1}^{T_j} \big\{(t,0,\bs_1,\bs_2): \bs_1\in\mathcal{S}, \bs_2 \in \mathcal{N}(\bs_1)\big\}$. We then define
$$\mathcal D_{k,\mathcal{N}}^{(j)}=\bigcup_{i=0}^k P_{i,\mathcal{N}}^{(j)},$$ 
which is the collection of pairs of distinct observations in $\bfX_j$ that are at most $k$ time units and $d$ spatial distance apart.}
The pairwise log-likelihood of $\bfX_j$ is then defined as
\begin{align}\label{CPL.k}
\mathrm{PL}(\psi_j;\bfX_j) 
= \sum_{(t,i,\bs_1,\bs_2)\in \mathcal{D}_{k,\mathcal{N}}^{(j)}} \log f(x_{t,\bs_1}^{(j)}, x_{t+i,\bs_2}^{(j)};\psi_j) 
=\sum_{(t,i,\bs_1,\bs_2)\in \mathcal{D}_{k,\mathcal{N}}^{(j)}} l_{pair}(\psi_j;x_{t,\bs_1}^{(j)}, x_{t+i,\bs_2}^{(j)})\,.
\end{align}

For the choices of $k$ and $\mathcal{N}$ (i.e.\ $d$), intuitively, \textcolor{black}{a large neighborhood} can be used if there exists strong spatial correlation across $\mathcal{S}$, and \textcolor{black}{a small neighborhood} should be favored if the spatial correlation is weak; see \cite{Varin2005Biometrika} and \cite{Bai2012}. On the other hand, similar to \cite{Ma2016}, if the main focus is estimating change-points rather than model parameters, it usually suffices to use the smallest $k$ and $d$ that ensure identifiability of the models in the candidate model set $\mathcal{M}$. See more discussions below Assumption \ref{ass.model.id} in Section \ref{sec:main}.

\subsection{Edge effect and a remedial composite likelihood}\label{subsec:edge_effect}
By the definition of $\mathcal D_{k,\mathcal{N}}^{(j)}$ in \eqref{CPL.k}, each data point in a generic segment $\bfX_j=\{\bfx_1^{(j)},\ldots,\bfx_{T_j}^{(j)}\}$ may {\textit{not}} appear in $\mathrm{PL}(\psi_j;\bfX_j)$ for the same number of times. For example, $\bfx_1^{(j)}$ can only be paired with $\{\bfx_t^{(j)}:t=1,\ldots,k+1\}$, and thus appears approximately half frequently compared to observations $\bfx_{\tilde{t}}^{(j)}$, $k+1\leq \tilde{t} \leq T_j-k$, which can be paired with $\{\bfx_t^{(j)}: \tilde{t}-k \leq t\leq \tilde{t}+k\}$. This can be viewed as that different weights are implicitly assigned to the observations in $\bfX_j$, and observations on the edge of a segment receive less weights. We refer to this phenomenon of the pairwise likelihood~(PL) defined in equation \eqref{CPL.k} as the \textit{edge effect}. 

To determine whether a change-point $\tau$ exists on a segment $\bfX_j$, we need to compare two \textcolor{black}{log-likelihood} quantities: the pairwise \textcolor{black}{log-}likelihood formed by $\{\bfx_t^{(j)}: 1 \leq t \leq T_j\}$, and the sum of pairwise \textcolor{black}{log-}likelihoods formed by $\{\bfx_t^{(j)}: 1 \leq t \leq \tau\}$ and $\{\bfx_t^{(j)}: \tau+1 \leq t < T_j\}$. For the latter quantity, all observations within $k$ time units from $\tau$ will suffer from the edge effect and receive less weights, which thus causes that the latter quantity has $O(S)$ fewer terms than the former one.

The edge effect is negligible if the spatial dimension $S$ is fixed since it is of order $O_p(1)$, as is in the multivariate time series setting. However, when $S\longrightarrow \infty$, the edge effect is non-ignorable and can cause inconsistency of the PL based method. In Section \ref{sec:edge_effect} of the supplement, through a simple and intuitive example, we show that due to the fact $S\longrightarrow \infty$ in the spatio-temporal setting, the edge effect can cause false positives in the PL based change-point estimation asymptotically with probability 1, \textcolor{black}{which is further confirmed by an accompanying simulation study.}

To correct the edge effect of PL and achieve consistency of change-point estimation, \textcolor{black}{we design a compensating mechanism for} the missing pairwise log-likelihoods on the edge of each segment, so that each data point appears the same number of times in the likelihood function. In particular, the mechanism introduces additional \textit{marginal} log-likelihoods for data points observed on the edge $t=(1,\ldots, k)\cup (T_j-k+1,\ldots,T_j)$. \textcolor{black}{Based on the compensating mechanism}, we propose a newly designed composite likelihood that takes the form 
\begin{align}\label{eq:CPL}
&L_{ST}^{(j)}(\psi_j; \bfX_j)= 
\mathrm{PL}(\psi_j;\bfX_j)
+ \sum_{(i,\bs)\in \mathcal E_{k,\mathcal{N}}}\log f(x_{i,\bs}^{(j)};{\psi_j}) + \sum_{(i,\bs) \in \mathcal E_{k,\mathcal{N}}}\log f(x_{T_j-i+1,\bs}^{(j)};{\psi_j}) \nonumber\\
:= &\sum_{(t,i,\bs_1,\bs_2)\in \mathcal{D}_{k,\mathcal{N}}^{(j)}} l_{pair}(\psi_j;x_{t,\bs_1}^{(j)}, x_{t+i,\bs_2}^{(j)})+ \sum_{(i,\bs)\in \mathcal E_{k,\mathcal{N}}}\left[l_{marg}(\psi_j;x_{i,\bs}^{(j)}) + l_{marg}(\psi_j;x_{T_j-i+1,\bs}^{(j)})\right]\,, 
\end{align}
where $l_{marg}(\psi;x)=\log f(x;\psi)$, $l_{pair}(\psi;x_1, x_2)=\log f(x_1,x_2,\psi)$, and $\mathcal E_{k,\mathcal{N}}= \bigcup_{i=1}^k \{(i,\bs):\bs\in\mathcal{S},\text{~repeat } (i,\bs) \text{ by } (k-i+1)(1+|\mathcal{N}(\bs)|) \text{ times}\}$ denotes the collection of marginal log-likelihoods used for correcting the edge effect.

\subsection{Derivation of the criterion} \label{section.mdl}
With the modified composite likelihood, in this section we derive a criterion function for estimating the change-points and model parameters in each segment. The criterion is based on the minimum description length (MDL) principle, which aims to select the best-fitting model that requires the minimum amount of code length to store the data \citep{Rissanen2012}, and has been shown to have promising performance for change-point estimation, see \cite{Davis2006,Davis2008,Lu2010}.
 
One classical way to construct the MDL is the two-stage approach \citep{Hansen2001,Lee2001}, which splits the code length, $\mathrm{CL}(\bfY)$, into two components: 
\begin{align*}
\mathrm{CL}(\bfY) = \mathrm{CL}(\widehat{\mathcal{M}}) + \mathrm{CL}( \bfY\mid\widehat{\mathcal{M}})\,,
\end{align*}
where $\mathrm{CL}(\widehat{\mathcal{M}})$ is the code length for the fitted model $\widehat{\mathcal{M}}$ and $\mathrm{CL}(\bfY\mid\widehat{\mathcal{M}})$ is the information in $\bfY$ unexplained by $\widehat{\mathcal{M}}$. Recall that the model parameter set of the $j$th segment is specified by $\psi_j=(\xi_j,\theta_j)$, where $\xi_j$ is the model order and $\theta_j$ is the model parameter. Given $\xi_j$, the composite likelihood estimator of $\theta_j$ is obtained by $\hat{\theta}_{j}=\argmax_{\theta\in\Theta(\xi_j)} 
L_{ST}^{(j)}\{(\xi_j,\theta); \bfX_j\}$, where $L_{ST}^{(j)}$ is defined in \eqref{eq:CPL}. Since the fitted model $\widehat{\mathcal{M}}$ can be completely described by $m$, $\lambda_j$'s and $\psi_j$'s, we have
\begin{align*}
	\mathrm{CL}(\widehat{\mathcal{M}}) =  \mathrm{CL}(m) + \mathrm{CL}(\lambda_1)+\cdots+ \mathrm{CL}(\lambda_{m})+ \mathrm{CL}(\xi_1)+\cdots+\mathrm{CL}(\xi_{m+1}) 
	  +  \mathrm{CL}(\hat{\theta}_1)+\cdots+\mathrm{CL}(\hat{\theta}_{m+1})\,.
\end{align*}

Note that $(\lambda_1,\ldots,\lambda_m)$ contains information equivalent to the integer-valued vector $(T_1,\ldots,T_{m+1})$, and the code length for an integer $I$ is \textcolor{black}{approximately} $\log_2 I$. From \cite{Hansen2001} and \cite{Lee2001},  
the code length for an estimate of a real-valued parameter depends on the precision by the optimal quantization of the parameter space, which is related to the standard error of the estimate. In particular, under the increasing domain asymptotics and some regularity conditions, a parameter estimator computed from $N$ observations is $1/ \sqrt{N}$-consistent, and hence the code length for an estimate is $(\log_2 N)/2$. Thus, we have
\begin{align}\label{eq:CLM}
\mathrm{CL}(\widehat{\mathcal{M}})= \log_2 (m+1) + \sum_{j=1}^{m+1}\log_2 T_j+\sum_{j=1}^{m+1}\sum_{i=1}^{c_j}\log_2 \xi_{i,j}+ \sum_{j=1}^{m+1} \frac{d_j}{2}(\log_2 T_j +\log_2 S)\,,
\end{align}
where $\xi_j=(\xi_{1,j},\ldots,\xi_{c_j,j})$. 

\textcolor{black}{We remark that since the segment length $T_j$ is upper bounded by $T$ and there are finite number of models in $\mathcal{M}$, the uniform code $({m+1})\log_2 T$ and $(m+1)\log_2 |\mathcal{M}|$ can instead be used for the second and third terms in \eqref{eq:CLM} and will lead to the same asymptotic results. However, in finite sample, we find that $\log_2 T_j$ achieves higher detection power due to its smaller magnitude, and $\sum_{i=1}^{c_j}\log_2 \xi_{i,j}$ encourages parsimony as lower-index models in $\mathcal{M}$ are less complex.}

As demonstrated by \cite{Rissanen2012}, $\mathrm{CL}(\bfY\mid\widehat{\mathcal{M}})=-\log_2 L$, where $L$ is the maximized full likelihood. By regarding the composite likelihood as a proxy to the full likelihood and using logarithm base $e$ rather than base $2$, the sum of the negative of \eqref{eq:CPL} can be used to define $\mathrm{CL}(\bfY\mid \widehat{\mathcal{M}})$. However, by construction, each data point may appear several times in the composite likelihood function. In particular, for any $\bfX_j$ with length $T_j\geq 2k$, the average number of times that an observation is used in the composite log-likelihood $L_{ST}^{(j)}(\psi_j, \bfX_j)$ equals to  
\begin{align}\label{eq:Ck}
C_{k,\mathcal{N}}=\frac{2 \text{Card}(\mathcal D_{k,\mathcal{N}}^{(j)})+2\text{Card}(\mathcal E_{k,\mathcal{N}})}{ST_j}=\frac{\sum_{\bs\in\mathcal{S}}(2k+(2k+2)|\mathcal{N}(\bs)|)}{S}\,,
\end{align}
where $\mbox{Card}(\cdot)$ denotes the cardinality of a set. Indeed, if the data are identically and independently distributed, the composite log-likelihood is a product of marginal densities and $L_{ST}^{(j)}(\psi_j, \bfX_j)$ is essentially $C_{k,\mathcal{N}}$ times the full log-likelihood. Thus, we compensate the code length by multiplying $\mathrm{CL}(\widehat{\mathcal{M}})$ by a factor $C_{k,\mathcal{N}}$ and define the composite likelihood-MDL~(CLMDL) criterion as  
\begin{eqnarray} \label{MDLform}
{\mathrm{CLMDL}}(m,\Lambda,\Psi) &=& C_{k,\mathcal{N}}\left\{ \log m +\sum_{j=1}^{m+1}\sum_{i=1}^{c_j}\log \xi_{i,j}+ \sum_{j=1}^{m+1} \left[\left(\frac{d_j}{2}+1\right)\log T_j + \frac{d_j}{2}\log S\right]\right\} \nonumber \\
&&- \sum_{j=1}^{m+1}L_{ST}^{(j)}(\psi_j;\bfX_j)\,.
\end{eqnarray}

{\color{black} We remark that unlike the classical MDL criterion, the proposed CLMDL criterion in \eqref{MDLform} may no longer be interpreted as a code-length function since the adjusted composite likelihood is used. Nevertheless, as will be seen, this construction still allows one to balance between the lack of fit and model complexity, and offers consistent estimation of change-points and model parameters.}

\textcolor{black}{\textbf{CLMDL-based estimation}: We estimate the unknown true parameters $(m_o,\Lambda^o,\Psi^o)$ by minimizing the CLMDL criterion $\eqref{MDLform}$}. To ensure identifiability of the change-points, we impose a tuning parameter $\epsilon_\lambda \in (0,1/2)$ such that $\min_{1\leq j \leq m+1}|\lambda_j - \lambda_{j-1} |\geq\epsilon_\lambda$. In other words, we consider
\begin{equation} \label{A}
A_{\epsilon_\lambda}^m=\{\Lambda\in (0,1)^m: 0=\lambda_0<\lambda_1<\cdots<\lambda_m<\lambda_{m+1}=1,~ \lambda_j-\lambda_{j-1}\geq\epsilon_\lambda, ~j=1,\ldots,m+1\}\,.
\end{equation}
Thus, the number of change-points is \textcolor{black}{upper bounded by $M_\lambda:=[1/\epsilon_\lambda-1]$. For theoretical validity, we require $\epsilon_\lambda\leq\epsilon_\lambda^o$, where $\epsilon_\lambda^o=\min_{1\leq j\leq m_o+1}(\lambda_{j}^o-\lambda_{j-1}^o)$ is the minimum spacing between true change-points. This is a common assumption in the change-point estimation literature for parametric models, see e.g.\ \colorlinks{black}{\cite{Andrews1993}, \cite{Davis2006},  \cite{Ling2014}, \cite{Ma2016}, and \cite{romano2022detecting}}. A sensitivity analysis is conducted in \Cref{sec:add_num}~(\Cref{add_simu_epsilon}) of the supplement, which shows CLMDL is robust to the choice of $\epsilon_\lambda$.} Denote $\mathcal{M}^{m+1}$ as the Cartesian product of $\mathcal{M}$.


The estimated number and locations of change-points and parameters in each segment are thus
\begin{equation} \label{minMDL}
(\hat{m},\hat{\Lambda}_{ST},\hat{\Psi}_{ST}) = \argmin_{m \leq M_\lambda, \Lambda\in A_{\epsilon_\lambda}^m, \Psi \in \mathcal{M}^{m+1}} {\mathrm{CLMDL}}(m,\Lambda,\Psi)\,,
\end{equation}
where $\hat{\Lambda}_{ST}=(\hat{\lambda}_1,\ldots,\hat{\lambda}_{\hat{m}})$ and 
$\hat{\Psi}_{ST}=(\hat{\psi}_1,\ldots,\hat{\psi}_{\hat{m}+1})$ with 
$\hat{\psi}_j=(\hat{\xi}_j,\hat{\theta}_j)$. Note that 
\begin{equation*}
\hat{\theta}_j=\argmax_{\theta\in\Theta(\hat{\xi}_j)} L_{ST}^{(j)}\{(\hat{\xi}_j,\theta);{\hat{\bfX}_j}\}
\end{equation*}
is the composite likelihood estimator of the model parameters of the $j$th estimated segment of the spatio-temporal process, i.e.\
$\hat{\bfX}_j= \left\{\bfy_t: [T\hat{\lambda}_{j-1}]+1\leq t\leq [T\hat{\lambda}_j]\right\}$.

\vspace{-1mm}
\section{Main Results under Increasing Domain Asymptotics}\label{sec:main}
In this section, we first impose some mild regularity conditions on the composite log-likelihood and the strong-mixing coefficients of the spatio-temporal process, and then present the main results.

For the asymptotic theory, we assume that the piecewise stationary spatio-temporal process is generated by a random field in a (possibly unevenly spaced) lattice $\mathcal{Z}=\mathbb{N}^+ \times \mathcal{S} \subset \mathbb{N}^+\times\mathbb{R}^2$. The data $\bfY$ is observed on $\mathcal{Z}_n=\mathcal{T}_n\times \mathcal{S}_n$ with $\mathcal{T}_n=[1,T_n]$ and $|\mathcal{S}_n|=S_n$. In other words, we have $\bfY=\{y_{t,\bs}: t\in [1,T_n], ~\bs\in \mathcal{S}_n\}$. The asymptotic theory is based on $n\longrightarrow\infty$, where the number of observations $|\mathcal{Z}_n|=S_nT_n > |\mathcal{Z}_{n'}|=S_{n'}T_{n'}$ whenever $n>n'$. For notational simplicity, in the following we use $(S,T)$ instead of $(S_n,T_n)$ when there is no possibility of confusion. 

We define a metric $\rho$ on $\mathcal{Z}$ by $\rho(\bd_1,\bd_2)=\max(|t_2-t_1|, |s_2^1-s_1^1|, |s_2^2-s_1^2|)$, where 
$\bd_i=(t_i,\bs_i)$, $\bs_i=(s_i^1,s_i^2)$, $i=1,2$, denote any two points in $\mathcal{Z}$. The distance between any two subsets $U,V\in \mathcal{Z} $ is further defined as $\rho(U,V)=\inf\{\rho(\bd_1,\bd_2): \bd_1\in U \text{ and } \bd_2\in V \}\,.$ In this section, the theoretical results are established under the increasing domain asymptotics framework as in \cite{Jenish2009} and \cite{Bai2012}, which is made explicit by Assumption \ref{ass.countable}. 
\begin{assumption} \label{ass.countable}
	The lattice $\mathcal{Z}\subset \mathbb{N}^+\times\mathbb{R}^2$ is countably infinite. All elements in $\mathcal{Z}$ are located at distances of at least $\rho_0>0$ from each other, i.e., for all $ \bd_1, \bd_2 \in \mathcal{Z},$ we have $\rho(\bd_1,\bd_2)\geq \rho_0$.
\end{assumption}
Assumption \ref{ass.countable} allows for unevenly spaced locations and general forms of sample regions, which is often encountered in real data. By Assumption \ref{ass.countable}, we can assume the maximum cardinality of the neighborhood set 
$\mathcal{N}(\bs)$ is bounded by a constant $B_{\mathcal{N}}$. Throughout this section, we assume the time lag used in CLMDL is $k$ and the maximum cardinality of $\mathcal{N}(\bs)$ in CLMDL is $B_{\mathcal{N}}$.


\begin{assumption}[$r$] \label{ass.mom} There exists an $\epsilon > 0$ such that for any fixed model order $\xi$, we have\\
	\indent (i) for $a=0$ and each stationary segment $\bfX_j^o$, $j=1,\ldots,m_o+1$,
	\begin{align*}
	&\sup_{S,T}\sup_{(t,i,\bs_1,\bs_2) \in \mathcal{D}^{(j)}_{k,\mathcal{N}}}\mathbb E\left[\sup_{\theta \in \Theta(\xi)} |l_{pair}^{[a]}\{(\xi,\theta);\dot x_{t,\bs_1}^{(j)}, \dot x_{t+i,\bs_2}^{(j)}\}|^{r+\epsilon}\right] < \infty  \,,\\
	&\sup_{S,T}\sup_{(t,\bs)\in \mathcal E_{k,\mathcal{N}}}\mathbb E\left[\sup_{\theta \in \Theta(\xi)} |l_{marg}^{[a]}\{(\xi,\theta);\dot x_{t,\bs}^{(j)}\}|^{r+\epsilon}\right] < \infty  \,,
	\end{align*}
	
	(ii) for $a=1,2$, the above moment conditions hold with $r=2$,\\	
	where $l_{marg}$ and $l_{pair}$ are defined in \eqref{eq:CPL} and $[a]$ stands for the $a$th order derivative w.r.t. $\theta$.
\end{assumption}
Assumption \ref{ass.mom}($r$) requires that the composite log-likelihood \eqref{eq:CPL} is twice continuously differentiable and has a finite $(r+\epsilon)$th moment, and its first and second order derivatives have a finite $(2+\epsilon)$th moment. Note that if the data $\bfY$ is observed on a regular lattice, by the piecewise stationarity assumption, 
the supremum w.r.t.\ $(S,T)$ can be dropped in Assumption \ref{ass.mom}($r$).

Given $\psi$, define the expected log-likelihood of each stationary segment as $\thickbar{L}_{ST}^{(j)}(\psi)=\mathbb E\{L_{ST}^{(j)}(\psi; \bfX_j^o)\}$. The derivatives $\thickbar{L}_{ST}^{'(j)}(\psi)=\mathbb E\{L_{ST}^{'(j)}(\psi; \bfX_j^o)\}$ and 
$\thickbar{L}_{ST}^{''(j)}(\psi)=\mathbb E\{L_{ST}^{''(j)}(\psi; \bfX_j^o)\}$ are defined similarly. 
\begin{assumption}\label{ass.model.id}
    For each stationary segment $\bfX_j^o$ of the random field, where $j=1,\ldots,m_o+1$,
	
	(i) there exists a model $\xi_j^o \in \mathcal{M}$ with a parameter 
	$\theta_j^o\in\mathbb{R}^{d_j^o}$ satisfying 
	$$\psi_{j}^{o}=(\xi_j^o,\theta_j^o)=\argmax_{\xi\in\mathcal M, ~\theta\in\Theta(\xi)}\thickbar{L}^{(j)}_{ST}\{(\xi,\theta)\}\,.$$ 
	The model order $\xi_j^o$ is uniquely identifiable in the sense that if there exists another model 
	$(\xi_j^*,\theta_j^*)  \neq (\xi_j^o, \theta_j^o)$ with
	$\theta_j^* \in \mathbb{R}^{d_j^*}$ and $\thickbar{L}^{(j)}_{ST}\{(\xi_j^*,\theta_j^*)\}= \thickbar{L}^{(j)}_{ST}\{(\xi_j^o,\theta_j^o)\}$, then $d_j^* > d_j^o$. 
	Moreover, for any $\delta>0$, we have 
	$\sup\limits_{S,T}\frac{1}{ST}\left(\sup_{\|\theta-\theta^{o}_{j}\|_2>\delta} \thickbar{L}^{(j)}_{ST}\{(\xi_j^o,\theta)\} -\thickbar{L}^{(j)}_{ST}\{(\xi_j^o,\theta_j^o)\}\right)< 0$. The same holds for $(\xi_j^*,\theta_j^*).$ 
	
	(ii) $\sup\limits_{S,T}\frac{1}{ST}\left(\thickbar{L}^{(j)}_{ST}(\psi_{j-1}^o) - \thickbar{L}^{(j)}_{ST}(\psi_j^o)\right)<0$ and $\sup\limits_{S,T}\frac{1}{ST}\left(\thickbar{L}^{(j-1)}_{ST}(\psi_j^o)-\thickbar{L}^{(j-1)}_{ST}(\psi_{j-1}^o)\right)<0$, 
	where $\psi_{j-1}^o$ and $\psi_j^o$ are defined in (i).  
\end{assumption}

Similar as above, if the data $\bf Y$ is observed on a regular lattice, then by the piecewise stationarity assumption, 
the supremum w.r.t.\ $(S,T)$ can be dropped. Note that Assumption \ref{ass.model.id} does not require that the stationary process is from the model class $\mathcal{M}$. Instead, Assumption \ref{ass.model.id}(i) only assumes the existence of a pseudo-true model $\psi_{j}^{o}$ in $\mathcal{M}$, which is of the simplest form. That is, the model cannot be expressed by another model $\psi_{j}^*=(\xi_{j}^*,\theta_{j}^*)$ in $\mathcal{M}$, where $\theta_j^{*}$ is of a smaller dimension. The last statement in Assumption \ref{ass.model.id}(i) asserts that for any model order $\xi^{*}_{j}$, the point $\theta^{*}_{j}$ is the unique parameter value that maximizes the expected composite likelihood.

Assumption \ref{ass.model.id}(ii) rules out the degenerate case that the $(j-1)$th and $j$th stationary segments are indistinguishable by the composite likelihood. Assumption \ref{ass.model.id}(ii) may fail if two adjacent segments follow different models but have the same expected composite likelihood. \textcolor{black}{In \Cref{subsec:pathological} of the supplement, we give two pathological examples where such scenarios may happen.} In principle, this situation can always be avoided by using a larger time lag $k$ and spatial neighborhood $\mathcal{N}$ when defining \eqref{eq:CPL}. As noted by \cite{Ma2016}, this situation seldom occurs in practice and $k=1$ or $2$ is usually sufficient to distinguish between stationary segments.

\textcolor{black}{For more intuition regarding \Cref{ass.model.id}(ii), consider the case where two adjacent segments share the same model order $\xi_{j-1}^o=\xi_j^o$, and thus $\theta_{j-1}^o$ and $\theta_j^o$ can be directly compared. In view of Assumption \ref{ass.model.id}(i) and boundedness of the first order derivatives in \Cref{ass.mom}(ii), for Assumption \ref{ass.model.id}(ii) to hold, it is equivalent to require $\|\theta_{j-1}^o-\theta_j^o\|_2>\epsilon$ for some $\epsilon>0$. In other words, the change size of the model parameter is non-vanishing. In \Cref{subsec:vanish_cp}, we further show that \Cref{ass.model.id}(ii) can indeed be relaxed to allow vanishing change sizes under additional conditions.}

The next two assumptions regulate the dependence structure of the spatio-temporal process by imposing mild $\alpha$-mixing conditions on the underlying random field. For the $j$th stationary segment $\bfX_j^o$ of the random field, denote $\sigma_{\bfX_j^o}(U)=\sigma(\dot x_{t,\bs}^{(j)}: (t,\bs)\in U)$ as the sigma field generated by the random variables in the index set $U \subset \mathbb{N}^+\times \mathcal{S}.$ Define
\begin{align*}
\alpha_{\bfX_j^o}(U,V)=\sup\big\{|P(A\cap B)-P(A)P(B)|: A\in\sigma_{\bfX_j^o}(U), B\in \sigma_{\bfX_j^o}(V) \big\}\,.
\end{align*}
The $\alpha$-mixing coefficient for the $j$th stationary segment $\bfX_j^o$ is defined as
\begin{align*}
\alpha_{\bfX_j^o}(d;u,v)=\sup\big\{\alpha_{\bfX_j^o}(U,V): |U|\leq u, ~|V|\leq v,~ \rho(U,V)\geq d\big\}\,.
\end{align*}
Also, denote $M=(1+k)(1+B_{\mathcal{N}})$, where $2M$ is an upper bound of the number of composite likelihood components $l_{pair}$ that any point $(t,\bs)$ can have in $\mathcal{D}_{k,\mathcal{N}}^{(j)}$.

\begin{assumption}[$r$] \label{ass.mix}
    For each stationary segment $\bfX_j^o$, where $j=1,\ldots,m_o+1$, there exist some $\epsilon>0$ and $c\in 2\mathbb{N}^+$ where $c>r$, such that for all $u,v \in \mathbb{N}^+$, $u+v\leq c$, $u,v\geq 2$, we have
    $$\sum_{d=1}^{\infty}(d+1)^{3(c-u+1)-1}[\alpha_{\bfX_j^o}(d;Mu,Mv)]^{\epsilon/(c+\epsilon)}<\infty\,.$$
\end{assumption}

Assumption \ref{ass.mix} requires a polynomial decay rate for the $\alpha$-mixing coefficient of the random field, which is mild and is used for invoking the moment inequality in \cite{Doukhan1994} to control the asymptotic size of the deviation of the composite likelihood from its expectation. Note that the mixing rate in Assumption \ref{ass.mix} depends on $M=(1+k)(1+B_{\mathcal{N}})$. 
This is intuitive since a longer time lag $k$ and a larger neighborhood size $B_{\mathcal{N}}$ induce a slightly higher dependence among the composite likelihood and thus require   stronger conditions on the mixing coefficients. Define also
\begin{align}\label{mixing.time}
\textcolor{black}{\alpha_{\bfX_j^o}^*(d)=\sup\Big\{\alpha_{\bfX_j^o}\big([t_1,t_2]\times \mathcal{S}', ~[t_2+d, t_3]\times \mathcal{S}' \big): 1\leq t_1\leq t_2,~ t_2+d\leq t_3<\infty,~ \mathcal{S}'\subset \mathcal{S} \Big\}},
\end{align}
which characterizes the dependence of the spatio-temporal process along the time dimension, and can be regarded as an analogue to the classical $\alpha$-mixing coefficient in the time series setting. 

\begin{assumption}\label{ass.mixtime}
For each stationary segment $\bfX_j^o$ of the random field, where $j=1,\ldots,m_o+1$, there exist $r>2$, $\delta>0$, $\tau>r(r+\delta)/(2\delta)$ and $C>0$, such that for any fixed model order $\xi$,
\begin{align*}
    (i)~\sup_{S,T}\sup_{(t,i,\bs_1,\bs_2) \in \mathcal{D}_{k,\mathcal{N}}^{(j)}}\mathbb E\left[\sup_{\theta \in \Theta(\xi)} |l_{pair}^{[1]}\{(\xi,\theta);\dot x_{t,\bs_1}^{(j)}, \dot x_{t+i,\bs_2}^{(j)}\}|^{r+\delta}\right] < \infty\,, \text{ and }
    (ii)~\alpha_{\bfX_j^o}^*(d)\leq C d^{-\tau}.
\end{align*}
\end{assumption}

Assumption \ref{ass.mixtime}(i) is a moment condition on the first order derivative of the composite likelihood. Assumption \ref{ass.mixtime}(ii) is a mild mixing condition along the time dimension and is used for invoking an maximal moment inequality in \cite{Yang2007} that controls the asymptotic size of the first order derivative of the composite likelihood at the true parameter value. Note that for a fixed moment condition $c=r+\delta$, the slowest polynomial decay rate for the mixing condition is $\tau>\frac{c}{c-2}$, which is achieved when $r \to 2$. Thus, a higher order moment condition on the first order derivative requires a weaker mixing condition. Next, we impose Assumptions \ref{ass.thetadistr} and \ref{ass.variance}, which are standard in establishing the asymptotic distribution of the parameter estimator $\hat{\theta}_j$ in a stationary segment; see Assumption 3 in \cite{Jenish2009} and Assumptions 6--8 in \cite{Bai2012}.
\begin{assumption}\label{ass.thetadistr}
	For each stationary segment $\bfX_j^o$ of the random field, where $j=1,\ldots,m_o+1$, there exists some $\delta>0$ such that (i) $\sum_{d=1}^{\infty}\alpha_{\bfX_j^o}(d;M, M)d^{3(2+\delta)/\delta-1}<\infty$, (ii) $\sum_{d=1}^{\infty}d^2\alpha_{\bfX_j^o}(d;Mu,Mv)<\infty$ for $u+v\leq 4$, and (iii) $\alpha_{\bfX_j^o}(d;M,\infty)=O(d^{-3-\epsilon})$ for some $\epsilon>0$.
\end{assumption}

\begin{assumption}\label{ass.variance}
	For each stationary segment $\bfX_j^o$ of the random field, we have
	\begin{align*}
	(i)~\frac{1}{ST}\mbox{Var}(L_{ST}^{'(j)}(\psi_j^o; \bfX_j^o)) \longrightarrow \Sigma_1^{(j)}\,,\hspace{0.3cm}
	(ii)~-\frac{1}{ST}\mathbb E(L_{ST}^{''(j)}(\psi_j^o; \bfX_j^o)) \longrightarrow \Sigma_2^{(j)}\,,
	\end{align*}
	where $j=1,\ldots,m_o+1$, and $\Sigma_1^{(j)}$ and $\Sigma_2^{(j)}$ are positive definite matrices.
\end{assumption}

{\color{black}
\begin{remark}[Mixing conditions]\label{remark:mixing}
Define $\alpha_{\bfX_j^o}(d):=\sup_{u\geq 1}\sup_{v\geq 1} \alpha_{\bfX_j^o}(d;u,v)$, which is a mixing coefficient commonly used in the random field literature, see \colorlinks{black}{\cite{BerkesMorrow1981} and \cite{Doukhan1994}}~(Section 1.3). Clearly, we have $\alpha_{\bfX_j^o}^*(d)\leq \alpha_{\bfX_j^o}(d).$ Thus, all mixing conditions in Assumptions \ref{ass.mix}, \ref{ass.mixtime}(ii) and \ref{ass.thetadistr} hold if $\alpha_{\bfX_j^o}(d)$ decays at a sufficiently fast \textit{polynomial} rate. This is a mild condition and is satisfied by many widely used spatio-temporal parametric models, such as a Gaussian random field with a separable space-time Mat\'ern covariance function or with other popular non-separable space-time covariance functions proposed in \colorlinks{black}{\cite{stein2005space} and \cite{fuentes2008class}}. To conserve space, we refer to \Cref{sec:alphMixing} of the supplement for more details. 
\end{remark}
}

\subsection{Consistency of CLMDL under increasing domain asymptotics}
We now present the theoretical results. We begin with a somewhat surprising finding on the consistency of change-point estimation using the composite likelihood function alone without any penalty for model complexity. It requires the existence of the $(2+\epsilon)$th moments of the composite likelihood in \eqref{eq:CPL} and its derivatives, and a mild divergence rate requirement between $T$ and $S$.

Analogous to the CLMDL based estimator in \eqref{minMDL}, define the CL based estimator 
    \begin{eqnarray}\label{minCL}
    (\tilde{m},\tilde{\Lambda}_{ST},\tilde{\Psi}_{ST}) = \argmin_{m \leq M_\lambda, \Lambda\in A_{\epsilon_\lambda}^m, \Psi \in \mathcal{M}^{m+1}} 
    - \sum_{j=1}^{m+1}L_{ST}^{(j)}(\psi_j;\bfX_j) \,, 
    \end{eqnarray}
where $\tilde{\Lambda}_{ST}=(\tilde{\lambda}_1,\ldots,\tilde{\lambda}_{\tilde{m}})$, $\tilde{\Psi}_{ST}=(\tilde{\psi}_1,\ldots,\tilde{\psi}_{\tilde{m}+1})$ with $\tilde{\psi}_j=(\tilde{\xi}_j,\tilde{\theta}_j)$, and the estimated model parameter is $\tilde{\theta}_j=\argmax_{\theta\in\Theta(\tilde{\xi}_j)} L_{ST}^{(j)}\{(\tilde{\xi}_j,\theta);{\tilde{\bfX}_j}\}$ with $\tilde{\bfX}_j= \Big\{\bfy_t: [T\tilde{\lambda}_{j-1}]+1\leq t\leq [T\tilde{\lambda}_j]\Big\}$. 

\begin{proposition}\label{prop.minCL}
    Let $\bfY$ be a piecewise stationary spatio-temporal process specified by $(m_o,\Lambda^o,\Psi^o)$, and Assumptions \ref{ass.countable}, \ref{ass.mom}($r$), \ref{ass.model.id} and \ref{ass.mix}($r$) hold with some $r>2$.

    \noindent (i).\ If $S,T\longrightarrow \infty$, \textcolor{black}{with probability going to 1, we have $\tilde m\geq m_o$} and for each $j=1,\ldots,m_o$, there exists a $\tilde{\lambda}_{i_j}\in \tilde{\Lambda}_{ST}, 1\leq i_j \leq \tilde{m}$, such that
    $[T\lambda_j^o] = [T\tilde{\lambda}_{i_j}]$.

    \noindent (ii).\ If, in addition, $T\cdot S^{-r/2}\longrightarrow 0$ and Assumption \ref{ass.mixtime} holds, with probability going to 1, we have
    \begin{align}\label{weak.consist.CL}
	\tilde{m} = m_o, \quad  [T\tilde{\Lambda}_{ST}]=[T\Lambda^o]\,,
    \end{align}
    and $\tilde\theta_j$ has a dimension greater than or equal to $d_j^o=d_{j}(\xi^o_j)$ for $j=1,\ldots,m_o+1$.
\end{proposition}

The consistency of the CL-based change-point estimator in Proposition \ref{prop.minCL}(ii) can be viewed as the opposite phenomenon to the inconsistency caused by the edge effect discussed in Section \ref{subsec:edge_effect}. \textcolor{black}{It is due to the compensating mechanism of the proposed composite likelihood~(which is designed to correct the edge effect) and the fact that the spatial dimension $S$ diverges~(i.e. $S\longrightarrow\infty$).}

The high-level intuition is as follows. Consider the $j$th stationary segment $\mathbf X_j^o$. By assigning one false change-point to $\mathbf X_j^o$, the difference introduced on the CL criterion function in \eqref{minCL} is the difference between the dropped pairwise likelihood and the corresponding compensating marginal likelihood evaluated at $\theta_j^o$~(and a $O_p(1)$ term due to the estimation error of model parameters). Compared to the dropped pairwise log-likelihood, the compensating marginal log-likelihood incorrectly imposes temporal independence on the stationary segment. Thus, by the definition of $\theta_j^o$ in Assumption \ref{ass.model.id}(i), the expected value of this difference is positive and is of order $O(S)$. \textcolor{black}{Based on a moment inequality, we can further show the actual value of this difference is concentrated around its expectation, and therefore is positive and of order $O_p(S)$.} Thus, a model with an over-estimated change-point will have a significantly larger criterion function value than the true model, and will not be selected. In contrast, in the classical time series setting, $S$ is fixed and the consistency of $m_o$ needs to be achieved by an extra penalty term in the criterion function to suppress over-estimation. 

Proposition \ref{prop.minCL} only guarantees consistency in change-point estimation but not model selection. Indeed, introducing the MDL penalty term ensures consistency of model selection in each segment and improves the finite sample performance in terms of reducing false positives when there is no change-point. Theorem \ref{unknownprob} states the consistency of our procedure based on the CLMDL criterion.

\begin{theorem} \label{unknownprob}
Let $\bfY$ be a piecewise stationary spatio-temporal process specified by $(m_o,\Lambda^o,\Psi^o)$, and Assumptions \ref{ass.countable}, \ref{ass.mom}($r$), \ref{ass.model.id} and \ref{ass.mix}($r$) hold with some $r>2$. Consider the CLMDL based estimator $(\hat{m},\hat{\Lambda}_{ST},\hat{\Psi}_{ST})$ defined in \eqref{minMDL}.

\noindent (i).\ If $S,T\longrightarrow \infty$, \textcolor{black}{with probability going to 1, we have $\hat m\geq m_o$} and for each $j=1,\ldots,m_o$, there exists a $\hat{\lambda}_{i_j}\in \hat{\Lambda}_{ST}, 1\leq i_j \leq \hat{m}$, such that $[T\lambda_j^o] =[T\hat{\lambda}_{i_j}]$. 

\noindent (ii).\ If, in addition, $T\cdot S^{-r/2}\longrightarrow 0$ and Assumption \ref{ass.mixtime} holds, we further have
	\begin{align}\label{weak.consist}
	\hat{m} = m_o, ~ 
	[T\hat{\Lambda}_{ST}]=[T\Lambda^o], ~
        \text{and} ~~  \hat{\xi}_j=\xi_j^o, ~\hat{\theta}_j \longrightarrow \theta_j^o ~ \text{ for all } j=1,\cdots,m_o+1,
	\end{align}
 with probability going to 1.
\end{theorem}

In Proposition \ref{prop.minCL} and Theorem \ref{unknownprob}, we have $\mathbb P([T\hat{\Lambda}_{ST}]=[T\Lambda^o])\longrightarrow 1$, which indicates that asymptotically we can recover the exact location of the change-point\textcolor{black}{s} without error. This result is different from the one under the multivariate time series setting, where an $O_p(1)$ error is observed asymptotically. This difference is due to the expansion of the spatial domain $\mathcal{S}$ under the spatio-temporal setting, which intuitively provides more information for the detection of change-points.

\begin{remark}[Infill asymptotics]
	An alternative asymptotic framework commonly used in the spatial statistics literature is the infill asymptotics, where the time domain is increasing while the spatial sampling domain is fixed. In \Cref{sec:main_infill} of the supplement, we modify the CLMDL criterion to tailor to the infill setting and further establish its estimation consistency in \Cref{asym_infill}. Intuitively, under the infill setting, since we are sampling from a bounded spatial domain, the increasing sample size $S$ along the space dimension may or may not accumulate more information for change-point estimation. In particular, we show that exact recovery of the change-point is possible if the change happens in a microergodic parameter. Roughly speaking, a parameter is microergodic if a higher spatial sampling resolution improves its estimation accuracy~(see \cite{stein1999interpolation} and \cite{Zhang2004}). For changes in a non-microergodic parameter, the localization error achieves the classical $O_p(1)$ rate. To conserve space, we refer to \Cref{sec:main_infill} of the supplement for more details.
\end{remark}

To guarantee the consistency of the estimated number of change-points, Proposition \ref{prop.minCL} and Theorem \ref{unknownprob} require an additional polynomial rate condition on the divergence rate between $S$ and $T$, i.e.\ $T\cdot S^{-r/2}\longrightarrow 0$. This technical condition is needed for controlling the asymptotic size of terms due to the compensating mechanism via a union bound. Note that a higher moment condition $r$ implies a less restrictive divergence rate requirement. When $\bfY$ is Gaussian, all moments of the composite likelihood exist, and the rate requirement becomes minimal. Indeed, if the spatio-temporal process $\bfY$ is observed on a regular lattice, then under an additional assumption on the mixing coefficients, a finer result in terms of the divergence rate between $S$ and $T$ can be obtained.

\begin{theorem}\label{finerate} 
Let $\bfY$ be a piecewise stationary spatio-temporal process specified by $(m_o,\Lambda^o,\Psi^o)$, and Assumptions \ref{ass.countable}, \ref{ass.mom}($r$), \ref{ass.model.id}, \ref{ass.mix}($r$) and \ref{ass.mixtime} hold with $r>2$. Moreover, assume that 
\begin{eqnarray}\label{eq.mix.space}
\alpha_{\bfX_j^o}(d):=\sup_{u\geq1}\sup_{v\geq1}\alpha_{\bfX_j^o}(d;u,v) \leq C d^{-\tau}\,, 
\end{eqnarray}
for some $C>0$, $\tau >2(1+\eta)(q+\delta)/(q-2+\delta)$ with some $q>2$, $\delta>0$ and $0<\eta<1/2$. 
Then, for the estimator $(\hat{m},\hat{\Lambda}_{ST},\hat{\Psi}_{ST})$ defined in \eqref{minMDL}, with probability going to 1, we have
    \begin{align}\label{weak.consist2}
	\hat{m} = m_o, ~ 
	[T\hat{\Lambda}_{ST}]=[T\Lambda^o], ~
        \text{and} ~~  \hat{\xi}_j=\xi_j^o, ~\hat{\theta}_j \longrightarrow \theta_j^o ~ \text{ for all } j=1,\cdots,m_o+1
    \end{align}
provided that $S,T\longrightarrow \infty$ with $\log T/S \longrightarrow 0$.
\end{theorem}

\begin{remark}\label{dep.seg}
The consistency results given in Proposition \ref{prop.minCL}, Theorems \ref{unknownprob} and \ref{finerate} are based on the independence assumption across stationary segments, which is commonly used in the change-point estimation literature. In \Cref{sec:relaxIS} of the supplement~(see \Cref{consistency_relaxIS}), we further relax this assumption and show that, similar to the classical time series setting, CLMDL can still achieve consistency with a $O_p(1)$ localization error rate when there is weak dependence across segments.
\end{remark}

\subsubsection{\textcolor{black}{{Vanishing change sizes}}}\label{subsec:vanish_cp}
{\color{black}
The exact recovery phenomenon in Theorems \ref{unknownprob} and \ref{finerate} suggest CLMDL may further work with vanishing change sizes under the increasing domain asymptotics, which is confirmed in this subsection.

To facilitate a transparent and intuitive definition of change sizes, in this subsection, we assume the model order of the pseudo-true parametric model in $\mathcal{M}$ are the same across the $m_o+1$ stationary segment, i.e.\ $\xi_1^o=\cdots=\xi_{m_o+1}^o=\xi^o$. Therefore, the pseudo-true model parameters $\theta_j^o$ of all $j=1,\cdots,m_o+1$ stationary segments share the \textit{same} meaning and same dimension $d^o=d(\xi^o)$, and can be directly compared. Specifically, the parameter change at the $j$th true change-point can be defined as $\Delta_j=\theta_{j+1}^o-\theta_j^o$ and the change size can be measured by its $l_2$-norm $\|\Delta_j\|_2.$ Otherwise, if $\xi_j^o\neq \xi_{j+1}^o$, the difference between $\theta_j^o$ and $\theta_{j+1}^o$ is not well defined and we need to quantify the change size via the expected composite log-likelihood, which is less intuitive and interpretable\footnote{\textcolor{black}{With tedious notations, we can extend our result to the case where the pseudo-true model orders $\{\xi_1^o,\cdots,\xi_{m_o+1}^o\}$ are different but consist of nested models, such as space-time AR($q$) models with $q=1,\cdots,p$. In such case, $\theta_j^o$ can be of different dimensions but we can define the change size using the pseudo-true parameter $\theta_j^*$ at the highest model order, i.e.\ $\max\{\xi_1^o,\cdots,\xi_{m_o+1}^o\}$, which shares the same meaning and dimension thanks to the nested model nature.}}.

To model the vanishing change size, we assume that $\Delta_j=\kappa \delta_j$, where $\delta_j \in\mathbb R^{d^o}$ is a $d^o$-dimensional vector with $\|\delta_j\|_2>0$ for $j=1,\cdots,m_o$. The decay rate of the change size is controlled by $\kappa \in \mathbb R^+$, which vanishes as $S,T$ increase, i.e.\ $\kappa=\kappa_{ST}\longrightarrow 0.$ Under this setting, it is easy to see that Assumption \ref{ass.model.id}(ii) no longer holds, as the difference between $\theta_j^o$ and $\theta_{j+1}^o$~(and thus $\psi_j^o $ and $\psi_{j+1}^o$) converges to 0 as $S,T$ increase. Thus, a more delicate technical argument based on Taylor expansion is needed to quantify the difference between the composite likelihood functions in CLMDL, which requires the following stronger moment conditions on its derivatives.

\begin{assumption}[$r$] \label{ass.mom2}
    \Cref{ass.mom}(ii) holds for $a=1,2$ with some $r>2.$
\end{assumption}

\Cref{thm_diminishing} states that the CLMDL based estimator $(\hat{m},\hat{\Lambda},\hat{\Psi})$ defined in \eqref{minMDL} can still achieve consistent estimation of change-points given that the change size $\kappa$ does not vanish too fast. 

\begin{theorem}\label{thm_diminishing}
Let $\bfY$ be a piecewise stationary spatio-temporal process specified by $(m_o,\Lambda^o,\Psi^o)$, and Assumptions \ref{ass.countable}, \ref{ass.mom}($r$), \ref{ass.model.id}(i), \ref{ass.mix}($r$), \ref{ass.mixtime} and \ref{ass.mom2}(r) hold with some $r>2$, and furthermore $T\cdot S^{-r/2}\longrightarrow 0$. Suppose the change size satisfies that $T\cdot (S\kappa^2)\longrightarrow\infty$.

\noindent (i).\ If $\liminf\limits_{S,T\to\infty} S\kappa^2 >0$, with probability going to 1, we have 
$$\hat m=m_o, ~ [T\hat{\Lambda}_{ST}]=[T\Lambda^o], ~\text{and} ~~  \hat{\xi}_j=\xi^o, ~\hat{\theta}_j \longrightarrow \theta_j^o ~ \text{ for all } j=1,\cdots,m_o+1.$$

\noindent (ii).\ If $S\kappa^2\longrightarrow 0$, with probability going to 1, we have
\begin{align}\label{eq:diminish_rate}
    \hat m=m_o, ~\text{and} ~~\max_{j=1,\cdots,m_o} \big| [T\hat\lambda_{j}] - [T\lambda_j^o] \big|=O_p\big((S\kappa^2)^{-r/(r-2)}\wedge \sqrt{T/(S\kappa^2)}\big),
\end{align}
and $\hat\theta_j$ has a dimension greater than or equal to $d^o=d(\xi^o)$ for $j=1,\cdots,m_o+1$. If in addition, $S=O(T)$, we further have $\hat{\xi}_j=\xi^o$ and $\hat{\theta}_j \longrightarrow \theta_j^o$ for $j=1,\cdots,m_o+1$ with probability going to 1.
\end{theorem}

\Cref{thm_diminishing} suggests that in some sense $S\kappa^2$ can be viewed as the effective (squared) change size of the problem. In particular, \Cref{thm_diminishing}(i) states that as long as $S\kappa^2$ does not vanish as $S,T$ increase, CLMDL can still provide consistent estimation for $m_o$ and achieve exact recovery of all change-points $[T\Lambda^o]$, which resembles \Cref{unknownprob}(i). Interestingly, \Cref{thm_diminishing}(i) indicates that exact recovery can be achieved not only for a diverging $S\kappa^2\longrightarrow\infty$ but also for a constant/converging $S\kappa^2 \longrightarrow c>0$. This in fact can be attributed to the compensating mechanism of the proposed composite likelihood, which is designed to correct the edge effect, and shares a similar intuition as that of \Cref{prop.minCL}. We refer to Remark \ref{remark:hdmean} of the supplement for more details.

\Cref{thm_diminishing}(ii) states that under a vanishing $S\kappa^2$, as long as $T\cdot (S\kappa^2)\longrightarrow \infty$, CLMDL again consistently estimates $m_o$, though no longer achieves exact recovery of $[T\Lambda^o].$ However, CLMDL still gives consistent estimation of the normalized change-points, as by \eqref{eq:diminish_rate}, we have $\big| \hat\lambda_{j} - \lambda_j^o \big|=O_p(\sqrt{1/(ST\kappa^2)})=o_p(1)$. Due to the bias caused by the localization error of $[T\hat\lambda_{j}]$, CLMDL may over-estimate the model order $\xi^o$ on each segment unless $S=O(T).$ 



\Cref{thm_diminishing}(ii) also indicates that the existence of a higher moment $r$ leads to a faster localization error rate. In particular, if the moment conditions hold for all $r>2$, \eqref{eq:diminish_rate} implies that
$\big| [T\hat\lambda_{j}] - [T\lambda_j^o] \big|=O_p((S\kappa^2)^{-1-\delta})$ for any $\delta>0.$ The presence of $r$ is due to the use of a moment inequality in \colorlinks{black}{\cite{Doukhan1994}}, which works for a general random field with mild $\alpha$-mixing conditions. Indeed, we show in \Cref{remark:sharper_rate} of the supplement that under some additional martingale difference or linear process assumptions on the score function, by invoking the generalized H\'ajek-R\'enyi inequality in \colorlinks{black}{\cite{Bai1994}}, the localization error rate \eqref{eq:diminish_rate} can be sharpened to $O_p((S\kappa^2)^{-1})$ as long as $r>2.$ However, it seems unnatural and difficult to cast a general spatio-temporal process into the framework of linear processes, and we thus opt to use a general (albeit less sharp) moment inequality for CLMDL.

To our best knowledge, the consistency and exact recovery property of CLMDL for multiple change-points estimation in a parametric spatio-temporal model are the first time seen in the literature. This is substantial as parametric modeling is one of the main workhorses for spatio-temporal analysis. For simplicity and clarity of presentation, we assume all changes $\Delta_j, ~j=1,\cdots, m_o$ vanish. Indeed, by combining the arguments in Theorems \ref{unknownprob} and \ref{thm_diminishing}, we can show CLMDL works when both vanishing and non-vanishing changes exist. We omit the details to conserve space.

We remark that the exact recovery phenomenon has been previously observed in the change-point literature. For example, \colorlinks{black}{\cite{bai2010common} and its extension \cite{bhattacharjee2019change}} study single change-point estimation for non-parametric high-dimensional mean change and show that exact recovery can be achieved when the change size is large in certain sense. However, the technical arguments used in the results for CLMDL are substantially different. In particular, unlike the mean change problem which retains linearity, due to the parametric nature of our problem, the objective function of CLMDL is more complex and does not have a closed-form solution as a result of its non-linearity. On the other hand, there is also some intuitive connection between the localization error rate of CLMDL in \Cref{thm_diminishing} and the rate derived in the high-dimensional mean change literature. We refer to \Cref{remark:hdmean} of the supplement for a more detailed discussion.
}

\subsection{Asymptotic distribution under increasing domain asymptotics}

We now investigate the asymptotic distribution of the change-point estimator. Theorems \ref{unknownprob}(ii), \ref{finerate} and \ref{thm_diminishing}(i) indicate that under the increasing domain asymptotics, the integer-valued change-points can be recovered exactly. Thus, there is no non-degenerate asymptotic distribution for the estimated change-points $\hat{\Lambda}_{ST}$. Nevertheless, the following Theorem \ref{asym.distr} provides some insights on the finite-sample behavior of $\hat{\Lambda}_{ST}$. For simplicity, we only present the result for $k=1$, the result for $k> 1$ is similar but notationally more complicated. Define $E_{1}=\{(\bs_1,\bs_2): \bs_1\in\mathcal{S}, \bs_2\in \bs_1\cup\mathcal{N}(\bs_1) \}$ and $E_2=\{\bs\in\mathcal{S}: \text{repeat } \bs \text{ by } 1+|\mathcal{N}(\bs)| \text{ times} \}$.

For $q>0$, define
\begin{align*}
&A_1(q, \psi_j^o, \psi_{j+1}^o)=\sum\limits_{\bs\in E_2}\left[ l_{marg}(\psi_j^o;\dot x_{q,\bs}^{(j+1)})
+ l_{marg}(\psi_{j+1}^o;\dot x_{q+1,\bs}^{(j+1)})\right]-
\sum\limits_{(\bs_1,\bs_2)\in E_1}l_{pair}(\psi_{j+1}^o;\dot x_{q,\bs_1}^{(j+1)}, \dot x_{q+1,\bs_2}^{(j+1)})\,,\\
&A_2(\psi_j^o, \psi_{j+1}^o)=\sum\limits_{(\bs_1,\bs_2)\in E_1}l_{pair}(\psi_{j}^o;\dot x_{T_j^o,\bs_1}^{(j)},\dot x_{1,\bs_2}^{(j+1)})-
\sum\limits_{\bs\in E_2} \left[ l_{marg}(\psi_{j}^o;\dot x_{T_j^o,\bs}^{(j)}) 
+ l_{marg}(\psi_{j+1}^o;\dot x_{1,\bs}^{(j+1)})\right]\,,\\
&A_3(q, \psi_j^o, \psi_{j+1}^o)=\sum\limits_{(t,i,\bs_1,\bs_2)\in D_1(q)}l_{pair}(\psi_{j}^o; \dot x_{t,\bs_1}^{(j+1)}, \dot x_{t+i,\bs_2}^{(j+1)})
- \sum\limits_{(t,i,\bs_1,\bs_2)\in D_1(q)} l_{pair}(\psi_{j+1}^o; \dot x_{t,\bs_1}^{(j+1)}, \dot x_{t+i,\bs_2}^{(j+1)})\,,
\end{align*}
where $D_1(q)= D_{10}(q)\cup D_{11}(q)$ with $D_{10}(q)=\bigcup_{t=1}^{q}\{(t,0,\bs_1,\bs_2):  \bs_1\in\mathcal{S}, \bs_2 \in \mathcal{N}(\bs_1) \}$ and $D_{11}(q)=\bigcup_{t=1}^{q-1}\{(t,1,\bs_1,\bs_2):  \bs_1\in\mathcal{S}, \bs_2 \in \bs_1\cup\mathcal{N}(\bs_1) \} $.

For $q<0$, define
\begin{align*}
&B_1(q, \psi_j^o, \psi_{j+1}^o)=\sum\limits_{\bs \in E_2}\left[ l_{marg}(\psi_j^o;\dot x_{{T_j^o+q},\bs}^{(j)})
+  l_{marg}(\psi_{j+1}^o;\dot x_{{T_j^o+q+1},\bs}^{(j)})\right] -
\sum\limits_{(\bs_1,\bs_2)\in E_1}l_{pair}(\psi_j^o;\dot x_{{T_j^o+q},\bs_1}^{(j)}, \dot x_{{T_j^o+q+1},\bs_2}^{(j)})\,,\\
&B_2(\psi_j^o, \psi_{j+1}^o)=\sum\limits_{(\bs_1,\bs_2)\in E_1}l_{pair}(\psi_{j+1}^o;\dot x_{T_j^o,\bs_1}^{(j)},\dot x_{1,\bs_2}^{(j+1)})-
\sum\limits_{\bs\in E_2} \left[ l_{marg}(\psi_{j}^o;\dot x_{T_j^o,\bs}^{(j)}) 
+ l_{marg}(\psi_{j+1}^o;\dot x_{1,\bs}^{(j+1)})\right]\,,\\
&B_3(q, \psi_j^o, \psi_{j+1}^o)=\sum\limits_{(t,i,\bs_1,\bs_2)\in D_2(q)}l_{pair}(\psi_{j+1}^o; \dot x_{t,\bs_1}^{(j)}, \dot x_{t+i,\bs_2}^{(j)})
- \sum\limits_{(t,i,\bs_1,\bs_2)\in D_2(q)} l_{pair}(\psi_{j}^o; \dot x_{t,\bs_1}^{(j)}, \dot x_{t+i,\bs_2}^{(j)})\,,
\end{align*}
where $D_2(q)= D_{20}(q)\cup D_{21}(q)$ with $D_{20}(q)=\bigcup_{t=T_j^o+q+1}^{T_j^o}\{(t,0,\bs_1,\bs_2):  \bs_1\in\mathcal{S}, \bs_2 \in \mathcal{N}(\bs_1) \}$ and $D_{21}(q)=\bigcup_{t=T_j^o+q+1}^{T_j-1}\{(t,1,\bs_1,\bs_2):  \bs_1\in\mathcal{S}, \bs_2 \in \bs_1\cup\mathcal{N}(\bs_1) \} $. Note that $A_i$'s and $B_i$'s quantify the effects of the estimation error $[T\hat{\lambda}_j]-[T\lambda_j^o]=q$ on the CLMDL. For $j=1,\ldots, m_o$, we define a double-sided random walk for the $j$th change-point,
\begin{align}\label{diff.seg}
W_{ST}^{(j)}(q; \psi_{j}^o, \psi_{j+1}^o) &=\begin{cases}
&A_1(q, \psi_j^o, \psi_{j+1}^o)+A_2(\psi_j^o, \psi_{j+1}^o)+A_3(q, \psi_j^o, \psi_{j+1}^o)\,,~~~q>0\,,\\
&0\,,~~~\hspace{7.8cm} q=0\,,\\
&B_1(q, \psi_j^o, \psi_{j+1}^o)+B_2(\psi_j^o, \psi_{j+1}^o)+B_3(q, \psi_j^o, \psi_{j+1}^o)\,,~~~q<0\,.
\end{cases}
\end{align}
Theorem \ref{asym.distr} gives an approximation of the finite-sample behavior of $\hat{\Lambda}_{ST}$ and the asymptotic distribution of the estimated parameters $\hat{\theta}_j$.

\begin{theorem} \label{asym.distr}
	Suppose that the conditions in Theorems \ref{unknownprob}(ii) or \ref{finerate} or \ref{thm_diminishing}(i) are satisfied, and Assumptions \ref{ass.thetadistr} and \ref{ass.variance} hold. We have that
	\begin{align}
	\sqrt{ST}(\hat{\theta}_j-\theta_j^o) &\longrightarrow N[0, \{\Sigma_2^{(j)}\}^{-1}\Sigma_1^{(j)}\{\Sigma_2^{(j)}\}^{-1}]~\text{in distribution}\,,  \label{asy.dist.theta}
	\end{align}
	as $S,T\longrightarrow \infty$. If, additionally, $S=o(T)$, we have for $j=1,\ldots,m_{o}+1$,
	\begin{align}	\label{asy.dist.lam}
		[T\hat{\lambda}_j]-[T\lambda_j^o] &= \arg\max_{q\in \mathbb{Z}} W_{ST}^{(j)}(q;\psi_{j}^o,\psi_{j+1}^o)+o_p(1)\,.
	\end{align}
 Moreover, $\{\hat{\lambda}_{1},\ldots,\hat{\lambda}_{m_{o}}$, $\hat{\theta}_{1},\ldots, \hat{\theta}_{m_{o}+1}\}$ are asymptotically independent.
\end{theorem}

From the proof of Theorems \ref{unknownprob} and \ref{thm_diminishing}, it can be shown that 
	\begin{equation} \label{w.to.1}
	P\left(\arg\max_{q\in \mathbb{Z}} W_{ST}^{(j)}(q;\psi_{j}^o,\psi_{j+1}^o)=0\right)\longrightarrow 1\,,
	\end{equation} 
which again indicates that the true change-points can be recovered without errors. 
Although $[T\hat{\lambda}_j]-[T\lambda_j^o]$
eventually converges to a degenerate distribution, as is shown by the numerical experiments in \Cref{sec:num}, $\arg\max_{q\in \mathbb{Z}} W_{ST}^{(j)}(q;\psi_{j}^o,\psi_{j+1}^o)$ can still give a reasonably accurate approximation to the finite-sample behavior of $\hat{\Lambda}_{ST}$. 
The finite-sample approximation of $[T\hat{\lambda}_j]-[T\lambda_j^o]$ in Theorem \ref{asym.distr} requires $S=o(T)$. For the case where $S$ is greater than $o(T)$, the approximation in \eqref{asy.dist.lam} becomes inaccurate. The intuitive reason is that 
the distribution of $[T\hat{\lambda}_j]-[T\lambda_j^o]$ converges too fast towards its degenerate limit when more information from the spatial dimension is available. 

Since a closed-form expression for the distribution function of $W_{ST}^{(j)}(\cdot,\psi^{o}_{j},\psi^{o}_{j+1})$ is unavailable, we need to simulate replicates of $W_{ST}^{(j)}(\cdot,\hat{\psi}_{j},\hat{\psi}_{j+1})$ to conduct inference. However, the double-sided random walk $W_{ST}^{(j)}(\cdot,\psi^{o}_{j},\psi^{o}_{j+1})$ depends not only on the pseudo-true parameters $\psi^{o}_{j}$ and $\psi^{o}_{j+1}$, but also on the true distributions of the $j$th and $(j+1)$th segments. Hence, this simulation procedure is valid only if the true models of the $j$th and $(j+1)$th segments are known. In other words, if the true model is not included in $\mathcal{M}$, $\hat{\Lambda}_{ST}$ is consistent, but inference cannot be made via Theorem \ref{asym.distr}.


\section{Numerical Experiments}\label{sec:num}
We begin this section by discussing the optimization algorithm for the minimization of CLMDL in \eqref{minMDL} and then present the extensive simulation studies and a real data application.

Due to the additive form of CLMDL in \eqref{MDLform}, its minimization can be performed via dynamic programming~\citep{Jackson2005}, which incurs a quadratic computational complexity $O(ST^2)$. To further lower the computational cost, we adapt the pruned exact linear time (\textsc{PELT}) algorithm proposed by \cite{Killick2012} (originally designed for univariate time series) to the spatio-temporal setting. A key component of PELT is to find a suitable threshold $K$, which is used to prune unnecessary candidates in recursive computation of the dynamic programming, and thus lowers the computational complexity to between $O(ST)$ and $O(ST^2)$. Under the spatio-temporal setting, the threshold $K$ is more challenging to derive due to the non-ignorable edge effect when spatial dimension $S\longrightarrow \infty$. Nevertheless, in \Cref{PELTK} of the supplement, we provide a suitable choice of $K$ and establish the asymptotic validity of applying PELT for the minimization of CLMDL. We refer to \Cref{sec:PELT} of the supplement for more details.

\subsection{Simulation studies}\label{subsec:simu}

Throughout the numerical experiments, we mainly consider the following four-parameter autoregressive spatial model,
\begin{equation} \label{sim.process}
\bfy_t - \mu= \phi (\bfy_{t-1}-\mu) + \boldsymbol{\varepsilon}_t\,,
\end{equation}
where $\bfy_t=\{ y_{t,\bs}: \bs \in \mathcal{S} \}$ is defined on a regular two-dimensional grid $\mathcal{S}$~(see definition later) and $\boldsymbol{\varepsilon}_t=\{ \varepsilon_{t,\bs}: \bs \in \mathcal{S} \}$ is a Gaussian process with exponential covariance function 
${\rm Cov}(\varepsilon_{t,\textbf{s}}, \varepsilon_{t,\textbf{s}'}) = \sigma^2\exp\{-\left\lVert \textbf{s}-\textbf{s}'\right\lVert_2/\rho\}$ and ${\rm Cov}(\varepsilon_{t,\textbf{s}}, \varepsilon_{t',\textbf{s}'}) =0$, when $t\neq t'$. The model is specified by $\theta=(\mu,\phi, \rho,\sigma^2)^\top$, where $\phi\in(-1,1)$, $\rho >0$ and $\sigma^2 >0$. Spatial and temporal dependence are determined by $\rho$ and $\phi$ respectively. Meanwhile, $\mu$ and $\sigma^2$ control the overall mean and variance.

We investigate the performance of CLMDL under the increasing domain setting and define the spatial domain as $\mathcal{S}=\{(s_1,s_2):s_1,s_2\in \{1,2,3\ldots,s\} \}$ where the spatial sample size $S$ grows by increasing $s$. For all simulation studies, we set $k=1$ and $d=2$ in defining the composite likelihood, set $\epsilon_\lambda=0.1$ in the optimization, and set the number of replications to be 1000.

\textbf{Competing methods}: To our knowledge, there is no natural competing method for CLMDL in the literature. \textcolor{black}{Nevertheless, for illustration purposes, we compare CLMDL with \colorlinks{black}{\cite{Davis2006}}, which is an important work for multiple change-point estimation in parametric models of univariate time series}, and with SBS in \cite{Cho2015} and DCBS in \cite{Cho2016}, which are \textcolor{black}{important works that} allow multiple change-point estimation in both the \textcolor{black}{(non-parametric)} mean and second-order structure of a high-dimensional time series. To conserve space, we refer to Section \ref{sec:add_num} of the supplement (\Cref{add_simu_hd} and \textcolor{black}{\ref{add_simu_univariate}}) for the detailed comparison. 


\textbf{Additional simulation studies}: In the supplement, we have further conducted numerical experiments examining the performance of CLMDL for multiple change-point estimation and model selection within each stationary segment, \textcolor{black}{for change-point estimation under partial changes, and} under the spatial infill setting, and for its robustness w.r.t.\ the tuning parameters $(k,d)$ and $\epsilon_\lambda$. We refer to Section \ref{sec:add_num} of the supplement (Simulation \ref{add_simu_multiCP}-\ref{add_simu_epsilon}) for more details.

\begin{simulation}\label{sim.1} In this simulation, we examine the estimation accuracy of CLMDL under various sample sizes and signal levels. The underlying data generating process~(DGP) in each stationary segment follows \eqref{sim.process} with $\mu=0$, hence the process is specified by $\theta=(\phi, \rho,\sigma^2)^\top$.

Let $\theta_1=(-0.5,0.6,1)^\top$ and $\theta_2=(-0.5+\delta_\phi,0.6+\delta_\rho,1)^\top$ be the underlying parameter vectors for the segments before and after the change-point, respectively. When there is no change-point (i.e. $\delta_\phi=\delta_\rho=0$), the entire process is simulated from $\theta_1$, otherwise there is a change-point at $\lambda_1^o=0.5$. We consider four scenarios corresponding to no change-point, change in temporal dependence ($\delta_\phi$), change in spatial dependence ($\delta_\rho$) and change in both spatial and temporal dependence. \textcolor{black}{Note that the signal levels $\delta_\phi, \delta_\rho$ are pre-fixed and do not vary with the sample sizes $S,T$~(i.e.\ non-vanishing).}

Table \ref{scp.table} reports the estimated number of change-points under various settings. Under the no-change scenario, there is no false positive even when the sample size is small. Some over-estimation is observed for small sample when there is change in spatial dependence, which is probably due to the larger variation in estimating $\rho$. The detection power improves when either $S$, $T$ or signal level $(\delta_\phi, \delta_\rho)$ increases. For example, for $(\delta_\phi=0.2$, $S=6^2$, $T=100)$, the detection power increases from 37\% to 81\% or 98\% respectively, when $S$ increases to $10^2$ or $T$ increases to 200. To be expected, the proposed procedure is most powerful when there is change in both $\delta_\phi$ and $\delta_\rho$.  

\begin{table}[h]
	\centering
	\caption{Percentage of estimated change-points $\hat m$ among 1000 replications under various spatial size $S$, temporal size $T$, and signal levels $(\delta_\phi, \delta_\rho)$.}\label{scp.table}
	\begin{tabular}{cccrrrrrrrrr}
		\hline \hline
		\multirow{3}{*}{$T$} & \multirow{3}{*}{\begin{tabular}[c]{@{}l@{}}$\delta_\phi\times 10$\end{tabular}} & \multirow{3}{*}{\begin{tabular}[c]{@{}l@{}}$\delta_\rho\times 10$\end{tabular}} & \multicolumn{9}{c}{\% of $\hat{m}$} \\
		&                                                                                     &                                                                                    & \multicolumn{3}{c}{$S=6^2$} & \multicolumn{3}{c}{$S=8^2$} & \multicolumn{3}{c}{$S=10^2$} \\ \cline{4-12} 
		&                                                                                     &                                                                                    & 0      & 1      & $\geq2$   & 0      & 1      & $\geq2$   & 0      & 1      & $\geq2$    \\ \hline
		100 & 0                                                                                   & 0                                                                                  & 100    & 0      & 0         & 100    & 0      & 0         & 100    & 0      & 0          \\
		& 2                                                                                   & 0                                                                                  & 63     & 37     & 0         & 29     & 71     & 1         & 2      & 98     & 0          \\
		& 3                                                                                   & 0                                                                                  & 22     & 79     & 0         & 1      & 99     & 1         & 0      & 100    & 0          \\
		& 0                                                                                   & 6                                                                                  & 81     & 18     & 2         & 39     & 60     & 2         & 7      & 91     & 2          \\
		& 0                                                                                   & 10                                                                                 & 16     & 81     & 3         & 1      & 95     & 4         & 1      & 96     & 3          \\
		& 2                                                                                   & 2                                                                                  & 54     & 47     & 0         & 14     & 86     & 1         & 0      & 100    & 0          \\
		& 3                                                                                   & 3                                                                                  & 11     & 89     & 1         & 1      & 99     & 1         & 0      & 100    & 0          \\\hline
		200 & 0                                                                                   & 0                                                                                  & 100    & 0      & 0         & 100    & 0      & 0         & 100    & 0      & 0          \\
		& 2                                                                                   & 0                                                                                  & 20     & 81     & 0         & 1      & 99     & 0         & 0      & 100    & 0          \\
		& 3                                                                                   & 0                                                                                  & 0      & 100    & 0         & 0      & 100    & 0         & 0      & 100    & 0          \\
		& 0                                                                                   & 6                                                                                  & 38     & 61     & 1         & 13     & 82     & 5         & 0      & 98     & 2          \\
		& 0                                                                                   & 10                                                                                 & 3      & 96     & 2         & 1      & 98     & 2         & 0      & 100    & 0          \\
		& 2                                                                                   & 2                                                                                  & 10     & 91     & 0         & 0      & 100    & 0         & 0      & 100    & 0          \\
		& 3                                                                                   & 3                                                                                  & 4      & 95     & 0         & 0      & 100    & 0         & 0      & 100    & 0    \\ \hline \hline  
	\end{tabular}
\end{table}
\end{simulation}

\begin{simulation} In this simulation, we compare the empirical distribution of the change-point estimator and its asymptotic distribution as stated in Theorem \ref{asym.distr}, and further illustrate the exact recovery property of CLMDL in \Cref{unknownprob} \textcolor{black}{under non-vanishing change sizes}. The underlying DGP follows \eqref{sim.process} with $\mu=0$ and $\theta_1=(-0.5,0.6,1)^\top$ and $\theta_2=(-0.5+\delta_\phi, 0.6+\delta_\rho,1)^\top$. We vary $S=6^2, 8^2, 10^2$ and fix $T_1^o=T_2^o=100$. When the number of change-points is correctly estimated, 100 replicates of $W_{ST}(\cdot)$ are simulated using $\hat{\psi}_1$ and $\hat{\psi}_2$ to compute 
$\arg\max_{q\in \mathbb{Z}} W_{ST}(q;\hat{\psi}_{1},\hat{\psi}_{2})$.

\begin{table}[h]
	\centering
	\caption{Percentage of $\hat{m}=1$, and percentage of $\hat{\lambda}=0.5$, mean, empirical standard deviation (esd), mean of 90\% confidence interval (CI) of $\hat{\lambda}$, and empirical coverage probability (CP) of the 90\% CI (given $\hat m=m_o=1)$ under various settings.}
	\label{lam.table}
	\begin{tabular}{cccrrcccc}
		\hline\hline
		$\delta_\phi\times 10$ & $\delta_\rho\times 10$ & $S$      & \% of $\hat{m}$ = 1 & \multicolumn{5}{c}{$\hat \lambda$ (given $\hat{m}=1)$}                                                                                                                                 \\ \cline{5-9} 
		&       &        &                     & \multicolumn{1}{l}{\% of $\hat{\lambda}$ = 0.5} & mean                       & esd                        & 90\% CI                      & CP                       \\ \hline
		2     & 0     & $6^2$  & 81                  & 37                                        & 0.4923                     & 0.0513                     & {[}0.4611, 0.5426{]} & 88.0                     \\
		&       & $8^2$  & 99                  & 63                                        & 0.4964                     & 0.0310                     & {[}0.4538, 0.5279{]} & 87.6                     \\
		&       & $10^2$ & 100                 & 79                                        & 0.4979                     & 0.0171                     & {[}0.4726, 0.5189{]} & 91.6                     \\\hline
		3 &   0    & $6^2$  & 100                 & 81                                        & 0.4944                     & 0.0326                     & {[}0.4695, 0.5368{]} & 92.4                     \\
		 &       & $8^2$  & 100                 & 81                                        & 0.5013                     & 0.0137                     & {[}0.4722, 0.5145{]} & 89.1                     \\
		&       & $10^2$ & 100                 & 100                                       & 0.5000                     & -                          & -                            & -                        \\\hline
		2      & 2       & $6^2$  & 91                 & 82                                       & \multicolumn{1}{l}{0.5046} & 0.0275                         & {[}0.4827, 0.5114{]}                            & 90.6                        \\
       &        & $8^2$  & 100                 & 100                                       & \multicolumn{1}{l}{0.5000} & -                          & -                            & -                        \\
	&       & $10^2$ & 100                 & 100                                       & 0.5000                     & -                          & -                            & -                        \\ \hline\hline
	\end{tabular}
\end{table}
Table \ref{lam.table} summarizes the detailed simulation result. Clearly, the percentage that the estimated change-points equal the true ones, i.e.\ $\{\hat{\lambda}=\lambda^o\}$, increases to 100\% when the sample size increases, which demonstrates the exact recovery of the true change-points. Table \ref{lam.table} further reports the performance of the 90\% confidence interval (CI) obtained from the quantiles of $\arg\max_{q\in \mathbb{Z}} W_{ST}(q;\hat{\psi}_{1},\hat{\psi}_{2})$. Both the width of CI and empirical standard deviation (esd) decrease as sample size increases, and the empirical coverage probabilities are close to the nominal level. For more intuition, Figure \ref{QQ_hist}~(top panel) gives the QQ plots of the empirical quantile of $\hat\lambda$ against its theoretical quantile based on $\arg\max_{q\in \mathbb{Z}} W_{ST}(q;\hat{\psi}_{1},\hat{\psi}_{2})$ in Theorem \ref{asym.distr}. Note that the QQ plot closely aligns with the 45 degree line. Figure \ref{QQ_hist}~(bottom panel) depicts the histograms of $\hat{\lambda}$. The distribution of $\hat{\lambda}$ is non-standard and becomes degenerate as the sample size increases, which aligns with the asymptotic results.

\begin{figure}[h]
	\begin{center}
		\includegraphics[scale=0.6]{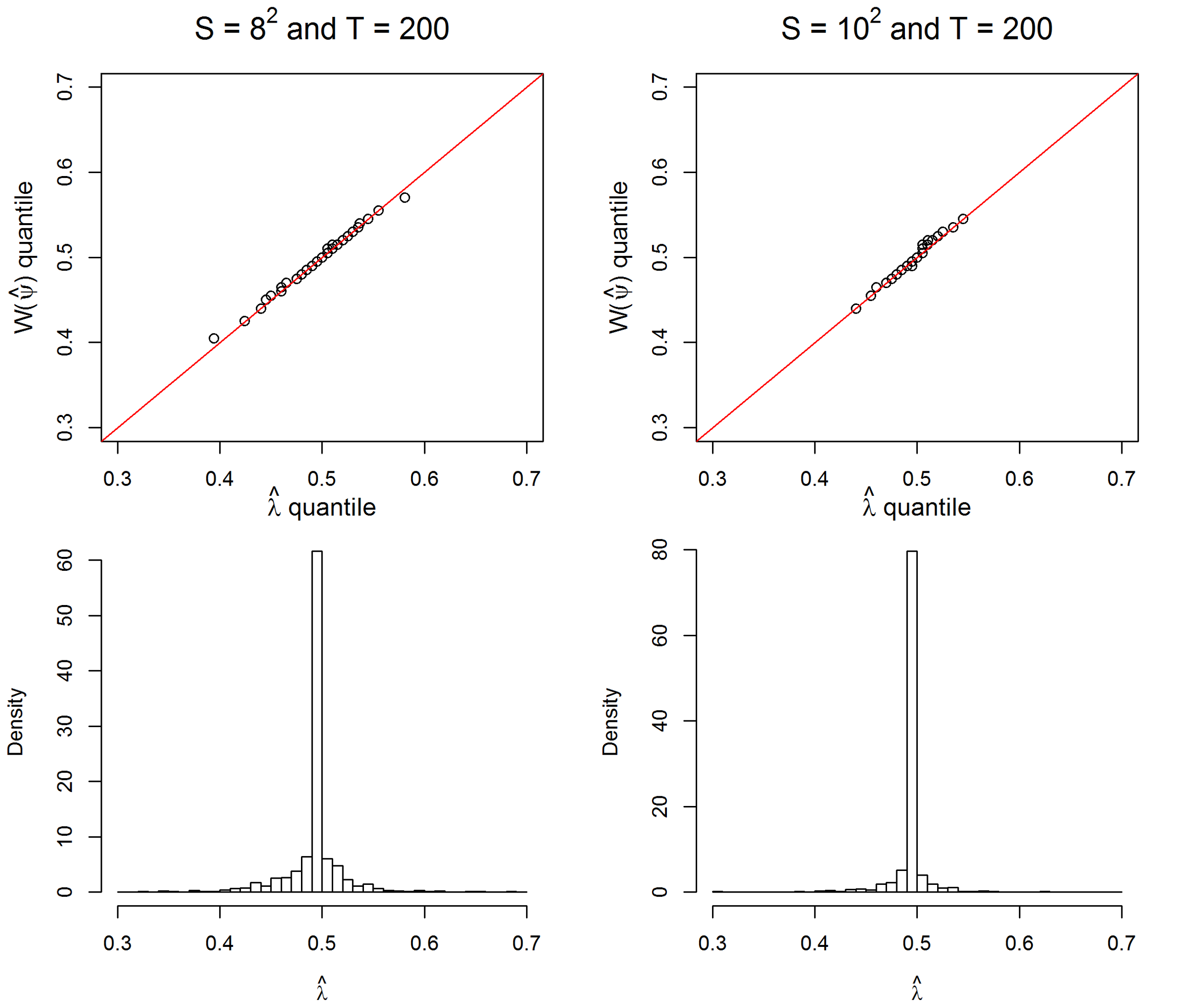}
		\caption{Top: QQ plots of sample quantiles of $\arg\max_{q\in \mathbb{Z}} W_{ST}(q;\hat{\psi}_{1},\hat{\psi}_{2})$ against the sample quantiles of empirical $\hat\lambda$; Bottom: Histogram of empirical $\hat\lambda$. The signal level is $\delta_\phi=0.2, \delta_\rho=0$.}
		\label{QQ_hist}
	\end{center}
    \vspace{-5mm}
\end{figure}
\end{simulation}

{\color{black}
\begin{simulation}\label{sim.vanish} In this simulation, we further examine the performance of CLMDL under vanishing change sizes as studied in \Cref{thm_diminishing}. The underlying DGP follows \eqref{sim.process} with $\mu=0$ and $\theta_1=(-0.5,0.6,1)^\top$ and $\theta_2=(-0.5+\delta_\phi,0.6,1)^\top$. We fix $T_1^o=T_2^o=50$ and vary $S=6^2,10^2,30^2$. We set the change size as $\delta_\phi=S^{-0.4}$ or $S^{-0.5}$, which is a function of the sample size $S$ and vanishes as $S$ increases. To conserve space, we refer to \Cref{add_simu_vanishing} in \Cref{sec:add_num} of the supplement for a more thorough numerical study regarding CLMDL under vanishing change sizes. 

Table \ref{table.vanish.CLMDL} summarizes the simulation results. As can be seen, for both $\delta_\phi=S^{-0.4}$ and $S^{-0.5}$, the detection power and estimation accuracy of CLMDL improve as $S$ increases (and the change size vanishes). Intuitively, CLMDL performs better under a slower decaying rate (i.e.\ $S^{-0.4}$) of change sizes. Note that exact recovery of $\lambda_1^o=0.5$ is possible even for $\delta_\phi=S^{-0.5}$ (though it requires a large sample size), which provides further numerical evidence for the theoretical results in \Cref{thm_diminishing}. 


\begin{table}[h]
\centering
{\color{black}
\caption{Percentage of $\hat{m}$ among 1000 replications, and percentage of $\hat{\lambda}=0.5$, mean and empirical standard deviation (esd) of $\hat{\lambda}$~(given $\hat m=m_o=1$) under various vanishing change sizes.
}\label{table.vanish.CLMDL}
\begin{tabular}{cccrrrrrrr}
\hline\hline
& $S$    & \begin{tabular}[c]{@{}c@{}}$\delta_\phi\times 10$\end{tabular}  & \multicolumn{3}{c}{\% $\hat{m}$} &  & \multicolumn{3}{c}{$\hat{\lambda}$ (given $\hat{m}=1)$}          \\ \cline{4-6} \cline{8-10} 
&   &        & 0       & 1       & $\geq 2$     &  & \% of $\hat{\lambda}= 0.5$ & mean   & esd    \\ \hline
$\delta_\phi=S^{-0.4}$ & $6^2$ & 2.38   & 40      & 60      & 0            &  & 43   & 0.4895 & 0.0314 \\
& $10^2$  & 1.58    & 25      & 75      & 0            &  & 68                         & 0.4960 & 0.0168 \\
& $30^2$ & 0.66    & 0       & 100     & 0            &  & 100                        & 0.5000 & -       \\ \hline
$\delta_\phi=S^{-0.5}$ & $6^2$ & 1.67   &    74     & 26      & 0            &  & 27  & 0.4773 & 0.0595 \\
& $10^2$ & 1.00   &     68    & 32      & 0            &  & 57                         & 0.5025 & 0.0314 \\
& $30^2$ & 0.33  &     14    & 86      & 0            &  & 100                        & 0.5000 & -       \\ \hline \hline
\end{tabular}}
\end{table}
\end{simulation}
}

\begin{simulation}\label{sim.CL}
    In this simulation, we examine the consistency of the CL based estimator defined in \eqref{minCL}, which has no penalty terms. The underlying DGP follows \eqref{sim.process} with $\mu=0$, and we set $\theta_1=(-0.5,0.6,1)^\top$ and $\theta_2=(-0.5+\delta_\phi,0.6,1)^\top$ as the true parameters for the segments before and after the change-point. We consider change sizes of both fixed $\delta_\phi=0.1,0.2$ \textcolor{black}{and vanishing $\delta_\phi=S^{-0.4}, S^{-0.5}$.} We fix $T=100$ and vary $S=30^2,60^2$. When there is no change-point (i.e.\ $\delta_\phi=0$), the data is simulated from $\theta_1$, otherwise there is a change-point at $\lambda_1^o=0.5$.
 
    Table \ref{nopen.table} reports the estimated number of change-points by CLMDL and CL. Due to the lack of penalty, CL does experience false positives for $S=30^2$. However, under both fixed \textcolor{black}{and vanishing change sizes, the false detection disappears as $S$ increases to $60^2$, which indicates the consistency of CL (without penalty) and aligns with the theoretical results in Proposition \ref{prop.minCL} and \Cref{lem_rate_m_unknown}. On the other hand, note that CL does require a very large sample size to achieve such consistency, which is often not reasonable in real data and thus makes CL less practical.} In contrast, thanks to the MDL penalty, CLMDL is robust to false positives and achieves superior performance over CL under constant change sizes. Compared with CL, it has slightly less power under the vanishing change sizes for $S=30^2$ when $\delta_\phi$ is extremely small (i.e.\ $\delta_\phi=S^{-0.5}$). However, such power loss disappears as $S$ increases to $60^2$. Thus, we prefer CLMDL as the MDL penalty can guard against false positives in finite sample without significant negative impact on its detection power. 
 	
\begin{table}[h]
\centering
\caption{Percentage of estimated change-points $\hat m$ among 1000 replications by the CL-based estimator (without penalty terms) and the CLMDL estimator under various change sizes.}\label{nopen.table}
\begin{tabular}{lcrrrrrr>{\color{black}}c>{\color{black}}r>{\color{black}}r>{\color{black}}r>{\color{black}}r>{\color{black}}r>{\color{black}}r}
\hline \hline
$S$    & \begin{tabular}[c]{@{}c@{}}$\delta_\phi$ \end{tabular} & \multicolumn{3}{c}{CL}        & \multicolumn{3}{c}{CLMDL}     & $\delta_\phi $ & \multicolumn{3}{c}{\color{black} CL}        & \multicolumn{3}{c}{\color{black} CLMDL}     \\
       &                                 $\times 10$                                   & \multicolumn{3}{c}{\% of $m$} & \multicolumn{3}{c}{\% of $m$} &          $\times 10$     & \multicolumn{3}{c}{\color{black} \% of $m$} & \multicolumn{3}{c}{\color{black} \% of $m$} \\ \cline{3-8} \cline{10-15} 
       &                                                                    & 0       & 1      & $\geq2$    & 0       & 1      & $\geq2$    &               & 0       & 1      & $\geq2$    & 0       & 1      & $\geq2$    \\ \hline
$30^2$ & 0                                                                  & 96      & 3      & 1          & 100     & 0      & 0          & 0             & 96      & 3      & 1          & 100     & 0      & 0          \\
       & 1                                                                  & 0       & 96     & 4          & 0       & 100    & 0          & 0.33~($S^{-0.5}$)    & 0       & 100    & 0          & 14       & 86    & 0          \\
       & 2                                                                  & 0       & 81     & 19         & 0       & 100    & 0          & 0.66~($S^{-0.4}$)    & 0       & 100    & 0          & 0       & 100    & 0          \\ \hline
$60^2$ & 0                                                                  & 100     & 0      & 0          & 100     & 0      & 0          & 0             & 100     & 0      & 0          & 100     & 0      & 0          \\
       & 1                                                                  & 0       & 100    & 0          & 0       & 100    & 0          & 0.17~($S^{-0.5}$)    & 0       & 100    & 0          & 0       & 100    & 0          \\
       & 2                                                                  & 0       & 100    & 0          & 0       & 100    & 0          & 0.38~($S^{-0.4}$)    & 0       & 100    & 0          & 0       & 100    & 0          \\ \hline\hline
\end{tabular}
\end{table}
\end{simulation}

\begin{simulation}\label{sim.robustness}
In this simulation, we examine the robustness of CLMDL against model misspecification. In particular, we consider the non-separable space-time covariance function in \cite{Cressie1999},
\begin{eqnarray} \label{crescov}
	C(h,u|\theta) = \left\{ \begin{array}{llc}
	&\frac{2\sigma^2c}{(a^2u^2+1)^\nu(a^2u^2 +c) \Gamma(\nu)} \left\{\frac{b}{2}(\frac{a^2u^2+1}{a^2u^2+c})^{1/2}h\right\}^\nu K_\nu\left(b(\frac{a^2u^2+1}{a^2u^2+c})^{1/2}h\right)\,, &\text{\hspace{5mm} if } h > 0\,,\\
	&\frac{\sigma^2c}{(a^2u^2+1)^\nu(a^2u^2 +c) }\,, &\text{\hspace{5mm} if } h = 0\,,
	\end{array}
	\right.
\end{eqnarray}
where $h$ and $u$ are the space and time distance, $\nu>0$ is the smoothness parameter, $K_\nu$ is the modified Bessel function, $a\geq 0$ is the time scaling parameter, $b \geq 0$ is the space scaling parameter, $c>0$ is the space-time interaction parameter, and $\sigma^2=C(0,0|\theta)>0$ is the variance. We collect the parameter $\theta=(a,b,c,\nu,\sigma^2)$. Note that $C(h,u|\theta)$ in \eqref{crescov} generalizes various popular covariance functions. For example, if $u=0$, $C(h,0|\theta)$ becomes the Mat\'{e}rn spatial covariance function. For $\nu=0.5$, it further reduces to the exponential covariance function. Moreover, a separable space-time covariance is obtained when $c=1$. See \cite{Cressie1999} for details. 


Let $\theta_1=(1,1,3,0.2,1)^\top$ and $\theta_2=(1+\delta,1+\delta,3,0.2,1)^\top$ be the parameters for the segments before and after the change-point, respectively. When there is no change-point~(i.e.\ $\delta=0$), the entire process is simulated from $\theta_1$, otherwise there is a change-point at $\lambda_1^o=0.5$. We set $T=100$ and vary $S=6^2,8^2$. We conduct the change-point estimation using both the separable model \eqref{sim.process} and the true model \eqref{crescov}. Table \ref{robust.table} summarizes the numerical result. It shows that CLMDL works well under model misspecification, with a low false positive rate when there is no change-point and high detection power when change-point exists. Moreover, compared to the true model, the loss in detection power or estimation accuracy due to model misspecification is small.


\begin{table}[h]
	\centering
	\caption{Percentage of estimated change-points $\hat m$ among 1000 replications, and mean, empirical standard deviation (esd) of $\hat{\lambda}$~(given $\hat{m}=m_o=1$) using misspecified model and true model.} \label{robust.table}
	\begin{tabular}{ccrrrlcclrrrlcc}
		\hline \hline
		$S$ & $\delta \times 10$ & \multicolumn{6}{c}{Misspecified model}                       &  & \multicolumn{6}{c}{True model}                              \\ 
		&       & \multicolumn{3}{c}{\% of $\hat{m}$} &   & \multicolumn{2}{c}{$\hat{\lambda}$ (given $\hat{m}=1)$} &  & \multicolumn{3}{c}{\% of $\hat{m}$} &  & \multicolumn{2}{c}{$\hat{\lambda}$ (given $\hat{m}=1)$} \\ \cline{3-5} \cline{7-8} \cline{10-12} \cline{14-15} 
		&       & 0        & 1       & $\geq 2$      &   & mean         & esd          &  & 0       & 1        & $\geq 2$      &  & mean         & esd          \\ \hline
		$6^2$ & 0     & 97       & 2       & 0      &   & -            & -           &  & 97      & 2        & 0      &  & -            & -           \\
		& 5     & 94       & 6       & 0      &   & 0.4244       & 0.1135      &  & 93      & 7        & 0      &  & 0.4237       & 0.1136      \\
		& 10    & 33       & 67      & 1      &   & 0.5109       & 0.0407      &  & 30      & 70       & 1      &  & 0.5097       & 0.0375      \\
		& 15    & 1        & 98      & 1      &   & 0.5039       & 0.0214      &  & 1       & 98       & 1      &  & 0.5037       & 0.0199      \\\hline    
		$8^2$ & 0     & 99       & 1       & 0      &   & -            & -           &  & 99      & 1        & 0      &  & -            & -           \\
		& 5     & 69       & 31      & 0      &   & 0.4886       & 0.0417      &  & 66      & 33       & 0      &  & 0.5104       & 0.0383      \\
		& 10    & 0        & 99      & 1      &   & 0.5019       & 0.0111      &  & 0       & 99       & 1      &  & 0.4983       & 0.0109      \\
		& 15    & 0        & 99      & 1      &   & 0.5008       & 0.0053      &  & 0       & 99       & 1      &  & 0.5007       & 0.0052      \\\hline \hline   
	\end{tabular}
\end{table}
\end{simulation}

\subsection{Application to the U.S. precipitation data}\label{subsec:realdata}
Change-point detection in the amount of precipitation has been recognized as an important problem in climate and environmental science, see \cite{Gallagher2012} for a review on some common approaches. However, existing literature, e.g.\ \cite{Gromenko2017}, seems to mainly consider the at-most one change-point scenario and requires space-time separability of the covariance function.


We consider the data from the Global historical climatological network database (GHCN),
which is a main database for global climate monitoring. In particular, some key climate variables such as the amount of precipitation are collected from stations located all over the world. The documentation and datasets are available from \cite{Menne2012} and GHCN official website\footnote{ftp://ftp.ncdc.noaa.gov/pub/data/ghcn}. 
Similar to \cite{Gromenko2017}, selected precipitation data from 
the Midwest region of the U.S., including Illinois, Indiana, Iowa, Kansas, Michigan, Minnesota, Missouri, Nebraska, North Dakota, Ohio, South Dakota and Wisconsin, are considered.

In the GHCN database, daily precipitation from each station is recorded. However, there are missing data in many stations. We focus on the  land surface stations that provide at least 5 daily records of in each month over the entire period, and compute the monthly average precipitation data for the analysis. In summary, we have the monthly average precipitation $\{y^*_{t,\bs}\}$ in tenths of a millimeter (mm) from January 1941 to December 2010 for 76 stations (i.e.\ $S=76$ and $T=720$). We refer to Figure \ref{US.map} of the supplement for the location of the 76 stations.

To alleviate the heavy tail behavior of the data, we log-transform the monthly average precipitation record. \textcolor{black}{In addition, to remove seasonality in the mean and variance of the data, we follow the stationarizing transform in \colorlinks{black}{\cite{bloomfield1994periodic} and \cite{lund1995climatological}} and set 
\begin{align*}
    y_{t,\bs}=\frac{\log (y^*_{t,\bs} +1 )- \hat{\mu}_{\nu(t),\bs}}{\hat\sigma_{\nu(t),\bs}}\,,
\end{align*}
where $\nu(t)\in \{1,\cdots,12\}$ denotes the month that time $t$ is in, and $\hat{\mu}_{\nu,\bs}$ and $\hat{\sigma}_{\nu,\bs}$ are the sample mean and standard deviation (over time) of the $\log (y^*_{t,\bs} +1 )$ value for month $\nu$ at station $\bs.$ To conserve space, we refer to \Cref{subsec:addrealdata} of the supplement for a sample plot of the data.}

For the mean function $\mathbb E(y_{t,\bs})$, a linear regression with \textcolor{black}{an intercept and three station-level covariates} is adopted, which includes the latitude, longitude and elevation of the corresponding station. See \cite{Erhardt2015} and references therein for similar treatment of spatial regression. For robustness, we employ the non-separable \textcolor{black}{Gaussian} space-time covariance function \eqref{crescov} of \cite{Cressie1999} with $\nu=0.5$ in the formulation of the composite likelihood. For change-point estimation, we employ the proposed CLMDL in the main text, which is derived under the increasing domain setting. We also implement the modified CLMDL derived in the supplement, which is tailored for the infill setting. We select all pairs of time lag within 3 months $(k=3)$ and spatial distance within 500 kilometers in the composite likelihood. For the modified CLMDL tailored for the infill setting, we further set $B_{\mathcal{N}}=4$. The geodesic distance \citep{Karney2013} is used as the spatial distance, which is the shortest path between two points on the WGS84 ellipsoid. We note that the estimation result remains similar for different choices of time lag and spatial distance for the neighborhood.

The proposed CLMDL detects two change-points at \textcolor{black}{June 1953} and March 1968. Interestingly, the same change-points are detected by the modified CLMDL tailored for the infill setting, suggesting the robustness of our finding. The first change-point is within the great drought and prolonged heatwave which had great impact on the Midwestern U.S. \citep{Mishra2010,Westcott2011}. The second change-point is close to the proposed change in climate (1970) in North America from \cite{Bartomeus2011}. Moreover, the second change-point matches the one detected by \cite{Gromenko2017}, which analyzes a similar dataset but under annual resolution and allows at most one change-point. Based on Theorem \ref{asym.distr}, the 90\% CIs for the two change-points are \textcolor{black}{(Nov.\ 1950, Sep.\ 1955) and (Dec.\ 1966, Nov.\ 1971)}.

\textcolor{black}{In \Cref{subsec:addrealdata} of the supplement, we report the estimated model parameters for each stationary segment and provide visualization based on the estimation result. In addition, a robustness check is conducted for the change-point estimation result, which suggests that the changes are mainly due to the mean function of the spatio-temporal precipitation data.} 


\vspace{-0.2cm}
\section{Conclusion}\label{sec:conclusion}
In this paper, by combining composite likelihood and the MDL principle, we propose CLMDL, a unified and computationally efficient method for multiple change-point estimation in a piecewise stationary spatio-temporal process. CLMDL allows for non-separable space-time covariance specification and can detect changes in both mean and covariance functions. Moreover, it works under both the increasing domain and infill asymptotics. We show that exact recovery of true change-points can be achieved in the spatio-temporal setting under mild conditions. Furthermore, the effectiveness and practicality of CLMDL are demonstrated via extensive numerical studies.



\textcolor{black}{For future research, one interesting direction is to further consider the setting of change-point estimation in a locally stationary environment, where on each segment, the model parameter is allowed to vary smoothly instead of being constant, see \colorlinks{black}{\cite{wu2007inference}} and \colorlinks{black}{\cite{chen2022inference}} for works in non-parametric mean change under such setting. In \Cref{subsec:localvarying} of the supplement, we give a road map of how to achieve so under the CLMDL framework, where preliminary results show promise. Another interesting direction is to allow the number of change-points to diverge in the theoretical result, which is feasible under stronger assumptions on the Hessian matrix of the log-likelihood function. We give a detailed discussion in \Cref{subsec:div_cp} of the supplement. Lastly, most MDL based change-point estimation procedures in the literature (including this work) use the two-part code of standard MDL, as it suffices the purpose of change-point estimation. It will be interesting to design an algorithm based on the one-part code of refined MDL~\colorlinks{black}{\citep{grunwald2007minimum}} and compare the two strategies in terms of theoretical, computational and numerical performance.}

\vspace{2mm}
\setlength{\bibsep}{2pt plus 1ex}
\begin{spacing}{1.05}
	\bibliographystyle{apalike}
	\bibliography{Reference}
\end{spacing}

\end{document}


\setlength{\abovedisplayskip}{4pt}
	\setlength{\belowdisplayskip}{4pt}
	\setlength{\abovedisplayshortskip}{4pt}
	\setlength{\belowdisplayshortskip}{4pt}
	
	\if1\blind
	{
		\title{Supplementary Material: A Composite Likelihood-based Approach for Change-point Detection in Spatio-temporal Processes}	
		\author{
		Zifeng Zhao$^{1}$, Ting Fung Ma$^{2}$, Wai Leong Ng$^{3}$, Chun Yip Yau$^{4}$\\
		University of Notre Dame$^{1}$, University of South Carolina$^{2}$,\\
		Hang Seng University of Hong Kong$^{3}$ and Chinese University of Hong Kong$^{4}$
		}
		\date{}	
		\maketitle
	} \fi
	
	\if0\blind
	{
		\title{Supplementary Material: A Composite Likelihood-based Approach for Change-point Detection in Spatio-temporal Processes}
		\author{}
		\date{}
		\maketitle
		\vspace{-2cm}
	} \fi

The supplementary material is organized as follows. Section \ref{sec:edge_effect} illustrates with a simple example the non-ignorable edge effect and the related pairwise likelihood based change-point estimation inconsistency in the spatio-temporal setting~($S\to \infty$). \Cref{sec:relaxIS} provides a relaxation of the independence assumption among stationary segments. \textcolor{black}{\Cref{sec:alphMixing} gives three commonly used spatio-temporal parametric models that satisfy all $\alpha$-mixing conditions required for theoretical guarantees of CLMDL. \Cref{sec:extensions} discusses the potential extension of CLMDL to the locally stationary setting and the partial change setting.} Section \ref{sec:main_infill} provides detailed results of CLMDL under the infill asymptotics setting. Section \ref{sec:PELT} contains derivation of the PELT algorithm under the spatio-temporal setting. Section \ref{sec:add_num} provides \textcolor{black}{additional simulation results} and real data illustration. Section \ref{sec:notation} gives the outline of the technical proofs and defines the notation used in the proofs. Sections \ref{sec:proof_increasing_domain} and \ref{sec:vanishing} contain the proof of theoretical results under the increasing domain asymptotics for the constant \textcolor{black}{and vanishing change sizes, respectively.} Section \ref{sec:proof_infill} contains the proof for theoretical results under the infill asymptotics. In Section \ref{PELT.issue}, we further provide the proof for the validity of the pruned dynamic programming algorithm.

\section{Example of the non-ignorable edge effect}\label{sec:edge_effect}
Consider the simple example of $k=1$ and let the criterion function be in the form of
\begin{eqnarray*}
	IC(m)=-\sum_{j=1}^{m+1}\mathrm{PL}_{ST}(\hat{\psi}_{j};\bfX_j)+C(m+1)\log(ST)\,,  
\end{eqnarray*} 
where $m$ is the number of change-points, $\mathrm{PL}_{ST}(\hat{\psi}_{j};\bfX_j)$ is the pairwise log-likelihood of the $j$th segment evaluated under a parameter estimate $\hat{\psi}_j$, and the penalty $C(m+1)\log(ST)$ is of similar magnitude as some common information criteria such as BIC. We compare between two scenarios:

\vspace{0.2cm}

$\bullet$ Scenario 1: No change point is assumed and the pairwise log-likelihood function gives an estimate $\hat{\psi}$ over the entire observations $\bfY$. We have $IC(0)= -\mathrm{PL}_{ST}(\hat{\psi};\bfY)+C\log(ST)$, where
\begin{align*}
	\mathrm{PL}_{ST}(\hat{\psi};\bfY)=\sum_{(t,i,s_1,s_2)\in \mathcal{D}_{k,\mathcal{N}}}\log f(y_{t,s_1},y_{t+i,s_2};\hat{\psi})\,,
\end{align*}
$\mathcal{D}_{k,\mathcal{N}}=\{(t,i,s_1,s_2): 1\leq t, t+i \leq T; 0\leq i\leq 1; 1\leq s_1 \leq S; s_2\in s_1\cup \mathcal{N}(s_1); \text{ if }i=0,s_1\neq s_2\}.$
\vspace{0.2cm}

$\bullet$ Scenario 2: A change point is fixed at ${T}/{2}$ and the pairwise log-likelihoods give the estimates $\hat{\psi}_1, \hat{\psi}_2$ in the two segments, respectively. Thus, $IC(1)=-\sum_{j=1}^{2}\mathrm{PL}_{ST}(\hat{\psi}_{j};\bfX_j)+2C\log(ST)$, where  
\begin{eqnarray*}
	\sum_{j=1}^{2}\mathrm{PL}_{ST}(\hat{\psi}_{j};\bfX_j) = \sum_{(t,i,s_1,s_2)\in \mathcal{D}_{k,\mathcal{N}}^1}\log f(y_{t,s_1},y_{t+i,s_2};\hat{\psi}_1)+\sum_{(t,i,s_1,s_2)\in \mathcal{D}_{k,\mathcal{N}}^2}\log f(y_{t,s_1},y_{t+i,s_2};\hat{\psi}_2)\,,
\end{eqnarray*}
$\mathcal{D}_{k,\mathcal{N}}^1=\{(t,i,s_1,s_2): 1\leq t, t+i \leq \frac{T}{2}; 0\leq i\leq 1; 1\leq s_1 \leq S; s_2\in s_1\cup \mathcal{N}(s_1); \text{ if }i=0,s_1\neq s_2\}$, and $\mathcal{D}_{k,\mathcal{N}}^2=\{(t,i,s_1,s_2): \frac{T}{2}+1\leq t, t+i \leq T; 0\leq i\leq 1; 1\leq s_1 \leq S; s_2\in s_1\cup \mathcal{N}(s_1); \text{ if }i=0,s_1\neq s_2\}.$

\vspace{0.3cm}

Suppose that there is no change-point in the data, i.e., Scenario 1 is true, and the true parameter value is ${\psi}_o$. Then, 
under some regularity conditions, the estimators $\hat{\psi}, \hat{\psi}_1$ and $\hat{\psi}_2$ converge in probability to ${\psi}_o$ with order $\sqrt{ST}$. 
Hence, by Taylor's expansion, it can be seen that
\begin{eqnarray}
	IC(0)-IC(1)
	&=& \sum_{j=1}^{2}\mathrm{PL}_{ST}({\psi}_o;\bfX_j)-\mathrm{PL}_{ST}({\psi}_o;\bfY) - C\log(ST) + O_p(1)\,. \label{eg.IC}
\end{eqnarray}
Compared to $\mathrm{PL}_{ST}({\psi}_o;\bfY)$, due to the edge effect, $\sum_{j=1}^{2}\mathrm{PL}_{ST}({\psi}_o;\bfX_j)$ does not contain the pairwise log-likelihoods at $\mathcal{D}_{k,\mathcal{N}}^c:=\{(\frac{T}{2},\frac{T}{2}+1,s_1,s_2)|1\leq s_1 \leq S, s_2\in s_1\cup \mathcal{N}(s_1)\}$, which is of order $O(S)$. In other words,
\begin{eqnarray}
	\sum_{j=1}^{2}\mathrm{PL}_{ST}({\psi}_o;\bfX_j)-\mathrm{PL}_{ST}({\psi}_o;\bfY) = -\sum_{(s_1,s_2)\in\mathcal{D}_{k,\mathcal{N}}^c}\log f\left(y_{\frac{T}{2},s_1},y_{\frac{T}{2}+1,s_2};{\psi}_o\right) \,. \label{eg.PL.diff}
\end{eqnarray}
Note that for a pairwise log-likelihood evaluated at $\bfy=(y_1,y_2)$, we have
\begin{eqnarray*}
	-\log f\left(y_{1},y_{2};{\psi}_o\right)&=& - \log \frac{1}{2\pi \sqrt{|\Sigma|}}\exp(-{(\bfy-\mu)'}{\Sigma}^{-1}(\bfy-\mu)/2)\\
	&=&\log 2\pi + 1/2 \log |\Sigma| +{(\bfy-\mu)'}{\Sigma}^{-1}(\bfy-\mu)/2\,. 
\end{eqnarray*}
If $|\Sigma|>1$, then the term in \eqref{eg.PL.diff} is strictly positive 
and is of order $O_p(S)$, which dominates the penalty term of order $O(\log ST)$. Combining with \eqref{eg.IC}, 
$IC(0)-IC(1)>0$ with probability approaching one, indicating that a change-point model is wrongly selected. 
\hfill $\square$

{\color{black}
\subsection{Numerical experiments for inconsistency due to the edge effect}
In this subsection, we further conduct numerical experiments to illustrate the inconsistency of change-point estimation based on the pairwise likelihood~(PL), which is not corrected for the edge effect. Specifically, we consider a PLMDL criterion by replacing the composite likelihood in our proposed CLMDL criterion (see \eqref{MDLform} of the main text) with the pairwise likelihood without edge correction (see \eqref{CPL.k} of the main text).

The underlying DGP is exactly the same as that of \Cref{sim.1} of the main text. In particular, we assume each stationary segment follows the four-parameter autoregressive spatial model \eqref{sim.process} with $\mu=0$, hence the process is specified by $\theta=(\phi, \rho,\sigma^2)^\top$. We set $\theta_1=(-0.5,0.6,1)^\top$ and $\theta_2=(-0.5+\delta_\phi,0.6+\delta_\rho,1)^\top$, where $\delta_\phi$ controls the change in the temporal dependence and $\delta_\rho$ controls the change in the spatial dependence. For $\delta_\phi=\delta_\rho=0$, there is no change-point (i.e.\ $m_o=0$), otherwise, there is one change-point at $\lambda_1^o=0.5$ (i.e.\ $m_o=1$). For more details of the simulation setting, we refer to \Cref{sim.1} of the main text.

\Cref{scp.table.PL.only} reports the estimated number of change-points $\hat m$ based on PLMDL under various settings, where it can be seen that the non edge-corrected PL based criterion clearly suffers from false positive estimation. In particular, under both the no change-point scenario where $m_o=0$ and the one change-point scenario where $m_o=1$, PLMDL always detects at least two change-points across all sample sizes of $S$ and $T$. Compared with \Cref{scp.table} in the main text, which reports the performance of the proposed CLMDL criterion under the same simulation setting, it is clear that edge correction is necessary for consistent change-point estimation.

\begin{table}[h]
	\centering
{\color{black}
	\caption{Percentage of estimated change-points $\hat m$ among 1000 replications under various spatial size $S$, temporal size $T$, and signal levels $(\delta_\phi, \delta_\rho)$ based on the PLMDL criterion.}\label{scp.table.PL.only}
	\begin{tabular}{cccrrrrrrrrr}
		\hline \hline
		\multirow{3}{*}{$T$} & \multirow{3}{*}{\begin{tabular}[c]{@{}l@{}}$\delta_\phi\times 10$\end{tabular}} & \multirow{3}{*}{\begin{tabular}[c]{@{}l@{}}$\delta_\rho\times 10$\end{tabular}} & \multicolumn{9}{c}{\% of $\hat{m}$} \\
		&                                                                                     &                                                                                    & \multicolumn{3}{c}{$S=6^2$} & \multicolumn{3}{c}{$S=8^2$} & \multicolumn{3}{c}{$S=10^2$} \\ \cline{4-12} 
		&                                                                                     &                                                                                    & 0      & 1      & $\geq2$   & 0      & 1      & $\geq2$   & 0      & 1      & $\geq2$    \\ \hline
		100 & 0 & 0  & 0 & 0 & 100     & 0 & 0 & 100     & 0 & 0 & 100     \\
    & 2 & 0  & 0 & 0 & 100     & 0 & 0 & 100     & 0 & 0 & 100     \\
    & 3 & 0  & 0 & 0 & 100     & 0 & 0 & 100     & 0 & 0 & 100     \\
    & 0 & 6  & 0 & 0 & 100     & 0 & 0 & 100     & 0 & 0 & 100     \\
    & 0 & 10 & 0 & 0 & 100     & 0 & 0 & 100     & 0 & 0 & 100     \\
    & 2 & 2  & 0 & 0 & 100     & 0 & 0 & 100     & 0 & 0 & 100     \\
    & 3 & 3  & 0 & 0 & 100     & 0 & 0 & 100     & 0 & 0 & 100     \\ \hline
200 & 0 & 0  & 0 & 0 & 100     & 0 & 0 & 100     & 0 & 0 & 100     \\   
    & 2 & 0  & 0 & 0 & 100     & 0 & 0 & 100     & 0 & 0 & 100     \\
    & 3 & 0  & 0 & 0 & 100     & 0 & 0 & 100     & 0 & 0 & 100     \\
    & 0 & 6  & 0 & 0 & 100     & 0 & 0 & 100     & 0 & 0 & 100     \\
    & 0 & 10 & 0 & 0 & 100     & 0 & 0 & 100     & 0 & 0 & 100     \\
    & 2 & 2  & 0 & 0 & 100     & 0 & 0 & 100     & 0 & 0 & 100     \\
    & 3 & 3  & 0 & 0 & 100     & 0 & 0 & 100     & 0 & 0 & 100     \\ \hline \hline  
	\end{tabular}}
\end{table}
}

\section{Relaxation of the independence assumption across segments}\label{sec:relaxIS}

The consistency results given in Proposition \ref{prop.minCL}, Theorems \ref{unknownprob} and \ref{finerate} are based on the independence assumption across segments, which we denote as IS. Based on IS, the likelihood function in the proposed CLMDL takes the additive form $\sum_{j=1}^{m+1}L_{ST}^{(j)}(\psi_j;\bfX_j)$, which is a sum of composite likelihoods across different segments, and thus can be solved via dynamic programming.

We show in this subsection that, similar to the classical time series setting, CLMDL still gives consistency result under relaxation of IS when there is weak dependence across segments. We quantify the relaxation with Assumption \ref{ass.relaxIS}, which is a spatio-temporal extension of Assumption 3.1 in \cite{Aue2009}.


\begin{assumption}\label{ass.relaxIS}
	The observed spatio-temporal process $\bfY$ can be written as $y_{t,\bs}=f(y_{t,\bs}^*, z_{t,\bs})$ with some measurable function $f$, where $\bfY^*$ is a piecewise stationary spatio-temporal process with IS and $\bfZ$ is another spatio-temporal process that introduces dependence across segments of $\bfY$. There further exists a constant $C_I$ such that
	\begin{align*}
	&\sup_{S,T}\sup_{\psi\in \mathcal{M},0\leq t_1<t_2\leq T}\mathbb E\left\{\frac{1}{S}\left|L_{ST}^{[a]}(\psi;\bfY_{(t_1+1):t_2})-L_{ST}^{[a]}(\psi;\bfY_{(t_1+1):t_2}^*)\right|\right\} \leq  C_I\,, \text{ 	for } a=0,1,2.
	\end{align*}
\end{assumption}

Assumption \ref{ass.relaxIS} is expected to hold under mild conditions. For example, let $y_{t,\bs}=y^*_{t,\bs}+z_{t,\bs}$ with $y^*_{t,\bs}$ being a Gaussian random field, then by Taylor expansion and Cauchy-Schwartz inequality, a sufficient condition for Assumption \ref{ass.relaxIS} is $\sup_{S,T} S^{-1}\sum_{\bs \in \mathcal{S}}\sum_{t=1}^T \sqrt{\mathbb E(z_{t,\bs}^2)}< \infty$, which requires $\sqrt{\mathbb E(z_{t,\bs}^2)}$ to be absolutely summable along the time dimension. For a more concrete example, let $y^{*(j)}_{t,\bs}=\alpha^{(j)} y^{*(j)}_{t-1,\bs}+\varepsilon^{*(j)}_{t,\bs}$ for $[T\lambda_{j-1}^o]+1\leq t\leq [T\lambda_{j}^o]$, $j=1,\ldots,m_o+1$ be the piecewise stationary spatio-temporal process with IS, where the superscript $j$ indicates the stationary segment that $y^*_{t,\bs}$ is from. To introduce dependence across segments, let $y_{1,\bs}=y^{*(1)}_{1,\bs}$ and $y_{t,\bs}=\alpha^{(j)} y_{t-1,\bs}+\varepsilon^{*(j)}_{t,\bs}$ for $t>1$. Note that the initial observations of segment $j$ depends on the last observations of segment $j-1$. Simple algebra yields $z_{t,\bs}=(\alpha^{(j)})^{t-[T\lambda_{j-1}^o]}(y_{[T\lambda_{j-1}^o]}-y_{[T\lambda_{j-1}^o]}^{*(j)})$ for $[T\lambda_{j-1}^o]+1\leq t\leq [T\lambda_{j}^o]$, which clearly satisfies the absolutely summable condition.

The consistency result of CLMDL under the relaxed IS is presented in Corollary \ref{consistency_relaxIS}.
\begin{corollary}\label{consistency_relaxIS}
Let Assumption \ref{ass.relaxIS} hold for the observed process $\bfY$, and $\bfY^*$ is a piecewise stationary spatio-temporal process specified by $(m_o,\Lambda^o,\Psi^o)$ and satisfies conditions in Theorems \ref{unknownprob} or \ref{finerate}. We have that $\hat{m} = m_o, ~[T\hat{\Lambda}_{ST}] -[T\Lambda^o] = O_p(1), \text { and }  \hat{\Psi}_{ST} \longrightarrow \Psi^o\,,$ in probability provided that $S,T\longrightarrow \infty$ and $S=O(T)$, with $T\cdot S^{-r/2}\longrightarrow 0$~(Theorem \ref{unknownprob}) or with $\log T/S \longrightarrow 0$~(Theorem \ref{finerate}).
\end{corollary}

By Corollary \ref{consistency_relaxIS}, the exact recovery property of the change-point estimator $[T\hat{\Lambda}_{ST}]$ in Theorems \ref{unknownprob} and \ref{finerate} no longer holds. However, the relative change-point estimation remains consistent, i.e, $\hat{\Lambda}_{ST}-\Lambda^o=O_{p}(T^{-1})$, which is common in the classical time series setting. Intuitively, the presence of $z_{t,\bs}$ induces an extra $O_p(S)$ term which blurs the observed $y_{t,\bs}$ process around the true change-point $[T\Lambda^o]$. Note that Corollary \ref{consistency_relaxIS} additionally requires $S=O(T)$ to limit the impact of the extra $O_p(S)$ term, indicating the difficulty of detecting change-points with relaxation of IS.

{\color{black}
\section{Discussions of the $\alpha$-mixing conditions}\label{sec:alphMixing}
In this section, we provide a detailed discussion on the $\alpha$-mixing conditions used in the main text and further give three parametric models widely used in the spatio-temporal literature that satisfy all the assumed mixing conditions.

In particular, the three examples are Gaussian random fields with space-time covariance functions specified by (1).\ a separable classical Mat\'ern covariance function, (2).\ a non-separable generalized Mat\'ern covariance function proposed in \cite{fuentes2008class} and (3).\ a non-separable space-time covariance function proposed in \cite{stein2005space}.
	
To start, recall the result in \Cref{remark:mixing} of the main text. In particular, define $$\alpha_{\bfX_j^o}(d)=\alpha_{\bfX_j^o}(d;\infty,\infty):=\sup_{u\geq 1}\sup_{v\geq 1} \alpha_{\bfX_j^o}(d;u,v),$$
where $\bfX_j^o$ is the $j$th stationary segment. This is a mixing coefficient commonly used in the random field literature, see \cite{BerkesMorrow1981} and \cite{Doukhan1994}~(Section 1.3). Thus, by definition, we have that $\alpha_{\bfX_j^o}(d)\geq \alpha_{\bfX_j^o}(d,u,v)$ for any $u\geq 1$ and $v\geq 1.$ In addition, recall in \eqref{mixing.time} of the main text, we define
$$\alpha_{\bfX_j^o}^*(d)=\sup\Big\{\alpha_{\bfX_j^o}\big([t_1,t_2]\times \mathcal{S}', ~[t_2+d, t_3]\times \mathcal{S}' \big): 1\leq t_1\leq t_2,~ t_2+d\leq t_3<\infty,~ \mathcal{S}'\subset \mathcal{S} \Big\},
$$
which by definition also satisfies $\alpha_{\bfX_j^o}^*(d)\leq \alpha_{\bfX_j^o}(d)$.
	
In other words, for any given $d>0$, all mixing coefficients used in Assumptions \ref{ass.mix}, \ref{ass.mixtime}, \ref{ass.thetadistr} and equation \eqref{eq.mix.space} in \Cref{finerate} are dominated by $\alpha_{\bfX_j^o}(d)$. Thus, all the $\alpha$-mixing related assumptions required for CLMDL hold if we can show that $\alpha_{\bfX_j^o}(d)$ decays at a sufficiently fast \textit{polynomial} rate.

Importantly, \Cref{prop:Doukhan} states that the polynomial decay rate of the $\alpha$-mixing coefficient $\alpha_{\bfX_j^o}(d)$ can be bounded via the decay rate of the covariance function of the random field $\bfX_j^o$. \Cref{prop:Doukhan} directly follows from Corollary 2 in Section 2.1 of \cite{Doukhan1994} and we thus omit the proof. Recall that the spatio-temporal process considered in our paper is observed on $\mathbb N^+ \times \mathcal{S}$, where $\mathcal{S} \subset \mathbb R^2.$ In \Cref{sec:main} of the main text, we define a distance metric such that $\rho((t_1,\bs_1),(t_2,\bs_2))=\max\{|t_2-t_1|,|s_2^1-s_1^1|,|s_2^2-s_1^2|\}$ for any two points $(t_1,\bs_1)$ and $(t_2,\bs_2)$ with $t_1,t_2\in \mathbb N^+$ and $\bs_1=(s_1^1,s_1^2),\bs_2=(s_2^1,s_2^2)\in \mathcal{S}$.

\begin{proposition}\label{prop:Doukhan}
Assume that $\bfX=\{x_{t,\bs}, \bs\in\mathcal{S}\}_{t\in\mathbb N^+}$ is a stationary Gaussian random field and $\mathcal{S}\subset\mathbb R^2$ is a regular lattice. Suppose there exists some $A>3$ and a constant $c>0$ such that
		\begin{align}\label{eq:Cov_decay}
			\mathrm{Cov}(x_{t_1,\bs_1},x_{t_2,\bs_2})\leq c\cdot \left\{\rho((t_1,\bs_1),(t_2,\bs_2))\right\}^{-A},
		\end{align}
for any $t_1,t_2\in \mathbb N^+$ and $\bs_1,\bs_2\in \mathcal{S}$. If in addition, the spectral density of $\bfX$ is bounded below, we then have that $\alpha_{\bfX}(d)=O(d^{3-A}).$
\end{proposition}
	
Based on \Cref{prop:Doukhan}, if we can show that a stationary Gaussian random field has a positive lower bounded spectral density and its covariance function decays \textit{exponentially} w.r.t.\ the distance $\rho((t_1,\bs_1),(t_2,\bs_2))$, then \eqref{eq:Cov_decay} holds for any $A>0$. As a result, $\alpha_{\bfX}(d)$ decays faster than any polynomial rate and all the $\alpha$-mixing related assumptions required for CLMDL hold.
 
Importantly, many widely used spatio-temporal parametric models for Gaussian random fields have an exponentially decaying covariance function as shown in \Cref{prop_example}. Without loss of generality, in the following, we assume that the stationary Gaussian random field is of mean zero, thus its behavior is completely specified by the space-time covariance function.
	
\begin{proposition}\label{prop_example}
    Denote $\bfX=\{x_{t,\bs}, \bs\in\mathcal{S}\}_{t\in\mathbb N^+}$ as a stationary Gaussian random field on a regular lattice $\mathcal{S}\subset \mathbb R^2$. Suppose $\bfX$ follows one of the spatio-temporal parametric models in Examples \ref{ex1} or \ref{ex2} or \ref{ex3}, we have that $\bfX$ satisfies all the $\alpha$-mixing related assumptions for CLMDL in the main text.
\end{proposition}

\begin{example}\label{ex1}
The Gaussian random field $\bfX$ follows a separable Mat\'ern covariance function of parameter $(\sigma,\rho_1,\rho_2,\nu_1,\nu_2)$, defined as
\begin{equation*}
    {\rm Cov}(x_{t_1,\bs_1}, x_{t_2,\bs_2}) = \sigma^2 {\rm C_{space}}(\|\bs_1-\bs_2\|_2){\rm C_{time}}(|t_1-t_2|)
\end{equation*}
with
\begin{align}\label{eq:Cspace}
    {\rm C_{space}}(\|\bs_1- \bs_2\|_2)=  \frac{1}{2^{\nu_1-1}\Gamma(\nu_1)}\left(\frac{\sqrt{2{\nu_1}}\|\bs_1- \bs_2\|_2}{\rho_1}\right)^{\nu_1} K_{\nu_1}\left(\frac{\sqrt{2{\nu_1}}\|\bs_1-\bs_2\|_2}{\rho_1}\right) 
\end{align}
and
\begin{align}\label{eq:Ctime}
    {\rm C_{time}}(|t_1-t_2|)=  \frac{1}{2^{{\nu_2}-1}\Gamma({\nu_2})}\left(\frac{\sqrt{2{\nu_2}}|t_1-t_2|}{\rho_2}\right)^{\nu_2} K_{\nu_2}\left(\frac{\sqrt{2{\nu_2}}|t_1-t_2|}{\rho_2}\right) 
\end{align}	
where $\|\cdot\|_2$ is the Euclidean norm, $\Gamma(\cdot)$ is the gamma function, $K_{\nu_1}(\cdot)$ and $K_{\nu_2}(\cdot)$  are the modified Bessel functions of the second kind, $\sigma>0$ is the scale parameter, $\rho_1>0$ and $\rho_2>0$ are the range parameters, and $\nu_1>0$ and $\nu_2>0$ are the smoothness parameters. 
\end{example}
	
In \Cref{ex1}, for the case of $\nu_1=\nu_2=1/2$, the space and time covariance in \eqref{eq:Cspace} and \eqref{eq:Ctime} reduce to the exponential covariance functions such that ${\rm C_{space}}(\|\bs_1- \bs_2\|_2)=\exp(-\|\bs_1-\bs_2\|_2/\rho_1)$ and ${\rm C_{time}}(|t_1-t_2|)=\exp(-|t_1-t_2|/\rho_2)=\{\exp(-1/\rho_2)\}^{|t_1-t_2|}$, respectively. This recovers the autoregressive spatial model \eqref{sim.process} defined in Section \ref{subsec:simu} of the main text, which is used throughout our numerical experiments.

Examples \ref{ex2} and \ref{ex3} are widely used (non-separable) space-time covariance functions proposed in \cite{fuentes2008class} and \cite{stein2005space}, which are constructed based on spectral densities via the celebrated Bochner's theorem (see e.g.\ \cite{chen2021space}). In particular, given a valid spectral density $f(\boldsymbol{\omega}, \tau)$ on $\mathbb R^2\times \mathbb R$, one can construct a space-time covariance function via
\begin{align*}
        {\rm Cov}(x_{t_1,\bs_1}, x_{t_2,\bs_2})=\int_{\mathbb R^2}\int_{\mathbb R} \exp(i\boldsymbol{\omega}^\top (\bs_1-\bs_2)+i\tau (t_1-t_2)) f(\boldsymbol{\omega}, \tau) d\boldsymbol{\omega}d\tau \,,
\end{align*}
where $i=\sqrt{-1}.$

\begin{example}\label{ex2}
    The Gaussian random field $\bfX$ follows the generalized Mat\'ern covariance function in \cite{fuentes2008class} of parameter $(\alpha,\beta,\gamma,\nu,\epsilon)$. In particular, we have 
    \begin{align*}
        &{\rm Cov}(x_{t_1,\bs_1}, x_{t_2,\bs_2}) =\int_{\mathbb R^2}\int_{\mathbb R} \exp(i\boldsymbol{\omega}^\top (\bs_1-\bs_2)+i\tau (t_1-t_2)) f(\boldsymbol{\omega}, \tau) d\boldsymbol{\omega}d\tau 
    \end{align*}
    and the spectral density $f(\boldsymbol{\omega}, \tau)$ takes the form
    \begin{align}\label{eq:fuentes}
        f(\boldsymbol{\omega}, \tau)=\gamma \left(\alpha^2\beta^2+\beta^2\|\boldsymbol{\omega}\|_2^2+\alpha^2\tau^2+\epsilon\|\boldsymbol{\omega}\|_2^2\tau^2\right)^{-\nu},
    \end{align}
    where $\alpha>0$ is the spatial range parameter, $\beta>0$ is the temporal range parameter, $\gamma>0$ is the scale parameter, $\nu>3/2$ is the smoothness parameter and $\epsilon\in[0,1].$
\end{example}

\cite{fuentes2008class} shows that the space-time covariance function specified in \Cref{ex2} is non-separable unless $\epsilon=1$ in \eqref{eq:fuentes}, in which case the covariance function reduces to a separable Mat\'ern space-time covariance function. We refer to \cite{fuentes2008class} for more details.

\begin{example}\label{ex3}
    The Gaussian random field $\bfX$ follows the space-time covariance function in \cite{stein2005space} of parameter $(c_1,c_2,a_1,a_2,\alpha_1,\nu)$. In particular, we have 
    \begin{align*}
        &{\rm Cov}(x_{t_1,\bs_1}, x_{t_2,\bs_2}) =\int_{\mathbb R^2}\int_{\mathbb R} \exp(i\boldsymbol{\omega}^\top (\bs_1-\bs_2)+i\tau (t_1-t_2)) f(\boldsymbol{\omega}, \tau) d\boldsymbol{\omega}d\tau 
    \end{align*}
    and the spectral density $f(\boldsymbol{\omega}, \tau)$ takes the form
    \begin{align}\label{eq:stein}
        f(\boldsymbol{\omega}, \tau)= \left\{c_1(a_1^2+\tau^2)^{\alpha_1}+c_2(a_2^2+\|\boldsymbol{\omega}\|_2^2)\right\}^{-\nu},
    \end{align}
    where $c_1,c_2>0$, $a_1^2+a_2^2>0$, $\alpha_1$ is a positive integer, and $\nu>1+1/(2\alpha_1).$
\end{example}
The spectral density in \eqref{eq:stein} can generate a rich class of space-time covariance functions such as non-separable Mat\'ern covariance functions. We refer to Section 4 of \cite{stein2005space} for more details.

\subsection{Proof of \Cref{prop_example}}
\begin{proof}[Proof of \Cref{prop_example}] 

To prove \Cref{prop_example}, we only need to verify the conditions in \Cref{prop:Doukhan} hold for Examples \ref{ex1}, \ref{ex2} and \ref{ex3}. In particular, we need to show that (a).\ its covariance function $\mathrm{Cov}(x_{t_1,\bs_1},x_{t_2,\bs_2})$ has a polynomial decay rate as in \eqref{eq:Cov_decay} for any $A>0$, and (b).\ its spectral density, denoted by $f^1(\boldsymbol{\omega},\tau)$, is lower bounded.

First, note that by the definition of the distance metric $\rho(\cdot,\cdot)$, if we can show that
\begin{align*}
    \mathrm{Cov}(x_{t_1,\bs_1},x_{t_2,\bs_2})\leq c_1\exp\left\{-c_2\left(|t_2-t_1|+\|\bs_2-\bs_1\|_2\right)\right\},
\end{align*}
for some $c_1,c_2>0$, then \eqref{eq:Cov_decay} holds for any $A>0$. In addition, by assumption, $\bfX$ is observed on a regular lattice $\mathcal{S}\subset \mathbb R^2$. In other word, $\mathcal{S}$ consists of uniformly spaced spatial locations that are $\Delta$-distance apart in $\mathbb R^2$.

In addition, in the proof, we make use of the following property of the spectral density~(see e.g.\ Section 2.1 in \cite{fuentes2007approximate}). Denote $f_d(\boldsymbol{\lambda}), \boldsymbol{\lambda}\in \mathbb R^d$ as the spectral density of a random field observed over $\mathbb R^d$. The spectral density of the same random field but only observed on a regular lattice in $\mathbb R^d$ that are $\Delta$-distance apart takes the form $f_d^1(\boldsymbol{\lambda})=\sum_{Q\in \mathbb Z^d} f(\boldsymbol{\lambda}+2\pi Q/\Delta)$ for $\boldsymbol{\lambda}\in [0,2\pi/\Delta]^d.$ Note that $f_d^1(\boldsymbol{\lambda})$ is a periodic function of period $2\pi/\Delta$ along each axis of $\mathbb R^d.$ Without loss of generality, in the following, we assume that $\Delta=1$ and thus $\mathcal{S}=\mathbb Z^2.$ The exact same arguments can be used to prove the result for any $\Delta>0.$

\textbf{Verification for \Cref{ex1}}: We first verify the conditions of \Cref{prop:Doukhan} for Example \ref{ex1}. Recall that the Mat\'ern covariance function with parameter $(\rho,\nu)$ is of the form: $${\rm C}(h)= \frac{1}{2^{\nu-1}\Gamma(\nu)}\left(\frac{\sqrt{2\nu}h}{\rho}\right)^\nu K_\nu\left(\frac{\sqrt{2\nu}h}{\rho}\right)\,,$$
where $K_\nu(\cdot)$ is the modified Bessel function of the second kind with an asymptotic expansion $K_\nu(h) \asymp (\pi/(2h))^{1/2}\exp(-h)$ as $h \rightarrow \infty$. Here, $f(x) \asymp g(x)$ denotes $\lim_{x \rightarrow \infty} f(x)/g(x) = c \in (0,\infty)$. Thus, using this asymptotic expansion of $K_\nu(h)$ for large $h$, the tail behavior of the Mat\'ern covariance function is given by
$$
{\rm C}(h) \asymp h^{\nu-1/2} \exp\left(-\frac{\sqrt{2\nu}h}{\rho}\right).
$$
In other words, ${\rm C}(h)$ decays exponentially w.r.t.\ $h.$

Since for \Cref{ex1}, we have that
$${\rm Cov}(x_{t_1,\bs_1}, x_{t_2,\bs_2}) = \sigma^2 {\rm C_{space}}(\|\bs_1-\bs_2\|_2){\rm C_{time}}(|t_1-t_2|),$$
it is then clear that ${\rm Cov}(x_{t_1,\bs_1}, x_{t_2,\bs_2})$ decay exponentially w.r.t.\ both $|t_1-t_2|$ and $\|\bs_1-\bs_2\|_2$. Therefore, we have that \eqref{eq:Cov_decay} holds for any $A>0.$

In addition, for a Gaussian random field observed over $\mathbb R^d$ with a Mat\'ern covariance function of parameter $(\rho,\nu)$, its spectral density $f_d(\boldsymbol{\lambda})$ takes the form
$$
f_d(\boldsymbol{\lambda}|\rho,\nu)=\frac{2^d \pi^{d/2} \Gamma(\nu+d/2) (2\nu)^{\nu}}{\Gamma(\nu) \rho^{2\nu}} \left(\frac{2\nu}{\rho^2}+4\pi^2 \|\boldsymbol{\lambda}\|_2^2\right)^{-(\nu+d/2)}, ~\text{ for all }\boldsymbol{\lambda}\in \mathbb R^d.
$$

Thus, the spectral density of a random field over $\mathbb R^2 \times \mathbb R$ with the separable space-time Mat\'ern covariance function in \Cref{ex1} takes the form
$f(\boldsymbol{\omega},\tau)=f_2(\boldsymbol{\omega}|\rho_1,\nu_1)\cdot f_1(\tau|\rho_2,\nu_2)$, where the equality of the product form follows due to the separability of the space-time covariance function~(see e.g.\ \cite{chen2021space}).

Therefore, by properties of the spectral density as discussed above, for a Gaussian random field observed over $\mathbb Z^2 \times \mathbb Z$ with the separable space-time Mat\'ern covariance function in \Cref{ex1}, its spectral density $f^1(\boldsymbol{\omega},\tau)$ takes the form
$f^1(\boldsymbol{\omega},\tau)=\sum_{Q_1\in\mathbb Z^2}\sum_{Q_2\in \mathbb Z} f_2(\boldsymbol{\omega}+2\pi Q_1|\rho_1,\nu_1)\cdot f_1(\tau+2\pi Q_2|\rho_2,\nu_2),$
which is clearly lower bounded over $[0,2\pi]^3$ and thus over $\mathbb R^3$. This completes the verification for \Cref{ex1}.

\textbf{Verification for \Cref{ex2}}: First, by the property of the spectral density as discussed above, we know that the spectral density $f^1(\boldsymbol{\omega},\tau)$ for the random field in \Cref{ex2} takes the form
$f^1(\boldsymbol{\omega},\tau)=\sum_{Q_1\in \mathbb Z^2}\sum_{Q_2\in \mathbb Z}f(\boldsymbol{\omega}+2\pi Q_1, \tau+2\pi Q_2)$, where $f(\boldsymbol{\omega},\tau)$ is defined in \eqref{eq:fuentes}. Thus $f^1(\boldsymbol{\omega},\tau)$ is clearly lower bounded over $[0,2\pi]^3$ and thus over $\mathbb R^3$.

We now focus on the covariance function ${\rm Cov}(x_{t_1,\bs_1}, x_{t_2,\bs_2})$. By \cite{fuentes2008class} (Section 4), the covariance function can be further written as
\begin{align*}
    {\rm Cov}(x_{t_1,\bs_1}, x_{t_2,\bs_2})=\int_{\mathbb R}\exp(i\tau (t_1-t_2))g(\bs_1-\bs_2,\tau)d\tau,
\end{align*}
where $g(\bs_1-\bs_2,\tau)$ takes the form
\begin{align*}
    g(\bs_1-\bs_2,\tau)=\frac{\pi\gamma}{2^{\nu-2}\Gamma(\nu)}(\beta^2+\epsilon\tau^2)^{-\nu}\left(\frac{\|\bs_1-\bs_2\|_2}{\theta(\tau)}\right)^{\nu-1}{\mathcal{K}}_{\nu-1}(\theta(\tau)\|\bs_1-\bs_2\|_2)
\end{align*}
with $\theta(\tau)=\sqrt{\alpha^2(\beta^2+\tau^2)/(\beta^2+\epsilon\tau^2)}$. Recall that $\nu>3/2.$

For $\epsilon=0$, \cite{ip2017some} show that ${\rm Cov}(x_{t_1,\bs_1}, x_{t_2,\bs_2})$ has a closed-form solution takes the form of $2^{5/2-\nu}/\Gamma(\nu-3/2)\left(\sqrt{\alpha^2\|\bs_1-\bs_2\|_2^2+\beta^2|t_1-t_2|^2}\right)^{\nu-3/2}\mathcal K_{\nu-3/2}(\sqrt{\alpha^2\|\bs_1-\bs_2\|_2^2+\beta^2|t_1-t_2|^2})$, which clearly decays exponentially w.r.t.\ both $\|\bs_1-\bs_2\|_2$ and $|t_1-t_2|.$

Thus, we focus on $\epsilon\in (0,1].$ In such case, note that $\theta(\tau)$ is clearly lower and upper bounded by some constants $c_1$ and $c_2$. In addition, we have that $K_\nu(h) \asymp (\pi/(2h))^{1/2}\exp(-h)$ as $h\to \infty$. Therefore, the part
$({\|\bs_1-\bs_2\|_2}/{\theta(\tau)})^{\nu-1}{\mathcal{K}}_{\nu-1}(\theta(\tau)\|\bs_1-\bs_2\|_2)$ will decay exponentially w.r.t.\ $\|\bs_1-\bs_2\|_2$ uniformly over all $\tau\in \mathbb R.$ Moreover, it is easy to see that for any $t_1-t_2,$ the integral
$|\int_{\mathbb R}\exp(i\tau (t_1-t_2))(\beta^2+\epsilon\tau^2)^{-\nu} d\tau|\leq \int_{\mathbb R}(\beta^2+\epsilon\tau^2)^{-\nu} d\tau <\infty $ since $\nu>3/2.$ Therefore, we have that  ${\rm Cov}(x_{t_1,\bs_1}, x_{t_2,\bs_2})$ decays exponentially w.r.t.\ $\|\bs_1-\bs_2\|_2$ uniformly for any $t_1-t_2.$

To prove the result w.r.t.\ $|t_1-t_2|$, note that the part $({\|\bs_1-\bs_2\|_2}/{\theta(\tau)})^{\nu-1}{\mathcal{K}}_{\nu-1}(\theta(\tau)\|\bs_1-\bs_2\|_2)$ is upper bounded uniformly over any $\|\bs_1-\bs_2\|_2$ and $\tau$. Thus, we only need to focus on the integral
\begin{align*}
    \int_{\mathbb R}\exp(i\tau (t_1-t_2))(\beta^2+\epsilon\tau^2)^{-\nu} d\tau
   =\int_{\mathbb R}\cos(\tau |t_1-t_2|)(\beta^2+\epsilon\tau^2)^{-\nu} d\tau.
\end{align*}
In particular, we need to show that $\int_{\mathbb R}\cos(\tau h)(\beta^2+\epsilon\tau^2)^{-\nu} d\tau=O(h^{-A})$ for any $A>0.$ Denote $m(\tau)=(\beta^2+\epsilon\tau^2)^{-\nu}$ and $m^{(l)}(\tau)$ as its $l$th derivative. First, it is easy to see that $|\int_{\mathbb R} m^{(l)}(\tau) d\tau |$ is uniformly upper bounded for all $l\in \mathbb N.$ In addition, by integration by parts, we have that
\begin{align*}
    &\left|\int_{\mathbb R}\cos(\tau h)(\beta^2+\epsilon\tau^2)^{-\nu} d\tau \right|\\
   =&h^{-l}\left|\int_{\mathbb R}m^{(l)}(\tau)\left[ \sin(\tau h)\cdot\mathbb I(l \text{ is odd}) + \cos(\tau h)\cdot\mathbb I(l \text{ is even})\right] d\tau \right| < c h^{-l},
\end{align*}
which hold for any $l>0.$ This completes the verification for \Cref{ex2}.

\textbf{Verification for \Cref{ex3}}: First, using the exact same argument as that for \Cref{ex2}, it is easy to prove that the spectral density for the random field in \Cref{ex3} is lower bounded.

By \cite{stein2005space} (Section 4), the covariance function can be further written as 
\begin{align*}
    {\rm Cov}(x_{t_1,\bs_1}, x_{t_2,\bs_2})=\int_{\mathbb R}\exp(i\tau (t_1-t_2))g(\bs_1-\bs_2,\tau)d\tau,
\end{align*}
where $g(\bs_1-\bs_2,\tau)$ takes the form
\begin{align*}
    g(\bs_1-\bs_2,\tau)=\frac{\pi}{2^{\nu-2}c_2^\nu\Gamma(\nu)}\left(\frac{\|\bs_1-\bs_2\|_2}{\theta(\tau)}\right)^{\nu-1}{\mathcal{K}}_{\nu-1}(\theta(\tau)\|\bs_1-\bs_2\|_2)
\end{align*}
with $\theta(\tau)=(c_1c_2^{-1}(a_1^2+\tau^2)^{\alpha_1}+a_2^2)^{1/2}$ and $\nu>1+1/(2\alpha_1).$ By the property of $\mathcal{K}_{\nu-1}$, we have $g(\bs_1-\bs_2,\tau) \asymp \left({\|\bs_1-\bs_2\|_2}/{\theta(\tau)}\right)^{\nu-1}(\pi/(2\theta(\tau)\|\bs_1-\bs_2\|_2))^{1/2}\exp(-\theta(\tau)\|\bs_1-\bs_2\|_2)\asymp 
\{\theta(\tau)\}^{-\nu+1/2}(\|\bs_1-\bs_2\|_2)^{\nu-1/2}\exp(-\theta(\tau)\|\bs_1-\bs_2\|_2)
$. In particular, note that $\{\theta(\tau)\}^{-\nu+1/2}$ and its $l$th derivative for any $l>0$ is an integrable function w.r.t.\ $\tau.$ Thus, using the same arguments as that for \Cref{ex2}, we can show that ${\rm Cov}(x_{t_1,\bs_1}, x_{t_2,\bs_2})=O(\|\bs_1-\bs_2\|_2^{-A}+|t_1-t_2|^{-A})$, for any $A>0$ and thus complete the verification for \Cref{ex3}.
\end{proof}


\subsection{Pathological examples where composite likelihood fails}\label{subsec:pathological}
As mentioned in the discussion regarding \Cref{ass.model.id} in the main text, a change-point becomes indistinguishable to CLMDL if the pre-change data and post-change data cannot be distinguished by the composite likelihood used in the CLMDL criterion function (see equation \eqref{eq:CPL}). Recall that the composite likelihood in CLMDL is defined on the index set $\mathcal{D}_{k,\mathcal{N}}$, which includes pairs of observations that are at most $k$ time unit apart and at most $d$ spatial distance apart.

As discussed in \cite{davis2011comments} and \cite{Ma2016}, which concerns the use of pairwise likelihood in univariate time series, such scenario is uncommon in practice and can usually be avoided by enlarging the collection of observations where the composite likelihood is defined. In the following, we first give two pathological examples where such scenario may happen for CLMDL and then discuss the strategy to fix it.

(I). In terms of temporal dependence, if the true pre- and post-change data follow different autoregressive models of order 2 but have the same lag-0 and lag-1 autocovariance (and also same spatial covariance), the composite likelihood may not detect this change if time lag $k=1$ is used in $\mathcal{D}_{k,\mathcal{N}}$, i.e.\ we only consider pairs of observations that are at most $k=1$ time unit apart when constructing the composite likelihood.

(II). In terms of spatial dependence, consider a Gaussian random field with a spatial correlation matrix of a banded structure, where for $\bf s\neq \bf s'$, we have $$\text{Cor}(x_{t,\bf s}, x_{t,\bf s'})=\rho\cdot\mathbb I(\|{\bf s}-{\bf s}'\|_2\leq d_0) + \gamma\cdot\mathbb I(d_0<\|{\bf s}-{\bf s}'\|_2\leq d_1)$$ for some distance $0<d_0<d_1.$ If the true pre- and post-change data follow such a model with the same $\rho$ but different $\gamma$ (and the same temporal dependence), the composite likelihood may not detect this change if a $d\leq d_0$ is used in $\mathcal{D}_{k,\mathcal{N}}$, i.e.\ we only consider pairs of observations that are at most $d\leq d_0$ spatial distance apart when constructing the composite likelihood.

To summarize, in general, CLMDL may fail to detect a change-point if the pre- and post-change data exhibit the exact same behavior in the collection of observations where the composite likelihood is defined~(i.e. $\mathcal{D}_{k,\mathcal{N}}$). However, such issue can typically be solved by increasing the time lag $k$ or the spatial distance $d$ when defining the index set $\mathcal{D}_{k,\mathcal{N}}$. Indeed, it is easy to see that CLMDL will work for both pathological examples once $k>1$ and $d>d_0$ is used in defining $\mathcal{D}_{k,\mathcal{N}}$.
}

{\color{black}
\section{Possible extensions of CLMDL}\label{sec:extensions}
\subsection{Extension to a locally stationary environment}\label{subsec:localvarying}
In this subsection, we provide a road map of how to extend the proposed CLMDL to the setting of change-point estimation in a locally stationary environment, where on each segment, the model parameter is allowed to vary smoothly with time instead of being constant. We refer to this setting as the ``piecewise smooth setting", which is more general than the ``piecewise constant setting" considered in the main text.

We remark that existing works for change-point estimation in a locally stationary environment focus on the \textit{non-parametric} mean or variance setting, see e.g.\ \cite{wu2007inference}, \cite{wu2019multiscale} and \cite{chen2022inference} for non-parametric mean change and \cite{casini2021change} for non-parametric variance change. In contrast, our proposed change-point estimation method~(i.e.\ CLMDL) is designed for a spatio-temporal \textit{parametric} model. 

To incorporate local stationarity, we first modify our model to a \textit{semi-parametric} setting. Denote $0=\tau_0^o<\tau_1^o<\cdots <\tau_{m_o}^o <\tau_{m_o+1}^o=T$ as the $m_o$ change-points. We assume that on each stationary segment $[\tau_{j-1}^o+1,\tau_{j}^o]$ for $j=1,\cdots,m_o+1$, given the pseudo-true model order $\xi_j^o$, the pseudo-true model parameter $\theta_j^o\in \mathbb R^{d_j^o}$ is not a constant but can vary smoothly with time. In particular, denote the model parameter at time $t$ as $\theta_j^o(t)$, we assume that
$$\theta_j^o(t)=\mathbf{g}_j(t/T), ~\text{ for } t \in [\tau_{j-1}^o+1,\tau_{j}^o],$$
where $\mathbf{g}_j(r)=[g_{j,1}(r), g_{j,2}(r),\cdots,g_{j,d_j^o}(r)]$ is a $d_j^o$-dimensional vector-valued smooth function, such as the Lipschitz or H\"older functions. At each change-point $\tau_j^o$, the smooth function $\mathbf{g}_j(\cdot)$ switches to another smooth function $\mathbf{g}_{j+1}(\cdot)$ and experiences an abrupt change such that
$$\mathbf{g}_j(\tau_j^o/T)\neq \mathbf{g}_{j+1}(\tau_j^o/T), \text{ for } j=1,\cdots,m_o.$$
The primary interest is to estimate the $m_o$ unknown change-points. 

To extend CLMDL to this piecewise smooth setting, we consider a basis expansion based approach, which is commonly used in the functional data analysis literature~\citep{ramsay2005}. In particular, given a generic segment $[t_1/T, t_2/T]$ and a model order $\xi$, we approximate the $d=d(\xi)$ dimensional vector-valued smooth function $\mathbf{g}(r)=[g_1(r),g_2(r),\cdots,g_d(r)]$ via a basis expansion such that $$g_i(r)\approx\sum_{k=1}^K h_{i,k} \phi_k(r),$$
where $\{\phi_k(\cdot)\}_{k=1}^\infty$ is a basis system on $[t_1/T, t_2/T]$, such as the Fourier series or B-splines, and $K$ is a truncation parameter that governs the bias-variance trade-off. The model parameter is $\{h_{i,k}, k=1,\cdots,K\}_{i=1}^d$, which can be estimated via the composite likelihood.

Therefore, methodologically speaking, this basis expansion based semi-parametric modeling framework can be seamlessly combined with the proposed CLMDL method to conduct change-point estimation in the piecewise smooth setting. In \Cref{subsec:num_localvarying}, we have further conducted numerical experiments to investigate the performance of such a strategy and found that it can accurately detect abrupt changes while being robust to smooth changes.

On the other hand, establishing theoretical guarantees for such a procedure seems to be highly non-trivial, as we are now in a \textit{semi-parametric} setting, which is notably different from our current parametric setting. Thus, the theoretical result will require substantially different technical assumptions and arguments. Some challenges could involve the functional class of $\mathbf{g}_j$, the asymptotic order of the truncation parameter $K$, and the choice of the MDL penalty. Indeed, theoretical results regarding locally stationary spatio-temporal models are scarce, even for the case \textit{without} change-points. Therefore, the theoretical analysis is out of the scope of our current manuscript and we leave it as a future research direction.}

{\color{black}
\subsubsection{Numerical experiments for the locally stationary setting}\label{subsec:num_localvarying}

In this subsection, we further conduct numerical experiments to illustrate the performance of the modified CLMDL proposed in \Cref{subsec:localvarying} under the scenario of local stationarity.

We remark that all simulation settings, including the tuning parameters of CLMDL, follow the ones used in \Cref{sec:num} of the main text unless otherwise noted. On time $t,$ the data $\bfy_t=\{ y_{t,\bs}: \bs \in \mathcal{S} \}$ is defined on a regular two-dimensional grid $\mathcal{S}=\{(s_1,s_2):s_1,s_2\in \{1,\cdots,s\}\}$. We fix $T=100$ with a single change-point at $\tau_1^o=50$ (i.e.\ $\lambda_1^o=0.5$) and vary the spatial dimension $S.$ 

For the data generating process~(DGP), we modify the four-parameter autoregressive spatial model \eqref{sim.process} in the simulation studies of the main text to generate local stationarity.  In particular, on each segment, the observation $\bfy_t$ follows
\begin{equation}\label{sim.process_restate_local}
\bfy_t - \mu(t)= \phi (\bfy_{t-1}-\mu(t-1)) + \boldsymbol{\varepsilon}_t\,,
\end{equation}
where $\mu(t)$ denotes the locally stationary mean function, and $\boldsymbol{\varepsilon}_t=\{ \varepsilon_{t,\bs}: \bs \in \mathcal{S} \}$ is a Gaussian process with an exponential covariance function 
${\rm Cov}(\varepsilon_{t,\textbf{s}}, \varepsilon_{t,\textbf{s}'}) = \sigma^2\exp\{-\left\lVert \textbf{s}-\textbf{s}'\right\lVert_2/\rho\}$. Following the main text, we set $\theta=(\phi, \rho,\sigma^2)^\top=(-0.5,0.6,1)$.

For the true mean function $\mu(t)$, using the Fourier basis, we first define a smooth function $g(r)$ on $r\in [0,1]$ such that
\begin{align*}
    g(r)=\sum_{i=1}^{10}i^{-3}\cos\left(2i\pi r\right) + \sum_{i=1}^{10}i^{-3}\sin\left(2i\pi r\right).
\end{align*}
To generate a piecewise smooth mean function, we set $\mu(t)$ as
\begin{align*}
    \mu(t)=\begin{cases}
        g(t/T), ~\text{ if } 1\leq t\leq 50, \\
        g(t/T)+\delta, ~\text{ if } 51 \leq t \leq 100,
    \end{cases}
\end{align*}
where $\delta$ regulates the change size. Clearly, if $\delta=0$, there is no change-point. Otherwise, there is one change-point at $\tau_1^o=50.$ For more intuition, \Cref{fig:functional} plots the mean function $\mu(t)$ with $\delta=0.5.$ As can be seen, the mean function $\mu(t)$ varies smoothly over time with an abrupt change of size $\delta$ at $t=50.$

\begin{figure}
    \centering
    {\color{black}
    \includegraphics[scale=0.4]{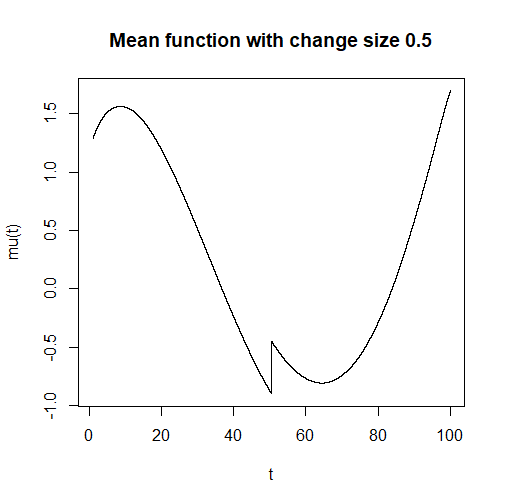}
    \vspace{-5mm}
    \caption{ The piecewise smooth mean function $\mu(t)$ with $\delta=0.5$ at $\tau_1^o=50$.}
    \label{fig:functional}}
\end{figure}

To implement the modified CLMDL proposed in \Cref{subsec:localvarying}, we use the autoregressive spatial model in \eqref{sim.process_restate_local} as the working model to compute the composite log-likelihood function, where we approximate the smooth mean function $\mu(t)$ via a Fourier basis such that
\begin{align*}
    \mu(t)=\beta_0 + \sum_{i=1}^3\beta_{cos,i}\cos\left(2i\pi \frac{t}{T}\right) + \sum_{i=1}^3\beta_{sin,i}\sin\left(2i\pi \frac{t}{T}\right).
\end{align*}
Importantly, note that the mean function of the working model admits an approximation error due to misspecification of the true mean function, as it only has \textit{three} instead of \textit{ten} Fourier basis functions $\cos(\cdot)$ and $\sin(\cdot)$.

We vary $\delta=0,0.5,1$ and vary $S=6^2,8^2,10^2,15^2,20^2,30^2$. For each simulation setting, we repeat the experiments 1000 times. \Cref{table.local.stationary} summarizes the detailed simulation result. As can be seen, the modified CLMDL proposed in \Cref{subsec:localvarying}  is robust to the approximation error due to model misspecification and indeed provides promising performance: when there is no abrupt change (i.e.\ $\delta=0$), the modified CLMDL is robust to false positive detection; when there is an abrupt change, the modified CLMDL can indeed detect its presence and the performance of CLMDL improves as the sample size $S$ grows or the change size $\delta$ increases. The modified CLMDL can still achieve exact recovery of the true change-points for sufficiently large $S$ or $\delta.$  To summarize, this numerical experiment suggests that the modified CLMDL can be used for change-point estimation for locally stationary processes.

\begin{table}[h]
\centering
{\color{black}
\caption{Percentage of $\hat{m} = 1$, and percentage of $\hat{\lambda}=0.5$, mean, empirical standard deviation (esd) of $\hat{\lambda}$ (given $\hat{m}=m_o=1$) under various signal levels ($\delta$) and sample sizes ($S$) with local stationarity.}\label{table.local.stationary}
\begin{tabular}{crrrrr}
\hline\hline
$S$    & $\delta$ & \% $\hat{m} = 1$ & \% $\hat{\lambda} =   0.5$ & mean & esd \\ \hline
$6^2$  & 0   & 0   & -   & -      & -      \\
$8^2$  &     & 0   & -   & -      & -      \\
$10^2$ &     & 0   & -   & -      & -      \\
$15^2$ &     & 0   & -   & -      & -      \\
$20^2$ &     & 0   & -   & -      & -      \\
$30^2$ &     & 0   & -   & -      & -      \\ \hline
$6^2$  & 0.5 & 5   & 80  & 0.4967 & 0.0208 \\
$8^2$  &     & 25  & 60  & 0.4980 & 0.0193 \\
$10^2$ &     & 55  & 73  & 0.4997 & 0.0066 \\
$15^2$ &     & 100 & 94  & 0.4996 & 0.0037 \\
$20^2$ &     & 100 & 99  & 0.4999 & 0.0001 \\
$30^2$ &     & 100 & 100 & 0.5000 & -      \\ \hline
$6^2$  & 1.0 & 100 & 100 & 0.5000 & -      \\
$8^2$  &     & 100 & 100 & 0.5000 & -      \\
$10^2$ &     & 100 & 100 & 0.5000 & -      \\
$15^2$ &     & 100 & 100 & 0.5000 & -      \\
$20^2$ &     & 100 & 100 & 0.5000 & -      \\
$30^2$ &     & 100 & 100 & 0.5000 & -      \\ \hline \hline
\end{tabular}}
\end{table}
}

{\color{black}
\subsection{Results under a partial change setting}\label{subsec:partialchange}
In this subsection, we investigate the performance of CLMDL under the scenario of partial change. We show that thanks to its ability to handle model misspecification, CLMDL is capable of detecting change-points where the observations only experience partial changes.

In particular, consider a spatio-temporal process $\bfX=\{\mathbf{x}_1,\cdots,\mathbf{x}_T\}$ observed over spatial locations $\mathcal{S}$ at time $t=1,\cdots, T$, where $\mathbf{x}_t=(x_{t,\bs})_{\bs\in\mathcal S}$. Suppose that among the $S=|\mathcal{S}|$ spatial locations, the ones in $\mathcal{S}_*$ follows a true parameter $\theta_2$ and the ones in $\mathcal{S}_*^c=\mathcal{S}\setminus \mathcal{S}_*$ follows a true parameter $\theta_1.$ Furthermore, we assume that $|\mathcal{S}_*|=rS$ and $|\mathcal{S}_*^c|=(1-r)S$ for some $r\in (0,1)$, where $r$ is allowed to vanish as $S,T\longrightarrow\infty.$ In other words, the data is a mixture of two stationary spatio-temporal processes with true parameters $\theta_1$ and $\theta_2.$

Denote $L_{ST}(\theta;\bfX)$ as the composite likelihood function for the entire data $\bfX$ as defined in \eqref{eq:CPL} of the main text. Denote $L_{\mathcal{S}_*T}(\theta;\bfX)$ as the composite log-likelihood function for the data in $\bfX$ observed on spatial locations $\mathcal{S}_*$. In particular, $L_{\mathcal{S}_*T}(\theta;\bfX)$ follows by replacing $\mathcal{S}$ with $\mathcal{S}_*$ in the definition of $D_{k,\mathcal{N}}$ and $E_{k,\mathcal{N}}$ for $L_{ST}(\theta;\bfX)$ in \eqref{eq:CPL}. Similarly, denote $L_{\mathcal{S}_*^cT}(\theta;\bfX)$ as the composite log-likelihood function for the data in $\bfX$ observed on spatial locations $\mathcal{S}_*^c.$ Note that we have $L_{ST}(\theta;\bfX)=L_{\mathcal{S}_*T}(\theta;\bfX)+L_{\mathcal{S}_*^cT}(\theta;\bfX)$. 

Denote the expected composite likelihood function as $\mathcal L(\theta)=\mathbb E(L_{ST}(\theta;\bfX))/(ST)$, $\mathcal L_{\mathcal{S}_*}(\theta)=\mathbb E(L_{\mathcal S_*T}(\theta;\bfX))/(|\mathcal S_*|T)$ and $\mathcal L_{\mathcal{S}_*^c}(\theta)=\mathbb E(L_{\mathcal{S}_*^cT}(\theta;\bfX))/(|\mathcal{S}_*^c|T)$. Denote $\theta^*$ as the maximizer of $\mathcal L(\theta)$, which serves as the pseudo-true parameter for $\bfX$. Recall $\bfX$ is a mixture of two stationary processes observed over $\mathcal{S}_*$ (with true parameter $\theta_2$) and $\mathcal{S}_*^c$ (with true parameter $\theta_1$).

\begin{proposition}\label{thm:partial}
    Suppose Assumptions \ref{ass.countable}, \ref{ass.mom}($r$) and \ref{ass.model.id} hold for both stationary processes observed over $\mathcal{S}_*$ and $\mathcal{S}_*^c.$ In addition, assume that for any $\epsilon>0$, it holds that
    \begin{align}\label{eq:first_order_bound}
        \inf_{\theta \in \{\|\theta-\theta_2\|>\epsilon\}\cap \Theta } \left\|\mathcal L_{\mathcal S_*}'(\theta) \right\|_2 >0.
    \end{align} 
    We have that there exists a constant $c>0$ such that $$\|\theta^*-\theta_1\|_2>cr\|\theta_1-\theta_2\|_2$$ for all sufficiently large $S,T.$
\end{proposition}

The assumption \eqref{eq:first_order_bound} is a mild condition as it requires that the expected composite log-likelihood function $\mathcal L_{\mathcal S_*}(\theta)$ has a non-zero gradient when $\theta$ is $\epsilon$-distance away from its unique maximizer $\theta_2$. A sufficient condition for $\eqref{eq:first_order_bound}$ is that $\mathcal L_{\mathcal S_*}(\theta)$ is a concave function.

\textbf{Implication of \Cref{thm:partial}}: Given a change-point $\tau^o$, suppose the pre-change parameter is $\theta_1$ and at $\tau^o$, the data experiences a partial change where only $r$-proportion of the spatial locations switch to the post-change parameter $\theta_2$. Denote $\theta^*$ as the pseudo-true parameter for the post-change segment. By \Cref{thm:partial}, we have that the \textit{effective} change size $\|\theta_1-\theta^*\|_2$ at $\tau^o$ can be lower bounded by $\|\theta_1-\theta^*\|_2>cr\|\theta_1-\theta_2\|_2$ for some $c>0.$ Importantly, combined with the results in Theorems \ref{unknownprob} and \ref{thm_diminishing}, this implies that CLMDL can still detect change-points where the observations only experience partial changes.

\textbf{An illustrative example}: For more intuition and clarity, consider the simple example of a mean change model. Assume the true pre-change data is generated as $y_{t,\bf s}=0 + \varepsilon_{t,\bf s}$, where $\varepsilon_{t,\bf s}$ is i.i.d.\ $N(0,1)$ across time and space. Assume the true post-change data is generated as $y_{t,\bf s}=\delta \cdot {\mathbb I}({\bf s}\in \mathcal{S}_*) + \varepsilon_{t,\bf s}$, where $\varepsilon_{t,\bf s}$ is i.i.d.\ $N(0,1)$ across time and space. Here $\mathcal{S}_*$ dictates the spatial locations that have mean change with a cardinality of $|\mathcal{S}_*|=rS$, where $r\in [0,1]$ indicates the proportion. If we use a parametric mean-variance model with $y_{t,\bf s}\overset{i.i.d.}{\sim} N(\mu, \sigma^2)$ to fit the data, it is easy to see that the \textit{pseudo-true} parameter will be $\theta^*_{pre}=(\mu_{pre},\sigma_{pre}^2)=(0,1)$ for the pre-change data and $\theta^*_{post}=(\mu_{post},\sigma_{post}^2)=(r\delta, 1+r(1-r)\delta^2)$ for the post-change data. 

Based on the notation in \Cref{thm:partial}, we have $\theta_1=(0,1)$ and $\theta_2=(\delta,1).$ In particular, the pseudo-true parameter $\theta^*_{post}$ for the post-change data follows the lower bound given in \Cref{thm:partial} such that $\|\theta^*_{post}-\theta_1\|_2>cr\|\theta_2-\theta_1\|_2$ with $c=1.$ The effective change size is $S\|\theta^*_{post}-\theta^*_{pre}\|^2_2=r^2S\delta^2(1+(1-r)^2\delta^2)\asymp r^2S$, where $\delta$ is a constant. By the result in Theorems \ref{unknownprob} and \ref{thm_diminishing}, we have that CLMDL can achieve exact recovery given that $r>c/\sqrt{S}$ for some $c>0$ and can achieve consistent change-point estimation given that $TSr^2 \to \infty.$ In particular, the proportion $r$ of spatial locations that experience changes is allowed to vanish as $S,T$ increase.

\subsubsection{Numerical experiments for the partial change setting}\label{subsec:num_partialchange}
In this subsection, we further conduct numerical experiments to illustrate the performance of CLMDL under the scenario of partial changes.

We remark that all simulation settings, including the tuning parameters of CLMDL, follow the ones used in \Cref{sec:num} of the main text unless otherwise noted. On time $t,$ the data $\bfy_t=\{ y_{t,\bs}: \bs \in \mathcal{S} \}$ is defined on a regular two-dimensional grid $\mathcal{S}=\{(s_1,s_2):s_1,s_2\in \{1,\cdots,s\}\}$. We fix $T=100$ with a single change-point at $\tau_1^o=50$ (i.e.\ $\lambda_1^o=0.5$) and vary the spatial dimension $S.$

For the pre-change data $\{\bfy_1,\cdots,\bfy_{50}\}$, we generate it using the four-parameter autoregressive spatial model \eqref{sim.process} in the simulation studies of the main text. In particular, the pre-change data follows 
\begin{equation}\label{sim.process_restate}
\bfy_t - \mu= \phi (\bfy_{t-1}-\mu) + \boldsymbol{\varepsilon}_t\,,
\end{equation}
where $\boldsymbol{\varepsilon}_t=\{ \varepsilon_{t,\bs}: \bs \in \mathcal{S} \}$ is a Gaussian process with an exponential covariance function 
${\rm Cov}(\varepsilon_{t,\textbf{s}}, \varepsilon_{t,\textbf{s}'}) = \sigma^2\exp\{-\left\lVert \textbf{s}-\textbf{s}'\right\lVert_2/\rho\}$. Following the main text, we set $\theta_1=(\mu,\phi, \rho,\sigma^2)^\top=(0,-0.5,0.6,1)$ for the pre-change data.

To generate partial changes, we modify the DGP in \eqref{sim.process_restate} and simulate the post-change data $\{\bfy_{51},\cdots,\bfy_{100}\}$ via
\begin{equation}\label{sim.process_restate2}
\bfy_{t,\bs} - \mu\cdot\mathbb{I}(\bs\in\mathcal{S}^*)= \phi (\bfy_{t-1,\bs}-\mu\cdot\mathbb{I}(\bs\in\mathcal{S}^*)) + \boldsymbol{\varepsilon}_{t,\bs}\,, \text{ for all } \bs\in\mathcal{S},
\end{equation}
where $\mathcal{S}^*$ is a randomly selected subset of $\mathcal{S}$ with cardinality $|\mathcal{S}^*|=rS.$ We set $\theta_2=(\mu,\phi, \rho,\sigma^2)^\top=(1,-0.5,0.6,1)$ for the post-change data.

To summarize, at the change-point $\tau^o_1=50$ (i.e. $\lambda_1^o=0.5$), there is a mean change from 0 to 1 but only for data observed over the spatial locations in $\mathcal S^*$. The spatio-temporal dependence remain unchanged. To implement CLMDL, we use the four-parameter autoregressive spatial model \eqref{sim.process_restate} as the working model to compute the composite log-likelihood function. Note that the model \eqref{sim.process_restate} mis-specifies the post-change data due to the partial change.

We vary $S=6^2,8^2,10^2,30^2,60^2$ and set the proportion of change as $r=S^{-0.4}$ or $r=S^{-0.5}.$ In other words, the proportion of spatial locations that experience a change vanishes while $S$ increases. For each simulation setting, we repeat the experiments 1000 times. Table \ref{table.partial.change} summarizes the detailed simulation result.  As can be seen, for both $r=S^{-0.4}$ and $S^{-0.5}$, the performance of CLMDL improves as $S$ grows. This suggests that CLMDL still works under the setting of partial changes, even for the case where the proportion of change vanishes as $S$ increases. Note that for the case $r=1/\sqrt{S}$, CLMDL can still achieve exact recovery of the true change-points for sufficiently large $S$, which supports the theoretical results we derived in \Cref{subsec:partialchange}.

\begin{table}[h]
\centering
{\color{black}
\caption{Percentage of $\hat{m} = 1$ among 1000 replications, and percentage of $\hat{\lambda}=0.5$, mean, empirical standard deviation (esd) of $\hat\lambda$~(given $\hat m=m_o=1$) under various partial change settings.}\label{table.partial.change}
\begin{tabular}{cccrrrr}
\hline\hline
           & $S$    & proportion of change & \% $\hat{m} = 1$ & \multicolumn{3}{c}{$\hat{\lambda}$}     \\ \cline{5-7}
           &        &           $(r\times 100)$           &                  & \% of $\hat{\lambda} = 0.5$  & \multicolumn{1}{c}{mean} & \multicolumn{1}{c}{esd}   \\ \hline
$r=S^{-0.4}$ & $6^2$  & 23.8                 & 100                                  & 49                                         & 0.5027                   & 0.0341                  \\
           & $8^2$  & 18.9                 & 99                                   & 62                                         & 0.4993                   & 0.0313                  \\
           & $10^2$ & 15.8                 & 99                                   & 85                                         & 0.5007                   & 0.0180                  \\
           & $30^2$ & 6.6                  & 100                                  & 100                                        & 0.5000                   & -                       \\
           & $60^2$ & 3.8                  & 100                                  & 100                                        & 0.5000                   & -                       \\ \hline
$r=S^{-0.5}$ & $6^2$  & 16.7                 & 33                                   & 42                                         & 0.5015                   & 0.0474                  \\
           & $8^2$  & 12.5                 & 39                                   & 67                                         & 0.4946                   & 0.0290                  \\
           & $10^2$ & 10.0                 & 43                                   & 79                                         & 0.5007                   & 0.0215                  \\
           & $30^2$ & 3.3                  & 90                                   & 100                                        & 0.5000                   & -                       \\
           & $60^2$ & 1.7                  & 100                                  & 100                                        & 0.5000                   & -                       \\ \hline\hline 
\end{tabular}}
\end{table}

\subsubsection{Proof of \Cref{thm:partial}}
\begin{proof}[Proof of \Cref{thm:partial}]
    Define $\delta=\|\theta_1-\theta_2\|_2>0$ as the difference between the two parameters. First, by triangle inequality, we have that
    \begin{align*}
        \|\theta^*-\theta_1\|_2+\|\theta^*-\theta_2\|_2\geq \|\theta_2-\theta_1\|_2=\delta,
    \end{align*}
    which implies that $\max\{\|\theta^*-\theta_1\|_2, \|\theta^*-\theta_2\|_2\}>\delta/2$. Suppose $\|\theta^*-\theta_1\|_2>\|\theta^*-\theta_2\|_2$, we then finish the proof. Thus, in the following, we assume that $\|\theta^*-\theta_1\|_2<\|\theta^*-\theta_2\|_2$, which implies that $\|\theta^*-\theta_2\|_2>\delta/2$ is a constant. Recall that we have $|\mathcal{S}_*|=rS$ and $|\mathcal{S}_*^c|=(1-r)S.$ 

    We first consider the scenario where $r$ is a fixed proportion in $(0,1)$. In such case, we only need to show that $\|\theta^*-\theta_1\|_2>c$ for some constant $c$. We prove by contradiction. Suppose that $\|\theta^*-\theta_1\|_2\to 0$ as $S,T$ grow. By the definition of $\theta^*$, $\theta_1$ and a Taylor expansion, we have that
    \begin{align*}
       0= \mathbb E(L_{ST}'(\theta^*;\bfX))= \mathbb E(L'_{\mathcal{S}_*T}(\theta^*;\bfX)) + \mathbb E(L'_{\mathcal{S}_*^cT}(\theta^*;\bfX))=  \mathbb E(L'_{\mathcal{S}_*T}(\theta^*;\bfX)) + \mathbb E(L''_{\mathcal{S}_*^cT}(\theta^+_1;\bfX))(\theta^*-\theta_1),
    \end{align*}
    where $\theta^+_1$ is between $\theta^*$ and $\theta_1$. In other words, we have that
    \begin{align*}
        -\mathcal{L}_{S_*}'(\theta^*)\cdot \frac{r}{1-r}=\mathcal{L}_{S_*^c}''(\theta_1^+)(\theta^*-\theta_1)
    \end{align*}
    
    By the assumption that $\|\theta^*-\theta_1\|_2\to 0$, we have $\|\theta^+_1-\theta_1\|_2\to 0$. Thus, by the definition of $\theta_1$, we have that $\mathcal{L}_{S_*^c}''(\theta_1^+)$ converges to a negative definite matrix. Thus, we have that for sufficiently large $S,T$,
    \begin{align*}
        \|\theta^*-\theta_1\|_2= \left\| -\frac{r}{1-r} \{\mathcal{L}_{S_*^c}''(\theta_1^+)\}^{-1} \mathcal{L}_{S_*}'(\theta^*)\right\|_2
        > \frac{c'r}{1-r}\|\mathcal{L}_{S_*}'(\theta^*)\|_2>c,
    \end{align*}
    for some constant $c>0$, where the inequality follows from \eqref{eq:first_order_bound} and the fact that $\|\theta_2-\theta^*\|>\delta/2$. Thus, there is a contradiction and we have that $\|\theta^*-\theta_1\|>c$ for some $c>0.$

    For the scenario where $r\to 0$, the proof is again by contradiction and follows the same argument. Assume that $\|\theta^*-\theta_1\|_2 /(r\delta) \to 0$, we have that $\|\theta^*-\theta_1\|_2\to 0$ since $r\delta\to 0.$ By using the same argument as above, we have that
    \begin{align*}
        \|\theta^*-\theta_1\|_2 =\left\| -\frac{r}{1-r} \{\mathcal{L}_{S_*^c}''(\theta_1^+)\}^{-1} \mathcal{L}_{S_*}'(\theta^*)\right\|_2>cr,
    \end{align*}
    for some $c>0$, which gives a contradiction. Thus, we finish the proof.
\end{proof}

\subsection{Extension to diverging number of change-points}\label{subsec:div_cp}
The asymptotic results of CLMDL operate under the classical infill framework (along the time dimension) widely used in the change-point literature. In particular, denote $\tau_j^o$ as the $j$th true change-point, we require $\tau_j^o/T\to \lambda_j^o \in (0,1)$ for $j=1,\cdots,m_o$ and $\min_{1\leq j\leq m_o+1}(\lambda_j^o-\lambda_{j-1}^o)= \epsilon_\lambda^o>0$, where we define $\lambda_0^o = 0$ and $\lambda_{m_o+1}^o= 1.$ Note that the minimum spacing $\epsilon_\lambda^o$ naturally imposes an upper bound of $[1/\epsilon_\lambda^o-1]$ on the number of true change-points.

This is a common assumption in the literature for change-point estimation, see for example\ \cite{Andrews1993}, \cite{Davis2006}, \cite{Aue2009}, \cite{Matteson2014}, \cite{Ma2016}, \cite{fearnhead2019changepoint}, \cite{romano2022detecting}, and \cite{zhao2022segmenting}.

A natural extension for the theoretical results of CLMDL is to allow the minimum spacing $\epsilon_\lambda^o$ to vanish as $S,T$ increase (i.e.\ $\epsilon_\lambda^o \to 0$). However, such an extension is challenging without additional strong technical assumptions on the behavior of the composite log-likelihood function. In particular, due to the complex nature of the composite log-likelihood based criterion function, our theoretical analysis of CLMDL in Sections \ref{sec:proof_increasing_domain}, \ref{sec:vanishing} and \ref{sec:proof_infill} relies on \textit{asymptotic} probability bounds instead of finite-sample high probability bounds. Thus, the finiteness of the number of change-points (and thus $\epsilon_\lambda^o>0$) is essential as only a \textit{finite} sum of probabilities that converge to zero is guaranteed to converge to zero.

In the following, we give a more detailed (though high level) explanation. Recall we denote $\tau_j^o=[T\lambda_j^o]$ and $\hat\tau_j=[T\hat\lambda_j].$ For simplicity of illustration, consider the case where $\tau_{j-1}^o\leq \hat\tau_j<\tau_j^o<\hat\tau_{j+1}\leq \tau_{j+1}^o$. To establish consistency of $\hat\tau_j$, we need to analyze log-likelihood terms such as
$$R(\hat\tau_j, \tau_j^o, \hat\tau_{j+1}) \coloneqq L(\hat\theta, \hat\tau_j+1,\hat\tau_{j+1})-L(\theta_j^o,\hat\tau_j+1,\tau_j^o)-L(\theta_{j+1}^o,\tau_j^o+1,\hat\tau_{j+1}),$$
where for notational simplicity, we denote $L(\theta,t_1,t_2)$ as the composite log-likelihood of data from time $t_1$ to $t_2$ and evaluated at $\theta$. In $L(\hat\theta, \hat\tau_j+1,\hat\tau_{j+1})$, $\hat\theta$ denotes its maximum composite likelihood estimator~(CMLE), i.e. $\hat\theta=\argmax_{\theta\in\Theta}L(\theta,\hat\tau_j+1,\hat\tau_{j+1})$. We need to show $R(\hat\tau_j, \tau_j^o, \hat\tau_{j+1})$ is of large negative order if $\hat\tau_j$ is far away from $\tau_j^o.$

For the mean problem, due to the linearity of mean, such a term takes the least squares form and has a simple closed-form solution. In contrast, we are dealing with a general spatio-temporal parametric model. The log-likelihood function $L(\theta,t_1,t_2)$ is nonlinear and does not admit a closed-form CMLE, making its analysis challenging.

In particular, to analyze $R(\hat\tau_j, \tau_j^o, \hat\tau_{j+1})$, a general strategy is Taylor expansion. However, note that $\hat\theta$ is estimated on a mixture of two stationary segments with \textit{different} parameters $\theta_j^o$ and $\theta_{j+1}^o$. Our analysis considers two scenarios.

(I).\ For the case where the change size $\|\theta_j^o-\theta_{j+1}^o\|_2$ is a constant (i.e.\ does not vanish), the Hessian matrices $\nabla_\theta^2 L(\tilde\theta_1, \hat\tau_j+1,\tau_j^o)$ and $\nabla_\theta^2 L(\tilde\theta_2, \tau_j^o+1, \hat\tau_{j+1})$ resulted from the Taylor expansion of $L(\hat\theta, \hat\tau_j+1,\hat\tau_{j+1})$ are \textit{not} guaranteed to be negative definite asymptotically, where $\tilde\theta_1$ is between $\hat\theta$ and $\theta_j^o$, and $\tilde\theta_2$ is between $\hat\theta$ and $\theta_{j+1}^o$. To overcome this difficulty, we proceed without Taylor expansion and use an almost sure argument, where we invoke Bolzano-Weierstrass theorem on $\hat\tau_j/T$ to obtain a convergent subsequence and directly analyze the behavior of $R(\hat\tau_j, \tau_j^o, \hat\tau_{j+1})$ on that subsequence (see e.g.\ the proof of \Cref{known} of the supplement). However, such an argument would fail if the number of change-points $m_o$ is unbounded, since a joint convergence of the $m_o$ subsequences resulted from each of the $m_o$ estimated change-points is then \textit{not} guaranteed. Moreover, if $\epsilon_\lambda^o \to 0$, some limits among $\tau_j^o/T$ will coincide, which further invalidates our arguments.

(II).\ For the case where the change size $\|\theta_j^o-\theta_{j+1}^o\|_2$ vanishes, via some technical arguments, we can indeed show that \textit{asymptotically}, the Hessian matrices $\nabla_\theta^2 L(\tilde\theta_1, \hat\tau_j+1,\tau_j^o)$ and $\nabla_\theta^2 L(\tilde\theta_2, \tau_j^o+1, \hat\tau_{j+1})$ become negative definite~(see e.g. the proof of \Cref{lem_rate} of the supplement). However, this is an asymptotic result and for this to hold for all $m_o$ change-points, we again require the finiteness of $m_o$ and thus $\epsilon_\lambda^o$ cannot vanish.

To summarize, there is no obvious way to modify our arguments to accommodate the case where $\epsilon_\lambda^o$ vanishes, unless we make some (strong) high-level assumptions on the behavior of the composite log-likelihood function, such as that the Hessian matrix $\nabla_\theta^2 L(\theta,t_1,t_2)$ is negative definite for any $\theta,t_1,t_2$. This indeed holds for the mean problem and linear regression problem, however, does not hold for a general spatio-temporal parametric model. Since we would like CLMDL to cover a wide range of spatio-temporal models, we opt not to pursue the extension of allowing $\epsilon_\lambda^o\to 0$ with additional assumptions. Indeed, to our best knowledge, most existing works for change-point estimation in a parametric model operate under the time infill setting with $\epsilon_\lambda^o>0$, see e.g.\ \cite{Andrews1993}, \cite{Davis2006},  \cite{Ling2014}, \cite{Ma2016}, and \cite{romano2022detecting}.
}

\section{Main results under infill asymptotics}\label{sec:main_infill}
In addition to the increasing domain asymptotics, an alternative asymptotic framework commonly used in the spatial statistics literature is the infill asymptotics, where the time domain is increasing while the spatial sampling domain is fixed. In this section, we modify the proposed CLMDL criterion to tailor to the infill setting and further establish its asymptotic properties.

We assume that the spatio-temporal process is generated by a piecewise stationary random field in $\mathcal{Z}=\mathbb{N} \times \mathcal{S} \subset \mathbb{N}\times\mathbb{R}^2$ where $\mathcal{S}$ is a bounded compact set in $\mathbb{R}^2$. With certain rescaling, we consider $\mathcal{S}=[0,1]^2$ without loss of generality. The data $\bfY$ is observed on $\mathcal{Z}_n=\mathcal{T}_n\times \mathcal{S}_n$ with $\mathcal{T}_n=[1,T_n]\cap \mathbb{N}$ and $\mathcal{S}_n\subset [0,1]^2$ 
with $|\mathcal{S}_n|=S_n$, in other words,
\begin{align*}
	\bfY=\{y_{t,\bs}: t\in [1,T_n]\cap \mathbb{N}, \bs\in \mathcal{S}_n, \mathcal{S}_n\subset [0,1]^2\}\,.
\end{align*}
Same as in Section \ref{sec:main}, the asymptotic theory is based on $n\longrightarrow\infty$, where the number of observations $|\mathcal{Z}_n|=S_nT_n > |\mathcal{Z}_{n'}|=S_{n'}T_{n'}$ whenever $n>n'$. For notational simplicity, we use $(S,T)$ instead of $(S_n,T_n)$ when there is no confusion. To our best knowledge, this is the first work in the change-point literature that considers a parametric spatio-temporal model under the spatial infill setting.

\subsection{Modified CLMDL for spatial infill asymptotics}\label{modifiedCLMDL}
To tailor to the infill setting where the spatial sampling domain is fixed, in the following, we reformulate the definition of the neighborhood set $\mathcal{N}(\bs)$, and further modify the CLMDL criterion proposed in Section \ref{section.mdl}. Recall the code length of $\bfY$ is $\mathrm{CL}(\bfY) = \mathrm{CL}(\widehat{\mathcal{M}}) + \mathrm{CL}( \bfY\mid\widehat{\mathcal{M}})$.

We first derive the $\mathrm{CL}(\bfY\mid\widehat{\mathcal{M}})$ part of CLMDL. Define $B_{\mathcal{N}}\in\mathbb{N}$ as a pre-specified constant. For each location $\bs \in \mathcal{S}$, denote $N_{B_{\mathcal{N}}}(\bs) \subset \mathcal{S}$ as the collection of the $B_{\mathcal{N}}$ closest points to $\bs$, and define $\mathcal{N}(\bs)=\{\bs'|\bs'\in\mathcal{S}, \bs' \neq \bs, \text{dist}(\bs,\bs')\leq d\} \cap N_{B_{\mathcal{N}}}(\bs)$ as the neighborhood set of $\bs$.	Note that $\mathcal{N}(\bs)$ depends on $n$, $B_{\mathcal{N}}$ and $d$ implicitly, and the maximum cardinality of $\mathcal{N}(\bs)$ is bounded by $B_{\mathcal{N}}$, which is a constant. The specification of $\mathrm{CL}(\bfY\mid \widehat{\mathcal{M}})$ remains the same as that in \eqref{MDLform}~(refer to \eqref{eq:CPL} for more details), except for the definition of the neighborhood set $\mathcal{N}(\bs)$.


Second, we derive the code length $\mathrm{CL}(\widehat{\mathcal{M}})$. Under the infill setting, a parameter estimator based on $ST$ observations is typically $1/ \sqrt{S^\delta T}$-consistent for some $\delta \in [0,1]$. Intuitively, due to the fixed domain, the space dimension may not contain as much information as the time dimension, and thus could result in $\delta < 1$; see Assumption \ref{ass.infill.consis} later for more details. Hence, the code length for an estimate should be $(\log_2 S^\delta T)/2$. However, since $\delta$ is generally unknown in practice, we use the upper bound $(\log_2 ST)/2$ to guard against false detection.

In addition, same as the increasing domain setting, each data point is used multiple times in $\mathrm{CL}(\bfY\mid\widehat{\mathcal{M}})$, and we thus need a compensating factor for $\mathrm{CL}(\widehat{\mathcal{M}})$. However, a key difference is that the correlation among spatial observations can grow to 1 under the infill asymptotics and we need to modify the compensating factor to account for that. Specifically, consider the extreme scenario where the spatio-temporal process $\bfY$ is generated by replicating a univariate time series $\{z_t\}$ along the spatial dimension for $S$ times, i.e., the spatio-temporal process $\bfY=\{y_{t,\bs}: 1\leq t\leq T, \bs\in \mathcal{S}\}$ satisfies  $y_{t,\bs}=z_t$ for all $\bs\in \mathcal{S}$. Intuitively, this spatio-temporal process resembles the situation under the infill asymptotics where the spatial domain $\mathcal{S}$ is densely sampled.

Note that for the case where the time series $\{z_t\}$ is temporally independent, the composite likelihood $L_{ST}(\psi,\bfY)$ is the same as $C_{k,\mathcal{N}}\times S$ times the full likelihood of $\{z_t\}$ where $C_{k,\mathcal{N}}$ is defined in \eqref{eq:Ck}. Thus, by regarding the composite likelihood as an approximation to the full likelihood, we can compensate the code length by magnifying $\mathrm{CL}(\widehat{\mathcal{M}})$ by a factor $C_{k,\mathcal{N}}^{\text{infill}}=C_{k,\mathcal{N}}S$ and define the modified CLMDL criterion under the infill asymptotic as  
\begin{eqnarray} \label{MDLform_infill}
	{\mathrm{CLMDL}}(m,\Lambda,\Psi) &=& C_{k,\mathcal{N}}^{\text{infill}}\left\{ \log m +\sum_{j=1}^{m+1}\sum_{i=1}^{c_j}\log \xi_{i,j}+ \sum_{j=1}^{m+1} \left[\left(\frac{d_j}{2}+1\right)\log T_j + \frac{d_j}{2}\log S\right]\right\} \nonumber \\
	&&- \sum_{j=1}^{m+1}L_{ST}^{(j)}(\psi_j;\bfX_j)\,.
\end{eqnarray}

Note that the differences between the CLMDL criterion in \eqref{MDLform_infill} and that in Section \ref{section.mdl}~(see equation \eqref{MDLform}) are solely in the definition of the neighborhood set $\mathcal{N}(\bs)$ in the specification of $\mathrm{CL}(\bfY\mid \widehat{\mathcal{M}})$, and the compensating factor $C_{k,\mathcal{N}}^{\text{infill}}$ of $\mathrm{CL}(\widehat{\mathcal{M}})$. The necessity of the modification is due to the fixed spatial sampling domain and the increasingly dependent observations under the infill asymptotics. The estimates of the number of change-points, the locations of the change-points and the parameters in each of the segments, given by the vector $(\hat{m},\hat{\Lambda}_{ST},\hat{\Psi}_{ST})$ defined in \eqref{minMDL}, can be obtained similarly 
by minimizing the modified CLMDL criterion \eqref{MDLform_infill}.

\begin{remark}\label{rm.compare.MDL}
	To handle the strong spatial dependence under the infill asymptotics, the compensating factor $C_{k,\mathcal{N}}^{\text{infill}}=C_{k,\mathcal{N}}S$ is $S$ times that under the increasing domain setting. Indeed, to establish theoretical guarantees for CLMDL under the infill asymptotics~(see Theorem \ref{asym_infill} later), we only require $C_{k,\mathcal{N}}^{\text{infill}}=c\cdot C_{k,\mathcal{N}}S$ for any constant $c>0$. In practice, we recommend setting $C_{k,\mathcal{N}}^{\text{infill}}=C_{k,\mathcal{N}}S/\text{ESS}$, where ESS is the so-called effective sample size in the spatial domain~\citep{Cressie_spatial_book}. We use $\text{ESS}=\trace(\widehat{V})/(\mathbf{1}^\top \widehat{V} \mathbf{1})$ as in \cite{Griffith2005}, where $\widehat{V}$ is the covariance matrix among the $S$ spatial locations derived based on the sample variogram.
\end{remark}

\subsection{Assumptions for spatial infill asymptotics}\label{assumption_infill} 
In this subsection, we impose conditions on the composite log-likelihood needed for establishing the main results. In the following, to ease presentation, we restrict the class of models, $\mathcal M$, to contain only one element, $\xi^o$ but allow the model to be mis-specified. Hence, we replace $\psi_j=(\xi_j,\theta_j)$ by $\theta_j$ for notational simplicity. For the class of models that contains more than one but finite elements, similar but lengthier assumptions to ensure model identifiability as Assumption \ref{ass.model.id} are required, and the generalization is straightforward.

\begin{assumption}\label{ass.infill.consis}
For each stationary segment $\bfX_j^o$ of the random field, where $j=1,\ldots,m_o+1$, there exists some $\delta_j \in [0,1]$ and parameters $\theta_j^o\in\Theta(\xi^o)$ such that for any $\epsilon>0$ and $\lambda_d, \lambda_u\in [0,1]$ with $\lambda_u>\lambda_d$, we have that 
\begin{align*}
    \lim_{S,T\to \infty} \frac{1}{S^{\delta_j} (\lambda_u-\lambda_d) T} \left(L_{ST}^{(j)}(\theta_j^o,\lambda_d,\lambda_u;\bfX_j^o)-\sup_{\theta_j \in \Theta(\xi_j^o), |\theta_j-\theta_j^o|>\epsilon}L_{ST}^{(j)}(\theta_j,\lambda_d,\lambda_u;\bfX_j^o)\right) >0 ,\text{ almost surely}\,.
\end{align*}
\end{assumption}

Assumption \ref{ass.infill.consis} asserts that with probability one, the point $\theta^{o}_{j}$ is asymptotically the unique parameter value that maximizes the composite likelihood. This type of divergence rate condition is commonly seen in the spatial statistics literature and is typically derived as a byproduct in the proof for consistency of parameter estimation in a time series or spatial context under infill asymptotics, see for example \cite{Ying1991}, \cite{Ying1993}, \cite{Zhang2004} and \cite{chang2017mixed}.

Note that the value of $\delta_j$ may depend on the parameterization of the model. For example, consider a Gaussian random field in $\mathbb{R}^d$ with a multiplicative exponential covariogram, and variance $\sigma^2$ and scale parameter $\varphi=(\varphi_1,\ldots,\varphi_d)\in\mathbb{R}^d_+$. If the parameter of interest is $\theta=(\sigma^2,\varphi)$ with parameter space $\Theta=[a,b]\times[w,v]^d$ for some constants $0 < a \leq b$ and $0 < w \leq v$, then we have $\delta=0$ as shown in \cite{Ying1991} for $d=1$, and  $\delta=1/2$ as shown in \cite{Ying1993} for $d=2$.
On the other hand, if one of the parameter is fixed, e.g., $\sigma^2=\sigma^2_*$ with $\sigma^2_*>0$ being a fixed constant or $\varphi=\varphi_*$ with $\varphi_*\in\mathbb{R}^d_+$ being a fixed constant, and the parameter of interest is $\theta=h(\sigma^2,\varphi)=\sigma^2\prod_{i=1}^d \varphi_i$ with parameter space $\Theta=\mathbb{R}^+$, then $\delta=1$ as shown in \cite{Ying1991,Ying1993}. We refer to \cite{chen2000infill} for other possible values of $\delta$ under different parameterizations of the model, and the presence of nugget effect.

\begin{assumption}\label{ass.diff.model}
	For each pair of consecutive stationary segments $\bfX_{j}^o$ and $\bfX_{j+1}^o$ of the random field, where $j=1,2,\ldots,m_o$, there exists some  $r_j \in [0,1]$ such that,
	\begin{align*}
		W_{ST}^{(j)}(q; \theta_{j}^o, \theta_{j+1}^o) &=\begin{cases}
			&O(q) \longrightarrow -\infty,\text{ as $|q| \longrightarrow \infty$, $q \in \mathbb{Z}$, if $r_j=0$\,,}\\
			&O(qS^{r_j})<0,\text{ almost surely, for any finite $q \in \mathbb{Z}\setminus\{0\}$, if $r_j>0$}\,,
		\end{cases}
	\end{align*}
	where $\theta_{j}^o$ and $\theta_{j+1}^o$ are defined in Assumption \ref{ass.infill.consis}, and  $W_{ST}^{(j)}(q; \theta_{j}^o, \theta_{j+1}^o)$ is defined in \eqref{diff.seg}.
\end{assumption}


Recall that $W_{ST}^{(j)}(q; \theta_{j}^o, \theta_{j+1}^o)$ quantifies the effects of the change-point estimation error $[T\hat{\lambda}_j]-[T\lambda_j^o]=q$ on the CLMDL. Assumption \ref{ass.diff.model} rules out the degenerate case that the $j$th and $(j+1)$th stationary segments are indistinguishable by the composite likelihood, and hence its role is similar to that of Assumption \ref{ass.model.id}(ii) in the increasing domain setting. Note that Assumption \ref{ass.model.id}(ii) can be conveniently expressed in terms of expectation as some form of strong law of large numbers can be applied under the increasing domain asymptotics. For reference, we remark that it can be shown Assumptions \ref{ass.infill.consis} and \ref{ass.diff.model} are satisfied with $\delta_j=1$ and $r_j=1$, for all $j=1,2,\ldots,m_o$ under the increasing domain setting given Assumptions \ref{ass.countable}, \ref{ass.mom}($r$), \ref{ass.model.id}, \ref{ass.mix}($r$) and \ref{ass.mixtime} in Section \ref{sec:main} hold with $r>2$.

Based on Assumption \ref{ass.diff.model}, we can classify the true change-points $\{\lambda_1^o,\cdots,\lambda^o_{m_o}\}$ into two types. The first type of change-points satisfy Assumption \ref{ass.diff.model} with $r_j>0$ and we denote them as
$$M_o=\left\{j \in \{1,2,\ldots,m_o\}: W_{ST}^{(j)}(q; \theta_{j}^o, \theta_{j+1}^o)=O(qS^{r_j})<0,\text{ a.s. for } q \in \mathbb{Z}\setminus\{0\}, \text{ with }  r_j>0\right\}.$$
The second type of change-points satisfy Assumption \ref{ass.diff.model} with $r_j=0$ and we denote them as ${M_o^c} =\{1,2,\ldots,m_o\} \setminus {M_o}$. Intuitively, under the infill setting, since we are sampling from a bounded spatial domain, the increasing sample size along the space dimension may or may not accumulate more information for change-point estimation. To distinguish the two cases, $M_o$ collects the change-points where more spatial observations do bring more information (with $r_j>0$) while $M_o^c$ collects the ones where more spatial observations do not (with $r_j=0$). As will be made clear in Theorem \ref{asym_infill} later, CLMDL can estimate $M_o$ more accurately than ${M_o^c}$. Note that for the increasing domain setting, all change-points are in $M_o$~(with $r_j=1$) as discussed above.

Define $B^{(j)}_{t,i,\bs_1,\bs_2}=l_{pair}(\theta_{j}^o; x_{t,\bs_1}^{(j)}, x_{t+i,\bs_2}^{(j)}) - l_{marg}({\theta}_{j}^o;x_{t,\bs_1}^{(j)}) - l_{marg}({\theta}_{j}^o;x_{t+i,\bs_2}^{(j)})$ for $j=1,2,\ldots,m_o+1$, $i=1,2,\ldots,k$, $1 \leq t,t+i \leq T_j$, $\bs_1 \in \mathcal{S}$ and $\bs_2 \in \bs_1\cup\mathcal{N}(\bs_1)$.
In the proof of consistency (see Theorem \ref{asym_infill} later), we need to control the difference of CLMDL caused by specifying an extra change-point, in order to show that the number of change-points will not be overestimated. 

To this end, Assumption \ref{ass.infill.mix.time} below requires that $B^{(j)}_{t,i,\bs_1,\bs_2}$ belongs to some exponential-type Orlicz space which leads to exponential-type inequalities related to the summation of $B^{(j)}_{t,i,\bs_1,\bs_2}$ over a spatial sampling region; see \cite{zajkowski2020norms}. For $p\geq1$, define the exponential-type Orlicz norm generated by the function $\psi_p(x)=\exp(|x|^p)-1$ as $\|X\|_{\psi_p}=\inf \{K>0 : \mathbb E\exp(|X/K|^p)\leq 2\}$, then the Orlicz space $L_{\psi_p}$ is defined as the set of all random variables $X$ with finite Orlicz norm, i.e., $L_{\psi_p}=\{X : \|X\|_{\psi_p} < \infty\}$. Define $K_{\psi_p}=\max\|B^{(j)}_{t,i,\bs_1,\bs_2}\|_{\psi_p}$, where the maximum is taken over all $(t,i,\bs_1,\bs_2)$ for $j=1,2,\ldots,m_o+1$, $i=1,2,\ldots,k$, $1 \leq t,t+i \leq T_j$, $\bs_1 \in \mathcal{S}$ and $\bs_2 \in \bs_1\cup\mathcal{N}(\bs_1)$. The value of $K_{\psi_p}$ determines the decay rate of the exponential-type tail probabilities of the summation of $B^{(j)}_{t,i,\bs_1,\bs_2}$ over a spatial sampling region, which is involved in the proof of consistency.

\begin{assumption}[$p$] \label{ass.infill.mix.time}
	For each stationary segment $\bfX_j^o$ of the random field, where $j=1,\ldots,m_o+1$,\\
	(i) for $i=1,2,\ldots,k, 1 \leq t,t+i \leq T_j, \bs_1 \in \mathcal{S}, \bs_2 \in \bs_1\cup\mathcal{N}(\bs_1)$, $B^{(j)}_{t,i,\bs_1,\bs_2} \in L_{\psi_p}$, for $p \geq 1$.
	
	\noindent (ii) the functions $l_{pair}$ and $l_{marg}$ are pointwise Lipschitz, i.e.,  there exist square-integrable functions $l_1^*$ and $l_2^*$ such that for every $\theta_1$ and $\theta_2$ in a neighborhood of $\theta_j^o$,
	$|l_{pair}(\theta_1; x_1, x_2)-l_{pair}(\theta_2; x_1, x_2)| \leq l_1^*(x_1, x_2) \|\theta_1-\theta_2\|$ and 	$|l_{marg}(\theta_1; x_1, x_2)-l_{marg}(\theta_2; x_1, x_2)| \leq l_2^*(x_1, x_2) \|\theta_1-\theta_2\|$ for all $x_1, x_2 \in \mathbb{R}^2$. Furthermore, the expectation of $l_{pair}$ and $l_{marg}$ are twice continuously differentiable at $\theta_j^o$.
\end{assumption}

To facilitate understanding and build intuition, we conclude this subsection with a concrete example. Specifically, consider the commonly used autoregressive spatial model in the literature,
\begin{equation*} 
	y_{t,\bs} - \mu= \phi (y_{t-1,\bs}-\mu) + \varepsilon_{t,\bs}\,,
\end{equation*}
where $\phi\in(-1,1)$, $t \in \{1,\ldots,T\}$, $\bs \in \mathcal{S}$ and $\{ \varepsilon_{t,\bs}: \bs \in \mathcal{S} \}$ is a zero-mean error process with covariance function ${\rm Cov}(\varepsilon_{t,\textbf{u}}, \varepsilon_{t,\textbf{v}})$. In Section \ref{infill_example} of the supplement, we provide two examples of the autoregressive spatial model that satisfy Assumptions \ref{ass.infill.consis}, \ref{ass.diff.model} and \ref{ass.infill.mix.time}($p$), where the error process is spatial Gaussian with isotropic or multiplicative exponential covariance functions.

\subsection{Consistency of CLMDL under spatial infill asymptotics}\label{consist_infill}
Theorem \ref{asym_infill} provides the theoretical guarantees for CLMDL under the infill setting. Compared to Theorems \ref{unknownprob} and \ref{finerate} for the increasing domain asymptotics, the main difference is that CLMDL exhibits different behavior for different change-points: for change-points in $M_o$, exact recovery can be achieved asymptotically, while for change-points in $M_o^c$, we obtain the classical $O_p(1)$ error.


\begin{theorem}\label{asym_infill}
	Let $\bfY$ be a piecewise stationary random field specified by the vector $(m_o,\Lambda^o,\Psi^o)$, and Assumptions \ref{ass.infill.consis}, \ref{ass.diff.model} and \ref{ass.infill.mix.time}($p$) hold for some $p \geq 1$.
	Then, for the estimator $(\hat{m},\hat{\Lambda}_{ST},\hat{\Psi}_{ST})$ defined in \eqref{minMDL} using the modified CLMDL criterion \eqref{MDLform_infill}, 
	we have
	\begin{align}\label{weak.consist_infill}
		\hat{m} \longrightarrow m_o, \quad 
		\hat{\Psi}_{ST} \longrightarrow \Psi^o, \quad 
		[T\hat{\lambda}_j]=[T\lambda_j^o] ~\text{ for }~ j \in M_o, \quad \text{and} \nonumber\\
		[T\hat{\lambda}_j]-[T\lambda_j^o] = O_p(1) ~\text{ for }~ j \in {M_o^c}\,,~~~~~
	\end{align}
	in probability, provided that $S,T\longrightarrow \infty$ and $S=O(T^{1/(\nu+1)})$ for any $\nu>-\min_{1\leq j\leq m_o+1}\delta_j$ if $p>1$, and in addition $T^{K_{\psi_p}-1}/S \longrightarrow 0$ if $p=1$, where $\delta_j$s are defined in Assumption \ref{ass.infill.consis}.
\end{theorem}

To better discuss and distinguish the exact recovery and the classical $O_p(1)$ localization error of the change-points in $M_o$ and $M_o^c$ under the infill setting, we first review the concept of microergodicity of a model parameter in measures for random fields observed on a bounded region under the infill asymptotics; see \cite{stein1999interpolation} and \cite{Zhang2004} for details.

We begin with the concept of equivalence and orthogonality of measures for random fields observed on a bounded region. Recall that for two probability measures $P_i$, $i=1,2$, defined on the same measurable space $(\Omega,\mathcal{F})$, $P_1$ is said to be absolutely continuous with respect to $P_2$, denoted by $P_1 \ll P_2$, if $P_1(A)=0$ for any $A\in \mathcal{F}$ such that $P_2(A)=0$. $P_1$ and $P_2$ are equivalent, denoted by $P_1 \equiv P_2$, if $P_1 \ll P_2$ and $P_2 \ll P_1$. If $P_1 \equiv P_2$ on $\mathcal{F}$ and $\mathcal{F}$ is the $\sigma$-algebra generated by a stochastic process $Y(\bs), \bs \in H$ for a given set $H$, we say that $P_i,~i=1,2$ are equivalent on the paths of $Y(\bs), \bs \in H$. Note that for two equivalent measures $P_1 \equiv P_2$, no matter what is observed, it is not possible to determine which measure is correct with probability one.

More specifically, consider a decision rule of the following form. For some event $A\in\mathcal{F}$, choose $P_1$ if $A$ occurs and choose $P_2$ otherwise. If $P_1 \equiv P_2$, then for any event $B\in\mathcal{F}$ such that $P_1(B) > 0$ (so that $P_2(B) > 0$), we cannot have both $P_1(A|B)=1$ and $P_2(A^c|B)=1$. Indeed, if $P_1(A|B)=1$ then $P_2(A^c|B)=0$. Thus, there is no event $B$ receiving positive probability under either measure such that, conditional on event $B$, perfect discrimination between the measures is possible; see Section 4.1 of \cite{stein1999interpolation}. Moreover, if $\{P_{\theta}, \theta\in\Theta\}$ is a family of equivalent measures and $\hat{\theta}_n,n\geq 1$ is a sequence of estimators, irrespective of what is observed, $\hat{\theta}_n$ cannot be a weakly consistent estimator of $\theta$ for all $\theta\in\Theta$; see Section 3 of \cite{Zhang2004}. On the other hand, two probability measures $P_i$, $i=1,2$, defined on the same measurable space $(\Omega,\mathcal{F})$, are said to be orthogonal, written $P_1 \perp P_2$, if there exists event $A\in\mathcal{F}$ such that $P_1(A)=1$ and $P_2(A)=0$. For $P_1 \perp P_2$, it is possible to determine which measure is correct with probability 1 based on observing $\omega \in \Omega$.

It is known that probability measures may be neither equivalent nor orthogonal in general, but for Gaussian measures on a separable Hilbert space, they are either equivalent or orthogonal; see Section 4.2 of \cite{stein1999interpolation} and Section 5 of \cite{Zhang2004}. Consider a class of Gaussian measures $\{P_{\phi}, \phi\in\Phi\}$ for a random field on a given bounded domain $\mathcal{S}$. A parameter $h(\phi)$ is defined to be microergodic if, for all $\phi$ and $\tilde{\phi}$ in the parameter space, $h(\phi) \neq h(\tilde{\phi})$ implies $P_\phi \perp P_{\tilde{\phi}}$, where $P_\phi$ denotes the Gaussian measure corresponding to the parameter $\phi$. 
For more details, we refer to Section 6.2 of \cite{stein1999interpolation}, \cite{stein2004equivalence} and \cite{Zhang2004}.

For stationary Gaussian random fields, it is easy to see that for a given change-point, Assumption \ref{ass.diff.model} is satisfied with $r_j>0$ if the change happens in a microergodic parameter that is consistently estimable. Moreover, we have that the composite likelihood estimator can consistently estimate the microergodic parameter using spatial data at only one time point. In other words, intuitively, by examining the spatial data at only one time point, we are able to correctly distinguish two Gaussian spatial models with different microergodic parameter values with probability one asymptotically. In this case, Theorem \ref{asym_infill} guarantees that an exact recovery of the change-point can be achieved. 

As a concrete example, Theorem 2 of \cite{Zhang2004} shows that for the isotropic Mat\'ern class in a bounded domain $\mathcal{S}\subset \mathbb{R}^d$ for $d=1,2$ or $3$ in the spatial context, the parameter $h(\sigma^2,\varphi)=\sigma^2\varphi^{2v}$ is microergodic where $\sigma^2$ is the variance of the process, $\varphi>0$ is the scale parameter and $v>0$ is the smoothness parameter. On the other hand, the parameter $(\sigma^2,\varphi)$ is non-microergodic. Hence, intuitively, the information contained in the bounded domain $\mathcal{S}$ is only rich enough to distinguish Mat\'ern models with different values of $h(\sigma^2,\varphi)=\sigma^2\varphi^{2v}$ but not $(\sigma^2,\varphi)$. In particular, when $v=1/2$, the Mat\'ern covariogram becomes the exponential covariogram, and we show that a change in the microergodic parameter $\sigma^2\varphi$ results in an exact recovery of the corresponding change-point in Section \ref{ill_example_exp} of the supplementary material. However, for changes in $\sigma^2$ and $\varphi$ that do not alter $\sigma^2\varphi$, Theorem \ref{asym_infill} states that we can only estimate the change-point with the classical $O_p(1)$ error with further help from information accumulated along the time dimension.

\section{Optimization of CLMDL using PELT}\label{sec:PELT}
The optimization problem in \eqref{minMDL} involves huge amount of combinatorial search for the best change-points configuration and thus is computationally intensive. Various searching algorithms have been proposed in the literature for this optimization problem.

In general, the searching algorithms can be classified as stochastic or deterministic.
Stochastic methods such as genetic algorithm \citep{Davis2006} require large scale simulation and give simulation-dependent solutions. Among deterministic methods, optimal partitioning \citep{Jackson2005}, which uses dynamic programming to simplify exhaustive search, can still be computationally intensive for large datasets. Recently, pruning-based dynamic programming algorithms \citep{Killick2012,Maidstone2017} are proposed to obtain exact optimization solution by discarding some unnecessary change-point configurations in dynamic programming.

In particular, the pruned exact linear time (\textsc{PELT}) algorithm proposed by \cite{Killick2012} is a modified dynamic programming algorithm that solves the minimization problem of the form 
\begin{equation} \label{costmin}
	F(T)=\sum_{j=1}^{m+1}\mathbb{C}( Y_{[(\tau_{j-1}+1):\tau_j]})+\beta f(m),
\end{equation}
where $ Y_{[a:b]}=\{y_t;t=a,\ldots,b\}$, $m$ is the number of change-points, $\mathbb{C}$ is the cost function for a segment and 
$\beta f(m)$ is a penalty function for the number of changes. 
If $f(m)=m+1$, \textsc{PELT} solves \eqref{costmin} by recursively computing $F(s)=\min_t\left\{F(t) + \mathbb{C}( Y_{[(t+1):s]})+\beta\right\}$ for $t<s<T.$ The minimization is taken over all $t<T$ excluding those satisfying the pruning condition
\begin{equation}\label{prunning}
	F(t) + \mathbb{C}( Y_{[(t+1):t']}) + K \geq F(t')\,, 
\end{equation}
for some $t'>t$ with $t'-t\geq T\epsilon_\lambda $.\footnote{Note that the additional requirement $t'-t\geq T\epsilon_\lambda$ is imposed by the identification condition in \eqref{A}.} The constant $K$ depends on $\mathbb{C}$ and is required to satisfy
\begin{equation}\label{prun.K}
	\mathbb{C}( Y_{[(t+1):s]})+\mathbb{C}( Y_{[(s+1):T]})+K\leq \mathbb{C}( Y_{[(t+1):T]})\,,  \ \ \mbox{for all $t<s<T$\,.} 
\end{equation}
In practice, a considerable number of time points can be pruned by \eqref{prunning} and hence the efficiency of the dynamic programming algorithm can be greatly improved.

To implement PELT in the current spatio-temporal setting, we express the CLMDL in \eqref{MDLform} for the increasing domain setting in the format of \eqref{costmin} as 
\begin{align*}
	\mathbb{C}(Y_{[(\tau_{j-1}+1):\tau_{j}]})= C_{k,\mathcal{N}} \left\{ \sum_{i=1}^{\hat{c}_j}\log \hat{\xi}_{i,j} 
	+ \left(\frac{\hat{d}_j}{2}+1\right)\log (\tau_j-\tau_{j-1})+\frac{\hat{d}_j}{2}\log S\right\} - L_{ST}^{(j)}(\hat{\psi}_j;Y_{[(\tau_{j-1}+1):\tau_{j}]})\,,
\end{align*}
where $\beta=C_{k,\mathcal{N}}$ and $f(m)=\log m$.

Compared to the classical univariate time series setting which PELT is designed for, under the spatio-temporal setting, the constant $K$ in \eqref{prun.K} is more challenging to derive due to the non-ignorable edge effect when spatial dimension $S\longrightarrow \infty$. If the condition \eqref{prun.K} for $K$ is not satisfied for all $t<s<T$, the exact minimizer of CLMDL cannot be guaranteed to be found by PELT. The following lemma provides a potential choice of $K$, which can guarantee the asymptotic validity of applying PELT for the minimization of CLMDL.



\begin{lemma} \label{PELTK}
	Consider all candidate model parameter sets $(\xi_{j},\theta_{j})$ in $\mathcal{M}$, $\xi_j=(\xi_{1,j},\ldots,\xi_{c_j,j})$ and $\theta_j\in \mathbb{R}^{d_j}$. 
	Let $d_{min}=\min_{j} d_{j}$, $d_{max}=\max_{j} d_{j}$, $\xi_{j}^{*}=\sum_{i=1}^{c_j}\log \xi_{i,j}$, 
	$\xi^{*}_{min}=\min_{j} \xi^{*}_{j}$ and $\xi^{*}_{max}=\max_{j} \xi^{*}_{j}$. Define
	\begin{equation*}
		K=C_{k,\mathcal{N}}\left \{ \left(\frac{d_{min}}{2}-d_{max}\right)\log ST+(2+d_{max})\log 2 + \xi_{min} - 2\xi_{max}-\log T\right\}.
	\end{equation*}
	The true change points $T\Lambda^o$ will not be pruned asymptotically as $S,T \longrightarrow \infty$ under the conditions of Theorem \ref{unknownprob} and $T^2\cdot S^{-r/2}\longrightarrow 0$, or the conditions of Theorem \ref{finerate} and $T\cdot S^{-r/2}\longrightarrow 0$. The same result holds under the conditions of Theorem \ref{asym_infill} if we replace $C_{k,\mathcal{N}}$ with $C_{k,\mathcal{N}}^{\text{infill}}$ in $K$.	
\end{lemma}

Since the true change-points $T\Lambda^o$ are not pruned asymptotically, PELT can find the minimizer of CLMDL as $S,T \longrightarrow \infty.$ Although there is no guarantee that PELT can obtain the exact solution of CLMDL when $S$ and $T$ are small, in practice, we find PELT works well as demonstrated by the numerical experiments. \textcolor{black}{In \Cref{add_simu_PELT} of \Cref{sec:add_num}, we compare the optimization performance of PELT with the optimal partitioning algorithm~(OP) in \cite{Jackson2005}}, which guarantees the exact minimization of CLMDL with a computational cost of $O(ST^2)$. \textcolor{black}{In particular, we find that PELT always returns the same minimizer of CLMDL as OP, even for small $S$ and $T$. We refer to \Cref{add_simu_PELT} for more details.
}

{
\color{black}
}

\section{Additional numerical studies}\label{sec:add_num}

\subsection{Additional simulation results}
\begin{simulation}\label{add_simu_hd} \textbf{(Comparison with high-dimensional methods)}

In this example, we compare the performance of CLMDL with existing methods designed for multiple change-point estimation in high-dimensional time series. Specifically, we consider the sparsified binary segmentation (SBS) method proposed by \cite{Cho2015} and the Double CUSUM Binary Segmentation algorithm (DCBS) proposed by \cite{Cho2016}. Both SBS and DCBS can estimate multiple change-points in the mean and second-order structure of a high-dimensional time series.

The underlying data generating process~(DGP) is the same as that of \Cref{sim.1} in Section \ref{subsec:simu} of the main text. Table \ref{scp.table.sbs} summarizes the performance of SBS under different settings. It can be seen that the detection power of SBS is rather low when $T$ is small even for a large spatial dimension $S=10^2$. The performance of SBS improves when $T=200$. Table \ref{scp.table.dcbs} shows the performance of DCBS, where the general observation is similar to that of SBS. 

Comparing Table \ref{scp.table} to Tables \ref{scp.table.sbs} and \ref{scp.table.dcbs}, it can be seen that CLMDL has a higher detection power compared to SBS and DCBS. As for the no-change scenario, all three methods are robust to false positive under different $S$ and $T$. Overall speaking, our proposed procedure outperforms SBS and DCBS. This is not surprising as CLMDL is specifically designed for spatio-temporal data while SBS and DCBS are generic algorithms designed for high-dimensional time series. 

\begin{table}[]
	\centering
	\caption{Percentage of estimated change-points $\hat m$ among 1000 replications under various spatial size $S$, temporal size $T$, and signal levels $(\delta_\phi, \delta_\rho)$ using SBS in \cite{Cho2015}.}\label{scp.table.sbs}
	\begin{tabular}{cccrrrrrrrrr}
		\hline \hline
		\multirow{3}{*}{$T$} & \multirow{3}{*}{\begin{tabular}[c]{@{}l@{}}$\delta_\phi\times 10$\end{tabular}} & \multirow{3}{*}{\begin{tabular}[c]{@{}l@{}}$\delta_\rho\times 10$\end{tabular}} & \multicolumn{9}{c}{\% of $\hat{m}$} \\
		&                                                                                     &                                                                                    & \multicolumn{3}{c}{$S=6^2$} & \multicolumn{3}{c}{$S=8^2$} & \multicolumn{3}{c}{$S=10^2$} \\ \cline{4-12} 
		&                                                                                     &                                                                                    & 0      & 1      & $\geq2$   & 0      & 1      & $\geq2$   & 0      & 1      & $\geq2$    \\ \hline
		100 & 0 & 0  & 100 & 0 & 0 & 100 & 0 & 0 & 100 & 0  & 0 \\
		& 2 & 0  & 100 & 0 & 0 & 100 & 0 & 0 & 100 & 0  & 0 \\
		& 3 & 0  & 100 & 0 & 0 & 100 & 0 & 0 & 100 & 0  & 0 \\
		& 0 & 6  & 100 & 0 & 0 & 100 & 0 & 0 & 100 & 0  & 0 \\
		& 0 & 10 & 100 & 0 & 0 & 100 & 0 & 0 & 100 & 0  & 0 \\
		& 2 & 2  & 100 & 0 & 0 & 100 & 0 & 0 & 100 & 0  & 0 \\
		& 3 & 3  & 100 & 0 & 0 & 99  & 1 & 0 & 97  & 3  & 0 \\\hline
		200 & 0 & 0  & 100 & 0 & 0 & 100 & 0 & 0 & 100 & 0  & 0 \\
		& 2 & 0  & 99  & 1 & 0 & 98  & 2 & 0 & 97  & 3  & 0 \\
		& 3 & 0  & 96  & 4 & 0 & 94  & 6 & 0 & 87  & 13 & 0 \\
		& 0 & 6  & 100 & 0 & 0 & 100 & 0 & 0 & 100 & 1  & 0 \\
		& 0 & 10 & 100 & 0 & 0 & 100 & 0 & 0 & 100 & 1  & 0 \\
		& 2 & 2  & 99  & 1 & 0 & 98  & 2 & 0 & 96  & 4  & 0 \\
		& 3 & 3  & 97  & 3 & 0 & 93  & 7 & 0 & 89  & 11 & 0 \\ \hline \hline  
	\end{tabular}
\end{table}

\begin{table}[]
	\centering
	\caption{Percentage of estimated change-points $\hat m$ among 1000 replications under various spatial size $S$, temporal size $T$, and signal levels $(\delta_\phi, \delta_\rho)$ using DCBS in \cite{Cho2016}.}\label{scp.table.dcbs}
	\begin{tabular}{cccrrrrrrrrr}
		\hline \hline
		\multirow{3}{*}{$T$} & \multirow{3}{*}{\begin{tabular}[c]{@{}l@{}}$\delta_\phi\times 10$\end{tabular}} & \multirow{3}{*}{\begin{tabular}[c]{@{}l@{}}$\delta_\rho\times 10$\end{tabular}} & \multicolumn{9}{c}{\% of $\hat{m}$} \\
		&                                                                                     &                                                                                    & \multicolumn{3}{c}{$S=6^2$} & \multicolumn{3}{c}{$S=8^2$} & \multicolumn{3}{c}{$S=10^2$} \\ \cline{4-12} 
		&                                                                                     &                                                                                    & 0      & 1      & $\geq2$   & 0      & 1      & $\geq2$   & 0      & 1      & $\geq2$    \\ \hline
		100 & 0 & 0  & 100 & 0  & 0 & 100 & 0  & 0  & 100 & 0  & 0  \\
		& 2 & 0  & 98  & 2  & 0 & 98  & 2  & 0  & 99  & 2  & 0  \\
		& 3 & 0  & 89  & 11 & 0 & 72  & 28 & 1  & 60  & 39 & 1  \\
		& 0 & 6  & 100 & 0  & 0 & 100 & 0  & 0  & 100 & 0  & 0  \\
		& 0 & 10 & 100 & 0  & 0 & 100 & 0  & 0  & 100 & 0  & 0  \\
		& 2 & 2  & 97  & 3  & 0 & 96  & 4  & 0  & 98  & 3  & 0  \\
		& 3 & 3  & 74  & 25 & 1 & 57  & 42 & 1  & 55  & 43 & 2  \\\hline
		200 & 0 & 0  & 100 & 0  & 0 & 100 & 0  & 0  & 100 & 0  & 0  \\
		& 2 & 0  & 93  & 7  & 0 & 64  & 35 & 1  & 51  & 49 & 0  \\
		& 3 & 0  & 45  & 52 & 3 & 14  & 83 & 2  & 14  & 82 & 5  \\
		& 0 & 6  & 99  & 1  & 0 & 100 & 0  & 0  & 100 & 0  & 0  \\
		& 0 & 10 & 99  & 1  & 0 & 100 & 0  & 0  & 100 & 0  & 0  \\
		& 2 & 2  & 81  & 19 & 1 & 49  & 50 & 1  & 44  & 55 & 1  \\
		& 3 & 3  & 20  & 72 & 8 & 8   & 83 & 10 & 9   & 70 & 21 \\ \hline \hline  
	\end{tabular}
\end{table}
\end{simulation}

\begin{simulation}\label{add_simu_multiCP}\textbf{(Model selection and multiple change-point detection)}

In this example, we study the performance of CLMDL for model selection with the presence of multiple change-points and model misspecification. Specifically, we simulate a spatio-temporal process with four stationary segments under the increasing domain setting.

The underlying process for the first, second and fourth segments follow the four-parameter autoregressive spatial model \eqref{sim.process} in the main text with $\theta_1=(0.0, -0.2, 0.6, 1.0)^\top$, $\theta_2=(0.0, -0.5,0.6,1.0)^\top$, $\theta_4=(0.3, -0.2., 0.9, 1.0)^\top$ and $T_1=T_2=T_4=50$. For the third segment, instead of exponential covariance function, Mat\'{e}rn spatial covariance function is used, i.e. 
\begin{equation*}
	{\rm Cov}(\varepsilon_{t,\textbf{u}}, \varepsilon_{t,\textbf{v}}) = \frac{\sigma^2}{2^{\nu-1}\Gamma(\nu)}\left(\sqrt{2\nu}\frac{\left\lVert \textbf{u}-\textbf{v}\right\lVert_2}{\rho}\right)^\nu K_\nu\left(\sqrt{2\nu}\frac{\left\lVert \textbf{u}-\textbf{v}\right\lVert_2}{\rho}\right),
\end{equation*}
\noindent where $K_\nu(\cdot)$ is the modified Bessel function of the second kind. We set $\phi=-0.5, \sigma^2=0.9,\mu=0.3$ in \eqref{sim.process} with $\nu=2$ and $\rho=0.9$ for the Mat\'{e}rn spatial covariance function and $T_3=50$. The three change-points correspond to change in $\phi$, change in $\mu$ and spatial covariance function, and change in $\phi$ and spatial covariance function respectively.

The candidate model class $\mathcal{M}$ consists of two models, $M_1$ and $M_2$, both in the form of \eqref{sim.process} with exponential spatial covariance function, while $M_1$ with $\mu=0$ fixed and $M_2$ with $\mu$ as a free parameter. The number of parameters of $M_1$ and $M_2$ are 3 and 4 respectively. Note that $M_1$ is nested within $M_2$ by setting $\mu=0$. For the third segment, the true model is not included in $\mathcal{M}$. Since $M_2$ allows non-zero $\mu$, it acts as the pseudo-true model. In summary, the (pseudo-) true models for the four segments are $M_1$, $M_1$, $M_2$ and $M_2$ respectively.

Table \ref{mcp.table} reports the results of the multiple change-point estimation. It can be seen that the proportion of correctly estimated number of change-point increases, and both over-fitting and under-fitting diminish as S increases. Moreover, the empirical standard deviations of the estimated location of change-point decrease gradually. Table \ref{model.select.table} reports the model selection results, which further shows CLMDL performs well for model selection and maintains its efficiency when the model is misspecified.

\begin{table}[h]
	\small
	\centering
	\caption{Distribution of $\hat m$, mean and empirical standard deviation (esd) of $\hat{\lambda}_j$, $j=1,2,3$, given $\hat m=m_{o}=3$.}
	\label{mcp.table}
	\begin{tabular}{lrrrrclrclrclrc} \hline \hline
		$S$ & \multicolumn{5}{c}{\% of $\hat m$} &  & \multicolumn{8}{c}{$\hat m =3$}                                                                      \\ \cline{2-6} \cline{8-15} 
		&        &        &        &         &        &  & \multicolumn{2}{c}{$\hat{\lambda}_1$} &  & \multicolumn{2}{c}{$\hat{\lambda}_2$} &  & \multicolumn{2}{c}{$\hat{\lambda}_3$} \\ \cline{8-9} \cline{11-12} \cline{14-15} 
		& 0      & 1      & 2      & 3       & $\geq 4$      &  & mean         & esd           &  & mean         & esd           &  & mean         & esd           \\ \hline
		$6^2$  & 0       & 4.7 & 39.6 & 55.2 & 0.5 &  & 0.2579            & 0.0304 &  & 0.4996           & 0.0133 &  & 0.7505            & 0.0314 \\
		$8^2$  & 0       & 0   & 2.3  & 96.9 & 0.8 &  & 0.2527            & 0.0158 &  & 0.4997           & 0.0040 &  & 0.7502            & 0.0143 \\
		$10^2$ & 0       & 0   & 0    & 99.6 & 0.4 &  & 0.2513            & 0.0065 &  & 0.5001            & 0.0023 &  & 0.7498           & 0.0083  \\ \hline\hline  
	\end{tabular}
\end{table}

\begin{table}[h]
	\small
	\centering
	\caption{Relative frequencies of the model selected given $\hat{m}=m_o=3$.} \label{model.select.table}
	\begin{tabular}{crrlrrlrrlrr}
		\hline \hline
		$S$  & \multicolumn{11}{c}{Segment}                                                                           \\ 
		& \multicolumn{2}{c}{1} &  & \multicolumn{2}{c}{2} &  & \multicolumn{2}{c}{3} &  & \multicolumn{2}{c}{4} \\ \cline{2-3} \cline{5-6} \cline{8-9} \cline{11-12}
		& $M_1$         & $M_2$       &  & $M_1$         & $M_2$       &  & $M_1$       & $M_2$         &  & $M_1$       & $M_2$         \\ \hline
		$6^2$  & 95.7       & 4.3      &  & 99.1       & 0.9      &  & 0.2      & 99.8       &  & 0.2      & 99.8       \\
		$8^2$  & 97.7       & 2.3      &  & 99.9       & 0.1      &  & 0.0      & 100.0      &  & 0.0      & 100.0      \\
		$10^2$ & 98.2       & 1.8      &  & 99.7       & 0.3      &  & 0.0      & 100.0      &  & 0.0      & 100.0     \\ \hline\hline
	\end{tabular}
\end{table}
\end{simulation}

\begin{simulation}\label{add_sim_fill} \textbf{(Infill setting with a fixed spatial domain)}

In this example, we study the performance of CLMDL under the infill setting with a fixed spatial domain. The underlying DGP follows \eqref{sim.process} with $\mu=0$ and $\theta_1=(-0.5,0.6,1)^\top$ and $\theta_2=(-0.5+\delta_\phi,0.6+\delta_\rho,1)^\top$. We set the two-dimensional spatial grid as $\mathcal{S}=\{(s_1,s_2): s_1,s_2 \in \{1,1+\Delta_s,1+2\Delta_s,\ldots; s_1,s_2\leq6\} \}$. We vary $\Delta_s=1,0.6$, $0.5$, $0.4$ and $0.2$ while fixing $T_1=T_2=50$. Note that spatial distance between the grid decreases and hence dependence between simulated data increases with $\Delta_s$, while the spatial domain remains fixed. For the choice of tuning parameters of CLMDL, same as that for the increasing domain simulation, we set $k=1$ and $d=2$ in defining the composite likelihood, $\epsilon_\lambda=0.1$ in the optimization. In addition, we set $B_{\mathcal{N}}=4$.
	
Table \ref{infill.tab} reports the estimated number of change-points under different settings using the infill CLMDL criterion \eqref{MDLform_infill}. For the case without change-point (i.e.\ $\delta_\phi=\delta_\rho=0$), the proposed procedure declares no false positive for all five levels of $\Delta_s$. As $\Delta_s$ decreases~(thus sample size increases), the detection power generally improves, similar to the case in the increasing domain setting. Note that for $\Delta_s=0.6$ and $0.5$, there are $S=9^2$ and $S=11^2$ spatial observations respectively, thus, the sample size is comparable to the case of $S=10^2$ and $T=100$ in Table \ref{scp.table} (Simulation \ref{sim.1}) of the main text. As can be seen, the detection power of CLMDL under the fixed spatial domain is lower than that under the increasing domain, which is intuitive as one has less information. In addition, under the infill setting, the estimation accuracy of CLMDL does improve as sample size increases, though the improvement is generally smaller than that under the increasing domain setting.
	
\begin{table}[]
		\centering
		\caption{Percentage of estimated change-points $\hat m$ among 1000 replications under various spatial resolutions $\Delta_s$ and signal levels $(\delta_\phi, \delta_\rho)$ using the modified CLMDL criterion \eqref{MDLform_infill}.}
		\label{infill.tab}
		\begin{tabular}{ccrrrrrrrrrrrrrrr}
			\hline \hline
			$\delta_\phi$ & $\delta_\rho$ & \multicolumn{15}{c}{\% of $\hat{m}$}  \\
			$\times 10$& $\times 10$   & \multicolumn{3}{c}{\begin{tabular}[c]{@{}c@{}} $S=6^2$\\ $\Delta_s=1$\end{tabular}} & \multicolumn{3}{c}{\begin{tabular}[c]{@{}c@{}} $S=9^2$\\ $\Delta_s=0.6$\end{tabular}} & \multicolumn{3}{c}{\begin{tabular}[c]{@{}c@{}} $S=11^2$\\ $\Delta_s=0.5$\end{tabular}} & \multicolumn{3}{c}{\begin{tabular}[c]{@{}c@{}} $S=13^2$\\ $\Delta_s=0.4$\end{tabular}}& \multicolumn{3}{c}{\begin{tabular}[c]{@{}c@{}} $S=26^2$\\ $\Delta_s=0.2$\end{tabular}} \\ \cline{3-17}
			&       & 0                        & 1                       & $\geq 2$                      & 0                         & 1                        & $\geq 2$                      & 0                         & 1                         & $\geq 2$                      & 0                         & 1                         & $\geq 2$& 0&  1& $\geq 2$                      \\ \hline
			0 & 0  & 100 & 0  & 0 & 100 & 0  & 0 & 100 & 0  & 0 & 100 & 0   & 0 & 100 & 0   & 0 \\
			2 & 0  & 74  & 25 & 1 & 69  & 31 & 0 & 58  & 40 & 3 & 55  & 45  & 0 & 39  & 61  & 0 \\
			3 & 0  & 32  & 66 & 3 & 8   & 91 & 1 & 3   & 96 & 1 & 3   & 97  & 0 & 0   & 100 & 0 \\
			0 & 6  & 92  & 8  & 0 & 79  & 20 & 1 & 61  & 39 & 0 & 45  & 55  & 0 & 12  & 88  & 0 \\
			0 & 10 & 30  & 69 & 1 & 23  & 77 & 0 & 10  & 90 & 0 & 4   & 96  & 0 & 0   & 100 & 0 \\
			2 & 2  & 67  & 32 & 1 & 39  & 61 & 0 & 28  & 72 & 0 & 19  & 81  & 0 & 6   & 94  & 0 \\
			3 & 3  & 23  & 78 & 0 & 6   & 94 & 0 & 2   & 97 & 1 & 0   & 100 & 0 & 0   & 100 & 0   \\ \hline \hline
		\end{tabular}
\end{table}
\end{simulation}

\begin{simulation}\label{add_simu_kd}\textbf{(Sensitivity analysis on tuning parameters $(k,d)$ and $B_{\mathcal{N}}$)}

In this example, we assess the sensitivity of the performance of CLMDL w.r.t.\ different values of tuning parameters $(k,d)$ and $B_{\mathcal{N}}$. The underlying DGP is the same as that of Simulation 1 in Section \ref{subsec:simu} of the main text.

Table \ref{scp.table.k2} and \ref{scp.table.k3} report the estimated number of change-points under various settings using $k=2$ and $3$ respectively. We can see that the performance of CLMDL using different $k$ is very similar to the case of $k=1$ reported in Table \ref{scp.table}. It indicates that CLMDL is robust to the choice of $k$. Table \ref{infill.tab.B8} reports the estimated number of change-points under different settings based on the infill CLMDL criterion \eqref{MDLform_infill} using $d=3$ and $B_{\mathcal{N}}=8$, where the DGP is the same as that in Simulation S.3. It can be seen that the performance of the proposed procedure is close to that in Table \ref{infill.tab}. These results together demonstrate that the proposed procedure is robust to the choices of tuning parameters $(k,d)$ and $B_{\mathcal{N}}$. 

\begin{table}[]
	\centering
	\caption{Percentage of estimated change-points $\hat m$ among 1000 replications under various spatial size $S$, temporal size $T$, and signal levels $(\delta_\phi, \delta_\rho)$ when $k=2$.}\label{scp.table.k2}
	\begin{tabular}{cccrrrrrrrrr}
		\hline \hline
		\multirow{3}{*}{$T$} & \multirow{3}{*}{\begin{tabular}[c]{@{}l@{}}$\delta_\phi\times 10$\end{tabular}} & \multirow{3}{*}{\begin{tabular}[c]{@{}l@{}}$\delta_\rho\times 10$\end{tabular}} & \multicolumn{9}{c}{\% of $\hat{m}$} \\
		&                                                                                     &                                                                                    & \multicolumn{3}{c}{$S=6^2$} & \multicolumn{3}{c}{$S=8^2$} & \multicolumn{3}{c}{$S=10^2$} \\ \cline{4-12} 
		&                                                                                     &                                                                                    & 0      & 1      & $\geq2$   & 0      & 1      & $\geq2$   & 0      & 1      & $\geq2$    \\ \hline
		100 & 0 & 0  & 100 & 0   & 0 & 100 & 0   & 0 & 100 & 0   & 0 \\
		& 2 & 0  & 64  & 36  & 0 & 35  & 65  & 0 & 4   & 96  & 0 \\
		& 3 & 0  & 19  & 81  & 0 & 2   & 98  & 0 & 0   & 100 & 0 \\
		& 0 & 6  & 82  & 16  & 2 & 35  & 64  & 1 & 9   & 91  & 0 \\
		& 0 & 10 & 17  & 80  & 3 & 4   & 93  & 3 & 3   & 95  & 2 \\
		& 2 & 2  & 53  & 48  & 0 & 15  & 84  & 1 & 0   & 100 & 0 \\
		& 3 & 3  & 11  & 88  & 2 & 1   & 99  & 0 & 0   & 100 & 0 \\\hline
		200 & 0 & 0  & 100 & 0   & 0 & 100 & 0   & 0 & 100 & 0   & 0 \\
		& 2 & 0  & 20  & 80  & 0 & 1   & 99  & 0 & 2   & 98  & 0 \\
		& 3 & 0  & 0   & 100 & 0 & 0   & 100 & 0 & 0   & 100 & 0 \\
		& 0 & 6  & 42  & 58  & 0 & 13  & 83  & 4 & 1   & 98  & 1 \\
		& 0 & 10 & 2   & 97  & 1 & 0   & 100 & 0 & 0   & 100 & 0 \\
		& 2 & 2  & 9   & 91  & 0 & 0   & 100 & 0 & 0   & 100 & 0 \\
		& 3 & 3  & 3   & 96  & 1 & 0   & 100 & 0 & 0   & 100 & 0 \\ \hline \hline 
	\end{tabular}
\end{table}

\begin{table}[]
	
	\centering
	\caption{Percentage of estimated change-points $\hat m$ among 1000 replications under various spatial size $S$, temporal size $T$, and signal levels $(\delta_\phi, \delta_\rho)$ when $k=3$.}\label{scp.table.k3}
	\begin{tabular}{cccrrrrrrrrr}
		\hline \hline
		\multirow{3}{*}{$T$} & \multirow{3}{*}{\begin{tabular}[c]{@{}l@{}}$\delta_\phi\times 10$\end{tabular}} & \multirow{3}{*}{\begin{tabular}[c]{@{}l@{}}$\delta_\rho\times 10$\end{tabular}} & \multicolumn{9}{c}{\% of $\hat{m}$} \\
		&                                                                                     &                                                                                    & \multicolumn{3}{c}{$S=6^2$} & \multicolumn{3}{c}{$S=8^2$} & \multicolumn{3}{c}{$S=10^2$} \\ \cline{4-12} 
		&                                                                                     &                                                                                    & 0      & 1      & $\geq2$   & 0      & 1      & $\geq2$   & 0      & 1      & $\geq2$    \\ \hline
		100 & 0 & 0  & 100 & 0   & 0 & 100 & 0   & 0 & 100 & 0   & 0 \\
		& 2 & 0  & 61  & 39  & 0 & 32  & 68  & 0 & 5   & 95  & 0 \\
		& 3 & 0  & 22  & 78  & 0 & 2   & 97  & 1 & 0   & 100 & 0 \\
		& 0 & 6  & 80  & 19  & 1 & 39  & 61  & 1 & 10  & 90  & 0 \\
		& 0 & 10 & 19  & 80  & 1 & 6   & 92  & 2 & 3   & 96  & 2 \\
		& 2 & 2  & 54  & 46  & 0 & 11  & 88  & 1 & 0   & 100 & 0 \\
		& 3 & 3  & 10  & 90  & 0 & 2   & 97  & 1 & 0   & 100 & 0 \\\hline
		200 & 0 & 0  & 100 & 0   & 0 & 100 & 0   & 0 & 100 & 0   & 0 \\
		& 2 & 0  & 21  & 79  & 0 & 1   & 97  & 2 & 0   & 99  & 1 \\
		& 3 & 0  & 0   & 100 & 0 & 0   & 100 & 0 & 0   & 100 & 0 \\
		& 0 & 6  & 36  & 63  & 1 & 19  & 77  & 4 & 1   & 97  & 2 \\
		& 0 & 10 & 2   & 96  & 2 & 0   & 100 & 0 & 0   & 100 & 0 \\
		& 2 & 2  & 11  & 89  & 0 & 0   & 100 & 0 & 0   & 100 & 0 \\
		& 3 & 3  & 7   & 93  & 0 & 0   & 100 & 0 & 0   & 100 & 0 \\ \hline \hline 
		
	\end{tabular}
\end{table}

\begin{table}[]
	
	\centering
	\caption{Percentage of estimated change-points $\hat m$ among 1000 replications under various spatial resolutions $\Delta_s$ and signal levels $(\delta_\phi, \delta_\rho)$ using the modified CLMDL criterion \eqref{MDLform_infill} with $d=3$ and $B_{\mathcal{N}}=8$.}
	\label{infill.tab.B8}
	\begin{tabular}{ccrrrrrrrrrrrrrrr}
		\hline \hline
		$\delta_\phi$ & $\delta_\rho$ & \multicolumn{15}{c}{\% of $\hat{m}$}  \\
		$\times 10$& $\times 10$   & \multicolumn{3}{c}{\begin{tabular}[c]{@{}c@{}} $S=6^2$\\ $\Delta_s=1$\end{tabular}} & \multicolumn{3}{c}{\begin{tabular}[c]{@{}c@{}} $S=9^2$\\ $\Delta_s=0.6$\end{tabular}} & \multicolumn{3}{c}{\begin{tabular}[c]{@{}c@{}} $S=11^2$\\ $\Delta_s=0.5$\end{tabular}} & \multicolumn{3}{c}{\begin{tabular}[c]{@{}c@{}} $S=13^2$\\ $\Delta_s=0.4$\end{tabular}}& \multicolumn{3}{c}{\begin{tabular}[c]{@{}c@{}} $S=26^2$\\ $\Delta_s=0.2$\end{tabular}} \\ \cline{3-17}
		&       & 0                        & 1                       & $\geq 2$                      & 0                         & 1                        & $\geq 2$                      & 0                         & 1                         & $\geq 2$                      & 0                         & 1                         & $\geq 2$& 0&  1& $\geq 2$                      \\ \hline
		0 & 0  & 100 & 0  & 0 & 100 & 0  & 0 & 100 & 0  & 0 & 100 & 0   & 0 & 100 & 0   & 0 \\
		2 & 0  & 71  & 29 & 0 & 66  & 34 & 0 & 58  & 41 & 2 & 54  & 46  & 0 & 37  & 63  & 0 \\
		3 & 0  & 3   & 69 & 3 & 15  & 85 & 1 & 17  & 93 & 0 & 2   & 98  & 0 & 0   & 100 & 0 \\
		0 & 6  & 90  & 10 & 0 & 76  & 22 & 2 & 59  & 41 & 0 & 43  & 57  & 0 & 14  & 86  & 0 \\
		0 & 10 & 28  & 72 & 0 & 25  & 75 & 0 & 15  & 85 & 0 & 0   & 94  & 0 & 0   & 100 & 0 \\
		2 & 2  & 65  & 34 & 1 & 44  & 56 & 0 & 26  & 74 & 0 & 18  & 81  & 1 & 4   & 96  & 0 \\
		3 & 3  & 20  & 80 & 0 & 11  & 89 & 0 & 5   & 95 & 0 & 0   & 100 & 0 & 0   & 100 & 0 \\ \hline \hline
	\end{tabular}
\end{table}    

\end{simulation}

\clearpage

{\color{black}

\begin{simulation}\label{add_simu_epsilon} \textbf{(Sensitivity analysis on the tuning parameter $\epsilon_\lambda$)}

In practice, the minimum spacing parameter $\epsilon_\lambda^o$ is typically unknown and thus for CLMDL to work, we need a lower bound of $\epsilon_\lambda^o$, denoted by $\epsilon_\lambda$ such that $\epsilon_\lambda\leq \epsilon_{\lambda}^o$. Note that $\epsilon_\lambda$ is a tuning parameter of CLMDL as it governs the set of candidate change-points $A_{\epsilon_\lambda}^m$ as defined in \eqref{A}. In this example, we conduct a sensitivity analysis of CLMDL w.r.t.\ $\epsilon_\lambda$.

The underlying DGP is the same as that of \Cref{sim.1} in \Cref{subsec:simu} of the main text. In particular, we assume each stationary segment follows the four-parameter autoregressive spatial model \eqref{sim.process} with $\mu=0$, hence the process is specified by $\theta=(\phi, \rho,\sigma^2)^\top$. We set $\theta_1=(-0.5,0.6,1)^\top$ and $\theta_2=(-0.5+\delta_\phi,0.6+\delta_\rho,1)^\top$, with $T_1^o=30$ and $T_2^o=70$. In other words, the true change-point is $\lambda^o_1=0.3$. Thus, the minimum spacing $\epsilon_\lambda^o=0.3$. We vary the spatial dimension $S=6^2,8^2,10^2.$ We consider both the scenarios of no change-point (i.e. $\delta_\phi=\delta_\rho=0$) and one change-point at $\lambda_1^o=0.3$ with $\delta_\phi=\delta_\rho=0.3.$

We implement CLMDL with $\epsilon_\lambda=0.05,0.1,0.2,0.3,0.35,0.4$. Note that for $\epsilon_\lambda=0.35,0.4$, the assumption that $\epsilon_\lambda\leq\epsilon_\lambda^o$ is violated. \Cref{table.elambda} summarizes the simulation results, which shows that the performance of CLMDL is robust w.r.t.\ the choice of $\epsilon_\lambda$. First, when there is no change-point, CLMDL rarely gives false positives across all ranges of $\epsilon_\lambda$.

Second, when there is change, both the detection power and estimation accuracy of CLMDL are quite stable across all $\epsilon_\lambda$ with $\epsilon_\lambda\leq \epsilon_\lambda^o=0.3.$ For $\epsilon_\lambda=0.35,0.4$, CLMDL does suffer from power loss due to the violation of $\epsilon_\lambda\leq \epsilon_\lambda^o.$ However, this power loss alleviates as the sample size $S$ increases. Intuitively, the sample mean of $\hat\lambda$ for CLMDL with $\epsilon_\lambda=0.35$ is around 0.35, as this is the closest point to $\lambda^o=0.3$ in the set of candidate change-points $A_{\epsilon_\lambda=0.35}^m$. The same holds for $\epsilon_\lambda=0.4.$

To summarize, the performance of CLMDL is robust w.r.t.\ the choice of $\epsilon_\lambda$, as long as the assumption that $\epsilon_\lambda\leq \epsilon_\lambda^o$ is not severely violated. Thus, similar to existing change-point literature, if no prior knowledge of $\epsilon_\lambda^o$ is available, we recommend setting $\epsilon_\lambda$ to be some small values such as $0.05$ or $0.1$, which helps minimize the possibility of $\epsilon_\lambda> \epsilon_\lambda^o$.

\begin{table}[h]
\centering
{\color{black}
\caption{Percentage of estimated change-points $\hat{m}$ among 1000 replications, and mean and empirical standard deviation (esd) of $\hat{\lambda}$ (given $\hat{m}=m_o=1$) with various $\epsilon_\lambda$.}\label{table.elambda}
\begin{tabular}{ccrrrrrrrr}
\hline\hline
$S$    & $\epsilon_\lambda$ & $\delta_\phi\times 10$ & $\delta_\rho \times 10$ & \multicolumn{3}{c}{\% $\hat{m}$} & & \multicolumn{2}{c}{$\hat{\lambda}$} \\ 
\cline{5-7} \cline{9-10}       &                    &                        &                         & 0    & 1    & $\geq 2$  && mean            & esd               \\ \hline
$6^2$  & 0.05               & 0                      & 0                       & 99   & 1    & 0                 & & -               & -                 \\
       & 0.1                & 0                      & 0                       & 99   & 1    & 0                  && -               & -                 \\
       & 0.2                & 0                      & 0                       & 100  & 0    & 0                  && -               & -                 \\
       & 0.3                & 0                      & 0                       & 100  & 0    & 0                  && -               & -                 \\
       & 0.35               & 0                      & 0                       & 100  & 0    & 0                  && -               & -                 \\
       & 0.4                & 0                      & 0                       & 100  & 0    & 0                  && -               & -                 \\\hdashline
       & 0.05               & 3                      & 3                       & 9    & 90   & 0                  && 0.2971          & 0.0315            \\
       & 0.1                & 3                      & 3                       & 9    & 90   & 0                  && 0.2971          & 0.0315            \\
       & 0.2                & 3                      & 3                       & 9    & 91   & 0                  && 0.2978          & 0.0320            \\
       & 0.3                & 3                      & 3                       & 10   & 90   & 0                  && 0.3083          & 0.0200            \\
       & 0.35               & 3                      & 3                       & 30   & 70   & 0                  && 0.3651          & 0.0318            \\
       & 0.4                & 3                      & 3                       & 45   & 55   & 0                  && 0.4175          & 0.0365            \\ \hline
$8^2$  & 0.05               & 0                      & 0                       & 99   & 1    & 0                  && -               & -                 \\
       & 0.1                & 0                      & 0                       & 100  & 0    & 0                  && -               & -                 \\
       & 0.2                & 0                      & 0                       & 100  & 0    & 0                  && -               & -                 \\
       & 0.3                & 0                      & 0                       & 100  & 0    & 0                  && -               & -                 \\
       & 0.35               & 0                      & 0                       & 100  & 0    & 0                  && -               & -                 \\
       & 0.4                & 0                      & 0                       & 100  & 0    & 0                  && -               & -                 \\\hdashline
       & 0.05               & 3                      & 3                       & 0    & 99   & 0                  && 0.2999          & 0.0339          \\
       & 0.1                & 3                      & 3                       & 0    & 99   & 0                  && 0.2999          & 0.0339          \\
       & 0.2                & 3                      & 3                       & 1    & 99   & 0                  && 0.3015          & 0.0297          \\
       & 0.3                & 3                      & 3                       & 1    & 99   & 0                  && 0.3081          & 0.0242           \\
       & 0.35               & 3                      & 3                       & 9    & 91   & 0                  && 0.3603          & 0.0222          \\
       & 0.4                & 3                      & 3                       & 20   & 80   & 0                  && 0.4136          & 0.0215          \\ \hline
$10^2$ & 0.05               & 0                      & 0                       & 99   & 1    & 0                  && -               & -                 \\
       & 0.1                & 0                      & 0                       & 100  & 0    & 0                  && -               & -                 \\
       & 0.2                & 0                      & 0                       & 100  & 0    & 0                  && -               & -                 \\
       & 0.3                & 0                      & 0                       & 100  & 0    & 0                  && -               & -                 \\
       & 0.35               & 0                      & 0                       & 100  & 0    & 0                  && -               & -                 \\
       & 0.4                & 0                      & 0                       & 100  & 0    & 0                  && -               & -                 \\\hdashline
       & 0.05               & 3                      & 3                       & 0    & 100  & 0                  && 0.2992          & 0.0159            \\
       & 0.1                & 3                      & 3                       & 0    & 100  & 0                  && 0.2992          & 0.0159            \\
       & 0.2                & 3                      & 3                       & 0    & 100  & 0                  && 0.2992          & 0.0159            \\
       & 0.3                & 3                      & 3                       & 0    & 100  & 0                  && 0.3023          & 0.0095            \\
       & 0.35               & 3                      & 3                       & 0    & 100  & 0                  && 0.3546          & 0.0105            \\
       & 0.4                & 3                      & 3                       & 1    & 99   & 0                  && 0.4127          & 0.0298 \\ \hline\hline           
\end{tabular}}
\end{table}
    
\end{simulation}

\clearpage

\begin{simulation}\label{add_simu_vanishing} \textbf{(Vanishing change sizes)}

In this example, we investigate the performance of CLMDL under vanishing change sizes. The underlying DGP is the same as that of \Cref{sim.vanish} in the main text. In particular, we assume each stationary segment follows the four-parameter autoregressive spatial model \eqref{sim.process} with $\mu=0$, hence the process is specified by $\theta=(\phi, \rho,\sigma^2)^\top$. We set $\theta_1=(-0.5,0.6,1)^\top$ and $\theta_2=(-0.5+\delta_\phi,0.6+\delta_\rho,1)^\top$, with $T_1^o=T_2^o=50$. In other words, the true change-point is $\lambda_1^o=0.5$. We vary $S=6^2,8^2,10^2,30^2,60^2$. We set the change size as $\delta_\phi=S^{-0.4}$ or $S^{-0.5}$ (with $\delta_\rho=0$), which vanishes as $S$ increases. We also consider a larger change size with $\delta_\phi=\delta_\rho=S^{-0.4}$ or $\delta_\phi=\delta_\rho=S^{-0.5}$.

Table \ref{table.vanish.CLMDL_full} summarizes the simulation results. For both $\delta_\phi=S^{-0.4}$ and $S^{-0.5}$, the detection power and estimation accuracy of CLMDL improve as $S$ increases. Intuitively, CLMDL performs better under a slower decaying rate (i.e.\ $S^{-0.4}$) of change sizes and under a larger change size~(i.e.\ both $\phi$ and $\rho$ change). Moreover, exact recovery of $\lambda_1^o$ is possible even for $\delta_\phi=S^{-0.5}$ (and $\delta_\phi=\delta_\rho=S^{-0.5}$), which provides further numerical evidence for the theoretical results in \Cref{thm_diminishing}.

\vskip -1mm
\begin{table}[h]
\centering
{\color{black}
\caption{Percentage of $\hat{m}$ among 1000 replications, and percentage of $\hat{\lambda}=0.5$, mean and empirical standard deviation (esd) of $\hat{\lambda}$~(given $\hat m=m_o=1$) under various vanishing change sizes. 
}\label{table.vanish.CLMDL_full}
\begin{tabular}{lcccrrrrrrr}
\hline\hline
& $S$ & \begin{tabular}[c]{@{}c@{}}$\delta_\phi$ $\times 10$\end{tabular} & \begin{tabular}[c]{@{}c@{}}$\delta_\rho$ $\times 10$\end{tabular}    & \multicolumn{3}{c}{\% $\hat{m}$} &  & \multicolumn{3}{c}{$\hat{\lambda}$}          \\ \cline{5-7} \cline{9-11} 
                               &                                                                     &                                                                     &        & 0       & 1       & $\geq 2$     &  & \% of $\hat{\lambda}= 0.5$ & mean   & esd    \\ \hline
$\delta_\phi=S^{-0.4}$ & $6^2$ & 2.38 & 0  & 40      & 60      & 0            &  & 43                         & 0.4895 & 0.0314 \\
 & $8^2$  & 1.89  & 0  & 29      & 71      & 0            &  & 47                         & 0.4954 & 0.0205 \\
 & $10^2$  & 1.58  & 0 & 25      & 75      & 0            &  & 68                         & 0.4960 & 0.0168 \\
 & $30^2$  & 0.66 & 0 & 0       & 100     & 0            &  & 100                        & 0.5000 & -      \\
 & $60^2$  & 0.38 & 0 & 0       & 100     & 0            &  & 100                        & 0.5000 & -      \\\hdashline
$\delta_\phi=\delta_\rho=S^{-0.4}$ & $6^2$  & 2.38 & 2.38 & 32      & 68      & 0            &  & 41                         & 0.4926 & 0.0295 \\
 & $8^2$  & 1.89 & 1.89  & 20      & 80      & 0            &  & 63                         & 0.4966 & 0.0207 \\
 & $10^2$  & 1.58 & 1.58 & 14      & 86      & 0            &  & 69                         & 0.5043 & 0.0195 \\
 & $30^2$  & 0.66 & 0.66  & 0       & 100     & 0            &  & 100                        & 0.5000 & -      \\
 & $60^2$  & 0.38 & 0.38 & 0       & 100     & 0            &  & 100                        & 0.5000 & -      \\ \hline
$\delta_\phi=S^{-0.5}$ & $6^2$ & 1.67  & 0  &    74     & 26      & 0            &  & 27                         & 0.4773 & 0.0595 \\
 & $8^2$  & 1.25  & 0 &     79    & 21      & 0            &  & 50                         & 0.4871 & 0.0436 \\
 & $10^2$ & 1.00  & 0 &     68    & 32      & 0            &  & 57                         & 0.5025 & 0.0314 \\
 & $30^2$ & 0.33 & 0 &     14    & 86      & 0            &  & 100                        & 0.5000 & -      \\
 & $60^2$ & 0.17 & 0  &     0    & 100     & 0            &  & 100                        & 0.5000 & -      \\\hdashline
$\delta_\phi=\delta_\rho=S^{-0.5}$ & $6^2$ & 1.67 & 1.67  & 68      & 33      & 0            &  & 19                         & 0.5021 & 0.0628 \\
 & $8^2$ & 1.25 & 1.25  & 66      & 34      & 0            &  & 56                         & 0.4985 & 0.0531 \\
 & $10^2$ & 1.00 & 1.00 & 60      & 40      & 0            &  & 75                         & 0.4997 & 0.0326 \\
 & $30^2$ & 0.33 & 0.33 & 7       & 93      & 0            &  & 100                        & 0.5000 & -      \\
 & $60^2$ & 0.17 & 0.17 & 0       & 100     & 0            &  & 100                        & 0.5000 & -      \\ \hline \hline
\end{tabular}}
\end{table}
\end{simulation}
\clearpage

\begin{simulation}\label{add_simu_univariate} \textbf{(Comparison with univariate methods)}

In this example, we compare the performance of CLMDL with existing methods designed for multiple change-point estimation in univariate time series. Specifically, we consider the important work of \cite{Davis2006}, which proposes an MDL based method for detecting change-points in parametric models of univariate time series. Hereafter, we refer to it as DLR.

The spatio-temporal data $\bfY$ can be viewed as $S$ univariate time series $\{\{y_{t,\bs}\}_{t=1}^T, \bs\in\mathcal{S}\}$ observed over $S$ different spatial locations. Thus, in principle, we can run DLR on each of the $S$ univariate time series to examine if change-points exist. However, intuitively such an approach may not be ideal for spatio-temporal data as it cannot efficiently combine information across the $S$ univariate time series to help better detect the change-points. In addition, this approach intrinsically \textit{cannot} detect any change in the spatial dependence among the $S$ univariate time series, as it can only analyze each time series \textit{individually}.

To illustrate the importance and advantage of treating the spatio-temporal data as one single entity instead of processing them separately, in the following we conduct two numerical studies to compare the performance of the proposed CLMDL and the DLR in \cite{Davis2006}.

\textsc{Non-vanishing change sizes}: We first consider the same DGP as the one in \Cref{sim.1} of the main text. In particular, we assume each stationary segment follows the four-parameter autoregressive spatial model \eqref{sim.process} with $\mu=0$, hence the process is specified by $\theta=(\phi, \rho,\sigma^2)^\top$. We set $\theta_1=(-0.5,0.6,1)^\top$ and $\theta_2=(-0.5+\delta_\phi,0.6,1)^\top$ and set the true change-point as $\lambda_1^o=0.5$. Note that we only consider change in the temporal dependence $\phi$, since as discussed above, the DLR based approach intrinsically cannot detect any change in the spatial dependence $\rho$.

We set the change size as $\delta_\phi=0,0.2,0.3$ and vary the sample size $T=100,200$ and $S=6^2,8^2,10^2$. For each simulation setting, we repeat the experiments 1000 times. \Cref{table.univariateMDL.DLR} summarizes the detection results of the DLR in \cite{Davis2006}. Note that for each experiment, given the simulated spatio-temporal data $\{\{y_{t,\bs}\}_{t=1}^T, \bs\in\mathcal{S}\}$, DLR is applied to the $S$ univariate time series \textit{individually} instead of treating them as one single entity. Therefore, for each experiment, DLR returns a set of change-point estimation result for each of the $S$ univariate time series. Thus, to facilitate comparison, we report the sample distribution of the estimated number of change-points $\hat m$ given by DLR among all $1000\times S$ univariate time series. For convenience of comparison, we further copy the detection results of the proposed CLMDL from \Cref{scp.table} of the main text here. 

Some observations are as follows. First, when there is no change-point, both DLR and our proposed CLMDL are robust to false positive detection. However, when there is a change-point, the detection power of DLR is significantly lower than that of CLMDL across all sample sizes and change sizes. In addition, note that the detection power of DLR is close to zero and does \textit{not} improve as the spatial dimension $S$ grows and only improves slightly as the time dimension $T$ grows. This is intuitive as DLR cannot combine information among the spatial dimension to achieve better change-point estimation. In contrast, the performance of the proposed CLMDL substantially improves whenever $T$ or $S$ increases. 

We remark that the comparison here is for illustrative purposes only and is mainly used to demonstrate the importance of \textit{jointly} analyzing all $S$ univariate time series for accurate change-point estimation in the spatio-temporal setting, which is the approach taken by CLMDL.

\begin{table}[h]
	\centering
{\color{black}
	\caption{ Percentage of estimated change-points $\hat m$ among 1000 replications under various spatial size $S$, temporal size $T$, and signal levels $(\delta_\phi, \delta_\rho)$ using \cite{Davis2006} and CLMDL.}\label{table.univariateMDL.DLR}
	\begin{tabular}{cccrrrrrrrrr}
		\hline \hline
            \multicolumn{12}{c}{DLR}\\\hline
		\multirow{3}{*}{$T$} & \multirow{3}{*}{\begin{tabular}[c]{@{}l@{}}$\delta_\phi\times 10$\end{tabular}} & \multirow{3}{*}{\begin{tabular}[c]{@{}l@{}}$\delta_\rho\times 10$\end{tabular}} & \multicolumn{9}{c}{\% of $\hat{m}$} \\
		&                                                                                     &                                                                                    & \multicolumn{3}{c}{$S=6^2$} & \multicolumn{3}{c}{$S=8^2$} & \multicolumn{3}{c}{$S=10^2$} \\ \cline{4-12} 
		&                                                                                     &                                                                                    & 0      & 1      & $\geq2$   & 0      & 1      & $\geq2$   & 0      & 1      & $\geq2$    \\ \hline
		100 & 0 & 0  & 100 & 0 & 0       & 100 & 0 & 0       & 100 & 0 & 0       \\
    & 2 & 0  & 97  & 3 & 0       & 98  & 2 & 0       & 97  & 3 & 0       \\
    & 3 & 0  & 93  & 7 & 0       & 93  & 7 & 0       & 93  & 7 & 0       \\\hline
200 & 0 & 0  & 100 & 0 & 0       & 100 & 0 & 0       & 100 & 0 & 0       \\
    & 2 & 0  & 96  & 4 & 0       & 97  & 3 & 0       & 96  & 4 & 0       \\
    & 3 & 0  & 91  & 9 & 0       & 92  & 8 & 0       & 91  & 9 & 0       \\\hline  
            \multicolumn{12}{c}{CLMDL}\\\hline
   		\multirow{3}{*}{$T$} & \multirow{3}{*}{\begin{tabular}[c]{@{}l@{}}$\delta_\phi\times 10$\end{tabular}} & \multirow{3}{*}{\begin{tabular}[c]{@{}l@{}}$\delta_\rho\times 10$\end{tabular}} & \multicolumn{9}{c}{\% of $\hat{m}$} \\
		&  &   & \multicolumn{3}{c}{$S=6^2$} & \multicolumn{3}{c}{$S=8^2$} & \multicolumn{3}{c}{$S=10^2$} \\ \cline{4-12} 
		&                                                                                     &                                                                                    & 0      & 1      & $\geq2$   & 0      & 1      & $\geq2$   & 0      & 1      & $\geq2$    \\ \hline
		100 & 0    & 0      & 100    & 0      & 0         & 100    & 0      & 0         & 100    & 0      & 0          \\
		& 2   & 0   & 63     & 37     & 0         & 29     & 71     & 1         & 2      & 98     & 0          \\
		& 3    & 0      & 22     & 79     & 0         & 1      & 99     & 1         & 0      & 100    & 0          \\\hline
		200 & 0  & 0    & 100    & 0      & 0         & 100    & 0      & 0         & 100    & 0      & 0          \\
		& 2    & 0    & 20     & 81     & 0         & 1      & 99     & 0         & 0      & 100    & 0          \\
		& 3    & 0     & 0      & 100    & 0         & 0      & 100    & 0         & 0      & 100    & 0          \\\hline \hline      
	\end{tabular}}
\end{table}

\textsc{Vanishing change sizes}: We further consider the same DGP as the one used in \Cref{add_simu_vanishing}. In particular, we assume each stationary segment follows the four-parameter autoregressive spatial model \eqref{sim.process} with $\mu=0$, hence the process is specified by $\theta=(\phi, \rho,\sigma^2)^\top$. We set $\theta_1=(-0.5,0.6,1)^\top$ and $\theta_2=(-0.5+\delta_\phi,0.6,1)^\top$, with $T_1^o=T_2^o=50$. We vary $S=6^2,8^2,10^2,30^2,60^2$. We set the change size as $\delta_\phi=S^{-0.4}$ or $S^{-0.5}$. Thus, for each of the $S$ univariate time series, its change size vanishes as $S$ increases. \Cref{table.univariateMDL.vanish.DLR} summarizes the detection results of the DLR based approach. Compared to \Cref{table.vanish.CLMDL_full}, which reports the performance of CLMDL under the same setting, it is clear that CLMDL works much better than DLR. In particular, as $S$ grows, DLR quickly becomes powerless, while in contrast, the power of CLMDL improves thanks to its ability to combine information across all $S$ univariate time series to achieve better change-point estimation.

\begin{table}[H]
\centering
{\color{black}
\caption{Percentage of estimated change-points $\hat{m}$ among 1000 replications using \cite{Davis2006} under vanishing change sizes.}\label{table.univariateMDL.vanish.DLR}
\begin{tabular}{ccrrr|ccrrr|ccrrr}
    \hline\hline
    $S$    & $\delta_\phi\times   10$ & \multicolumn{3}{c|}{$\hat{m}$} & & $\delta_\phi\times 10$ & \multicolumn{3}{c|}{$\hat{m}$} & & $\delta_\phi\times   10$ & \multicolumn{3}{c}{$\hat{m}$}  \\
           &   & 0         & 1       & 2       &                        &                          & 0         & 1       & 2    & & & 0         & 1       & 2    \\ \hline
    $6^2$  & 0  & 100  & 0  & 0  & $\delta_\phi=S^{-0.4}$   & 2.38  & 96   & 4   & 0   & $\delta_\phi=S^{-0.5}$  & 1.67 & 98 & 2 & 0  \\
    $8^2$  & 0   & 100       & 0       & 0       &     &      1.89   & 96       & 4       & 0  & & 1.25 & 99 & 1 & 0     \\
    $10^2$ & 0    & 100       & 0       & 0       &   &  1.58    & 98       & 2       & 0   & & 1.00 & 100 & 0 & 0    \\
    $30^2$ & 0 & 100       & 0       & 0    &    &   0.66    & 99       & 1       & 0    & & 0.33 & 100 & 0 & 0     \\
    $60^2$ & 0  & 100       & 0       & 0       &    &   0.38  & 100       & 0       & 0    & & 0.17 & 100 & 0 & 0     \\ \hline\hline
\end{tabular}}
\end{table}
\end{simulation}

\begin{simulation}\label{add_simu_PELT}\textbf{(Performance of PELT and OP)}

In this example, we examine the optimization performance and computational time of PELT proposed in \Cref{sec:PELT} and compare it with the optimal partitioning~(OP) algorithm in \cite{Jackson2005}. Recall that the OP algorithm guarantees the exact minimization of CLMDL at a higher computational cost, while thanks to an additional pruning step, the PELT algorithm achieves a lower computational cost but only guarantees to find the exact minimizer of CLMDL asymptotically (\Cref{PELTK}) under the spatio-temporal setting.

The simulation setting is the same as the one in \Cref{sim.1} of the main text. In particular, we assume each stationary segment follows the four-parameter autoregressive spatial model \eqref{sim.process} with $\mu=0$, hence the process is specified by $\theta=(\phi, \rho,\sigma^2)^\top$. We set $\theta_1=(-0.5,0.6,1)^\top$ and $\theta_2=(-0.5+\delta_\phi,0.6+\delta_\rho,1)^\top$, with $T_1^o=T_2^o=50$ (and thus $T=100$). We vary $S=6^2,8^2,10^2$ and vary the change sizes $(\delta_\phi,\delta_\rho).$ We skip $T=200$ as OP requires too much running time which makes the comparison computationally forbidding. Note that to implement OP, we simply remove the pruning step (see \eqref{prunning}) in PELT.

For each simulation setting, we repeat the experiments 100 times. The computational hardware that we run the experiments on is an Intel Xeon Platinum 8358 32-Core cluster. First, interestingly, for all the experiments that we run, we find that PELT and OP return \textit{exactly} the same minimizers of the CLMDL criterion. In other words, the pruning step does not seem to affect the optimization performance of PELT under the settings that we consider, even for small sample sizes $S$ and $T.$

Table \ref{pelt.op.time} further reports the average computational time of PELT and OP. As can be seen clearly, PELT is in general much faster and requires three times less computational time than OP. When there is no change-point ($\delta_\phi=\delta_\rho=0)$ or small scale change ($\delta_\phi=0.1)$, the improvement of PELT over OP is to a lesser degree, as in such case the pruning step is less effective in removing abundant candidates in the dynamic programming, see similar observations on PELT in change-point estimation for univariate time series in \cite{Killick2012}.

To summarize, the numerical experiments indicate that compared to OP, PELT provides notable computational gain while at the same time delivering a similar level of optimization performance.

\begin{table}[H]
\centering
{\color{black}
\caption{Average computational time (in minutes) of PELT and OP, and the relative computational gain by PELT (in percentage) under various spatial sizes $S$ and signal levels $(\delta_\phi, \delta_\rho)$. }\label{pelt.op.time}
\begin{tabular}{rr|rrr|rrr|rrr}
\hline\hline
 &   & \multicolumn{3}{c|}{PELT} & \multicolumn{3}{c|}{OP} & \multicolumn{3}{c}{\begin{tabular}[c]{@{}c@{}}OP/PELT \\ $\times 100$\end{tabular}} \\\hline
$\delta_\phi\times 10$ & $\delta_\rho\times 10$ & $S=6^2$     & $8^2$     & $10^2$    & $S=6^2$     & $8^2$     & $10^2$    & $S=6^2$   & $8^2$  & $10^2$ \\ \hline
0 & 0  & 166 & 289 & 378 & 204 & 371 & 954 & 123 & 128 & 252 \\
1 & 0  & 146 & 238 & 325 & 235 & 452 & 711 & 161 & 190 & 219 \\
2 & 0  & 77  & 151 & 234 & 260 & 473 & 797 & 337 & 313 & 341 \\
3 & 0  & 83  & 115 & 251 & 290 & 509 & 815 & 349 & 444 & 324 \\
2 & 2  & 73  & 143 & 216 & 244 & 457 & 734 & 335 & 319 & 339 \\
3 & 3  & 82  & 148 & 249 & 342 & 635 & 748 & 416 & 428 & 301 \\
0 & 6  & 86  & 156 & 258 & 264 & 499 & 789 & 307 & 319 & 306 \\
0 & 8  & 83  & 136 & 237 & 265 & 478 & 830 & 319 & 352 & 350 \\
0 & 10 & 81  & 158 & 230 & 256 & 485 & 788 & 315 & 308 & 342 \\ \hline\hline
\end{tabular}}
\end{table}     
\end{simulation}

}

\clearpage
\subsection{Additional real data analysis of the U.S. precipitation data}\label{subsec:addrealdata}
\begin{figure}[h]
	\begin{center}
		\includegraphics[scale=0.4]{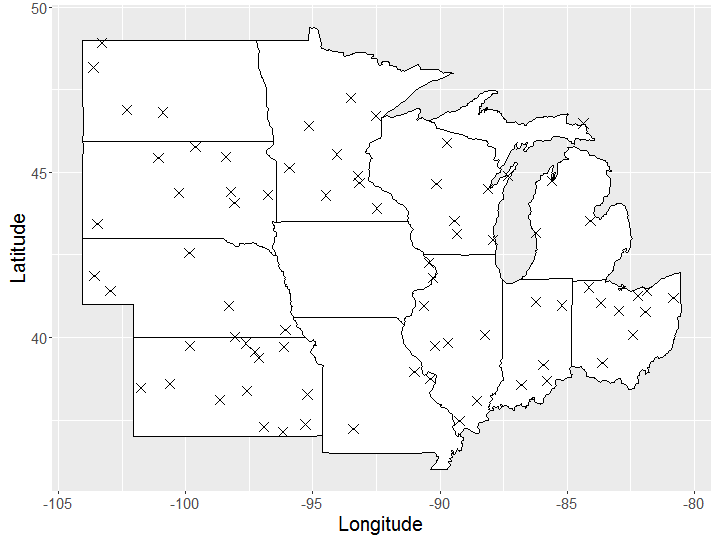} \vspace{-0.3cm}
	\end{center}
	\caption{Location of 76 land surface stations selected.}
	\label{US.map}
\end{figure}

\Cref{US.map} plots the locations of the 76 land surface stations used in the real data example.

{\color{black}
Recall that $\{\{y_{t,\bs}\}_{t=1}^T,\bs \in \mathcal{S}\}$ denotes the precipitation data after the log-transform and the stationarizing transform observed over the $S=76$ stations. \Cref{month.avg} plots the transformed precipitation data averaged over the 76 stations, i.e.\ $\big\{ \sum_{\bs\in \mathcal{S}} y_{t,\bs}/S\big\}_{t=1}^T$, along with the estimated change-points (solid vertical line) and the 90\% CIs (dashed vertical line) given by CLMDL. On each segment, based on the estimated linear regression for the mean function, we can further compute the fitted mean precipitation $\{\{\hat y_{t,\bs}\}_{t=1}^T,\bs \in \mathcal{S}\}$ for each of the 76 stations. Note that for a given station $\bs$ within each stationary segment, the fitted mean precipitation $\hat{y}_{t,\bs}$ does not vary over time. 

\Cref{month.avg} plots the fitted mean precipitation averaged over the 76 stations, i.e.\ $\big\{ \sum_{\bs} \hat y_{t,\bs}/S\big\}_{t=1}^T$, on each segment~(red dashed line). In addition, \Cref{station.avg} gives the histogram of the fitted mean precipitation of the 76 stations on each segment. As can be seen, the mean precipitation across the stations seems to shift down and then up over time. 

\begin{figure}[h]
    \begin{center}
        \includegraphics[scale=0.75]{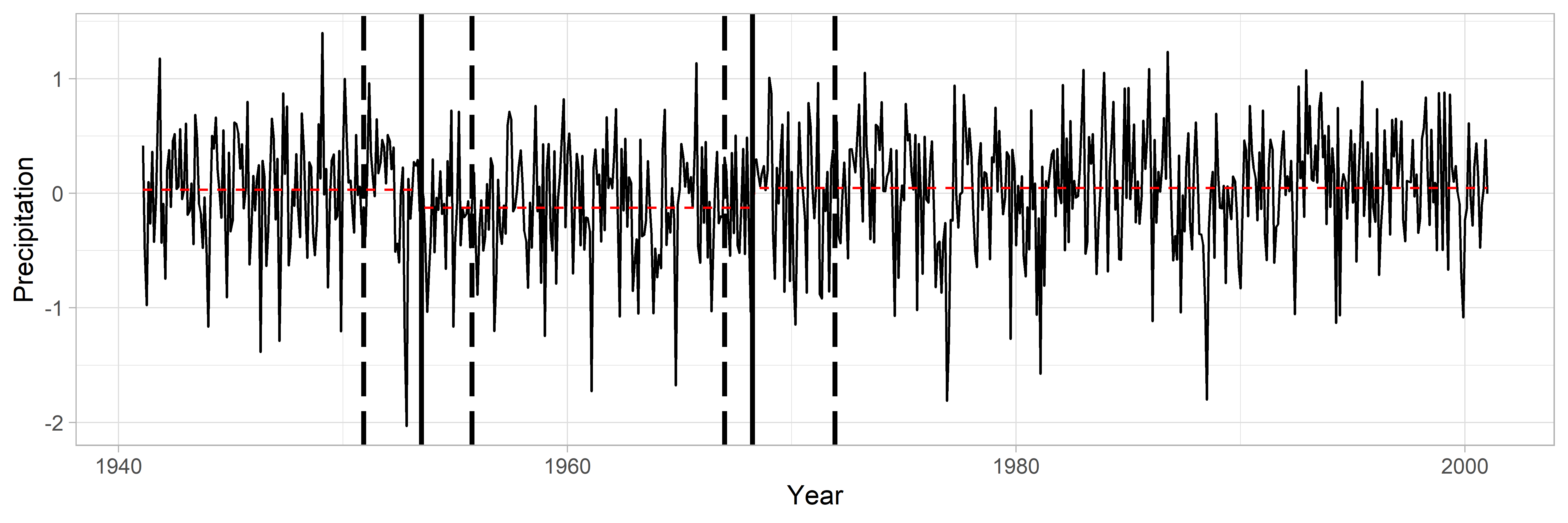}
    \end{center}
	\caption{\color{black} This figure plots the transformed precipitation data averaged over the 76 stations (black curve) and the fitted mean precipitation averaged over the 76 stations (red dashed line). In addition, it plots the two estimated change-points (solid vertical lines) and their corresponding 90\% CIs (dashed vertical lines).}
	\label{month.avg}
\end{figure}

\begin{figure}[h]
    \begin{center}
        \includegraphics[scale=1]{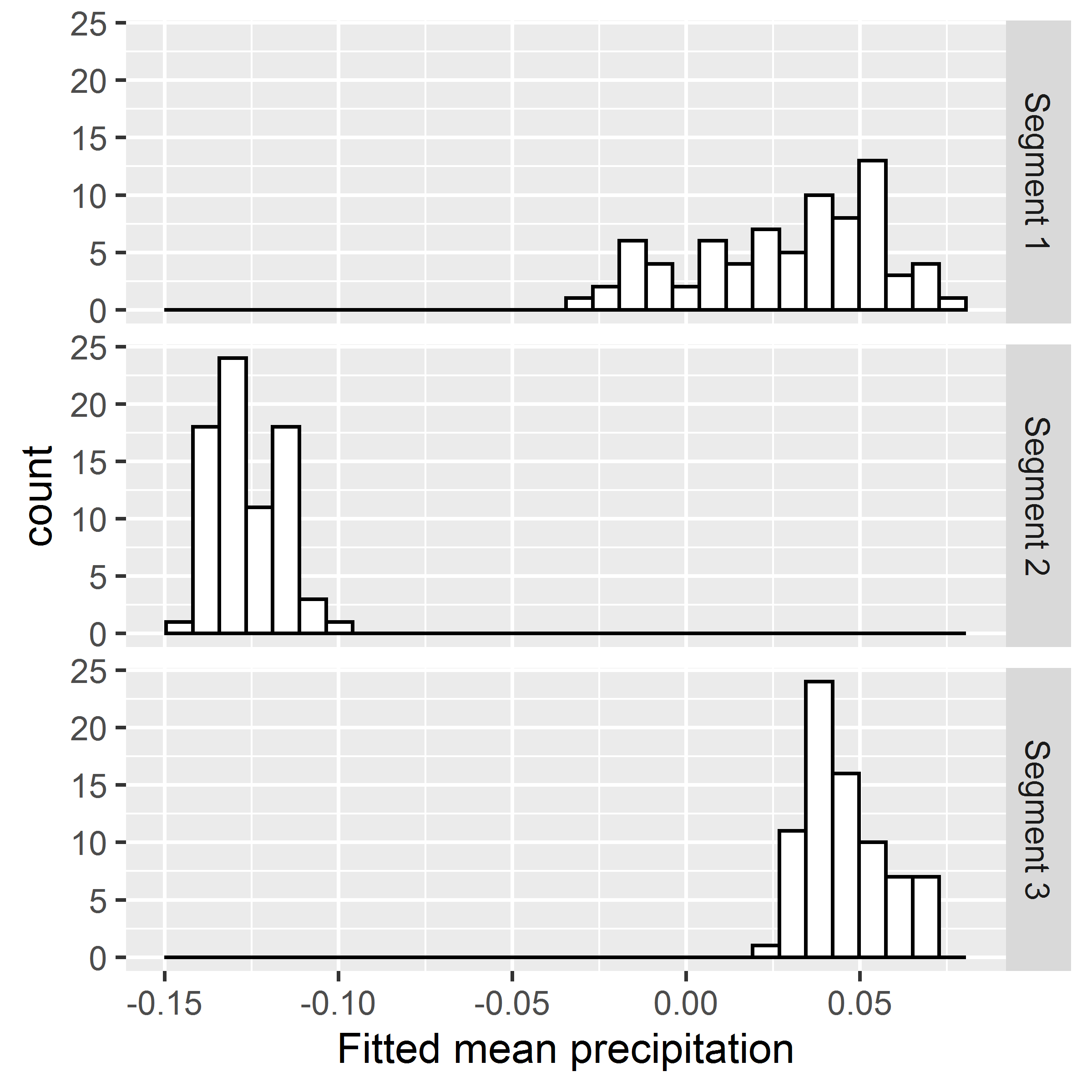}
    \end{center}
	\caption{\color{black} This figure gives the histogram of the fitted mean precipitation of the 76 stations on the three estimated segments.}
	\label{station.avg}
\end{figure}

For reference, \Cref{tab:para_est_USdata} reports the estimated parameters of the parametric model on each segment, where it is seen that the parameters of the space-time covariance function $(a,b,c,\sigma^2)$ seem to be similar across segments. To further investigate the root of the changes and examine the robustness of our findings, we conduct an additional analysis which is admittedly heuristic. First, based on the estimated mean function on each segment, we demean the observed spatio-temporal data and obtain $\{\{\tilde y_{t,\bs}=y_{t,\bs}-\hat{y}_{t,\bs}\}_{t=1}^T, \bs\in\mathcal{S}\}.$ We then estimate the parameters of the space-time covariance function $(\hat{a},\hat b,\hat c,\hat \sigma^2)$ based on the entire demeaned data  $\{\{\tilde y_{t,\bs}\}_{t=1}^T, \bs\in\mathcal{S}\}$. 

Now, we run CLMDL on the observed spatio-temporal data $\{\{ y_{t,\bs}\}_{t=1}^T,\bs\in\mathcal{S}\}$ but only allow changes in the linear regression based mean function while fixing the parameters of the space-time covariance function at $(\hat{a},\hat b,\hat c,\hat \sigma^2)$. Interestingly, we get the exact same change-points as before at 1953 and 1968. In addition, we further run CLMDL on the demeaned data $\{\{\tilde y_{t,\bs}\}_{t=1}^T, \bs\in\mathcal{S}\}$ but only allow changes in the space-time covariance function while fixing the mean function at 0. No change-point is detected this time. Therefore, this heuristic analysis suggests the robustness of the estimated change-points and indicates that the changes seem to stem from the mean function.

\begin{table}[H]
\centering
{\color{black}
\caption{Parameter estimates on the three estimated segments for the U.S. precipitation data .}\label{tab:para_est_USdata}
\begin{tabular}{rrrr}
\hline\hline
          & Segment 1       & Segment 2      & Segment 3      \\ \hline
intercept & -0.38    & -0.30    & 0.28     \\
latitude  & -2.74$\times 10^{-3}$ & 2.90$\times 10^{-3}$  & -2.62$\times 10^{-4}$ \\
longitude & -6.18$\times 10^{-3}$ & -6.30$\times 10^{-4}$ & 2.64$\times 10^{-3}$  \\
elevation & -1.25$\times 10^{-4}$ & -1.79$\times 10^{-5}$ & 5.58$\times 10^{-5}$  \\
$a$         & 2.81     & 3.02     & 2.93     \\
$b$         & 0.48    & 0.49     & 0.48     \\
$c$         & 1.85     & 1.48     & 1.64     \\
$\sigma^2$    & 0.99     & 0.96     & 0.97 \\ \hline\hline   
\end{tabular}}
\end{table}


}

\clearpage
\section{Notation and outline for technical proofs}\label{sec:notation}

Sections \ref{sec:proof_increasing_domain}, \ref{sec:vanishing}, \ref{sec:proof_infill} and \ref{PELT.issue} contain the proofs of all the theoretical results presented in \Cref{sec:main} of the main text and the ones presented in Sections \ref{sec:main_infill} and \ref{sec:PELT} of the supplement. For notational simplicity, in the following we use $(S,T)$ instead of $(S_n,T_n)$ when there is no possibility of confusion. The asymptotics should be understood as $n\to\infty,$ where we also interchangeably use the notation $(S,T)\to\infty$. 

For simplicity of presentation, in the following, we give the detailed proof for $k=1$, where $k$ is the time lag of the composite likelihood used in CLMDL. We remark that the proof for \textcolor{black}{any finite $k>1$ follows the exact same argument. The only difference is that the index set used in the definition of CLMDL is algebraically more tedious for $k>1$. We refer to \Cref{remark:kmore1} at the end of this section for a detailed discussion on the modification needed to cover the proof for $k>1$.}

Compared to time series settings such as in \cite{Davis2013}, \cite{Ling2014} and \cite{Ma2016}, the proof is theoretically more complicated due to the extension from time series~(random sequence) to spatio-temporal process~(random field). In addition, it is algebraically more complicated due to the explicit treatment of the edge effect and the cross-over terms between different segments, which are $O_p(1)$ terms under the time series setting (and thus can be ignored).

Given the spatial locations $\mathcal{S}\subset \mathbb{R}^2$ with cardinality $S$, we can always reindex $\bfy_t=\{y_{t,\bs}:\bs\in \mathcal{S}\}$ by $\bfy_t=\{y_{t,s}:s=1,\ldots,S\}$. Thus, we can reindex the spatio-temporal process $\bfY=\{y_{t,\bs}: t\in [1,T]\cap \mathbb{N}, \bs\in \mathcal{S}, \mathcal{S}\subset\mathbb{R}^2\}$ by $\bfY=\{y_{t,s}: 1\leq t\leq T, 1\leq s\leq S\}$. In the following, for notational simplicity, we mainly use the reindexed notation $\bfY=\{y_{t,s}: 1\leq t\leq T, 1\leq s\leq S\}$.

Denote the normalized true change points as $\{\lambda_0^o, \lambda_1^o,\ldots, \lambda_{m_o}^o, \lambda_{m_o+1}^o \}$ with $\lambda_0^o=0$ and $\lambda_{m_o+1}^o=1$. Denote the true change points as $\{\tau_0^o, \tau_1^o,\ldots, \tau_{m_o}^o, \tau_{m_o+1}^o \}$ where $\tau_j^o=[T\lambda_{j}^o]$ and denote $T_j=\tau_j^o-\tau_{j-1}^o$ for $j=1,\ldots,m_o+1$. Denote the piecewise stationary spatio-temporal process as $\bfY=\{y_{t,s}: 1\leq t\leq T, 1\leq s\leq S\}$, we have that the $j$th segment of $\bfY$ is generated from a stationary random field $\bfX_j^o=\{ x_{t,s}^{(j)}: 1\leq t\leq T_j, 1\leq s\leq S\}$ such that 
\begin{align*}
 x_{t-\tau_{j-1}^o, s}^{(j)}=y_{t,s}^{(j)}~~~~\text{for}~~\tau_{j-1}^o+1\leq t\leq \tau_j^o,
\end{align*}
where $j=1,\ldots, m_o+1.$ The superscripts for $y_{t,s}$ and $ x_{t-\tau_{j-1}^o,s}$ indicate the segment they come from. Note that the superscript for $y_{t,s}$ is unnecessary, however, for clearer presentation, we keep the superscript for $y_{t,s}$ when needed.

Let $\lambda_u$ and $\lambda_d$ be in $[0,1]$ satisfying $\lambda_d<\lambda_u$ and $\lambda_u-\lambda_d >\epsilon_\lambda$. Denote
\begin{align} \label{supdef}
\sup_{\lambda_d,\lambda_u}= \sup_{\substack{ \lambda_u\in[0,1], \lambda_d\in[0,1]\\ \lambda_u-\lambda_d >\epsilon_\lambda}}\,.
\end{align}

In addition, similar to $\mathcal D_{k,\mathcal{N}}$ and $\mathcal E_{k,\mathcal{N}}$ defined in Sections \ref{subsec:CLPL} and \ref{subsec:edge_effect} of the main text, we define several index sets that will be used throughout the proof. \textcolor{black}{In particular, these index sets provide more convenient references to the index sets $\mathcal D_{k,\mathcal{N}}$ and $\mathcal E_{k,\mathcal{N}}$ for $k=1$.} Given $T$ and $S$, define
\begin{align}\label{eq:index_sets}
E_{1ST}^{\lambda}&=\{(t,s):t=[T\lambda]+1, 1\leq s\leq S, \text{ each } s  \text{ repeats } (1+|\mathcal{N}(s)|) \text{ times} \},\nonumber\\
E_{2ST}^{\lambda}&=\{(t,s):t=[T\lambda], 1\leq s\leq S, \text{ each } s  \text{ repeats } (1+|\mathcal{N}(s)|) \text{ times} \},\nonumber\\
D^{\lambda_d,\lambda_u}_{ST}&=\{(t,i,s_1,s_2): [T\lambda_d]+1\leq t, t+i \leq [T\lambda_u], 0\leq i\leq 1, 1\leq s_1 \leq S, s_2 \in s_1\cup \mathcal{N}(s_1), \text{if }i=0, s_1\neq s_2\},\nonumber\\
D_{cST}^\lambda&=\{([T\lambda],[T\lambda]+1,s_1,s_2): 1\leq s_1 \leq S, s_2 \in s_1\cup \mathcal{N}(s_1)\}.
\end{align}

The sets $E_{1ST}^{\lambda}$ and $E_{2ST}^{\lambda}$ are used to index the marginal loglikelihood in CLMDL for compensating the edge effect. The set $D_{cST}^\lambda$ is used to index the pairwise loglikelihood across $[T\lambda]$. The set $D^{\lambda_d,\lambda_u}_{ST}$ is used to index the pairwise loglikelihood between $[T\lambda_d]+1$ and $[T\lambda_u]$. 

Moreover, denote a proportion of the observed composite log-likelihood for the $j$th stationary segment $\bfX_j^o$ ($j=1,\ldots,m_o+1$) as
\begin{align} \label{PL.prop}
&L_{ST}^{(j)}(\psi_j, \lambda_d,\lambda_u; \bfX_j^o)=L_{ST}^{(j)}\{(\xi_j,\theta_j), \lambda_d,\lambda_u; \bfX_j^o\}\nonumber\\=&\sum_{(t,i,s_1,s_2)\in D^{\lambda_d,\lambda_u}_{ST_j}}\log f\left(x_{t,s_1}^{(j)},x_{t+i,s_2}^{(j)};\psi_j\right)
+ \sum_{(t,s)\in E_{1ST_j}^{\lambda_d}}\log f\left(x_{t,s}^{(j)};{\psi}_j\right) + \sum_{(t,s)\in E_{2ST_j}^{\lambda_u}}\log f\left(x_{t,s}^{(j)};{\psi}_j\right)\nonumber\\
=&\sum_{(t,i,s_1,s_2)\in D^{\lambda_d,\lambda_u}_{ST_j}}l_{pair}\left(\psi_j; x_{t,s_1}^{(j)},x_{t+i,s_2}^{(j)}\right)
+ \sum_{(t,s)\in E_{1ST_j}^{\lambda_d}} l_{marg}\left({\psi}_j; x_{t,s}^{(j)}\right) + \sum_{(t,s)\in E_{2ST_j}^{\lambda_u}}l_{marg}\left({\psi}_j; x_{t,s}^{(j)}\right).
\end{align}
Here, for indexing observations on $\bfX_j^o,$ we replace $T$ with $T_j$ in the index sets $E_{1ST}^{\lambda}$, $E_{2ST}^{\lambda}$ and $D^{\lambda_d,\lambda_u}_{ST}$. Note that the parametric form of the likelihood functions $l_{marg}$ and $l_{pair}$ are fully determined by $\psi_j=(\xi_j, \theta_j)$, where $\xi_j \in \mathcal{M}$ determines the model order and $\theta_j\in \Theta_j(\xi_j)$ is the model parameter. For notational simplicity, we replace $\psi_j=(\xi_j,\theta_j)$ by $\theta_j$ in the likelihoods $l_{marg}$ and $l_{pair}$ when there is no confusion about the model order $\xi_j$.

Define
\begin{align*}
&L_{STpair}^{(j)}(\psi_j, \lambda_d,\lambda_u; \bfX_j^o)=\sum_{(t,i,s_1,s_2)\in D^{\lambda_d,\lambda_u}_{ST_j}}l_{pair}\left(\psi_j; x_{t,s_1}^{(j)},x_{t+i,s_2}^{(j)}\right),\\
&L_{STmarg}^{(j)}(\psi_j, \lambda_d,\lambda_u; \bfX_j^o)=\sum_{(t,s)\in E_{1ST_j}^{\lambda_d}} l_{marg}\left({\psi}_j; x_{t,s}^{(j)}\right) + \sum_{(t,s)\in E_{2ST_j}^{\lambda_u}}l_{marg}\left({\psi}_j; x_{t,s}^{(j)}\right).
\end{align*}
A useful decomposition of the composite loglikelihood is then
\begin{align*}
L_{ST}^{(j)}(\psi_j, \lambda_d,\lambda_u; \bfX_j^o)&=L_{STpair}^{(j)}(\psi_j, \lambda_d,\lambda_u; \bfX_j^o)+L_{STmarg}^{(j)}(\psi_j, \lambda_d,\lambda_u; \bfX_j^o).
\end{align*}
Define the expected values for the composite likelihood as
\begin{align*}
&\thickbar{L}_{STmarg}^{(j)}\{(\xi_j,\theta_j)\}=\mathbb E(L_{STmarg}^{(j)}(\psi_j, 0,1; \bfX_j^o)),\\
&\thickbar{L}_{STpair}^{(j)}\{(\xi_j,\theta_j)\}=\mathbb E(L_{STpair}^{(j)}(\psi_j, 0,1; \bfX_j^o)),\\
&\thickbar{L}_{ST}^{(j)}\{(\xi_j,\theta_j)\}=\thickbar{L}_{STmarg}^{(j)}\{(\xi_j,\theta_j)\}+\thickbar{L}_{STpair}^{(j)}\{(\xi_j,\theta_j)\}.
\end{align*}
Note that the definition here is consistent with the definition of $\thickbar{L}_{ST}^{(j)}(\psi_j)$ in the main text. By the strictly stationarity of the random field and the compactness of $\Theta$, we know that
\begin{align*}
\sup_{\theta\in \Theta}\frac{1}{ST}\left|\mathbb E(L_{ST}^{(j)}(\psi_j, \lambda_d,\lambda_u; \bfX_j^o))-(\lambda_u-\lambda_d)\thickbar{L}_{ST}^{(j)}\{(\xi_j,\theta_j)\}\right|=o(1),
\end{align*}
for all $j=1,\ldots, m_o+1$.

{\color{black}
\begin{remark}[Proof for time lag $k>1$ in CLMDL]\label{remark:kmore1}
    In the following, we provide a detailed discussion of how to modify the current proof for $k=1$ to cover the case where $k>1.$ In short, the proof for $k>1$ follows the exact same argument as the one for $k=1$. The only modification we need is to redefine the index sets $E_{1ST}^\lambda, E_{2ST}^\lambda$, $D^{\lambda_d,\lambda_u}_{ST}$ and $D_{cST}^\lambda$ in \eqref{eq:index_sets} to accommodate $k>1$. We remark that the technical arguments in the proof do not rely on the specific form of $E_{1ST}^\lambda, E_{2ST}^\lambda$, $D^{\lambda_d,\lambda_u}_{ST}$ and $D_{cST}^\lambda$. In particular, all arguments will go through for any finite time lag $k>1.$   
 
    To accommodate the case of $k>1$, we redefine
    \begin{align}\label{eq:index_sets_k}
    E_{1ST}^{\lambda}&=\bigcup_{i=1}^k\{(t,s):t=[T\lambda]+i, 1\leq s\leq S, \text{ each } s  \text{ repeats } (k-i+1)(1+|\mathcal{N}(s)|) \text{ times} \},\nonumber\\
    E_{2ST}^{\lambda}&=\bigcup_{i=1}^k\{(t,s):t=[T\lambda]-i+1, 1\leq s\leq S, \text{ each } s  \text{ repeats } (k-i+1)(1+|\mathcal{N}(s)|) \text{ times} \},\nonumber\\
    D^{\lambda_d,\lambda_u}_{ST}&=\bigcup_{i=0}^k\{(t,i,s_1,s_2): [T\lambda_d]+1\leq t, t+i \leq [T\lambda_u],  1\leq s_1 \leq S, s_2 \in s_1\cup \mathcal{N}(s_1), \text{if }i=0, s_1\neq s_2\},\nonumber\\
    D_{cST}^\lambda&= \bigcup_{i=0}^{k-1}\bigcup_{j=1}^{k-i} \{([T\lambda]-i,[T\lambda]+j,s_1,s_2): 1\leq s_1 \leq S, s_2 \in s_1\cup \mathcal{N}(s_1)\}.
    \end{align}
    Note that \eqref{eq:index_sets_k} reduces to \eqref{eq:index_sets} when $k=1.$ 

    In addition, in the proof of exact recovery for a true change-point, say $[T\lambda_1^o]$, we analyze the difference of two log-likelihoods that takes a generic form of
    \begin{align}\label{eq:loglik_comp_generic}
        \big[ L_{ST}(\hat\psi_1,\hat\lambda_0,\hat\lambda_1;\bfY)+L_{ST}(\hat\psi_2,\hat\lambda_1,\hat\lambda_2;\bfY) \big] - \big[L_{ST}(\hat\psi_1,\hat\lambda_0,\lambda_1^o;\bfY)+L_{ST}(\hat\psi_2,\lambda_1^o,\hat\lambda_2;\bfY)\big],
    \end{align}
    where $\hat\psi_1$ and $\hat\psi_2$ are the estimated parameters based on data between $[T\hat\lambda_0]+1$ to $[T\hat\lambda_1]$ and between $[T\hat\lambda_1]+1$ to $[T\hat\lambda_2]$, respectively. By symmetry, in the proof, we analyze \eqref{eq:loglik_comp_generic} for the case where the localization error is $[T\lambda_1^o]-[T\hat\lambda_1]=l\geq 1.$

    In particular, our analysis decomposes \eqref{eq:loglik_comp_generic} into three parts and analyzes each one individually. For the case of $k=1$, the decomposition of \eqref{eq:loglik_comp_generic} is algebraically simpler, where for any $[T\lambda_1^o]-[T\hat\lambda_1]=l\geq 1$, \eqref{eq:loglik_comp_generic} can be decomposed into
    \begin{align}\label{eq:loglik_comp_generic_k1}
        \eqref{eq:loglik_comp_generic}=&\sum_{(t,i,s_1,s_2) \in D_{ST}^{\hat\lambda_1,\lambda_1^o}} l(\hat\psi_2,y_{t,s_1},y_{t+i,s_2})-l(\hat\psi_1,y_{t,s_1},y_{t+i,s_2})\nonumber\\
        +&\sum_{(t,s)\in E_{2ST}^{\hat\lambda_1}}l(\hat\psi_1,y_{t,s})+\sum_{(t,s)\in E_{1ST}^{\hat\lambda_1}}l(\hat\psi_2,y_{t,s})-\sum_{(t_1,t_2,s_1,s_2)\in D_{cST}^{\hat\lambda_1}}l(\hat\psi_1,y_{t_1,s_1}, y_{t_2,s_2})\nonumber\\
        +&\sum_{(t_1,t_2,s_1,s_2)\in D_{cST}^{\lambda_1^o}}l(\hat\psi_2,y_{t_1,s_1}, y_{t_2,s_2})-\sum_{(t,s)\in E_{2ST}^{\lambda_1^o}}l(\hat\psi_1,y_{t,s})-\sum_{(t,s)\in E_{1ST}^{\lambda_1^o}}l(\hat\psi_2,y_{t,s}).
    \end{align}

    For $k>1$, the decomposition of \eqref{eq:loglik_comp_generic} is algebraically more complicated. Recall $[T\lambda_1^o]-[T\hat\lambda_1]=l\geq 1$. In particular, for the case where the localization error $l\geq k$, \eqref{eq:loglik_comp_generic_k1} still holds. However, when the localization error is small with $1\leq l< k$, we need to adjust the cross-over index set $D_{cST}^\lambda$ in \eqref{eq:loglik_comp_generic_k1} due to the fact that $[T\lambda_1^o]$ and $[T\hat\lambda_1]$ are within $k$ time lags. In particular, define two new index sets
    \begin{align*}
        D_{cST}^{\lambda,+l}&= \bigcup_{i=0}^{l-1}\bigcup_{j=1}^{k-i} \{([T\lambda]-i,[T\lambda]+j,s_1,s_2): 1\leq s_1 \leq S, s_2 \in s_1\cup \mathcal{N}(s_1)\},\\
        D_{cST}^{\lambda,-l}&= \bigcup_{i=0}^{k-1}\bigcup_{j=1}^{\min\{k-i,l\}} \{([T\lambda]-i,[T\lambda]+j,s_1,s_2): 1\leq s_1 \leq S, s_2 \in s_1\cup \mathcal{N}(s_1)\},
    \end{align*}
    We have that for $1\leq l<k$, the decomposition of \eqref{eq:loglik_comp_generic} takes the form    
    \begin{align}\label{eq:loglik_comp_generic_k2}
        \eqref{eq:loglik_comp_generic}=&\sum_{(t,i,s_1,s_2) \in D_{ST}^{\hat\lambda_1,\lambda_1^o}} l(\hat\psi_1,y_{t,s_1},y_{t+i,s_2})-l(\hat\psi_2,y_{t,s_1},y_{t+i,s_2})\nonumber\\
        +&\sum_{(t,s)\in E_{2ST}^{\hat\lambda_1}}l(\hat\psi_1,y_{t,s})+\sum_{(t,s)\in E_{1ST}^{\hat\lambda_1}}l(\hat\psi_2,y_{t,s})-\sum_{(t_1,t_2,s_1,s_2)\in D_{cST}^{\hat\lambda_1,-l}}l(\hat\psi_1,y_{t_1,s_1}, y_{t_2,s_2})\nonumber\\
        +&\sum_{(t_1,t_2,s_1,s_2)\in D_{cST}^{\lambda_1^o,+l}}l(\hat\psi_2,y_{t_1,s_1}, y_{t_2,s_2})-\sum_{(t,s)\in E_{2ST}^{\lambda_1^o}}l(\hat\psi_1,y_{t,s})-\sum_{(t,s)\in E_{1ST}^{\lambda_1^o}}l(\hat\psi_2,y_{t,s}).
    \end{align}

    We remark that for any finite $k>1$, the new term \eqref{eq:loglik_comp_generic_k2} can be analyzed in the exact same way as \eqref{eq:loglik_comp_generic_k1}. To summarize, the proof of $k=1$ can be readily modified to accommodate $k>1.$ $\hfill\square$
\end{remark}
}

\section{Proof for increasing domain asymptotics in Section \ref{sec:main}}\label{sec:proof_increasing_domain}
\subsection{Technical propositions and lemmas for the main theorems}

This section states some propositions and lemmas that will be used in the proof of the main results for the increasing domain setting in Section \ref{sec:main} of the main text. The essential theoretical results needed for the proof of main results are stated in Lemma \ref{thm2} and Lemma \ref{lil}. The proofs of these propositions and lemmas are deferred to Section \ref{proof.prop.lem}. 

The following moment inequality follows from Theorem 1 in Section 1.4 of \cite{Doukhan1994}.
\begin{proposition}\label{s_converge}
	Under Assumptions \ref{ass.countable}, \ref{ass.mom}($r$) and \ref{ass.mix}($r$) with $r>2$, there exists a constant $K$~(depending on $r$) and a constant $\epsilon>0$ such that, for any $\psi=(\xi, \theta), \theta\in\Theta(\xi)$ with a fixed model order $\xi \in \mathcal{M}$, any $\lambda \in [0,1]$ and $0\leq \lambda_d < \lambda_u \leq 1$, we have
	
	(i) for $a=0$,
	\begin{align*}
	&\mathbb E\left|\sum_{(t,s)\in E_{1ST}^{\lambda}} (l_{marg}^{[a]}\left(\psi; y_{t,s}\right)-\mathbb E(l_{marg}^{[a]}\left(\psi; y_{t,s}\right)))\right|^r \leq K S^{r/2},\\
	&\mathbb E\left|\sum_{(t,s)\in E_{2ST}^{\lambda}} (l_{marg}^{[a]}\left(\psi; y_{t,s}\right)-\mathbb E(l_{marg}^{[a]}\left(\psi; y_{t,s}\right)))\right|^r \leq K S^{r/2},\\
	&\mathbb E\left|\sum_{(t,s_1,s_2)\in D_{cST}^{\lambda}} (l_{pair}^{[a]}\left(\psi; y_{t,s_1},y_{t+1,s_2}\right)-
	\mathbb E(l_{pair}^{[a]}\left(\psi; y_{t,s_1},y_{t+1,s_2}\right)))\right|^r \leq K S^{r/2}\\
	&\mathbb E\left|\sum_{(t,s)\in E_{ST}^{\lambda_d,\lambda_u}} (l_{marg}^{[a]}\left(\psi; y_{t,s}\right)-\mathbb E(l_{marg}^{[a]}\left(\psi; y_{t,s}\right)))\right|^r \leq K ((\lambda_u-\lambda_d)TS)^{r/2},\\
	&\mathbb E\left|\sum_{(t,i,s_1,s_2)\in D^{\lambda_d,\lambda_u}_{ST}}(l_{pair}^{[a]}\left(\psi; y_{t,s_1},y_{t+i,s_2}\right)-
	\mathbb E(l_{pair}^{[a]}\left(\psi; y_{t,s_1},y_{t+i,s_2}\right)))\right|^r \leq K ((\lambda_u-\lambda_d)TS)^{r/2},
	\end{align*}
    \indent(ii) for $a=1,2$, the above inequalities hold with $r$ replaced by $2+\epsilon$, 
	where $a=0,1,2$ in $[a]$ stands for the $a$th derivative w.r.t. $\theta$.
\end{proposition}

Based on Proposition \ref{s_converge}, the following proposition gives a strong uniform law of large numbers~(ULLN) for the marginal and pairwise loglikelihood functions.
\begin{proposition}\label{s_converge2}
	Under the conditions of Proposition \ref{s_converge}, for any $\psi=(\xi, \theta), \theta\in\Theta(\xi)$ with a fixed model order $\xi \in \mathcal{M}$ and any $0\leq \lambda_d < \lambda_u \leq 1$, we have 
	\begin{align*}
	&\sup_{\theta \in \Theta}\left|\frac{1}{S([\lambda_uT]-[\lambda_dT])}\sum_{(t,s)\in E_{ST}^{\lambda_d,\lambda_u}} (l_{marg}^{[a]}\left({\psi}; y_{t,s}\right)-\mathbb E(l_{marg}^{[a]}\left({\psi}; y_{t,s}\right)))\right|\longrightarrow 0 ,\\
	&\sup_{\theta \in \Theta}\left|\frac{1}{S([\lambda_uT]-[\lambda_dT])}\sum_{(t,i,s_1,s_2)\in D^{\lambda_d,\lambda_u}_{ST}}(l_{pair}^{[a]}\left(\psi; y_{t,s_1},y_{t+i,s_2}\right)-
	\mathbb E(l_{pair}^{[a]}\left(\psi; y_{t,s_1},y_{t+i,s_2}\right)))\right|\longrightarrow 0,
	\end{align*}
almost surely, where $a=0,1,2$ in $[a]$ stands for the $a$th derivative w.r.t. $\theta$.
\end{proposition}

Proposition \ref{edge_effect} controls the size of the marginal loglikelihood used for compensating the edge effect and the size of the cross-over pairwise loglikelihood between different segments.
\begin{proposition}\label{edge_effect}
Under the conditions of Proposition \ref{s_converge}, for any $\psi=(\xi, \theta), \theta\in\Theta(\xi)$ with a fixed model order $\xi \in \mathcal{M}$, we have
\begin{align*}
&\frac{1}{ST}\sup_{\lambda\in[0,1]} \sup_{\theta\in\Theta(\xi)}\sum_{(t,s)\in E_{2ST}^\lambda}\left|l_{marg}\left(\psi; y_{t, s}\right)\right| \longrightarrow 0,\\
&\frac{1}{ST}\sup_{\lambda\in[0,1]} \sup_{\theta\in\Theta(\xi)}\sum_{(t,s_1,s_2)\in D_{cST}^{\lambda}}\left|l_{pair}\left(\psi; y_{t, s_1}, y_{t+1, s_2}\right)\right| \longrightarrow 0, \text{ almost surely.}
\end{align*}
\end{proposition}

The following proposition establishes the uniform convergence of $L_{ST}^{(j)}(\psi, \lambda_d,\lambda_u; \bfX_j^o)$ in \eqref{PL.prop}.
\begin{proposition} \label{partdata} Under the conditions of Proposition \ref{s_converge}, for all $j=1,\ldots, m_o+1$ and any fixed model order $\xi$, we have 
	\begin{equation}
	\sup_{\lambda_d,\lambda_u}\sup_{\theta\in\Theta(\xi)}\frac{1}{ST}\left| L_{ST}^{(j)[a]}\{(\xi,\theta),\lambda_d,\lambda_u; \bfX_j^o\}-(\lambda_u-\lambda_d)\thickbar{L}_{ST}^{(j)[a]}\{(\xi,\theta)\}\right| \longrightarrow 0  \label{prop1.eq} \,,\\
	\end{equation} 
	almost surely, where the supremum notation is defined in \eqref{supdef} and $a=0,1,2$ in $[a]$ stands for the $a$th derivative w.r.t. $\theta.$
\end{proposition}

In Proposition \ref{partdata}, $\lambda_d$ and $\lambda_u$ are restricted to $[0,1]$, i.e.,  
we search for the maximum value of the loglikelihood using data within the stationary piece.
An extension has to be made so that $\lambda_d$ and $\lambda_u$ are allowed to be slightly outside $[0,1]$ 
since the $j$th estimated segment may cover part of the $(j-1)$th or the $(j+1)$th true segment. 
For any real-valued function $g_{ST}(\lambda_d,\lambda_u)$ on $\mathbb{R}^2$, we use
\begin{equation} \label{supfdef}
\sup_{\underline{\lambda_d},\overline{\lambda_u}} g_{ST}(\lambda_d,\lambda_u) \longrightarrow 0 \,,
\end{equation}
to denote the supremum of $g_{ST}(\lambda_d,\lambda_u)$ over 
all $\lambda_{d}, \lambda_{u}$ satisfying 
$-h_T < \lambda_d < \lambda_u < 1+ k_T$ and $\lambda_u-\lambda_d > \epsilon_\lambda$, where $h_T$ and $k_T$ are pre-specified positive-valued sequences converging to $0$ as $T \rightarrow \infty$. 

\begin{proposition} \label{proprep}
	Proposition \ref{partdata} holds when $\sup_{\lambda_d,\lambda_u}$ is replaced by $\sup_{\underline{\lambda_d},\overline{\lambda_u}}$.
\end{proposition}

The following proposition shows the consistency of the parameter estimation within the $j$th segment. 
\begin{proposition} \label{MPLE}
	Let $\psi_j^o = (\xi_j^o,\theta_j^o)$ be the true model parameter for the $j$th segment. Suppose that a model $\xi_j$ is specified for estimation. Define
	\begin{align}\label{eq:hat.theta}
	\hat{\theta}_{ST} &\equiv \hat{\theta}_{ST}^{(j)}(\lambda_d,\lambda_u)=\arg\max_{\theta_j\in\Theta_j(\xi_j)}L_{ST}^{(j)}\{(\xi_j,\theta_j),\lambda_d,\lambda_u;\bfX_j^o\} \text{ ,} \\\
	\theta_j^* &= \arg\max_{\theta_j\in\Theta_j(\xi_j)}\thickbar{L}_{ST}^{(j)}\{(\xi_j,\theta_j)\}\text{ .}\nonumber
	\end{align}
    Under the conditions of Proposition \ref{s_converge} and Assumption \ref{ass.model.id}, for all $j=1,\ldots, m_o+1$, we have 
	\begin{align} \label{suppartcon}
	\sup_{\underline{\lambda_d},\overline{\lambda_u}} \left|\hat{\theta}_{ST}^{(j)}(\lambda_d,\lambda_u)-\theta_j^*\right|\longrightarrow 0 \,,
	\end{align}
	almost surely.
\end{proposition}

The following lemma provides a preliminary result for the convergence of the change-point estimators and the model order estimators when the number of change-points is known.

\begin{lemma} 
	\label{known}
	Let $\bfY$ be observations from a piecewise stationary random field specified by the vector $(m_o,\Lambda^o,\Psi^o)$ and Assumptions \ref{ass.countable}, \ref{ass.mom}($r$), \ref{ass.model.id} and \ref{ass.mix}($r$) hold with some $r>2$. Suppose that the number of change-points $m_o$ is known. 
	We estimate the change-points and model parameters by 
	\begin{align}
	\{\hat{\Lambda}_{ST},\hat{\Psi}_{ST}\}=\arg\min_{\substack{ \Psi\in \mathcal{M}^{m_o+1}\\ \Lambda\in A_{\epsilon_\lambda}^{m_o}}}\frac{1}{ST}\mathrm{CLMDL}(m_o,\Lambda,\Psi)\,,
	\end{align}
	where $A_{\epsilon_\lambda}^{m_o}$ is defined in \eqref{A}. Then, $\hat{\Lambda}_{ST}\longrightarrow\Lambda^o$ almost surely, and for each 
	segment the estimated model does not underestimate the true model, i.e., $\hat{\xi}_{j}\geq \xi^o_{j}$.
\end{lemma}

The following lemma states some properties of the proposed procedure when the number of change-points is unknown.

\begin{lemma} \label{coro1}
	Under the conditions of Lemma \ref{known}, if the number of change-points $m$ is unknown and is estimated from the data using \eqref{minMDL}, then we have the following.
	
	(a) The number of change-points cannot be underestimated. In other words, $\hat{m}\geq m_o$ for sufficiently large $(S,T)$, almost surely.
	
	(b) When $\hat{m}\geq m_o$, then $\Lambda^o$ must be a subset of the limit points of $\hat{\Lambda}_{ST}$, i.e., for any  $\epsilon>0$ and $\lambda_j^o\in\Lambda^o$, there exists a $\hat{\lambda}_k\in \hat{\Lambda}_{ST}$ such that $|\lambda_j^o-\hat{\lambda}_k| < \epsilon$ for sufficiently large $(S,T)$, almost surely. 
	
	(c) The order of the model in each segment cannot be underestimated.
	
\end{lemma}

Lemma \ref{coro1} is refined in the following Lemma \ref{thm2}, where an exact recovery property of the change-point estimator is established. 
\begin{lemma}
	\label{thm2}
	Under the conditions of Lemma \ref{known}, with $(\hat{m},\hat{\Lambda}_{ST})$ estimated from the data using \eqref{minMDL}, we have that, for each $j=1,\ldots,m_o$, there exists a 
	$\hat{\lambda}_{i_j}\in \hat{\Lambda}_{ST}, 1\leq i_j \leq \hat{m}$, such that
	$$\mathbb P\left(\left|[T\lambda_j^o] - [T\hat{\lambda}_{i_j}]\right| = 0\right) \longrightarrow 1$$ as $(S,T)\longrightarrow \infty$.
\end{lemma}

The following result gives the asymptotic order of the composite likelihood estimator for the model parameters. 
\begin{lemma}\label{lil}
	Under the conditions of Lemma \ref{known} and Assumption \ref{ass.mixtime}, we have for each $j=1,\ldots,m_o+1$,
	\begin{align} \label{liltheta}
	\sup_{\lambda_d,\lambda_u}\left|\hat{\theta}_{ST}^{(j)}(\lambda_d,\lambda_u)-\theta_j^*\right|=O_p\left\{(ST)^{-1/2} \right\}\,,
	\end{align}
	where $\theta_j^*$ is the pseudo-true parameter vector and $\hat{\theta}_{ST}^{(j)}(\lambda_d,\lambda_u)$ is defined in 
	\eqref{eq:hat.theta} with $\xi_j=\xi_j^*$.
\end{lemma}

\subsection{Proofs of main theorems}
\begin{proof}[Proof of Proposition \ref{prop.minCL}]
	The proof of Proposition \ref{prop.minCL} follows the same argument as the proof of Theorem \ref{unknownprob} by noticing that \eqref{exponentialinequality} still holds when replacing $-(\log S+ \log T)$ by 0. Thus we omit its proof and refer the readers to the proof of Theorem \ref{unknownprob}.
\end{proof}

\begin{proof}[Proof of Theorem \ref{unknownprob}]
	The first part of the Theorem \ref{unknownprob} follows directly from Lemma \ref{thm2}. We give the proof for the second part of Theorem \ref{unknownprob}. Following Lemmas \ref{coro1} and \ref{thm2}, it suffices to prove that for any integer $d=1,\ldots,M-m_o$, 
	\begin{align} \label{argminprob}
	\min_{\substack{\Psi\in\mathcal{M}^{m_o+d+1},\Lambda\in A_{\epsilon_\lambda}^{(m_o+d)}\\ {\Lambda}^o \subset \Lambda}}\left\{ \frac{1}{ST}{\mathrm{CLMDL}}(m_o+d,\Lambda,\Psi)\right\}-\frac{1}{ST}{\mathrm{CLMDL}}(m_o,{\Lambda}^o,\Psi^o)\,,
	\end{align}
	is positive with probability going to 1. We prove the result for $d=1$, the proof for $d>1$ is the same. Denote $\hat{\Lambda}_{ST}=(\hat{\lambda}_1,\ldots,\hat{\lambda}_{m_o+1})$ as the 
	minimizer of the first term in \eqref{argminprob}. Note that ${\Lambda}^{o}\subset \hat{\Lambda}_{ST}$ by construction. By a Taylor's expansion of the loglikelihood functions, the quantity 
	in \eqref{argminprob} can be expressed as
	\begin{align}
	&C_1' - C_2' \nonumber\\
	&+ \frac{1}{ST}\left[\sum_{j=1}^{m_o+1}L_{ST}^{(j)}\{(\xi_j^o,\theta_j^o),{\lambda}_{j-1}^o,{\lambda}_j^o; \bfY\}-\sum_{l=1}^{m_o+d+1}L_{ST}^{(j)}\{(\hat{\xi}_l,\theta_l^*),\hat{\lambda}_{l-1},\hat{\lambda}_l; \bfY\}\right] \label{sum2}\\
	& - \frac{1}{2}\sum_{l=1}^{m_o+d+1}(\hat{\theta}_{ST}^{(l)}-\theta_l^*)^\T \frac{1}{ST} L_{ST}^{''(l)}\{(\hat{\xi}_l,\theta_l^+),\hat{\lambda}_{l-1},\hat{\lambda}_l; \bfY\}(\hat{\theta}_{ST}^{(l)}-\theta_l^*) \label{taylor2sum},
	\end{align}
	where $C_1'-C_2'$ is positive and of order $O(\log ST /ST)$, and $|\theta_l^+-\theta_l^*|<|\hat{\theta}_{ST}^{(l)}-\theta_l^*|$.
    
    By Lemma \ref{lil}, $\hat{\theta}_{ST}^{(j)}- \theta_j^* = O_p\{(ST)^{-1/2}\}$. Together with Proposition \ref{partdata}, the summation in \eqref{taylor2sum} is of order $O_p\{(ST)^{-1}\}$. Note that the quantity $C_1'-C_2'$, 
    which is of order $O(\log ST/ST)$, 
    dominates the expression.

    Since for all segments the model order cannot be underestimated for sufficiently large $(S,T)$, by Assumption \ref{ass.model.id} the bracketed quantity in \eqref{sum2} is almost exactly zero. Due to the presence of the one extra pseudo-change point, say $\hat{\lambda}_l$, \eqref{sum2} is
	\begin{align}
	\frac{1}{ST}\left(\sum_{D_{cST}^{\hat{\lambda}_l}}l_{pair}(\theta_{l}^o; y_{t,s_1}^{(l)}, y_{t+1,s_2}^{(l)})- \sum_{E_{2ST}^{\hat{\lambda}_l}} l_{marg}({\theta}_{l}^o;y_{t,s}^{(l)})
	-\sum_{E_{1ST}^{\hat{\lambda}_l}} l_{marg}({\theta}_{l}^o;y_{t,s}^{(l)})\right),\label{cross-overterm}
	\end{align}
 which is the difference of the dropped pairwise likelihood due to the presence of $\hat{\lambda}_l$ and the corresponding compensating marginal likelihood. Note that by Assumption \ref{ass.model.id}, WLOG, we can assume the model orders of different terms in \eqref{cross-overterm} are the same and are at $\xi_l^o$. Thus we skip the model order $\xi_l^o$ in \eqref{cross-overterm} and only use $\theta_l^o$.
    
	Denote
	\begin{align*}
	B_{ \lambda ST}=
	\left(\sum_{D_{cST}^{\lambda}}l_{pair}(\theta_{l}^o; y_{t,s_1}^{(l)}, y_{t+1,s_2}^{(l)})- \sum_{E_{2ST}^{\lambda}} l_{marg}({\theta}_{l}^o;y_{t,s}^{(l)})
	-\sum_{E_{1ST}^{\lambda}} l_{marg}({\theta}_{l}^o;y_{t,s}^{(l)})\right).
	\end{align*}
	If we can show that
	\begin{align}
	\mathbb P\left(\inf_{\lambda^o_{l-1}+\epsilon_\lambda<\lambda<\lambda^o_l-\epsilon_\lambda}B_{\lambda ST}< -(\log S + \log T)\right)\longrightarrow 0,
	\label{exponentialinequality}
	\end{align}
	then we know the quantity in \eqref{argminprob} is indeed positive with probability going to $1$.

	By union bound and stationarity of the data, we have that
	\begin{align*}
	&\mathbb P\left(\inf_{\lambda^o_{l-1}+\epsilon_\lambda<\lambda<\lambda^o_l-\epsilon_\lambda}B_{ \lambda ST}< -(\log S + \log T)\right)\\
	\leq &T(\lambda_l^o-\lambda_{l-1}^o)\mathbb P\left(B_{ \lambda ST}< -(\log S + \log T)\right)\\
	\leq &TP\left(B_{ \lambda ST}-\mathbb E(B_{ \lambda ST}) <-\mathbb E(B_{ \lambda ST})-\log S -\log T \right)\\
	\leq &TP\left(\left|B_{ \lambda ST}-\mathbb E(B_{ \lambda ST})\right| >C_1\cdot S \right) \leq C_2\cdot T\cdot S^{-r/2},
	\end{align*}
	where $C_1$ and $C_2$ denote some positive numbers, the second to last inequality follows from information inequality, and the last inequality follows from Markov inequality and the moment inequality in Proposition \ref{s_converge}. Thus, \eqref{exponentialinequality} holds as long as $T\cdot S^{-r/2}\to 0$.
	
	When the number of change-points is correctly estimated but some $\xi_j^*$ is a strictly bigger model than $\xi_j^o$, we have that \eqref{sum2} is exactly zero, and the same argument shows that \eqref{taylor2sum} is of order $O_p(1/ST)$ and $C_1'-C_2'$ is positive and of order $O(\log ST /ST)$ respectively. Thus, the same consistency conclusion follows. 
	
	Lastly, the convergence rate follows from Lemma \ref{thm2} and the weak consistency of the estimated number of change-points. Hence the proof of \eqref{weak.consist} is complete. 
\end{proof}

\begin{proof}[Proof of Theorem \ref{finerate}]
	The proof of Theorem \ref{finerate} follows the same logic as that of Theorem \ref{unknownprob} but with a finer rate of the union bound. Following Lemmas \ref{coro1} and \ref{thm2}, it suffices to prove that for any integer $d=1,\ldots,M-m_o$,  
	\begin{align} \label{argminprob2}
	\min_{\substack{\psi_j\in\mathcal{M},\Lambda\in A_{\epsilon_\lambda}^{(m_o+d)}\\ {\Lambda}^o \subset \Lambda}}\left\{ \frac{1}{T}{\mathrm{CLMDL}}(m_o+d,\Lambda,\Psi)\right\}-\frac{1}{T}{\mathrm{CLMDL}}(m_o,{\Lambda}^o,\Psi^o)\,,
	\end{align}
	is positive with probability going to 1. We prove the result for $d=1$, and the proof for $d>1$ follows similarly. Denote $\hat{\Lambda}_{ST}=(\hat{\lambda}_1,\ldots,\hat{\lambda}_{m_o+1})$ as the 
	minimizer of the first term in \eqref{argminprob2}. Note that ${\Lambda}^{o}\subset \hat{\Lambda}_{ST}$ by construction. By a Taylor's expansion on the log-likelihood functions, the quantity 
	in \eqref{argminprob2} can be expressed as
	\begin{align}
	&C_1' - C_2' \nonumber\\
	&+ \frac{1}{ST}\left[\sum_{j=1}^{m_o+1}L_{ST}^{(j)}\{(\xi_j^o,\theta_j^o),{\lambda}_{j-1}^o,{\lambda}_j^o; \bfY\}-\sum_{l=1}^{m_o+d+1}L_{ST}^{(j)}\{(\hat{\xi}_l,\theta_l^*),\hat{\lambda}_{l-1},\hat{\lambda}_l; \bfY\}\right] \label{sum22}\\
	& - \frac{1}{2}\sum_{l=1}^{m_o+d+1}(\hat{\theta}_{ST}^{(l)}-\theta_l^*)^\T \frac{1}{ST} L_{ST}^{''(l)}\{(\hat{\xi}_l,\theta_l^+),\hat{\lambda}_{l-1},\hat{\lambda}_l; \bfY\}(\hat{\theta}_{ST}^{(l)}-\theta_l^*) \label{taylor2sum2}\,,
	\end{align}
	where $C_1'-C_2'$ is positive and of order $O(\log ST /ST)$, and $|\theta_l^+-\theta_l^*|<|\hat{\theta}_{ST}^{(l)}-\theta_l^*|$. Since for all segments the model order cannot be underestimated for sufficiently large $(S,T)$, the two terms in the bracketed quantity in \eqref{sum22} are almost identical. Due to the presence of the extra pseudo-change point, say $\hat{\lambda}_l$,  \eqref{sum22} is
 \begin{align}\label{cross-overterm2}
	\frac{1}{ST}\left(\sum_{D_{cST}^{\hat{\lambda}_l}}l_{pair}(\theta_{l}^o; y_{t,s_1}^{(l)}, y_{t+1,s_2}^{(l)})- \sum_{E_{2ST}^{\hat{\lambda}_l}} l_{marg}({\theta}_{l}^o;y_{t,s}^{(l)})
	-\sum_{E_{1ST}^{\hat{\lambda}_l}} l_{marg}({\theta}_{l}^o;y_{t,s}^{(l)})\right).
	\end{align}
	If there is no temporal dependence for the true data, then \eqref{cross-overterm2} is exactly zero; otherwise it is not. By Lemma \ref{lil}, $\hat{\theta}_{ST}^{(j)}(\lambda_d,\lambda_u)- \theta_j^o = O_p\{(ST)^{-1/2}\}$. Together with Proposition \ref{partdata}, the summation in \eqref{taylor2sum2} is of order $O_p\{(ST)^{-1}\}$. Note that the quantity $C_1'-C_2'$, 
	which is of order $O(\log ST/ST)$, 
	dominates the expression.
	
	For each $\lambda$, $T$ and $S$, denote
	\begin{align*}
	B_{ \lambda ST}=
	\left(\sum_{D_{cST}^{\lambda}}l_{pair}(\theta_{l}^o; y_{t,s_1}^{(l)}, y_{t+1,s_2}^{(l)})- \sum_{E_{2ST}^{\lambda}} l_{marg}({\theta}_{l}^o;y_{t,s}^{(l)})
	-\sum_{E_{1ST}^{\lambda}} l_{marg}({\theta}_{l}^o;y_{t,s}^{(l)})\right).
	\end{align*}
	If we can show that, for any $\epsilon>0$,
	\begin{align}\label{exponentialinequality2}
	\mathbb P\left(\inf_{\lambda^o_{l-1}+\epsilon_\lambda<\lambda<\lambda^o_l-\epsilon_\lambda}B_{ \lambda ST}< -\epsilon(\log S + \log T)\right) &\longrightarrow 0.
	\end{align}
	Then we know the quantity in \eqref{argminprob2} is indeed positive with probability going to $1$. Note that we have
		\begin{align*}
			 & \mathbb P\left(\inf_{\lambda^o_{l-1}+\epsilon_\lambda<\lambda<\lambda^o_l-\epsilon_\lambda}B_{ \lambda ST}< -\epsilon(\log S + \log T)\right)\\
			 \leq & \mathbb P\left(\cup_{\lambda^o_{l-1}+\epsilon_\lambda<\lambda<\lambda^o_l-\epsilon_\lambda} \{B_{ \lambda ST}< -\epsilon(\log S + \log T)\}\right)	\\
			 \leq & \sum_{\lambda^o_{l-1}+\epsilon_\lambda<\lambda<\lambda^o_l-\epsilon_\lambda} \mathbb P\left(B_{ \lambda ST}< -\epsilon(\log S + \log T)\right)\,.
		\end{align*}	
By the strong invariance principles for $\alpha$-mixing random fields in \cite{BerkesMorrow1981}, we can refine the field on a richer probability space together with a field of i.i.d. centered Gaussian random vectors such that for some $\gamma>0$, we have
		$$
		\|B_{ \lambda ST}-\mathbb{E}(B_{ \lambda ST})-N_{ \lambda ST}\|=O(|S|^{\frac{1}{2}-\gamma})\,,
		$$
		where $N_{ \lambda ST}=\sum_{D_{cST}^{\lambda}} N_{t,s_1,s_2}$ and $N_{t,s_1,s_2}$ are i.i.d. centered Gaussian random vectors.
		
		Consider 	
			\begin{eqnarray*}
			B_{ \lambda ST} & = & 
			\sum_{D_{cST}^{\lambda}}l_{pair}(\theta_{l}^o; y_{t,s_1}^{(l)}, y_{t+1,s_2}^{(l)})- \sum_{E_{2ST}^{\lambda}} l_{marg}({\theta}_{l}^o;y_{t,s}^{(l)})
			-\sum_{E_{1ST}^{\lambda}} l_{marg}({\theta}_{l}^o;y_{t,s}^{(l)})\\
			& = & 			
			\sum_{D_{cST}^{\lambda}}\left(l_{pair}(\theta_{l}^o; y_{t,s_1}^{(l)}, y_{t+1,s_2}^{(l)}) - l_{marg}({\theta}_{l}^o;y_{t,s_1}^{(l)})
			- l_{marg}({\theta}_{l}^o;y_{t+1,s_2}^{(l)})\right)\\
			& = & \sum_{D_{cST}^{\lambda}} B_{t,s_1,s_2} \,.
			\end{eqnarray*}
		 Under the regular lattice assumption, the random field $\{N_{t,s_1,s_2}: (t,s_1,s_2) \in D_{cST}^{\lambda}\}$ approximates the random field $\left\{B_{t,s_1,s_2}: (t,s_1,s_2) \in D_{cST}^{\lambda}\right\}$. Also, let $\Sigma_D={\rm Var}(N_{t,s_1,s_2})$, then $|D_{cST}^{\lambda}| \Sigma_D$ is the variance of $B_{ \lambda ST}$. 
Note that $\Sigma_D$ can be regarded as the long run variance term. Next, we have
		\begin{align*}
			 & \mathbb P\left(B_{ \lambda ST}< -\epsilon(\log S + \log T)\right)\\
			 = & \mathbb P\left(B_{ \lambda ST}-\mathbb{E}(B_{ \lambda ST})< -\epsilon(\log S + \log T)-\mathbb{E}(B_{ \lambda ST})\right)\\
			 \leq & \mathbb P\left( N_{ \lambda ST}< -\epsilon(\log S + \log T)-\mathbb{E}(B_{ \lambda ST})-O(|S|^{\frac{1}{2}-\gamma})\right)\,.
		\end{align*}
		Since $\mathbb{E}(B_{ \lambda ST})=O(|D_{cST}^{\lambda}|)=O(|S|)>0$, which dominates the right-hand side of the inequality if $\log T/S \rightarrow 0$, we can use the tail probability bound for i.i.d. Gaussian random variables
		\begin{align*}
			\mathbb P\left(|N_{ \lambda ST}| > x\right) \leq \frac{\sqrt{2|D_{cST}^{\lambda}| \Sigma_D}}{x\sqrt{\pi}} \exp\left(-\frac{x^2}{2|D_{cST}^{\lambda}| \Sigma_D}\right)\,.
		\end{align*}
		Hence, the result follows for $\log T/S \rightarrow 0$. 
		Lastly, the convergence rate follows from Lemma \ref{thm2} and the weak consistency of the estimated number of change-points. Hence the proof of \eqref{weak.consist2} is complete.
\end{proof}

\begin{proof}[Proof of Corollary \ref{consistency_relaxIS}]
	It is easy to see that Lemmas \ref{known} and \ref{coro1} still hold for process $\bfY$ with relaxed IS condition quantified by the additional Assumption \ref{ass.relaxIS}. With the additional condition that $S=O(T)$, we can easily prove that Lemma \ref{lil} also holds for $\bfY$.
	
	The exact recovery property in Lemma \ref{thm2} only holds for $\bfY^*$ but not for $\bfY$. Instead, we have that 
	$$[T\lambda_j^o] - [T\hat{\lambda}_{i_j}]= O_p(1)\,.$$ 
	The proof follows the same argument as Lemma \ref{thm2} and thus is omitted. The intuition of why we cannot achieve exact recovery for $[T{\Lambda}^o]$ but instead a $O_p(1)$ error is that the extra $O_p(S)$ term caused by the presence of $Z$ around the true change-points is able to compensate/counter the information inequality loss caused by $\left|[T\lambda_j^o] - [T\hat{\lambda}_{i_j}]\right|$ unless $\left|[T\lambda_j^o] - [T\hat{\lambda}_{i_j}]\right|$ is larger than a certain threshold $C$. The value of the threshold $C$ needed depends on the value of the $O_p(S)$ terms and thus leads to a $O_p(1)$ error term of $T\hat{\lambda}_{i_j}$.	Together, this proves the first part of Corollary \ref{consistency_relaxIS}.
	
	We give the proof for the second part of Corollary \ref{consistency_relaxIS}. Define $T\Lambda^o\pm C=[T\lambda^o-C, T\lambda^o+C]$. Following Lemma \ref{coro1} and the result that $[T\lambda_j^o] - [T\hat{\lambda}_{i_j}]= O_p(1)$, it suffices to prove that for any constant $C>0$ and integer $d=1,\ldots,M-m_o$,
	\begin{align} \label{argminprob_dependent}
	&\min_{\substack{\psi_j\in\mathcal{M},
			T\Lambda_1 \subset T\Lambda^o\pm C, T\Lambda_2 \not\subset T\Lambda^o\pm C\nonumber\\
			\Lambda=\Lambda_1\cup\Lambda_2\in A_{\epsilon_\lambda}^{(m_o+d)} }}\left\{ \frac{1}{ST}{\mathrm{CLMDL}}(m_o+d,\Lambda_1\cup\Lambda_2,\Psi)\right\}\\
	-&\min_{\substack{\psi_j\in\mathcal{M},
			T\Lambda_1 \subset T\Lambda^o\pm C}}\frac{1}{ST}{\mathrm{CLMDL}}(m_o,{\Lambda}_1,\Psi^o)>0\,,
	\end{align}
	with probability going to 1. We prove the result for $d=1$, the proof for $d>1$ is the same.
	
	Note that \eqref{argminprob_dependent} is implied by
	\begin{align}\label{argminprob_dependent1}
	\min_{\substack{\psi_j\in\mathcal{M},
			T\Lambda_1 \subset T\Lambda^o\pm C, T\Lambda_2 \not\subset T\Lambda^o\pm C\\
			\Lambda=\Lambda_1\cup\Lambda_2\in A_{\epsilon_\lambda}^{(m_o+d)} }}\left\{ \frac{1}{ST}{\mathrm{CLMDL}}(m_o+d,\Lambda_1\cup\Lambda_2,\Psi)\right\}-\frac{1}{ST}{\mathrm{CLMDL}}(m_o,{\Lambda}_1,\Psi^o)>0\,.
	\end{align}
	
	Thus, we work with \eqref{argminprob_dependent1}. Denote $\hat{\Lambda}_{ST}=\hat{\Lambda}_1\cup\hat{\Lambda}_2=(\hat{\lambda}_1,\ldots,\hat{\lambda}_{m_o+1})$ as the 
	minimizer of \eqref{argminprob_dependent1}. WLOG, we assume that $\hat{\Lambda}_2=\{\hat{\lambda}_{m_o+1}\}$ since $d=1$. Note that $T\hat{\Lambda}_1=\{T(\hat{\lambda}_1,\ldots,\hat{\lambda}_{m_o})\}\subset T\Lambda^o\pm C$ by construction.
	
	By a Taylor expansion of the loglikelihood functions, the quantity 
	in \eqref{argminprob_dependent1} can be expressed as
	\begin{align}
	&C_1' - C_2' \nonumber\\
	&+ \frac{1}{ST}\left[\sum_{j=1}^{m_o+1}L_{ST}^{(j)}\{(\xi_j^o,\theta_j^o),\hat{\lambda}_{j-1},\hat{\lambda}_j; \bfY\}-\sum_{l=1}^{m_o+d+1}L_{ST}^{(j)}\{(\hat{\xi}_l,\theta_l^*),\hat{\lambda}_{l-1},\hat{\lambda}_l; \bfY\}\right] \label{sum2_dependent}\\
	& - \frac{1}{2}\sum_{l=1}^{m_o+d+1}(\hat{\theta}_{ST}^{(l)}-\theta_l^*)^\T \frac{1}{ST} L_{ST}^{''(l)}\{(\hat{\xi}_l,\theta_l^+),\hat{\lambda}_{l-1},\hat{\lambda}_l; \bfY\}(\hat{\theta}_{ST}^{(l)}-\theta_l^*) \label{taylor2sum_dependent},
	\end{align}
	where $C_1'-C_2'$ is positive and of order $O(\log ST /ST)$, and $|\theta_l^+-\theta_l^*|<|\hat{\theta}_{ST}^{(l)}-\theta_l^*|$.
	
	By Lemma \ref{lil}, $\hat{\theta}_{ST}^{(j)}- \theta_j^* = O_p\{(ST)^{-1/2}\}$. Together with Proposition \ref{partdata}, the summation in \eqref{taylor2sum_dependent} is of order $O_p\{(ST)^{-1}\}$. Note that the quantity $C_1'-C_2'$, 
	which is of order $O(\log ST/ST)$, 
	dominates the expression.

	Since for all segments the model order cannot be underestimated for sufficiently large $(S,T)$, by Assumption \ref{ass.model.id} the bracketed quantity in \eqref{sum2_dependent} is almost exactly zero. Due to the presence of the one extra pseudo-change point, say $\hat{\lambda}_l$~(from above, $l=m_o+1$), \eqref{sum2_dependent} is
	\begin{align}
	\frac{1}{ST}\left(\sum_{D_{cST}^{\hat{\lambda}_l}}l_{pair}(\theta_{l}^o; y_{t,s_1}^{(l)}, y_{t+1,s_2}^{(l)})- \sum_{E_{2ST}^{\hat{\lambda}_l}} l_{marg}({\theta}_{l}^o;y_{t,s}^{(l)})
	-\sum_{E_{1ST}^{\hat{\lambda}_l}} l_{marg}({\theta}_{l}^o;y_{t,s}^{(l)})\right).\label{cross-overterm_dependent}
	\end{align}
	Note that by Assumption \ref{ass.model.id}, WLOG, we can assume the model orders of different terms in \eqref{cross-overterm_dependent} are the same and are at $\xi_l^o$. Thus we skip the model order $\xi_l^o$ in \eqref{cross-overterm_dependent} and only use $\theta_l^o$.
	
	Denote
	\begin{align*}
	B_{ \lambda ST}=
	\left(\sum_{D_{cST}^{\lambda}}l_{pair}(\theta_{l}^o; y_{t,s_1}^{(l)}, y_{t+1,s_2}^{(l)})- \sum_{E_{2ST}^{\lambda}} l_{marg}({\theta}_{l}^o;y_{t,s}^{(l)})
	-\sum_{E_{1ST}^{\lambda}} l_{marg}({\theta}_{l}^o;y_{t,s}^{(l)})\right).
	\end{align*}
	If we can show that
	\begin{align}
	\mathbb P\left(\inf_{\lambda^o_{l-1}+\epsilon_\lambda<\lambda<\lambda^o_l-\epsilon_\lambda}B_{\lambda ST}< -(\log S + \log T)\right)\longrightarrow 0,
	\label{exponentialinequality_dependent}
	\end{align}
	then we know the quantity in \eqref{argminprob_dependent} is indeed positive with probability going to $1$.
	
	Note that by the definition of the identification condition $A_{\epsilon_\lambda}^{(m_o+d)}$, asymptotically, as $T\to \infty$, we can treat the observed process $\{y_t|T(\lambda^o_{l-1}+\epsilon_\lambda)<t<[T(\lambda^o_l-\epsilon_\lambda)]\}$ as the stationary spatio-temporal process $\{y_t^*|T(\lambda^o_{l-1}+\epsilon_\lambda)<t<[T(\lambda^o_l-\epsilon_\lambda)]\}$ with IS, since it is as least $\epsilon_\lambda T$ away from the true change-point. Thus, by union bound and stationarity of the data, we have that
	\begin{align*}
	&\mathbb P\left(\inf_{\lambda^o_{l-1}+\epsilon_\lambda<\lambda<\lambda^o_l-\epsilon_\lambda}B_{ \lambda ST}< -(\log S + \log T)\right)\\
	\leq &T(\lambda_l^o-\lambda_{l-1}^o)\mathbb P\left(B_{ \lambda ST}< -(\log S + \log T)\right)\\
	\leq &TP\left(B_{ \lambda ST}-\mathbb E(B_{ \lambda ST}) <-\mathbb E(B_{ \lambda ST})-\log S -\log T \right)\\
	\leq &TP\left(\left|B_{ \lambda ST}-\mathbb E(B_{ \lambda ST})\right| >C_1\cdot S \right) \leq C_2\cdot T\cdot S^{-r/2},
	\end{align*}
	where $C_1$ and $C_2$ denote some positive numbers, the second to last inequality follows from information inequality, and the last inequality follows from Markov inequality and the moment inequality in Proposition \ref{s_converge}. Thus, \eqref{exponentialinequality_dependent} holds as long as $T\cdot S^{-r/2}\to 0$.
	
	When the number of change-points is correctly estimated but some $\xi_j^*$ is a strictly bigger model than $\xi_j^o$, we have that \eqref{sum2_dependent} is exactly zero, and the same argument shows that \eqref{taylor2sum_dependent} is of order $O_p(1/ST)$ and $C_1'-C_2'$ is positive and of order $O(\log ST /ST)$ respectively. Thus, the same consistency conclusion follows. Similar arguments applies to the regular lattice case of Theorem \ref{finerate}.
	
	Lastly, the convergence rate follows from the first part of Corollary \ref{consistency_relaxIS} and the weak consistency of the estimated number of change-points. Hence the proof of Corollary \ref{consistency_relaxIS} is complete.
\end{proof}

\begin{proof}[Proof of Theorem \ref{asym.distr}]		
    We first prove the result for the approximate distribution of $\hat{\lambda}_j$ in \eqref{asy.dist.lam}. To lighten the notation, denote $L_{\psi_j^o}^{'}=L_{ST}^{'(j)}(\psi_j^o,\lambda^{o}_{j-1},\lambda^{o}_j; \bfY)$ and 
	$L_{\psi_j^o}^{''}=L_{ST}^{''(j)}(\psi_j^o,\lambda^{o}_{j-1},\lambda^{o}_j; \bfY)$, for $j=1,\ldots,m_{o}+1$. By Taylor expansion, we have that
	\begin{align*}
	0=L_{ST}'^{(j)}(\hat{\theta}_{j},\hat{\lambda}_{j-1},\hat{\lambda}_j; \bfY)
	=L_{ST}'^{(j)}({\theta}^{o}_j, \hat{\lambda}_{j-1},\hat{\lambda}_j; \bfY) 
	+ L_{ST}''^{(j)}({\theta}^{+}_j, \hat{\lambda}_{j-1},\hat{\lambda}_j)(\hat{\theta}_{j}-\theta^{o}_j),
	\end{align*}
	where ${\theta}^{+}_j$ is between ${\theta}^{o}_j$ and $\hat{\theta}^{(j)}_{ST}$. Together with $[T\hat{\lambda}_{j}]-[T\lambda_j^o]=O_p(1)$ and $S=o(T)$, we have
	\begin{align*}
	\frac{1}{\sqrt{ST}}L_{ST}'^{(j)}({\theta}^{o}_j, \hat{\lambda}_{j-1},\hat{\lambda}_j; \bfY)&=
    \frac{1}{\sqrt{ST}}L_{ST}'^{(j)}({\theta}^{o}_j, {\lambda}_{j-1}^o,{\lambda}_j^o; \bfY) + \frac{1}{\sqrt{ST}}O_p(S)([T\hat{\lambda}_{j}]-[T{\lambda}_{j}^o]+[T\hat{\lambda}_{j-1}]-[T{\lambda}_{j-1}^o])\\
    &= \frac{1}{\sqrt{ST}}L_{ST}'^{(j)}({\theta}^{o}_j, {\lambda}_{j-1}^o,{\lambda}_j^o; \bfY) + o_p(1),
	\end{align*}
	thus together with the consistency of $\hat{\theta}_{j}$'s and Proposition \ref{proprep} we have
	\begin{align*}
	\sqrt{ST}(\hat{\theta}_{j}-\theta^{o}_j)=(L_{\psi_j^o}^{''})^{-1} \sqrt{ST}L_{\psi_j^o}^{'}+o_p(1).
	\end{align*}
	
	By two Taylor's expansions and similar arguments, we have
	\begin{align}
		L_{ST}^{(j)}(\hat{\psi}_j,\hat{\lambda}_{j-1},\hat{\lambda}_j;\bfY)-L_{ST}^{(j)}(\psi_j^o,\hat{\lambda}_{j-1},\hat{\lambda}_j;\bfY) 
		&=-\frac{1}{2}(\hat{\theta}_{j}-\theta^{o}_{j})^\T L_{\psi_j^o}^{''}(\hat{\theta}_j-\theta_j^o)+o_p(1)\nonumber\\
		&=-\frac{1}{2} L_{\psi_j^o}^{'}\left(L_{\psi_j^o}^{''}\right)^{-1}L_{\psi_j^o}^{'} +o_p(1)\,,\label{Taylor2}
	\end{align}
	which is independent of $\hat{\lambda}_{j}$'s. 
	
    Therefore, on the set $\{\hat{m}=m_{o}\}$, we have 
	\begin{align}\label{eq:pf.thm3.1}
	&\sum_{j=1}^{m_{o}+1}\{L_{ST}^{(j)}(\hat{\psi}_j,\hat{\lambda}_{j-1},\hat{\lambda}_j;\bfY)-L_{ST}^{(j)}(\psi_j^o,\lambda^{o}_{j-1},\lambda^{o}_j;\bfY)\} \nonumber  \\
	=&-\frac{1}{2} \sum_{j=1}^{m_{o}+1} L_{\psi_j^o}^{'}\left(L_{\psi_j^o}^{''}\right)^{-1}L_{\psi_j^o}^{'} +o_p(1)
    +\sum_{j=1}^{m_{o}+1}\{L_{ST}^{(j)}({\psi}_j^o,\hat{\lambda}_{j-1},\hat{\lambda}_j;\bfY)-L_{ST}^{(j)}(\psi_j^o,\lambda^{o}_{j-1},\lambda^{o}_j;\bfY)\} \nonumber \\
    =&-\frac{1}{2} \sum_{j=1}^{m_{o}+1} L_{\psi_j^o}^{'}\left(L_{\psi_j^o}^{''}\right)^{-1}L_{\psi_j^o}^{'} +o_p(1)
    +\sum_{j=1}^{m_{o}} W_{ST}^{(j)}([T\hat{\lambda}_{j}]-[T\lambda_j^o];\psi_{j}^o,\psi_{j+1}^o),
	\end{align}
	where the last equality purely follows from algebra manipulation. Since $(\hat{\lambda}_1,\ldots,\hat{\lambda}_{m_{o}})$ maximize the left hand side of \eqref{eq:pf.thm3.1}, we have that for $j=1,\ldots,m_o$,
	\begin{align}
	[T\hat{\lambda}_j]-[T\lambda_j^o] &= \arg\max_{q\in \mathbb{Z}} W_{ST}^{(j)}(q;\psi_{j}^o,\psi_{j+1}^o)+o_p(1)\,.\label{asym.changepoint}
	\end{align}
	Note that to derive \eqref{eq:pf.thm3.1}, we only use the fact that $[T\hat{\lambda}_j]-[T\lambda_j^o]=O_p(1)$, so \eqref{asym.changepoint} can serve as an approximation for the finite sample behavior of $\hat{\Lambda}_{ST}$.

	We now prove the result for the asymptotic distribution of $\hat{\theta}_j$ in \eqref{asy.dist.theta}. Since $\max_{j=1,\ldots,m_{o}}|[T\hat{\lambda}_{j}]-[T\lambda_{j}^{o}]|=0$ in probability, on the set $\{\hat{m}=m_{o}\}$, we can assume the change points are known. Under Assumptions \ref{ass.thetadistr}, the condition of Theorem 1 in \cite{Jenish2009} is satisfied. Together with a standard Taylor expansion and Assumption \ref{ass.variance}, \eqref{asy.dist.theta} can be shown. The asymptotic independence follows from the strong mixing assumption along the time dimension.
\end{proof}

\subsection{Proofs of propositions and lemmas} \label{proof.prop.lem}
\begin{proof}[Proof of Lemma \ref{known}]
	Let $B$ be the probability one set on which Propositions \ref{proprep} and \ref{MPLE} hold. We will show that for each $\tilde{\omega} \in B$ and $\hat{\Lambda}_{ST}\rightarrow \Lambda^o$ as $(S,T)\rightarrow \infty$. To begin, for any $\tilde{\omega} \in B$, suppose on the contrary that $\hat{\Lambda}_{ST}\nrightarrow \Lambda^o$. Since the values of $\hat{\Lambda}_{ST}$ are bounded, there exists a subsequence $\{(S_l,T_l)\}$ such that $\hat{\Lambda}_{ST} \rightarrow \Lambda^*\neq\Lambda^o$ along the subsequence. Note that $\Lambda^* \in A_{\epsilon_\lambda}^{m_o}$ for all $(S,T)$. Recall that $(\hat{\xi}_j,\hat{\theta}_{ST}^{(j)})$ contains the estimators of the model order and model parameters for the $j$th segment. Since $\mathcal{M}$ is a finite set, without loss of generality, we can assume that $\hat{\xi}_j \rightarrow \xi_j^*$ along $\{(S_l, T_l)\}$. Similarly, since $\Theta_j\equiv\Theta_j(\xi_j)$ is compact for every $\xi_j$, we can assume that $\hat{\theta}_{ST}^{(j)}\rightarrow \theta_j^*$ along $\{(S_l, T_l)\}$. To lighten the notation, we replace $(S_l, T_l)$ by $(S,T)$ and it follows that for all sufficiently large $(S,T)$,
	\begin{equation}
	\frac{1}{ST}{\mathrm{CLMDL}} (m_o, \hat{\Lambda}_{ST},\hat{\Psi}_{ST})=c_{ST} -\frac{1}{ST}\sum_{j=1}^{m_o+1}L_{ST}^{(j)}\{(\xi_j^*, \hat{\theta}_{ST}^{(j)}),\hat{\lambda}_{j-1},\hat{\lambda}_j; \bfY\},
	\end{equation}
	where $c_{ST}$ is deterministic, positive and of order $O(\log ST/ST)$. 
	
	For each limiting estimated segment $I_j^*=(\lambda_{j-1}^*,\lambda_j^*) ,j=1,\ldots,m+1$, 
	there are two possible cases. First, $I_j^*$ is nested in the $i$th segment $(\lambda_{i-1}^o, \lambda_i^o)$. 
	Second, $I_j^*$ fully or partly covers $k+2$ $(k\geq0)$ true intervals: $(\lambda_{i-1}^o,\lambda_i^o),\ldots,(\lambda_{i+k}^o,\lambda_{i+k+1}^o)$. 
	We consider these two cases separately.\\
	Case 1. If $\lambda_{i-1}^o\leq\lambda_{j-1}^* < \lambda_j^* \leq \lambda_i^o$, in particular, 
	if $\lambda_{i-1}^o < \lambda_{j-1}^* < \lambda_j^* <\lambda_i^o$, then for sufficiently large $(S,T)$, 
	the estimated $j$th segment is a proportion of the stationary process from the true $i$th segment. 
	If $\lambda_j^*=\lambda_i^o$ or $\lambda_{j-1}^*=\lambda_{i-1}^o$, then as $\hat{\lambda}_j \rightarrow \lambda_i^o$ or $\hat{\lambda}_{j-1}\rightarrow \lambda_{i-1}^o$,
	the estimated segment includes only a decreasing proportion of observations from 
	an adjacent segment. Taking $\max(\hat{\lambda}_j-\lambda_i^o,0)$ and $\max(\lambda_{i-1}^o-\hat{\lambda}_{j-1},0)$ as the $h_T$ 
	and $k_T$ in \eqref{supfdef}, Proposition \ref{proprep} implies that
	\begin{equation}\label{case1con}
	\frac{1}{ST}L_{ST}^{(j)}\{(\xi^{*}_{j},\hat{\theta}_{ST}^{(j)}), \hat{\lambda}_{j-1}, \hat{\lambda}_j; \bfY\}=\frac{1}{ST} (\lambda_j^*-\lambda_{j-1}^*)\thickbar{L}_{ST}^{(i)}\{(\xi_j^*,\theta_j^*)\} +o(1)\,, \text{ \ almost surely}.
	\end{equation}
	In particular, if $\xi_j^*=\xi_i^o$, then $\theta_j^*$ is in fact $\theta_i^o$, which is the true parameter value of the $i$th segment. 
	Thus, the last quantity of \eqref{case1con} is in fact $\frac{1}{ST}(\lambda_j^*-\lambda_{j-1}^*)\thickbar{L}_{ST}^{(i)}\{(\xi^o_i,\theta_i^o)\}$.
	If $\xi_j ^*$ underestimates $\xi_i^o$, then information inequality implies that
	\begin{equation} \label{case1lik}
	\thickbar{L}_{ST}^{(i)}\{(\xi_i^o,\theta_i^o)\} > \thickbar{L}_{ST}^{(i)}\{(\xi_j^*,\theta_j^*)\}.
	\end{equation}
	Case 2. If $\lambda_{i-1}^o\leq\lambda_{j-1}^* < \lambda_i^o <\cdots<\lambda_{i+k}^o < \lambda_j^*\leq \lambda_{i+k+1}^o$ for some $k \geq0$, then for sufficiently large $(S,T)$, the estimated $j$th segment contains observations from at 
	least two pieces of different stationary processes. Partitioning the loglikelihood by 
	the true configuration of the series, we have
	\begin{align} \label{partition}
	& \frac{1}{ST}L_{ST}^{(j)}\{(\xi^{*}_{j},\hat{\theta}_{ST}^{(j)}), \hat{\lambda}_{j-1},\hat{\lambda}_j;\bfY\} \nonumber \\
	=&\frac{1}{ST}L_{STpair}^{(j)}\{(\xi^{*}_{j},\hat{\theta}_{ST}^{(j)}), \hat{\lambda}_{j-1},\lambda_i^o; \bfY\}+\frac{1}{ST}\sum_{l=i}^{i+k-1}L_{STpair}^{(j)}\{(\xi^{*}_{j},\hat{\theta}_{ST}^{(j)}), \lambda_l^o,\lambda_{l+1}^o; \bfY\} \nonumber \\
	+& \frac{1}{ST}L_{STpair}^{(j)}\{(\xi^{*}_{j},\hat{\theta}_{ST}^{(j)}), \lambda_{i+k}^o, \hat{\lambda}_j; \bfY\}
	+ \frac{1}{ST}L_{STmarg}^{(j)}\{(\xi^{*}_{j},\hat{\theta}_{ST}^{(j)}), \hat{\lambda}_{j-1},\hat{\lambda}_j;\bfY\}\nonumber\\
	+& \frac{1}{ST}\sum_{l=i}^{i+k}\sum_{D_{cST}^{\lambda_l^o}}l_{pair}\{(\xi^{*}_{j},\hat{\theta}_{ST}^{(j)}); y_{[\lambda_l^oT],s_1}, y_{[\lambda_l^oT]+1,s_2})\}
	\end{align}
	By Proposition \ref{edge_effect}, the cross-over pairwise loglikelihood and the marginal loglikelihood used for compensating the edge effect converge to 0 almost surely. The rest terms in \eqref{partition} involves observations from one piece of the stationary process. 
	From the proof of Proposition \ref{partdata} and the fact that $\thickbar{L}_{ST}^{(l)}\{(\xi_l^o,\theta_l^o)\} \geq \thickbar{L}_{ST}^{(j)}\{(\xi_j^*,\theta_j^*)\}$ for $l=i,\ldots,i+k+1$, we have for sufficiently large $(S,T)$
	\begin{align*}
		\frac{1}{ST}L_{ST}^{(j)}\{(\xi^{*}_{j},\hat{\theta}_{ST}^{(j)}), \hat{\lambda}_{j-1},\lambda_i^o;\bfY\}\leq&\frac{1}{ST} (\lambda_i^o-\lambda_{j-1}^*)\thickbar{L}_{ST}^{(i)}\{(\xi_i^o,\theta_i^o)\} \,,\\
		\frac{1}{ST}L_{ST}^{(j)}\{(\xi^{*}_{j},\hat{\theta}_{ST}^{(j)}), \lambda_l^o,\lambda_{l+1}^o;\bfY\}\leq&\frac{1}{ST} (\lambda_{l+1}^o-\lambda_l^o)\thickbar{L}_{ST}^{(l+1)}\{(\xi_{l+1}^o,\theta_{l+1}^o)\}\,,\\
		\frac{1}{ST}L_{ST}^{(j)}\{(\xi^{*}_{j},\hat{\theta}_{ST}^{(j)}),\lambda_{i+k}^o, \hat{\lambda}_j; \bfY\}\leq&\frac{1}{ST} (\lambda_j^*-\lambda_{i+k}^o)\thickbar{L}_{ST}^{(i+k+1)}\{(\xi_{i+k+1}^o,\theta_{i+k+1}^o)\}\,,
	\end{align*}
	almost surely. Note that strict inequalities hold for at least one of the preceding equations, since by Assumption \ref{ass.model.id}, $(\xi_j^*,\theta_j^*)$ cannot correctly specify the model for all of the different segments. Thus, for sufficiently large $(S,T)$,
	\begin{align} \label{limineqcase2}
	&\frac{1}{ST}L_{ST}^{(j)}\{(\xi^{*}_{j},\hat{\theta}_{ST}^{(j)}),\hat{\lambda}_{j-1},\hat{\lambda}_j; \bfY\} \nonumber \\
	<& \frac{1}{ST}(\lambda_i^o-\lambda_{j-1}^*)\thickbar{L}_{ST}^{(i)}\{(\xi_i^o,\theta_i^o)\}	 +\frac{1}{ST}\sum_{l=i}^{i+k-1}(\lambda_{l+1}^o-\lambda_l^o)\thickbar{L}_{ST}^{(l+1)}\{(\xi_{l+1}^o,\theta_{l+1}^o)\} \nonumber \\
	 +&\frac{1}{ST}(\lambda_j^*-\lambda_{i+k}^o)\thickbar{L}_{ST}^{(i+k+1)}\{(\xi_{i+k+1}^o,\theta_{i+k+1}^o)\}\,,
	\end{align}
	almost surely. As the number of estimated segments is equal to the true number of segments and $\lambda^*\neq\lambda^o$, there is at least one segment in which Case 2 applies. Thus for sufficiently large $(S,T)$,
	\begin{align} \label{case2con}
	0&\geq \frac{1}{ST}{\mathrm{CLMDL}}(m_o,\hat{\Lambda}_{ST},\hat{\Psi}_{ST})- \frac{1}{ST}{\mathrm{CLMDL}}(m_o,\Lambda^o,\Psi^o)\text{\hspace{ 10mm} [property of the estimator]}\nonumber \\
	&> 0, \text{\quad [\eqref{limineqcase2} holds for at least one piece]} 
	\end{align}
	which is a contradiction. Hence $\hat{\Lambda}_{ST}\rightarrow\Lambda^o$ for each $\tilde{\omega}\in B$.
	
	On the other hand, if $\hat{\Lambda}_{ST}\rightarrow\Lambda^o$ but in some of the segments, 
	the estimated model underestimates the true model, 
	then in Case 1, \eqref{case1lik} holds for those segments and the contradiction in \eqref{case2con} still occurs. 
	Thus the estimated model does not underestimate the true model. This completes the proof of Lemma \ref{known}.
\end{proof}

\begin{proof}[Proof of Lemma \ref{coro1}]
	In the proof of Lemma \ref{known}, the assumption of a known number of change-points is used only to ensure that Case 2 applies for at least one $j$. 
	The contradiction in the proof of Lemma \ref{known} arises whenever Case 2 applies. 
	By observing this fact, (a) and (b) follow. Thus we can assume that only Case 1 applies. 
	If any of the model segments is underestimated, then 
	\eqref{case1lik} holds with $i=j$ for some $j=1,\ldots, m+1$, and the contradiction in \eqref{case2con} still occurs, yielding (c).
\end{proof}

\begin{proof}[Proof of Lemma \ref{thm2}]
	From Lemma \ref{coro1}, we can assume that $\hat{m}\geq m_o$ and for each $\lambda_j^o$ there exists a $\hat{\lambda}_{i_j}$ such that 
	$|\lambda_j^o - \hat{\lambda}_{i_j}| = o_p(1)$. This Lemma states an exact recovery property of the change-point estimators. 
	Suppose that for some $\lambda_l^o$, there does not exist an $\hat{\lambda}_{i_l}$ such that 
	$\mathbb P\left(\left|[T\lambda_l^o] - [T\hat{\lambda}_{i_l}]\right|=0\right) \longrightarrow 1$, then either one of the events
	\begin{eqnarray} \label{thm2case}
	&& \text{i) \quad } [T\lambda_l^o] - [T\hat{\lambda}_{i_l}] \geq 1, \nonumber  \text{\quad or} \\
	&& \text{ii) \quad } [T\hat{\lambda}_{i_l}] - [T\lambda_l^o]\leq 1\,,
	\end{eqnarray}
	holds with nonzero probability, where $\hat{\lambda}_{i_l}$ is the estimated change-point closest to $\lambda_l^o$. From Lemma \ref{coro1}, we know that $\hat{\lambda}_{i_l}\to \lambda_l^o$ in probability. We show that $\mathbb P\left([T\lambda_l^o]-[T\hat{\lambda}_{i_l}]\geq 1\right) \longrightarrow 0$. The proof for the other direction is the same.
    %

For notational simplicity, in the following we skip the subscripts $marg$ and $pair$ for the loglikelihood functions. By Lemma \ref{coro1}, we know that the order of the model will not be underestimated. Since $\mathcal{M}$ is finite, WLOG, we can further assume the order of the model is fixed for the two segments around $\hat{\lambda}_{i_l}$ and thus replace $\psi$ by $\theta$ in the notation. The superscript of $y_{t,s}$ emphasizes the true segment that it comes from. Similar to the argument in \cite{Ling2014}, the proof relies on analysis of the behavior of
{\small
\begin{align*}
&\frac{2}{S(T\lambda_l^o-T\hat{\lambda}_{i_l})}\left(\sup_{\Theta}\left|\sum_{D_{cST}^{\hat{\lambda}_{i_l}}}[l(\theta; y_{t,s_1}^{(l)}, y_{t+1,s_2}^{(l)})-\mathbb E(l(\theta; y_{t,s_1}^{(l)}, y_{t+1,s_2}^{(l)}))]\right|\right.\nonumber\\
&\left.+ \sup_{\Theta}\left|\sum_{D_{cST}^{\lambda_l^o}}[l(\theta; y_{t,s_1}^{(l)}, y_{t+1,s_2}^{(l+1)})-\mathbb E(l(\theta; y_{t,s_1}^{(l)}, y_{t+1,s_2}^{(l+1)}))]\right|
+ \sup_{\Theta}\left|\sum_{D_{ST}^{\hat{\lambda}_{i_l}, \lambda_l^o}}[l(\theta; y_{t,s_1}^{(l)}, y_{t+i,s_2}^{(l)})- \mathbb E(l(\theta; y_{t,s_1}^{(l)}, y_{t+i,s_2}^{(l)}))]\right|\right),
\end{align*}}
on the event $\{T\lambda_l^o-T\hat{\lambda}_{i_l}\geq 1\}$.

By a slight abuse of notation, here we use $\Theta$ to denote $\Theta_l(\xi_l)\cup \Theta_{l+1}(\xi_{l+1})$, indicating that the data can be either modeled by the parametric model denoted by $\xi_l$ of the $l$th segment or $\xi_{l+1}$ of the $(l+1)$th segment. For the cross-over terms in $D_{cST}^{\hat{\lambda}_{i_l}}$ and $D_{cST}^{\lambda_l^o}$, $\Theta$ further contains the case where $l(\theta; y_{[T\hat{\lambda}_{i_l}],s_1}^{(l)}, y_{[T\hat{\lambda}_{i_l}]+1,s_2}^{(l)})=l(\theta_l; y_{[T\hat{\lambda}_{i_l}],s_1}^{(l)})+l(\theta_{l+1};y_{[T\hat{\lambda}_{i_l}]+1,s_2}^{(l)})$ and $l(\theta; y_{[T\lambda_l^o],s_1}^{(l)}, y_{[T\lambda_l^o]+1,s_2}^{(l+1)})=l(\theta_l; y_{[T\lambda_l^o],s_1}^{(l)})+l(\theta_{l+1}; y_{[T\lambda_l^o]+1,s_2}^{(l+1)})$ with $\theta_l\in\Theta_l(\xi_l)$ and $\theta_{l+1}\in\Theta_{l+1}(\xi_{l+1})$. This corresponds to the case where $\hat{\lambda}_{i_l}$ or $\lambda_l^o$ is the assigned change point and the pairwise likelihood of $([T\hat{\lambda}_{i_l}], [T\hat{\lambda}_{i_l}]+1)$ or $([T\lambda_l^o], [T\lambda_l^o]+1)$ breaks into two marginal likelihoods with different parameters.

On the event $\{T\lambda_l^o-T\hat{\lambda}_{i_l}\geq 1\}$ we have that, 
{\small
\begin{align}
&\frac{2}{S(T\lambda_l^o-T\hat{\lambda}_{i_l})}\left(\sup_{\Theta}\left|\sum_{D_{cST}^{\hat{\lambda}_{i_l}}}[l(\theta; y_{t,s_1}^{(l)}, y_{t+1,s_2}^{(l)})-\mathbb E(l(\theta; y_{t,s_1}^{(l)}, y_{t+1,s_2}^{(l)}))]\right|\right.\nonumber\\
&\left.+ \sup_{\Theta}\left|\sum_{D_{cST}^{\lambda_l^o}}[l(\theta; y_{t,s_1}^{(l)}, y_{t+1,s_2}^{(l+1)})-\mathbb E(l(\theta; y_{t,s_1}^{(l)}, y_{t+1,s_2}^{(l+1)}))]\right|
+ \sup_{\Theta}\left|\sum_{D_{ST}^{\hat{\lambda}_{i_l}, \lambda_l^o}}[l(\theta; y_{t,s_1}^{(l)}, y_{t+i,s_2}^{(l)})- \mathbb E(l(\theta; y_{t,s_1}^{(l)}, y_{t+i,s_2}^{(l)}))]\right|\right)\nonumber\\
\geq &\frac{1}{S(T\lambda_l^o-T\hat{\lambda}_{i_l})}\Big\{ \nonumber\\
 &\sum_{E_{2ST}^{\hat{\lambda}_{i_l}}} l(\hat{\theta}_{i_l};y_{t,s}^{(l)})
+ \sum_{E_{1ST}^{\hat{\lambda}_{i_l}}} l(\hat{\theta}_{i_l+1};y_{t,s}^{(l)})
+ \sum_{D_{cST}^{\lambda_l^o}}l(\hat{\theta}_{i_l+1}; y_{t,s_1}^{(l)}, y_{t+1,s_2}^{(l+1)})
+ \sum_{D_{ST}^{\hat{\lambda}_{i_l}, \lambda_l^o}}l(\hat{\theta}_{i_l+1}; y_{t,s_1}^{(l)}, y_{t+i,s_2}^{(l)})\nonumber\\
&-\sum_{E_{2ST}^{\hat{\lambda}_{i_l}}} \mathbb E(l(\hat{\theta}_{i_l};y_{t,s}^{(l)}))
- \sum_{E_{1ST}^{\hat{\lambda}_{i_l}}} \mathbb E(l(\hat{\theta}_{i_l+1};y_{t,s}^{(l)}))
- \sum_{D_{cST}^{\lambda_l^o}}\mathbb E(l(\hat{\theta}_{i_l+1}; y_{t,s_1}^{(l)}, y_{t+1,s_2}^{(l+1)}))
- \sum_{D_{ST}^{\hat{\lambda}_{i_l}, \lambda_l^o}}\mathbb E(l(\hat{\theta}_{i_l+1}; y_{t,s_1}^{(l)}, y_{t+i,s_2}^{(l)}))\nonumber\\
&-\sum_{E_{2ST}^{\lambda_l^o}} l(\hat{\theta}_{i_l};y_{t,s}^{(l)})
- \sum_{E_{1ST}^{\lambda_l^o}} l(\hat{\theta}_{i_l+1};y_{t,s}^{(l+1)})
- \sum_{D_{cST}^{\hat{\lambda}_{i_l}}}l(\hat{\theta}_{i_l}; y_{t,s_1}^{(l)}, y_{t+1,s_2}^{(l)})
- \sum_{D_{ST}^{\hat{\lambda}_{i_l}, \lambda_l^o}}l(\hat{\theta}_{i_l}; y_{t,s_1}^{(l)}, y_{t+i,s_2}^{(l)})\nonumber\\
&+\sum_{E_{2ST}^{\lambda_l^o}} \mathbb E(l(\hat{\theta}_{i_l};y_{t,s}^{(l)}))
+ \sum_{E_{1ST}^{\lambda_l^o}} \mathbb E(l(\hat{\theta}_{i_l+1};y_{t,s}^{(l+1)}))
+ \sum_{D_{cST}^{\hat{\lambda}_{i_l}}}\mathbb E(l(\hat{\theta}_{i_l}; y_{t,s_1}^{(l)}, y_{t+1,s_2}^{(l)}))
+ \sum_{D_{ST}^{\hat{\lambda}_{i_l}, \lambda_l^o}}\mathbb E(l(\hat{\theta}_{i_l}; y_{t,s_1}^{(l)}, y_{t+i,s_2}^{(l)}))\nonumber\\
\Big\}\nonumber\\
\geq &\frac{1}{S(T\lambda_l^o-T\hat{\lambda}_{i_l})}\Big\{\nonumber\\
&\sum_{E_{2ST}^{\lambda_l^o}} \mathbb E(l(\hat{\theta}_{i_l};y_{t,s}^{(l)}))
+ \sum_{E_{1ST}^{\lambda_l^o}} \mathbb E(l(\hat{\theta}_{i_l+1};y_{t,s}^{(l+1)}))
+ \sum_{D_{cST}^{\hat{\lambda}_{i_l}}}\mathbb E(l(\hat{\theta}_{i_l}; y_{t,s_1}^{(l)}, y_{t+1,s_2}^{(l)}))
+ \sum_{D_{ST}^{\hat{\lambda}_{i_l}, \lambda_l^o}}\mathbb E(l(\hat{\theta}_{i_l}; y_{t,s_1}^{(l)}, y_{t+i,s_2}^{(l)}))\nonumber\\
&-\sum_{E_{2ST}^{\hat{\lambda}_{i_l}}} \mathbb E(l(\hat{\theta}_{i_l};y_{t,s}^{(l)}))
- \sum_{E_{1ST}^{\hat{\lambda}_{i_l}}} \mathbb E(l(\hat{\theta}_{i_l+1};y_{t,s}^{(l)}))
- \sum_{D_{cST}^{\lambda_l^o}}\mathbb E(l(\hat{\theta}_{i_l+1}; y_{t,s_1}^{(l)}, y_{t+1,s_2}^{(l+1)}))
- \sum_{D_{ST}^{\hat{\lambda}_{i_l}, \lambda_l^o}}\mathbb E(l(\hat{\theta}_{i_l+1}; y_{t,s_1}^{(l)}, y_{t+i,s_2}^{(l)}))\nonumber\\
\Big\}\nonumber\\
>& C +o(1), \text{~~~~for some $C >0$}. \label{sup_inequality}
\end{align}}
\vspace{-0.9cm}

Note that in \eqref{sup_inequality}, the expectation is taken w.r.t. $y_{t,s}$ but not $\theta$ or $\hat{\theta}.$ The first inequality follows from the definition of $\sup\Theta$. The second inequality follows from the fact that $\hat{\lambda}_{i_l}$ is the estimated change point and thus
\begin{align*}
L_{ST}^{(j)}(\hat{\theta}_{i_l}, \hat{\lambda}_{i_l-1},\hat{\lambda}_{i_l}; \bfY) + 
L_{ST}^{(j)}(\hat{\theta}_{i_l+1}, \hat{\lambda}_{i_l},\hat{\lambda}_{i_l+1}; \bfY) 
\geq L_{ST}^{(j)}(\hat{\theta}_{i_l}, \hat{\lambda}_{i_l-1},\lambda_l^o; \bfY) + 
L_{ST}^{(j)}(\hat{\theta}_{i_l+1}, \lambda_l^o,\hat{\lambda}_{i_l+1}; \bfY).
\end{align*}
The last inequality follows from the consistency of $\hat{\theta}$ in Proposition \ref{MPLE} and information inequality since $[T\lambda_l^o]$ is the true change point but not $[T\hat{\lambda}_{i_l}]$.

By Markov inequality and the moment inequality in Proposition \ref{s_converge}, there exists $r>2$ such that for any fixed $\theta\in \Theta$ and positive integer $T\lambda_{l-1}^o<k<T\lambda_l^o$, such that
\begin{align*}
\mathbb P\left\{\frac{1}{S(T\lambda_l^o-k)}\left|\sum_{D_{ST}^{k/T,\lambda_l^o}}[l(\theta; y_{t,s_1}^{(l)}, y_{t+i,s_2}^{(l)})- \mathbb E(l(\theta; y_{t,s_1}^{(l)}, y_{t+i,s_2}^{(l)}))]\right|>\epsilon\right\}\leq \frac{C}{S^{r/2}(T\lambda_l^o-k)^{r/2}},
\end{align*}
where $C$ is a constant independent of $k.$

By Lemma \ref{coro1} and the definition of $\hat{\lambda}_{i_l}$, we know that $\hat{\lambda}_{i_l}>\lambda_{l-1}^o+\epsilon_\lambda$ almost surely. Thus, by union bound, on the event $\{T\lambda_l^o-T\hat{\lambda}_{i_l}\geq 1\}$, we also have that for any fixed $\theta$,
\begin{align*}
&\mathbb P\left\{\frac{1}{S(T\lambda_l^o-T\hat{\lambda}_{i_l})}\left|\sum_{D_{ST}^{\hat{\lambda}_{i_l}, \lambda_l^o}}[l(\theta; y_{t,s_1}^{(l)}, y_{t+i,s_2}^{(l)})- \mathbb E(l(\theta; y_{t,s_1}^{(l)}, y_{t+i,s_2}^{(l)}))]\right|>\epsilon\right\}\\
\leq& \sum_{k=T\lambda_{l-1}^o+1}^{T\lambda_l^o-1}\mathbb P\left\{\frac{1}{S(T\lambda_l^o-k)}\left|\sum_{D_{ST}^{k/T,\lambda_l^o}}[l(\theta; y_{t,s_1}^{(l)}, y_{t+i,s_2}^{(l)})- \mathbb E(l(\theta; y_{t,s_1}^{(l)}, y_{t+i,s_2}^{(l)}))]\right|>\epsilon\right\}\\
\leq& C \sum_{k=1}^{T\lambda_l^o-1} S^{-r/2}k^{-r/2}\longrightarrow 0,
\end{align*}
as $(S, T) \to \infty,$ since $r>2.$ Note that this probability goes to zero as long as $S \to \infty$, which is the essential ingredient for the exact recovery property of the estimated change-points, and is the difference between the multivariate time series setting and the spatio-temporal setting.

By the compactness of $\Theta$, using a standard ULLN type of argument, we can show that
\begin{align}
\mathbb P\left\{\frac{1}{S(T\lambda_l^o-T\hat{\lambda}_{i_l})}\sup_{\Theta}\left|\sum_{D_{ST}^{\hat{\lambda}_{i_l}, \lambda_l^o}}[l(\theta; y_{t,s_1}^{(l)}, y_{t+i,s_2}^{(l)})- \mathbb E(l(\theta; y_{t,s_1}^{(l)}, y_{t+i,s_2}^{(l)}))]\right|>\epsilon\right\}\longrightarrow 0.
\label{changepoint_ULLN}
\end{align}

Similarly, we can show that
\begin{align}
&\mathbb P\left\{\frac{1}{S(T\lambda_l^o-T\hat{\lambda}_{i_l})}\sup_{\Theta}\left|\sum_{D_{cST}^{\hat{\lambda}_{i_l}}}[l(\theta; y_{t,s_1}^{(l)}, y_{t+1,s_2}^{(l)})-\mathbb E(l(\theta; y_{t,s_1}^{(l)}, y_{t+1,s_2}^{(l)}))]\right|>\epsilon\right\}\longrightarrow 0,
\label{changepoint_ULLN1}\\
&\mathbb P\left\{\frac{1}{S(T\lambda_l^o-T\hat{\lambda}_{i_l})}\sup_{\Theta}\left| \sum_{D_{cST}^{\lambda_l^o}}[l(\theta; y_{t,s_1}^{(l)}, y_{t+1,s_2}^{(l+1)})-\mathbb E(l(\theta; y_{t,s_1}^{(l)}, y_{t+1,s_2}^{(l+1)}))]\right|>\epsilon\right\}\longrightarrow 0.
\label{changepoint_ULLN2}
\end{align}

Thus we have
\begin{align*}
&\mathbb P\left( [T\lambda_l^o] - [T\hat{\lambda}_{i_l}] \geq 1\right)\\
=&\mathbb P\Bigg\{[T\lambda_l^o] - [T\hat{\lambda}_{i_l}] \geq 1, \frac{2}{S(T\lambda_l^o-T\hat{\lambda}_{i_l})}\left(\sup_{\Theta}\left|\sum_{D_{cST}^{\hat{\lambda}_{i_l}}}[l(\theta; y_{t,s_1}^{(l)}, y_{t+1,s_2}^{(l)})-\mathbb E(l(\theta; y_{t,s_1}^{(l)}, y_{t+1,s_2}^{(l)}))]\right|\right.\nonumber\\
&\left.+ \sup_{\Theta}\left|\sum_{D_{cST}^{\lambda_l^o}}[l(\theta; y_{t,s_1}^{(l)}, y_{t+1,s_2}^{(l+1)})-\mathbb E(l(\theta; y_{t,s_1}^{(l)}, y_{t+1,s_2}^{(l+1)}))]\right|
+ \sup_{\Theta}\left|\sum_{D_{ST}^{\hat{\lambda}_{i_l}, \lambda_l^o}}[l(\theta; y_{t,s_1}^{(l)}, y_{t+i,s_2}^{(l)})- \mathbb E(l(\theta; y_{t,s_1}^{(l)}, y_{t+i,s_2}^{(l)}))]\right|\right)\\
&>C+o(1)\Bigg\} \longrightarrow 0,
\end{align*}

where the convergence follows from \eqref{changepoint_ULLN}, \eqref{changepoint_ULLN1} and \eqref{changepoint_ULLN2}.
\end{proof}

\begin{proof}[Proof of Lemma \ref{lil}]
	By a Taylor expansion, we have that
	\begin{align*}
	0=L_{ST}'^{(j)}(\hat{\theta}^{(j)}_{ST},\lambda_d,\lambda_u; \bfX_j^o)
	=L_{ST}'^{(j)}({\theta}^{*}_j, \lambda_d,\lambda_u; \bfX_j^o) 
	+ L_{ST}''^{(j)}({\theta}^{+}_j, \lambda_d,\lambda_u)(\hat{\theta}^{(j)}_{ST}-\theta^{*}_j),
	\end{align*}
	where ${\theta}^{+}_j$ is between ${\theta}^{*}_j$ and $\hat{\theta}^{(j)}_{ST}$. Thus we have
	\begin{align*}
	-\frac{1}{\sqrt{ST}}L_{ST}'^{(j)}({\theta}^{*}_j, \lambda_d,\lambda_u; \bfX_j^o) 
	= \frac{1}{ST} L_{ST}''^{(j)}({\theta}^{+}_j, \lambda_d,\lambda_u; \bfX_j^o) \cdot \sqrt{ST}(\hat{\theta}^{(j)}_{ST}-\theta^{*}_j),
	\end{align*}
	By Proposition \ref{MPLE} and Proposition \ref{partdata}, we have that
	\begin{align*}
    \frac{1}{ST(\lambda_u-\lambda_d)} L_{ST}''^{(j)}({\theta}^{+}_j, \lambda_d,\lambda_u; \bfX_j^o)=
    \frac{1}{ST}\thickbar{L}_{ST}''^{(j)}(\theta_j^*)+o_p(1),
	\end{align*}
	where $\frac{1}{ST}\thickbar{L}_{ST}''^{(j)}(\theta_j^*)$ is a negative definite matrix by the definition of $\theta_j^*$. Note that $\lambda_u-\lambda_d>\epsilon_\lambda$, thus if we can show that $\sup_{\lambda_d,\lambda_u}\frac{1}{\sqrt{ST}}L_{ST}'^{(j)}({\theta}^{*}_j, \lambda_d,\lambda_u; \bfX_j^o)=O_p(1)$, the proof is complete.
	
	For any $\lambda_d, \lambda_u$, we have that
	\begin{align*}
	&\frac{1}{\sqrt{ST}}L_{ST}'^{(j)}(\theta^*_j, \lambda_d, \lambda_u; \bfX_j^o)\\
	\leq&
	\frac{1}{\sqrt{ST}}\left(L_{STpair}'^{(j)}(\theta^*_j, 0,\lambda_u; \bfX_j^o)-L_{STpair}'^{(j)}(\theta^*_j, 0,\lambda_d; \bfX_j^o)\right)+\frac{2}{\sqrt{ST}}\sup_{\lambda\in[0,1]}\sum_{s=1}^S l_{marg}'\left(\theta_j^*; x_{[\lambda T_j], s}^{(j)}\right).
	\end{align*}
	By the definition of $\theta^*$, all the first order derivatives at $\theta^*$ are mean-zero. By Proposition \ref{s_converge} and union bound, we can easily show that $\frac{1}{\sqrt{ST}}\sup_{\lambda\in[0,1]}\sum_{s=1}^S l_{marg}'\left(\theta_j^*; x_{[\lambda T_j], s}^{(j)}\right)=o_p(1)$. Thus, we can bound $\frac{1}{\sqrt{ST}}L_{ST}'^{(j)}({\theta}^{*}_j, \lambda_d,\lambda_u; \bfX_j^o)$ by
	\begin{align}
	\frac{1}{\sqrt{ST}}\left(\sup_{\lambda\in [0,1]}L_{STpair}'^{(j)}(\theta^*_j, 0,\lambda; \bfX_j^o)-
	\inf_{\lambda\in [0,1]}L_{STpair}'^{(j)}(\theta^*_j, 0,\lambda; \bfX_j^o)\right).
	\label{BrownianSheet}	
	\end{align}
	By Assumption \ref{ass.mom}($r$), Assumption \ref{ass.mixtime} and Theorem 2.2 in \cite{Yang2007}, we can show that
    $\eqref{BrownianSheet}$ is $O_p(1)$, thus complete the proof.
\end{proof}

\begin{proof}[Proof of Proposition \ref{s_converge}]
	We prove the result for $a=0$ and $$\mathbb E\left|\sum_{(t,i,s_1,s_2)\in D^{\lambda_d,\lambda_u}_{ST}}(l_{pair}\left(\psi; y_{t,s_1},y_{t+i,s_2}\right)-
	\mathbb E(l_{pair}\left(\psi; y_{t,s_1},y_{t+i,s_2}\right)))\right|^r \leq K ((\lambda_u-\lambda_d)TS)^{r/2},$$
	the proof for the rest results follows the same argument. 
	
	Define $Z_{t,s}=\sum\limits_{s_2\in \mathcal{N}(s)}l_{pair}\left(\psi; y_{t,s},y_{t,s_2}\right)+
	\sum\limits_{s_2\in \mathcal{N}(s)\cup s }l_{pair}\left(\psi; y_{t,s},y_{t+1,s_2}\right)$. Thus $Z_{t,s}$ can be seen as a new random field indexed by $\bd=(t,s)$. We need to show that the induced random field $\{Z_{t,s}\}$ satisfies the mixing conditions of the moment inequality of Theorem 1 in Section 1.4 of \cite{Doukhan1994}. Note that the metric $\rho(\cdot)$ defined in the main text can induce a valid metric $\rho_z(\cdot)$ for $Z_{t,s}$ such that $\rho_z(\bd_1,\bd_2)=\rho(\bd_1,\bd_2)$. For $\bd=(t,s)$, define $P_z(\bd)=\{(t+i,s'), 0\leq i\leq 1, s'\in s\cup\mathcal{N}(s) \}$, which includes the indices of all the $y_{t,s}$'s that are used in $Z_{t,s}$.
	
	By the definition of CLMDL~(the neighbor $\mathcal{N}(s)$ and time lag $k=1$), there exists a constant $c^*$ such that for any $(\bd_1,\bd_2)$, we have $\rho_z(\bd_1,\bd_2)\leq \rho(P_z(\bd_1),P_z(\bd_2))+ c^*$. By the definition of $\alpha$-mixing, it is easy to see that for $m> c^*$, the mixing coefficients for $\{Z_{t,s}\}$ satisfies
	\begin{align}
	\alpha_Z(m;u,v)&=\sup(\alpha_Z(U,V), |U|\leq u, |V|\leq v, \rho_z(U,V)\geq m, U,V\subset \mathcal{Z})\nonumber\\
	&\leq \sup(\alpha_Y(U,V), |U|\leq 2(1+|\mathcal{N}(s)|)u, |V|\leq 2(1+|\mathcal{N}(s)|)v, \rho(U,V)\geq m-c^*, U,V\subset \mathcal{Z})\nonumber\\
	&=\alpha_Y(m-c^*;2(1+|\mathcal{N}(s)|)u,2(1+|\mathcal{N}(s)|)v).\label{alpha_mixing}
	\end{align}
	By the independence between segments, Assumption \ref{ass.mix}($r$) and \eqref{alpha_mixing}, we have that there exists $\epsilon>0$ and $c\in 2\mathbb{N}$ where $c>r$, such that for all $u,v \in \mathbb{N}^+$, $u+v\leq c$, $u,v\geq 2$,
	$$\sum_{m=1}^{\infty}(m+1)^{3(c-u+1)-1}[\alpha_{Z}(m;u,v)]^{\epsilon/(c+\epsilon)} <\infty.$$
	By Lemma A.1 (ii) and (iii) of \cite{Jenish2009} and Assumption \ref{ass.countable}, we can show that the maximal value of the cardinal number of a ring with thickness 1 and radius $d$ or of a ball with radius $d$ in $\mathcal{Z}$ can be bound $Cd^2$ and $Cd^3$ respectively for some $C>0$.
	
    Together, we can show that the condition of Theorem 1 in Section 1.4 of \cite{Doukhan1994} is satisfied\footnote{Theorem 1 in Section 1.4 of \cite{Doukhan1994} states the result for the case where the index set is $\mathbb{Z}^d$. However, using Lemma A.1 of \cite{Jenish2009}, the proof can easily be adapted to cover the case where Assumption \ref{ass.countable} holds.}. Thus we know that 
	$$\mathbb E\left|\sum_{(t,i,s_1,s_2)\in D^{\lambda_d,\lambda_u}_{ST}}(l_{pair}\left(\psi; y_{t,s_1},y_{t+i,s_2}\right)-
	\mathbb E(l_{pair}\left(\psi; y_{t,s_1},y_{t+i,s_2}\right)))\right|^r \leq K M_{r,\epsilon} ((\lambda_u-\lambda_d)TS)^{r/2},$$
	where $K$ is a constant only depending on $r$ and on the mixing coefficients of $\bfY$ and
	$$M_{r,\epsilon}=\sup_{(t,s)} \mathbb E\left|Z_{t,s}-\mathbb E(Z_{t,s})\right|^{r+\epsilon}.$$
	By Assumption \ref{ass.mom}($r$), we know that $M_{r,\epsilon} < C$, and thus we complete the proof.
\end{proof}

\begin{proof}[Proof of Proposition \ref{s_converge2}]
	We prove the result for $a=0$ and
	$$\sup_{\theta \in \Theta}\left|\frac{1}{S([\lambda_uT]-[\lambda_dT])}\sum_{(t,i,s_1,s_2)\in D^{\lambda_d,\lambda_u}_{ST}}(l_{pair}^{[a]}\left(\psi; y_{t,s_1},y_{t+i,s_2}\right)-
	\mathbb E(l_{pair}^{[a]}\left(\psi; y_{t,s_1},y_{t+i,s_2}\right)))\right| \longrightarrow 0,$$
	the proof for the rest results follows the same argument.
	
    By Proposition \ref{s_converge} and Markov inequality, we know that, for any fixed $\psi$,
    \begin{align*}
    &\mathbb P\left(\left|\frac{1}{S([\lambda_uT]-[\lambda_dT])}\sum_{(t,i,s_1,s_2)\in D^{\lambda_d,\lambda_u}_{ST}}(l_{pair}^{[a]}\left(\psi; y_{t,s_1},y_{t+i,s_2}\right)-
    \mathbb E(l_{pair}^{[a]}\left(\psi; y_{t,s_1},y_{t+i,s_2}\right)))\right|>\epsilon\right)\\
    &\leq K (S([\lambda_uT]-[\lambda_dT]))^{-r/2}=K(\lambda_u-\lambda_d)(ST)^{-r/2}.
    \end{align*}
    
    By Borel-Cantelli lemma, we have that 
    \begin{align*}
    &\sum_{n=1}^{\infty}\mathbb P\left(\left|\frac{1}{S_n([\lambda_uT_n]-[\lambda_dT_n])}\sum_{(t,i,s_1,s_2)\in D^{\lambda_d,\lambda_u}_{S_nT_n}}(l_{pair}^{[a]}\left(\psi; y_{t,s_1},y_{t+i,s_2}\right)-
    \mathbb E(l_{pair}^{[a]}\left(\psi; y_{t,s_1},y_{t+i,s_2}\right)))\right|>\epsilon\right)\\
    &\leq \sum_{n=1}^{\infty} K(\lambda_u-\lambda_d)(S_nT_n)^{-r/2} <\infty,  
    \end{align*}
    where the last inequality follows from the fact that $S_nT_n < S_{n+1}T_{n+1}$ for any $n$.
    
    Thus, we have a pointwise SLLN. The uniform convergence result then follows from the compactness of $\Theta$ and the standard argument for establishing ULLN.
\end{proof}

\begin{proof}[Proof of Proposition \ref{edge_effect}]
	We prove that $$\frac{1}{ST}\sup_{\lambda\in[0,1]} \sup_{\theta\in\Theta(\xi)}\sum_{(t,s)\in E_{2ST}^\lambda}\left|l_{marg}\left(\psi; y_{t, s}\right)\right| =\frac{1}{ST}\sup_{\lambda\in[0,1]} \sup_{\theta\in\Theta(\xi)}\sum_{s=1}^{S}\left|l_{marg}\left(\psi; y_{[\lambda T], s}\right)\right| \longrightarrow 0$$ almost surely, the proof for the cross-over pairwise loglikelihood is the same. For any $\epsilon>0$, we can find a set of rational number ($r_1,\ldots, r_{N_\epsilon}$) in $[0,1]$ such that for any $\lambda \in [0,1]$, there exists a $k$ such that $r_k\leq \lambda\leq r_{k+1}$ and $r_{k+1}-r_k<\epsilon$.
	
	Similar to Proposition \ref{s_converge2}, we can prove that for sufficiently large $(S,T)$,
	\begin{align*}
		\frac{1}{ST}\sup_{\theta\in\Theta(\xi)}\sum_{s=1}^{S}\left|l_{marg}\left(\psi; y_{[\lambda T], s}\right)\right|&\leq
		\frac{1}{ST}\sup_{\theta\in\Theta(\xi)}\sum_{t=[r_k T]}^{[r_{k+1}T]}\sum_{s=1}^{S}\left|l_{marg}\left(\psi; y_{t, s}\right)\right| \\
		&< C(r_{k+1}-r_k)<C\epsilon, \text{ almost surely,}
	\end{align*}
	where $C=\mathbb E\left(\sup_{\theta\in\Theta(\xi)}\left|l_{marg}\left(\psi; y_{t, s}\right)\right|\right)$. The result follows from the fact that $\epsilon$ is independent of $\lambda$ and the rational number cover for $\lambda\in[0,1]$ is countable.
\end{proof}

\begin{proof}[Proof of Proposition \ref{partdata}]
	We prove only the case where $a=0$; the cases where $a=1$ and $2$ are handled similarly. It is easy to see that
	\begin{align*}
	\sup_{\lambda_d,\lambda_u}\sup_{\theta \in \Theta(\xi)} \frac{1}{ST}\left|\thickbar{L}^{(j)}_{ST}\{(\xi,\theta)\}-\thickbar{L}^{(j)}_{STpair}\{(\xi,\theta)\}\right|=o(1),
	\end{align*}
	thus, we only need to prove the result for $\thickbar{L}^{(j)}_{STpair}\{(\xi,\theta)\}$.
	
	Let $\mathbb{Q}_{[0,1]}$ be the set of rational numbers in $[0,1]$. For any fixed pair $r_1,r_2 \in \mathbb{Q}_{[0,1]}$ with $r_1 < r_2$, we have
	\begin{align}\label{appprop1}
	&\sup_{\theta \in \Theta(\xi)} \frac{1}{ST}\left|L_{ST}^{(j)}\{(\xi,\theta),r_1,r_2;\bfX_j^o\} - (r_2-r_1)\thickbar{L}^{(j)}_{STpair}\{(\xi,\theta)\}\right|\nonumber\\
    \leq &\sup_{\theta \in \Theta(\xi)}\frac{1}{ST}\sum_{(t,s)\in E_{1ST_j}^{r_1}}\left| l_{marg}\left({\psi}; x_{t,s}^{(j)}\right) \right| + \sup_{\theta \in \Theta(\xi)}\frac{1}{ST} \sum_{(t,s)\in E_{2ST_j}^{r_2}}\left|l_{marg}\left({\psi}; x_{t,s}^{(j)}\right)   \right|\nonumber\\
    +&\sup_{\theta \in \Theta(\xi)}\frac{1}{ST}\left|L_{STpair}^{(j)}\{(\xi,\theta),r_1,r_2;\bfX_j^o\} - \mathbb E(L_{STpair}^{(j)}\{(\xi,\theta),r_1,r_2;\bfX_j^o\})\right|\nonumber\\
    +& \sup_{\theta \in \Theta(\xi)}\frac{1}{ST}\left|\mathbb E(L_{STpair}^{(j)}\{(\xi,\theta),r_1,r_2;\bfX_j^o\}) - (r_2-r_1)\thickbar{L}^{(j)}_{STpair}\{(\xi,\theta)\}\right|\nonumber\\
	&\longrightarrow 0 \, \text{, almost surely}.
	\end{align}
	By Proposition \ref{edge_effect}, the first two terms converge to 0 almost surely. The third term converges to 0 by the ULLN in Proposition \ref{s_converge2} and the fourth term converges to 0 by the strictly stationarity of the random field and the compactness of $\Theta$.
	
	Next, let $B_{r_1,r_2}$ be the probability one set for which \eqref{appprop1} holds. Set
	\begin{equation*}
	B= \bigcap_{r_1,r_2 \in \mathbb{Q}_{[0,1]}} B_{r_1,r_2} \text{ ,}
	\end{equation*}
	so that 
	$\mathbb P(B)=1$. Now we show that the convergence is uniform in $\lambda_d,\lambda_u$ with $\lambda_u-\lambda_d > \epsilon_\lambda$. 
	
	For any fixed positive $\epsilon < \epsilon_\lambda$, we choose a large $m_1$ with $r_0,\ldots,r_{m_1}\in \mathbb{Q}_{[0,1]}$ such that 
	$0=r_0<r_1<\cdots<r_{m_1}=1$ and $\max_{i=1,\ldots,m_1}(r_i-r_{i-1})\leq \epsilon$. Then for any $\lambda_d,\lambda_u\in [0,1]$, 
	we can find integers $j'$ and $k$ such that $j'<k,r_{j'-1}<\lambda_d<r_{j'}$ and $r_{k-1}<\lambda_u < r_k$. Thus, we have
	\begin{align}\label{eq:propA1}
	&\frac{1}{ST}\left|L_{ST}^{(j)}(\psi, \lambda_d,\lambda_u; \bfX_j^o)-(\lambda_u-\lambda_d)\thickbar{L}_{STpair}^{(j)}(\psi)\right| \nonumber \\
	\leq&\frac{1}{ST}\left|L_{ST}^{(j)}(\psi,\lambda_d,\lambda_u; \bfX_j^o)-L_{ST}^{(j)}(\psi,r_{j'-1},r_k;\bfX_j^o)\right| \nonumber \\
	+& \frac{1}{ST}\left|L_{ST}^{(j)}(\psi, r_{j'-1},r_k; \bfX_j^o)-(r_k-r_{j'-1})\thickbar{L}_{STpair}^{(j)}(\psi)\right| \nonumber \\
	+&\frac{1}{ST}\left|(r_k-r_{j'-1})\thickbar{L}_{STpair}^{(j)}(\psi)-(\lambda_u-\lambda_d)\thickbar{L}_{STpair}^{(j)}(\psi)\right|\,.
	\end{align}
	Denote $D=D_{ST}^{r_{j'-1}, r_{j'}}\cup D_{ST}^{r_{k-1},r_k}$ and $E=E_{ST}^{r_{j'-1},r_{j'}}\cup E_{ST}^{r_{k-1},r_k}$. The first term on the right hand side is bounded by 
	\begin{align*}
	&\sup_{\theta \in \Theta(\xi)}\frac{1}{ST}\sum_{(t,i,s_1,s_2)\in D}\left|l_{pair}\left(\psi;  x_{t,s_1}^{(j)},x_{t+i,s_2}^{(j)}\right)\right| + 
	\sup_{\theta \in \Theta(\xi)}\frac{1}{ST}\sum_{(t,s)\in E} \left|l_{marg}\left({\psi}; x_{t,s}^{(j)}\right) \right|\\
    =&\frac{r_{j'}-r_{j'-1}+r_k-r_{k-1}}{ST}\sup_{\theta \in \Theta(\xi)} \mathbb E\left(\sum_{(t,i,s_1,s_2)\in D}\left|l_{pair}\left(\psi;  x_{t,s_1}^{(j)},x_{t+i,s_2}^{(j)}\right)\right|\right)\\
    +&\frac{r_{j'}-r_{j'-1}+r_k-r_{k-1}}{ST}\sup_{\theta \in \Theta(\xi)} \mathbb E\left(\sum_{(t,s)\in E} \left|l_{marg}\left({\psi}; x_{t,s}^{(j)}\right)\right|\right)+o(1),\\
    \leq & 4C\epsilon +o(1),
	\end{align*}
	where the last inequality follows from Assumption \ref{ass.mom}.
	
    Similarly, for sufficiently large $(S,T)$, the third term on the right hand side of \eqref{eq:propA1} is bounded by
	\begin{align*}
	\frac{r_k-r_{k-1}}{ST}\sup_{\theta\in\Theta}|\thickbar{L}_{STpair}^{(j)}\{(\xi,\theta)\}|+\frac{r_{j'}-r_{j'-1}}{ST}\sup_{\theta\in\Theta}|\thickbar{L}_{STpair}^{(j)}\{(\xi,\theta)\}| < 2\epsilon C.
	\end{align*}
	By \eqref{appprop1}, the second term is bounded by $\epsilon$ for sufficiently large $(S,T)$. It follows that
	\begin{align*}
		\sup_{\lambda_d,\lambda_u}\sup_{\theta\in\Theta(\xi)}\left|\frac{1}{ST}L_{ST}^{(j)}\{(\xi,\theta), \lambda_d,\lambda_u;\bfX_j^o\}
		-(\lambda_u-\lambda_d)\thickbar{L}_{STpair}^{(j)}\{(\xi,\theta)\}\right| < 6\epsilon C+ \epsilon \,,
	\end{align*}
	for sufficiently large $(S,T)$. As $\epsilon$ is arbitrary and independent of $\lambda_d$ and $\lambda_u$, the proof is complete.
\end{proof}

\begin{proof}[Proof of Proposition \ref{proprep}]
	Let $\tilde{B}$ be the probability one set that on which Proposition \ref{partdata} holds for 
	all segments. Fix an $\tilde{w}\in \tilde{B}$. By setting
	\begin{equation} \label{lambdadef}
	\grave{\lambda}_d=\max(0,\lambda_d),\quad
	\ddot{\lambda}_d=\min(0,\lambda_d),\quad
	\grave{\lambda}_u=\min(1,\lambda_u),\quad
	\ddot{\lambda}_u=\max(1,\lambda_u),
	\end{equation}
	then we can consider the stationary and leftover pieces from the adjacent segments separately. Similar to the proof of Proposition \ref{partdata}, we only need to prove the result for $\thickbar{L}^{(j)}_{STpair}\{(\xi,\theta)\}$. By elementary algebra,
	\begin{align} \label{approprep}
	&\frac{1}{ST}\left|L_{ST}^{(j)}(\psi, \lambda_d,\lambda_u; \bfX_j^o) - (\lambda_u-\lambda_d)\thickbar{L}_{STpair}^{(j)}(\psi)\right| \\
	\leq&\frac{1}{ST}\left|L_{ST}^{(j)}(\psi, \grave{\lambda}_d,\grave{\lambda}_u;\bfX_j^o) - (\grave{\lambda}_u-\grave{\lambda}_d)\thickbar{L}_{STpair}^{(j)}(\psi)\right|
	+\frac{1}{ST}\left|(\ddot{\lambda}_u-1-\ddot{\lambda}_d)\thickbar{L}_{STpair}^{(j)}(\psi) \right|\nonumber \\
	+&\left[\frac{1}{ST}\sum_{(t,i,s_1,s_2)\in D_{ST_{j-1}}^{(1+\ddot{\lambda}_d),1}}\left|l_{pair}(\psi;x^{(j-1)}_{t,s_1},x^{(j-1)}_{t+i,s_2})\right|
    +\frac{1}{ST}\sum_{(t,i,s_1,s_2)\in D_{ST_{j+1}}^{0,(\ddot{\lambda}_u-1)}}\left|l_{pair}(\psi;x^{(j+1)}_{t,s_1},x^{(j+1)}_{t+i,s_2})\right|\right]\nonumber \\
	+&\left[\frac{1}{ST}\sum_{(t,s)\in E_{ST_{j-1}}^{(1+\ddot{\lambda}_d),1}} \left|l_{marg}\left({\psi}; x_{t,s}^{(j-1)}\right)\right| 
	+ \frac{1}{ST}\sum_{(t,s)\in E_{ST_{j+1}}^{0,(\ddot{\lambda}_u-1)}} \left|l_{marg}\left({\psi}; x_{t,s}^{(j+1)}\right)\right|\right] \nonumber \\ 
    +&\left[\frac{1}{ST}\sum_{1\leq s_1 \leq S}\sum_{s_2\in s_1\cup\mathcal{N}(s_1)}\left|l_{pair}(\psi;x^{(j-1)}_{T_{j-1},s_1},x^{(j)}_{1,s_2})\right|
	+\frac{1}{ST}\sum_{1\leq s_1 \leq S}\sum_{s_2\in s_1\cup\mathcal{N}(s_1)}\left|l_{pair}(\psi;x^{(j)}_{T_j,s_1},x^{(j+1)}_{1,s_2})\right|\right]\nonumber
	\end{align}
	Note that on the right side of \eqref{approprep}, the first term involves observations from a stationary segment, and 
	the last three terms are the leftover and crossover pieces from adjacent segments.
	Since $0\leq\grave{\lambda}_d < \grave{\lambda}_u\leq 1$, 
	the first term converges to zero almost surely by Proposition \ref{partdata}. 
	Moreover, for any $\delta > 0,\max(|\ddot{\lambda}_d|,|\ddot{\lambda}_u-1|)<\delta$ for sufficiently large $(S,T)$. Thus, the second term is bounded by 
	$\frac{2\delta}{ST}|\thickbar{L}_{STpair}^{(j)}(\psi)|$, and the third term is bounded by
	\begin{align*}
	\frac{1}{ST}\sum_{(t,i,s_1,s_2)\in D_{ST_{j-1}}^{(1-\delta),1}}\left|l_{pair}(\psi;x^{(j-1)}_{t,s_1},x^{(j-1)}_{t+i,s_2})\right|+
    \frac{1}{ST}\sum_{(t,i,s_1,s_2)\in D_{ST_{j+1}}^{0,\delta}}\left|l_{pair}(\psi;x^{(j+1)}_{t,s_1},x^{(j+1)}_{t+i,s_2})\right|,
	\end{align*}
	which is asymptotically bounded by $2\delta C$ almost surely as is shown in the proof of Proposition \ref{partdata}. A similar bound can be established for the fourth term. By Proposition \ref{edge_effect}, the fifth term converge to 0 almost surely. Since $\delta$ is arbitrary, the left side of \eqref{approprep} converges to zero uniformly in $\lambda_d$ and $\lambda_u$ in the sense of \eqref{supfdef}.
\end{proof}

\begin{proof}[Proof of Proposition \ref{MPLE}]
	By the definition of $\hat{\theta}_{ST}$, we have 
	$$L_{ST}^{(j)}\{(\xi_j,\hat{\theta}_{ST}),\lambda_d,\lambda_u;\bfX_j^o\} \geq L_{ST}^{(j)}\{(\xi_j,\theta_j^*),\lambda_d,\lambda_u;\bfX_j^o\}$$
	for every $\lambda_d,\lambda_u$ and $(S,T)$.
	 
	With the uniform convergence from Proposition \ref{proprep}, we have
	\begin{align*}
		& \frac{\lambda_u-\lambda_d}{ST}[\thickbar{L}_{ST}^{(j)}\{(\xi_j,\theta_j ^*)\}-\thickbar{L}_{ST}^{(j)}\{(\xi_j,\hat{\theta}_{ST})\}] \\
		\leq& \sup_{\underline{\lambda_d},\overline{\lambda_u}}\left[\frac{\lambda_u-\lambda_d}{ST}\thickbar{L}_{ST}^{(j)}\{(\xi_j,\theta_j^*)\}-\frac{1}{ST}L_{ST}^{(j)}\{(\xi_j,\theta_j^*),\lambda_d,\lambda_u; \bfX_j^o\}\right]  + o(1)\\
		\leq& \sup_{\underline{\lambda_d},\overline{\lambda_u}}\sup_{\theta_j\in \Theta_j(\xi_j)}\left[\frac{\lambda_u-\lambda_d}{ST}\thickbar{L}_{ST}^{(j)}\{(\xi_j,\theta_j)\}-\frac{1}{ST}L_{ST}^{(j)}\{(\xi_j,\theta_j),\lambda_d,\lambda_u; \bfX_j^o\}\right] +o(1)\\
		\longrightarrow& 0 \text{ ,   almost surely}.
	\end{align*}
	Since $\thickbar{L}_{ST}^{(j)}\{(\xi_j,\theta_j^*)\}$ is the maximum value over $\thickbar{L}_{ST}^{(j)}\{(\xi_j,\cdot)\}$ and $\lambda_u-\lambda_d >\epsilon_\lambda$, we have
	\begin{equation} \label{MPLEcon}
	\frac{1}{ST}|\thickbar{L}_{ST}^{(j)}\{(\xi_j,\hat{\theta}_{ST})\} - \thickbar{L}_{ST}^{(j)}\{(\xi_j,\theta_j^*)\}| \longrightarrow 0\text{ , almost surely}.
	\end{equation}
	From Assumption \ref{ass.model.id}, it follows that $\frac{1}{ST}\thickbar{L}_{ST}^{(j)}\{(\xi_j,\theta_j)\}$ has a unique maximum at $\theta_j^*$. 
	Thus, it follows from \eqref{MPLEcon} and Assumption \ref{ass.model.id} that \eqref{suppartcon} holds.
\end{proof}

\section{\textcolor{black}{Proof for vanishing change sizes in \Cref{subsec:vanish_cp}}}\label{sec:vanishing}

{\color{black}
\subsection{Notations}

We remark that the technical arguments used in the proof for localization error rate results under vanishing change sizes are substantially different than the ones in \Cref{sec:proof_increasing_domain} for constant change sizes. Thus, for expositional simplicity, in this subsection, we further introduce some additional notations will be used in the following proof in \Cref{sec:vanishing}.

Recall that we denote the true change-points as $\{\tau_1^o,\ldots, \tau_{m_o}^o\}$ where $\tau_j^o=[T\lambda_{j}^o]$. Define $\tau_0^o=0$ and $\tau_{m_o+1}^o=T$, denote $T_j=\tau_j^o-\tau_{j-1}^o$ for $j=1,\ldots,m_o+1$. We first define several index sets that will be used throughout the proof, where
\begin{align*}
E_t&=\{(t,s):1\leq s\leq S, \text{ each } s  \text{ repeats } (1+|\mathcal{N}(s)|) \text{ times} \},\\
D^c_t&=\{(t,s_1,s_2): 1\leq s_1 \leq S, s_2 \in s_1\cup \mathcal{N}(s_1)\},\\
D_{t_1,t_2}&=\{(t,i,s_1,s_2): t_1\leq t, t+i \leq t_2, 0\leq i\leq 1, 1\leq s_1 \leq S, s_2 \in s_1\cup \mathcal{N}(s_1), \text{if }i=0, s_1\neq s_2\}.
\end{align*}
The set $E_t$ is used to index the marginal loglikelihood in CLMDL for compensating the edge effect. The set $D^c_t$ is used to index the pairwise loglikelihood across time $t$, and the set $D_{t_1,t_2}$ is used to index the pairwise loglikelihood between $t_1$ and $t_2.$

Recall $\psi=(\xi,\theta)$, where $\xi$ is the model index that denotes the form of the parametric model and $\theta=\theta(\xi)$ denotes the model parameter given $\xi$. We further define
\begin{align*}
    &L_p(\psi,t_1,t_2)=\sum_{(t,i,s_1,s_2)\in D_{t_1,t_2}} l(\psi;y_{t,s_1},y_{t+i,s_2}),\\
    &L_{c}(\psi,t)=\sum_{(t,s_1,s_2)\in D^c_t} l(\psi;y_{t,s_1},y_{t+1,s_2}),\\
    &L_{m}(\psi,t)=\sum_{(t,s)\in E_t} l(\psi;y_{t,s}),
\end{align*}
where for notational simplicity, we skip the subscripts $marg$ and $pair$ in the loglikelihood functions. It is easy to see that the composite log-likelihood function for observations $\{y_{t,s}: t_1\leq t\leq t_2, 1\leq s\leq S\}$ takes the form
\begin{align*}
    L(\theta,t_1,t_2)=L_p(\theta,t_1,t_2)+L_m(\theta,t_1)+L_m(\theta,t_2).
\end{align*}
In the following, we denote the first and second order derivative of $L$ by $L'$ and $L''$. Same applies to $L_p, L_c$, $L_m$, and $l$. 

Recall that the pseudo-true model order is $\xi^o$ for all the stationary segments with $\theta_j^o$ being the pseudo-true model parameter for the $j$th stationary segment for $j=1,\cdots,m_o+1.$ Note that for notational simplicity, in the following, we replace $\psi$ with $\theta$ when no confusion arises.

\subsection{Technical propositions and lemmas}

\Cref{lem_newULLN} provides a uniform convergence result for the log-likelihood function and its second order derivative, which will be used throughout the later proofs.
\begin{proposition}\label{lem_newULLN}
Under Assumptions \ref{ass.countable}, \ref{ass.mom}($r$), \ref{ass.mix}($r$) and \ref{ass.mom2}($r$) with some $r>2$, suppose $TS^{-r/2}\to 0$, we have for any $\epsilon>0$, and any $\psi=(\xi,\theta), \theta\in\Theta(\xi)$, it holds that
    \begin{align*}
       \text{(i) \hspace{5mm}} &\mathbb P \left(\sup_{\substack{t_1,t_2,\\ t_2-t_1\geq 0}}\sup_{\theta\in\Theta(\xi)} \frac{1}{(t_2-t_1+1)S}
       \left|L_p(\psi,t_1,t_2)-\mathbb E(L_p(\psi,t_1,t_2))\right|>\epsilon\right)\to 0,\\
       \text{(ii) \hspace{5mm}} &\mathbb P \left( \sup_{\substack{t_1,t_2,\\ t_2-t_1\geq 0}}\sup_{\theta\in\Theta(\xi)} \frac{1}{(t_2-t_1+1)S}\left|L_p''(\psi,t_1,t_2)-\mathbb E (L_p''(\psi,t_1,t_2)) \right| > \epsilon \right) \to 0,\\
       \text{(iii) \hspace{5mm}} &\mathbb P \left(\sup_{t}\sup_{\theta\in\Theta(\xi)} \frac{1}{S}
       \left|L_c(\psi,t)-\mathbb E(L_c(\psi,t))\right|>\epsilon\right)\to 0,\\
       \text{(iv) \hspace{5mm}} &\mathbb P \left(\sup_{t}\sup_{\theta\in\Theta(\xi)} \frac{1}{S}
       \left|L_m(\psi,t)-\mathbb E(L_m(\psi,t))\right|>\epsilon\right)\to 0.
    \end{align*}
\end{proposition}

\Cref{lem_BM} further provides a uniform bound on the asymptotic order of the first order derivative of the log-likelihood function when evaluated at the true model parameter on each stationary segment, which will be used throughout the later proofs.
\begin{proposition}\label{lem_BM}
    Under the conditions of \Cref{lem_newULLN}, and Assumptions \ref{ass.model.id}(i) and \ref{ass.mixtime}, we have that for each $j=1,2,\cdots,m_o+1$, we have
    \begin{align*}
        \sup_{\tau_{j-1}^o+1\leq t_1\leq t_2 \leq \tau_j^o} \left| L'(\theta_j^o, t_1, t_2)\right|=O_p(\sqrt{ST}),
    \end{align*}
    where $\theta_j^o$ is the pseudo-true parameter of the $j$th stationary segment. Same applies to $L_p'$, $L_c'$, $L_m'.$
\end{proposition}

\Cref{lem_theta_star} provides a uniform bound on the distance between the pseudo-true parameter on an arbitrary mixture of stationary segments $\theta_{t_1,t_2}^*$ and the pseudo-true parameter of each stationary segment $\theta_1^o,\cdots,\theta_{m_o+1}^o$, which will be used throughout the later proofs to establish negative definiteness of the Hessian matrices.
\begin{proposition}\label{lem_theta_star}
Denote $\theta_{t_1,t_2}^*$ as the maximizer of the expected composite log-likelihood function $\mathbb E(L((\xi^o, \theta), t_1,t_2)).$ Under the vanishing change size where $\kappa\to 0$ and suppose Assumptions \ref{ass.mom}(i) and \ref{ass.model.id}(i) hold, we have that 
\begin{align*}
    \sup_{\substack{t_1,t_2,\\t_2-t_1\geq \epsilon_\lambda^o T}}\max_{1\leq l\leq m_o+1}\|\theta_{t_1,t_2}^*-\theta_l^o\|_2 \to 0, \text{ as } S, T \text{ increase.}
\end{align*}
\end{proposition}

\Cref{lem_theta_hat} provides a uniform bound on the asymptotic deviation of the estimated model parameter $\hat\theta_{t_1,t_2}$ from $\theta_l^o, l=1,\cdots,m_o+1$ on an arbitrary mixture of stationary segments, which will be used in the proofs to help establish preliminary bounds on the localization error rate.
\begin{proposition}\label{lem_theta_hat}
Denote $\hat\theta_{t_1,t_2}$ as the maximizer of the composite log-likelihood function $L((\xi^o,\theta), t_1,t_2).$ Under the conditions of \Cref{lem_BM}, we have that
$$\sup_{\substack{t_1,t_2,\\t_2-t_1\geq \epsilon_\lambda^o T}}\max_{1\leq l\leq m_o+1}\|\hat\theta_{t_1,t_2}-\theta_l^o\|_2=O_p(1/\sqrt{ST})+O_p(1/T)+O_p(\kappa).$$
\end{proposition}

\Cref{lem_rate} provides the result for the model order selection and localization error rate of the CL-based change-point estimators defined in \eqref{minCL} of the main text, when the number of change-points $m_o$ is known. It serves as an important building block for the proof of \Cref{thm_diminishing}.
\begin{lemma} \label{lem_rate}
     Let $\bfY$ be a piecewise stationary spatio-temporal process specified by $(m_o,\Lambda^o,\Psi^o)$ and Assumptions \ref{ass.countable}, \ref{ass.mom}($r$), \ref{ass.model.id}(i), \ref{ass.mix}($r$), \ref{ass.mixtime} and \ref{ass.mom2}($r$) hold with some $r>2$ and furthermore $T\cdot S^{-r/2}\to 0$. Suppose that the number of change-points $m_o$ is known. We estimate the change-points and model parameters  by the CL based estimator in \eqref{minCL}, i.e.\
     \begin{align}
     \{\tilde{\Lambda}_{ST},\tilde{\Psi}_{ST}\}=\argmin_{\substack{ \Lambda\in A_{\epsilon}^{m_o}, \Psi\in \mathcal{M}^{m_o+1}}}-\sum_{j=1}^{m_o+1}L_{ST}^{(j)}(\psi_j;\bfX_j):=\argmax_{\substack{ \Lambda\in A_{\epsilon}^{m_o}, \Psi\in \mathcal{M}^{m_o+1}}}\mathrm{CL}(m_o,\Lambda,\Psi)\,,
     \end{align}
     where $A_{\epsilon}^{m_o}$ is defined in \eqref{A}. Suppose the change size satisfies $ T\cdot (S\kappa^2)\to \infty$.
     
     \noindent (i).\ If $\liminf\limits_{S,T\to\infty}S\kappa^2>0$, with probability going to 1, we have
     \begin{align*}
         [T\tilde{\lambda}_j]=[T\lambda_j^o] \text{ for all } j=1,\cdots, m_o.
     \end{align*}
     
     \noindent (ii).\ If $S\kappa^2 \to 0$, we have
     \begin{align}\label{eq:diminish_rate_CL}
         \max_{j=1,\cdots,m_o}\big|[T\tilde{\lambda}_j]-[T\lambda_j^o]\big| = O_p( \sqrt{T/(S\kappa^2}) \wedge (S\kappa^2)^{-r/(r-2)} ),
     \end{align}
     which directly implies that $\tilde \Lambda_{ST}$ is consistent, i.e.\
     \begin{align*}
       \max_{j=1,\cdots,m_o} |\tilde\lambda_j-\lambda_j^o| \to_p 0,
     \end{align*}
     In addition, for each segment, the estimated model does not underestimate the order of the true model, i.e., $\tilde{\xi}_{j}\geq \xi^o_{j}$ for $j=1,\cdots,m_o+1.$
\end{lemma}

\Cref{lem_rate_m_unknown} gives the localization error rate of the CL-based change-point estimator $(\tilde m, \tilde{\Lambda}_{ST},\tilde{\Psi}_{ST})$ defined in \eqref{minCL} of the main text when the number of change-points $m_o$ is unknown. In particular, the number of change-points $m_o$ can be correctly estimated with probability going to 1, and thus the results in \Cref{lem_rate} still hold. Therefore, \Cref{lem_rate_m_unknown} indicates that the phenomenon of consistency of the CL-based change-point estimator~(without penalty) in \Cref{prop.minCL} still holds under vanishing change sizes. In other words, under the spatio-temporal setting we consider, consistent change-point estimation can be achieved without any penalty term in the criterion function.

\begin{lemma} \label{lem_rate_m_unknown}
     Under the same condition as that in \Cref{lem_rate} except that the number of change-points $m_o$ is unknown, we estimate the number and locations of change-points and model parameters by the CL based estimator in \eqref{minCL}, i.e.\
     \begin{align}
     \{\tilde m, \tilde{\Lambda}_{ST},\tilde{\Psi}_{ST}\}=\argmin_{\substack{m\leq M_\lambda, \Lambda\in A_{\epsilon_\lambda}^{m}}, \Psi\in \mathcal{M}^{m+1}}-\sum_{j=1}^{m+1}L_{ST}^{(j)}(\psi_j;\bfX_j)=\argmax_{\substack{m\leq M_\lambda, \Lambda\in A_{\epsilon_\lambda}^{m}}, \Psi\in \mathcal{M}^{m+1}} \mathrm{CL}(m, \Lambda, \Psi).
     \end{align}
     Suppose the change size satisfies $ T\cdot (S\kappa^2)\to \infty$.

     \noindent (i).\ With probability going to 1, we have $\tilde{m}= m_o$ and the estimated model does not underestimate the order of the true model, i.e., $\tilde{\xi}_{j}\geq \xi^o_{j}$ for $j=1,\cdots,m_o+1.$ 

     \noindent (ii).\ Moreover, the localization error rate result in \Cref{lem_rate}(i) and (ii) still holds.
\end{lemma}

\subsection{Discussions on the localization error rate of CLMDL}\label{subsec:remarks}
\begin{remark}[The role of the moment condition $r$]\label{remark:sharper_rate}
The presence of $r$ in the localization error rate of CLMDL under the scenario where $S\kappa^2 \to 0$~(see \eqref{eq:diminish_rate_CL} in \Cref{lem_rate}(ii) and \eqref{eq:diminish_rate} in \Cref{thm_diminishing}(ii)) is due to a deviation bound in the proof that takes a generic form of
\begin{align}\label{eq:prob_bound}
    \mathbb P\left(\max_{0\leq t\leq \tau_1^o-M}\frac{1}{\tau_1^o-t}\bigg|\frac{1}{\sqrt{S\kappa^2}}L_p'(\theta_1^o,t+1,\tau_1^o)(\theta_2^o-\theta_1^o)\bigg|>\epsilon \sqrt{S\kappa^2} \right) \leq \frac{1}{\epsilon^r (S\kappa^2)^{r/2}} M^{-r/2+1},
\end{align}
where $L_p'(\theta^o_1,t+1,\tau_1^o)$ is the first order derivative of the composite log-likelihood~(i.e.\ score function) from time $t+1$ to $\tau_1^o$ and evaluated at the pseudo-true parameter $\theta_1^o$ (thus is of mean zero), and $M\in\mathbb N^+$ quantifies the localization error. To bound the l.h.s.\ of \eqref{eq:prob_bound}, we use a moment inequality (i.e.\ \Cref{s_converge}) and get the r.h.s.\ of \eqref{eq:prob_bound}. The moment inequality is derived from \cite{Doukhan1994}  and holds for a general spatio-temporal process~(i.e.\ random field). We refer to \eqref{eq:key} in the proof of \Cref{lem_rate} for more details of the origin of \eqref{eq:prob_bound}.

We remark that \eqref{eq:prob_bound} can be more tightly bounded given the availability of a sharper inequality~(in particular, the generalized H\'ajek-R\'enyi inequality). Define 
\begin{align*}
    &A_q=\frac{1}{\sqrt{S\kappa^2}}\sum_{s_1=1}^S\sum_{s_2\in\mathcal{N}(s_1)}l'(\theta_1^o;y_{q,s_1},y_{q,s_2})(\theta_2^o-\theta_1^o),\\
    &B_q=\frac{1}{\sqrt{S\kappa^2}}\sum_{s_1=1}^S\sum_{s_2\in\mathcal{N}(s_1)\cup \{s_1\}}l'(\theta_1^o;y_{q,s_1},y_{q+1,s_2})(\theta_2^o-\theta_1^o).
\end{align*}
It is easy to see that by definition of $L_p$, we have 
\begin{align*}
    \frac{1}{(\tau_1^o-t)\sqrt{S\kappa^2}}L_p'(\theta^o_1,t+1,\tau_1^o)(\theta_2^o-\theta_1^o)=\frac{1}{\tau_1^o-t}\sum_{q=t+1}^{\tau_1^o}A_q + \frac{1}{\tau_1^o-t}\sum_{q=t+1}^{\tau_1^o-1}B_q,
\end{align*}
and thus the l.h.s.\ of \eqref{eq:prob_bound} becomes
\begin{align*}
    \mathbb P\left(\max_{0\leq t\leq \tau_1^o-M}\bigg|\frac{1}{\tau_1^o-t}\sum_{q=t+1}^{\tau_1^o}A_q + \frac{1}{\tau_1^o-t}\sum_{q=t+1}^{\tau_1^o-1}B_q\bigg|>\epsilon \sqrt{S\kappa^2} \right).
\end{align*}

Note that $\{A_q\}_{q=1}^{\tau_1^o}$ and $\{B_q\}_{q=1}^{\tau_1^o-1}$ are two mean-zero  stationary univariate time series and by \Cref{s_converge}(ii) and Jensen's inequality, both have bounded variance as long as $r>2$. Thus, if we further assume that $\{A_q\}_{q=1}^{\tau_1^o}$ and $\{B_q\}_{q=1}^{\tau_1^o-1}$ are martingale difference sequences or linear processes, we can apply the generalized H\'ajek-R\'enyi inequality in \cite{Bai1994}~(see Proposition 1 and Assumption (B) therein) and show that
\begin{align*}
    \mathbb P\left(\max_{0\leq t\leq \tau_1^o-M}\bigg|\frac{1}{\tau_1^o-t}\sum_{q=t+1}^{\tau_1^o}A_q\bigg| >\epsilon \sqrt{S\kappa^2} \right)\leq C_1/(\epsilon^2 S\kappa^2 M),
\end{align*}
for some $C_1>0$. Same applies to $\{B_q\}.$ Thus, we can sharpen the deviation bound in \eqref{eq:prob_bound} to
\begin{align*}
    \mathbb P\left(\max_{0\leq t\leq \tau_1^o-M}\frac{1}{\tau_1^o-t}\bigg|\frac{1}{\sqrt{S\kappa^2}}L_p'(\theta_1^o,t+1,\tau_1^o)(\theta_2^o-\theta_1^o)\bigg|>\epsilon \sqrt{S\kappa^2} \right) \leq C/(\epsilon^2 S\kappa^2 M),    
\end{align*}
for some $C>0$. Combined with the rest of the proof, we can show that the localization error rate of CLMDL in \Cref{thm_diminishing}(ii) under the scenario where $S\kappa^2\to 0$ can be made sharper as
\begin{align*}
    \max_{j=1,\cdots,m_o} \big| [T\hat\lambda_{j}] - [T\lambda_j^o] \big|=O_p\big(1/(S\kappa^2)\big),
\end{align*}
as long as $r>2$.

However, it seems that the martingale difference or linear processes assumption is not able to cover a general spatio-temporal process. Thus, we opt not to impose such assumptions and rely on the less sharp moment inequality in \Cref{s_converge}, which gives a localization error rate that depends on the moment condition $r$ as in \Cref{lem_rate} and \Cref{thm_diminishing}. $\hfill \square$
\end{remark}
\vskip 2mm

\begin{remark}[Comparison with the high-dimensional mean change literature]\label{remark:hdmean}
    As discussed in the main text, the exact recovery phenomenon has also been observed in the literature for non-parametric high-dimensional mean change. In particular, \cite{bai2010common} and its extension \cite{bhattacharjee2019change} consider the single change-point model (using our notations)
    \begin{align}\label{eq:model_hdmean}
        \bfy_t={\boldsymbol\mu}_t+\boldsymbol{\varepsilon}_t, ~t=1,\cdots, T, \text{ with } \boldsymbol\mu_t=\boldsymbol\mu_{pre} \cdot \mathbb I(t\leq \tau^o)+\boldsymbol\mu_{post} \cdot \mathbb I(t>\tau^o),
    \end{align}
    where $\tau^o=[T\lambda^o]$ is the change-point, $\bfy_t,\boldsymbol\mu_t,\boldsymbol\varepsilon_t\in \mathbb R^S$ and $\{\boldsymbol{\varepsilon}_t\}$ is a mean-zero linear process, and both $S,T$ are allowed to increase. In addition, suitable assumptions are imposed on the coefficient matrices of the linear process $\{\boldsymbol{\varepsilon}_t\}$ to regulate the temporal and cross-sectional dependence.
    
    Given the knowledge of one and only one change-point, they show that a least squares based estimator $\hat\tau$ achieves the localization error of 
    \begin{align}\label{eq:rate_hdmean}
        |\hat\tau-\tau^o|=O_p(1/\|{\boldsymbol\mu}_{pre}-{\boldsymbol\mu}_{post}\|_2^2) \text{ given that } T/S\cdot \|{\boldsymbol\mu}_{pre}-{\boldsymbol\mu}_{post}\|_2^2\to\infty.
    \end{align}
    We refer to \cite{bai2010common}~(e.g.\ conditions (4) and (5) and Theorem 3.2) and \cite{bhattacharjee2019change}~(e.g.\ condition SNR$^*$ and Theorem 2.1) for more details. In particular, \eqref{eq:rate_hdmean} implies that exact recovery of $\tau^o=[T\lambda^o]$ can be achieved if one further has $\|{\boldsymbol\mu}_{pre}-{\boldsymbol\mu}_{post}\|_2^2\to \infty.$ We note that similar phenomenon has also been observed in the recent works of \cite{chen2022inference} and \cite{li2022ell} under more general settings for high-dimensional mean change.

    Clearly, CLMDL is substantially different from the high-dimensional mean change literature in terms of problem setting, methodology and technical arguments. In particular, CLMDL focuses on change-point estimation in a general spatio-temporal parametric model (which covers change in mean and space-time covariance functions), while the high-dimensional literature focuses on non-parametric mean change. Thus, the two are not directly comparable.

    On the other hand, there is also some intuitive connection between the localization error rate \eqref{eq:rate_hdmean} with the rate of CLMDL in \Cref{thm_diminishing}. In particular, due to the parametric nature of our setting and the accumulation of information along the spatial dimension $S$, intuitively, the quantity $S\kappa^2$ can be viewed as the ``effective (squared) change size" of our problem. In some sense, the quantity $S\kappa^2$ plays a similar role as the quantity $\|{\boldsymbol\mu}_{pre}-{\boldsymbol\mu}_{post}\|_2^2$ in \eqref{eq:rate_hdmean}. Therefore, \eqref{eq:rate_hdmean} resembles the localization error rate of CLMDL in \Cref{thm_diminishing}. In particular, both suggest that exact recovery can be achieved if $S\kappa^2\to\infty$ or $\|{\boldsymbol\mu}_{pre}-{\boldsymbol\mu}_{post}\|_2^2\to \infty$.
    
    Interestingly, \Cref{thm_diminishing}(i) further suggests that CLMDL can achieve exact recovery even if $S\kappa^2 \to c$ for some $c>0$, which is different than \eqref{eq:rate_hdmean}. This phenomenon can be attributed to the compensating mechanism of the proposed composite likelihood~(which is designed to correct the edge effect) and the fact that the spatial dimension $S$ diverges. In particular, we show in the proof that as long as there is localization error, the term resulted from the compensating mechanism of the composite likelihood (i.e.\ the difference between the dropped pairwise likelihood and the corresponding compensating marginal likelihood) will be asymptotically positive of order $O_p(S)$. Moreover, we can further show that once the localization error is $O_p(1)$, this term can further help eliminate any localization error of CLMDL at all. Indeed, this shares a similar intuition as the phenomenon of consistency of the CL-based change-point estimator discussed below \Cref{prop.minCL} in the main text. We refer to scenario (C) in the proof of \Cref{lem_rate} for more technical details.
    
    For the case where $\|{\boldsymbol\mu}_{pre}-{\boldsymbol\mu}_{post}\|_2 \to 0$, under some suitable linear processes assumptions of $\{\boldsymbol{\varepsilon}_t\}$ in \eqref{eq:model_hdmean}, \cite{bai2010common} and \cite{bhattacharjee2019change} give the sharp rate of $O_p(1/\|{\boldsymbol\mu}_{pre}-{\boldsymbol\mu}_{post}\|_2^2)$. As discussed in \Cref{remark:sharper_rate} above, CLMDL can also achieve the rate of $O_p(1/(S\kappa^2))$ with a linear process assumption of the score function. In addition, for the general case, if the moment condition holds for all $r>2$, \Cref{thm_diminishing}(ii) implies $\big| [T\hat\lambda_{j}] - [T\lambda_j^o] \big|=O_p((S\kappa^2)^{-1-\delta})$ for any $\delta>0$, which almost matches \eqref{eq:rate_hdmean}.
    
    Note that Theorem \ref{thm_diminishing} of CLMDL requires the change size to satisfy $T\cdot (S\kappa^2)\to \infty$. In particular, it takes the form of (spacing between change-points) $\times$ (squared change size) $\to \infty,$ which resembles the form of requirement for change sizes in the classical change-point literature in univariate/multivariate time series (see e.g.\ \cite{Csoergo1997}). This is weaker than the requirement of change size for the high-dimensional mean change in \eqref{eq:rate_hdmean}, which is $T/S\cdot \|{\boldsymbol\mu}_{pre}-{\boldsymbol\mu}_{post}\|_2^2\to\infty$ and is due to the difficulty brought by the non-parametric and high-dimensional nature of the problem. $\hfill \square$
\end{remark}

\subsection{Proof of the main theorem}
\begin{proof}[Proof of \Cref{thm_diminishing}]

    First, note that the consistency and localization error rate result of the CL-based change-point estimator in \Cref{lem_rate_m_unknown} still hold for the CLMDL-based estimator. To see that, note that the only difference between the CL and CLMDL criterion is due to the MDL part. The additional (positive) MDL part will further penalize and prevent the over-estimation of the number of change-points. Furthermore, note that the MDL part is uniformly upper bounded by $C\log(ST)$ for some positive $C>0$, which is clearly of order $o_p(S)$ due to the requirement that $T\cdot S^{-r/2} \to 0.$ Thus, all arguments in the proof of \Cref{lem_rate_m_unknown} still go through.

    Thus, in the following, we only need to focus on the proof of the consistency of the estimated model order. In other words, CLMDL does not over-estimate the model order $\xi^o.$

    (I). For the case where $\liminf\limits_{S,T\to\infty}S\kappa^2>0$, we have exact recovery of the true change-points, i.e.\
    $\mathbb P( [T\hat\lambda_j]=[T\lambda_j^o] \text{ for } j=1,\cdots,m_o) \to 1. $ Thus, the exact same argument as that in the proof of \Cref{unknownprob}(ii) can be used to show that the model order cannot be over-estimated, i.e. with probability going to 1, we have $\hat\xi_j=\xi_j^o=\xi^o$ for all $j=1,\cdots,m_o+1.$ 

    (II). Therefore, we only need to prove the case where $S\kappa^2 \to 0.$ The proof relies on a careful analysis of $\hat\theta_j=\hat{\theta}_j(\hat\xi_j)$ for $j=1,\cdots, m_o+1.$ Note that without exact recovery of the true change-points, it is possible that $\hat\theta_j=\hat{\theta}_j(\hat\xi_j)$ is estimated based on mixture of two or three stationary segments, which creates additional technical challenge.

    In the following, for notational simplicity, we denote $\hat{\tau}_j=[T\hat\lambda_j]$ and $\tau_j^o=[T\lambda_j^o]$. In addition, we replace $\psi$ with $\theta$ and $\hat{\psi}$ with $\hat\theta$ when no confusion arises. By the localization error rate, we have that $|\hat\tau_j-\tau_j^o|/T=O_p(\sqrt{T/(S\kappa^2)}/T) = O_p(1)\sqrt{1/(TS\kappa^2)} =o_p(1)$ since $TS\kappa^2\to \infty.$ In other words, $\hat\tau_j$ is consistent for $\tau_j^o$ for $j=1,\cdots,m_o.$ Thus, for a given stationary segment from $\tau_{l-1}^o+1$ to $\tau_{l}^o$, the most complicated scenario is that $\tau_{l-2}^o<\hat\tau_{l-1}<\tau_{l-1}^o$ and $\tau_{l}^o<\hat\tau_{l}<\tau_{l+1}^o.$
    
    In such case, $\hat{\theta}_l=\hat\theta_l(\hat\xi_l)$ is estimated based on three stationary segments, one is from $\hat\tau_{{l-1}}+1$ to $\tau_{l-1}^o$, one is from $\tau_{l-1}^o+1$ to $\tau_l^o$, and one is from $\tau_l^o+1$ to $\hat\tau_{l}$. We need to control the bias due to the first and third segment, whose presence is caused by the localization error of the estimated change-points. Recall by \Cref{lem_rate_m_unknown}, we have that $\hat\xi_l\geq \xi_l^o=\xi^o$ with probability going to 1. We first analyze $\hat{\theta}_l=\hat\theta_l(\hat\xi_l)$ under the case where $\hat\xi_l=\xi^o$ and thus the pseudo-true parameter is $\theta_l^o.$
    
    Note that by the preliminary localization error rate, the first and third segment length $\hat\tau_{i_l}-\tau_l^o$ is upper bounded by $O_p(\sqrt{T/(S\kappa^2)})=o(T)$, while the second segment is of length $O(T).$ By Taylor expansion, we have that
    \begin{align*}    0=&L'(\hat{\theta}_l, \hat{\tau}_{{l-1}}+1, \hat{\tau}_{{l}})\\
    =&L'(\theta_{l}^o,\hat{\tau}_{{l-1}}+1, \hat{\tau}_{{l}}) + L''(\theta^+,\hat{\tau}_{{l-1}}+1, \hat{\tau}_{{l}})(\hat{\theta}_l-\theta_{l}^o)\\
    =&L_p'(\theta_{l}^o,\hat\tau_{{l-1}}+1,\tau_{l-1}^o) +L_p'(\theta_{l}^o,\tau_{{l-1}}^o+1,\tau_l^o) + L_p'(\theta_{l}^o,\tau_l^o+1, \hat\tau_{l}) + L''(\theta^+,\hat\tau_{{l-1}}+1,\hat\tau_{l})(\hat{\theta}_l-\theta_{l}^o)+O_p(S)\\
    =&L_p'(\theta_{l-1}^o,\hat\tau_{{l-1}}+1,\tau_{l-1}^o)  + L_p''(\theta_{l-1}^+,\hat\tau_{{l-1}}+1,\tau_{l-1}^o)(\theta_{l}^o-\theta_{l-1}^o)\\
    +&
    L_p'(\theta_{l}^o,\tau_{{l-1}}^o+1,\tau_l^o) \\
    + & L_p'(\theta_{l+1}^o,\tau_l^o+1, \hat\tau_{l})
    + L_p''(\theta_{l+1}^+,\tau_l^o+1, \hat\tau_{l})(\theta_{l}^o-\theta_{l+1}^o)\\
     + & L''(\theta^+,\hat\tau_{{l-1}}+1,\hat\tau_{l})(\hat{\theta}_l-\theta_{l}^o)+O_p(S)\\
    =& O_p(\sqrt{ST}) + O_p( (\tau_{l-1}^o-\hat\tau_{l-1})S\kappa) + O_p( (\hat\tau_{l}-\tau_l^o)S\kappa) + O_p(ST)(\hat{\theta}-\theta_{l}^o)+O_p(S),
    \end{align*}
    where the first equality follows from the definition of $\hat{\theta}_l$ and the second equality follows from a Taylor expansion where $\theta^+$ is between $\hat{\theta}_l$ and $\theta^o_{l}$. The third equality follows by decomposing $L'(\theta_{l}^o,\hat\tau_{{l-1}}+1,\hat\tau_{l})$ into $L_p'(\theta_{l}^o,\hat\tau_{{l-1}}+1,\tau_{l-1}^o)$, $L_p'(\theta_{l}^o,\tau_{{l-1}}^o+1,\tau_l^o)$ and $L_p'(\theta_{l}^o,\tau_l^o+1, \hat\tau_{l})$ plus the residual term, which is uniformly $O_p(S)$ over any $\hat\tau_{l-1}$ and $\hat\tau_l$ by \Cref{lem_newULLN} and \Cref{ass.mom}($r$).

    The fourth equality follows from a Taylor expansion of $L_p'(\theta_{l}^o,\hat\tau_{{l-1}}+1,\tau_{l-1}^o)$ and $L_p'(\theta_{l}^o,\tau_l^o+1, \hat\tau_{l})$ where $\theta_{l-1}^+$ is between $\theta_{l-1}^o$ and $\theta_{l}^o$, and $\theta_{l+1}^+$ is between $\theta_{l}^o$ and $\theta_{l+1}^o.$ For the last equality, by \Cref{lem_BM}, we have $L_p'(\theta_{l-1}^o,\hat\tau_{{l-1}}+1,\tau_{l-1}^o)=O_p(\sqrt{ST})$, $L_p'(\theta_{l}^o,\tau_{{l-1}}^o+1,\tau_l^o)=O_p(\sqrt{ST})$ and $L_p'(\theta_{l+1}^o,\tau_l^o+1, \hat\tau_{l})=O_p(\sqrt{ST})$, as all three terms are mean zero. Furthermore, due to the vanishing change size, we have that $\|\theta_{l-1}^+-\theta_{l-1}^o\|_2\to 0$ and $\|\theta_{l+1}^+-\theta_{l+1}^o\|_2\to 0$, and by \Cref{lem_theta_hat}, we further have $\|\hat\theta_l-\theta_l^o\|_2\to_p 0$. Combined with \Cref{ass.mom}(ii), \Cref{ass.model.id}(i) and the uniform convergence result in \Cref{lem_newULLN}(ii), we have the three terms $L_p''(\theta_{l-1}^+,\hat\tau_{{l-1}}+1,\tau_{l-1}^o)/(\tau_{l-1}^o-\hat\tau_{l-1})$, $L_p''(\theta_{l+1}^+,\tau_l^o+1, \hat\tau_{l})/(\hat\tau_l-\tau_l^o)$, and $L''(\theta^+,\hat\tau_{{l-1}}+1,\hat\tau_{l})/(\hat\tau_l-\hat\tau_{l-1})$ asymptotically converge to some non-negative matrices. Thus, we have that $L_p''(\theta_{l-1}^+,\hat\tau_{{l-1}}+1,\tau_{l-1}^o)(\theta_{l}^o-\theta_{l-1}^o)=O_p( (\tau_{l-1}^o-\hat\tau_{l-1})S\kappa)$, $L_p''(\theta_{l+1}^+,\tau_l^o+1, \hat\tau_{l})(\theta_{l}^o-\theta_{l+1}^o)=O_p( (\hat\tau_{l}-\tau_l^o)S\kappa)$ and $L''(\theta^+,\hat\tau_{{l-1}}+1,\hat\tau_{l})=O_p(ST)$.
    
    Thus, together it implies that
    \begin{align*}
        \hat{\theta}_l-\theta_{l}^o= O_p(1/\sqrt{ST})+ O_p( (\tau_{l-1}^o-\hat\tau_{l-1})\kappa/T)+  O_p( (\hat\tau_{l}-\tau_l^o)\kappa/T)+O_p(1/T),
    \end{align*}
    where the first term can be viewed as the variance and the second, third and fourth term can be viewed as the bias due to the localization error of change-point detection. Plug in the preliminary localization error rate, this implies that 
    \begin{align}\label{eq:theta_rate_model}
        \hat{\theta}_l-\theta_{l}^o=O_p(1/\sqrt{ST})+O_p(\sqrt{T/(S\kappa^2)}\kappa /T)+O_p(1/T)=O_p(1/\sqrt{ST}+1/T).
    \end{align}
    Importantly, by the assumption that $S=O(T)$, \eqref{eq:theta_rate_model} implies that $\hat\theta_l-\theta_l^o=O_p(1/\sqrt{ST}).$

    For the case where $\hat\xi_l=\xi_l^*>\xi^o$, note that by the definition of $\theta_l^*=\theta_l^*(\xi_l^*)$ in \Cref{ass.model.id}(i), all the above Taylor expansion based analysis still goes through and we have that
    $\hat\theta_l^*-\theta_l^*=O_p(1/\sqrt{ST}).$ Thus, by Taylor expansion, we have that
    \begin{align*}
      &L(\hat{\theta}_l, \hat{\tau}_{{l-1}}+1, \hat{\tau}_{{l}})-L(\hat{\theta}_l^*, \hat{\tau}_{{l-1}}+1, \hat{\tau}_{{l}})\\
     =&L(\theta_l^o, \hat{\tau}_{{l-1}}+1, \hat{\tau}_{{l}})-L({\theta}_l^*, \hat{\tau}_{{l-1}}+1, \hat{\tau}_{{l}}) \\
     +&(\hat\theta_l-\theta_l^o)L''(\theta^+,  \hat{\tau}_{{l-1}}+1, \hat{\tau}_{{l}})(\hat\theta_l-\theta_l^o) + (\hat\theta_l^*-\theta_l^*)L''(\theta^{*+},  \hat{\tau}_{{l-1}}+1, \hat{\tau}_{{l}})(\hat\theta_l^*-\theta_l^*)=O_p(1),
    \end{align*}
    where the last equality follows from the fact that $L(\theta_l^o, \hat{\tau}_{{l-1}}+1, \hat{\tau}_{{l}})=L({\theta}_l^*, \hat{\tau}_{{l-1}}+1, \hat{\tau}_{{l}})$ by the definition of $\theta_l^*$ and that $\hat\theta_l-\theta_l^o=O_p(1/\sqrt{ST})$ and $\hat\theta_l^*-\theta_l^*=O_p(1/\sqrt{ST}).$  

    In other words, the log-likelihood part of the CLMDL will only be impacted up to an $O_p(1)$ term due to over-estimation of the model order. On the other hand, by definition, the MDL part of the CLMDL will increase by at least $\log(ST)$, which then dominate the $O_p(1)$ term and thus prevent the over-estimation of $\xi^o.$ This completes the proof.
\end{proof}

\subsection{Proofs of propositions and lemmas}
\begin{proof}[Proof of \Cref{lem_rate}]
   For notational convenience, in the proof, we use $\hat{ \cdot }$ instead of $\tilde{ \cdot }$ to denote the CL-based estimators. We first give the proof for $m_o=1$ and $m_o=2$, which serves as the building block of the proof for $m_o\geq 3.$ 

   \textbf{[I. $m_o=1$]}. We start with the case of one change-point. There is only one change-point $\tau_1^o$ and we denote its estimator as $\hat{\tau}_1.$ By symmetry, we assume $\hat\tau_1< \tau_1^o.$ We compare CL($m_o, \hat{\Lambda}, \hat{\Psi}$) and CL($m_o, {\Lambda}^o, {\Psi}^o$). 
   
   \textit{[Model order]} We first show that the model order on each segment cannot be under-estimated. Recall that all segments have the same optimal model order $\xi^o.$ By elementary algebra, we have
   \begin{align*}
       &\text{CL}(m_o, \hat{\Lambda}, \hat{\Psi})-\text{CL}(m_o, {\Lambda}^o, {\Psi}^o)\\
     =&L(\hat{\psi}_1, 1,\hat{\tau}_1)+L(\hat{\psi}_2, \hat{\tau}_1+1,T)-L({\psi}_1^o, 1,{\tau}_1^o)+L({\psi}_2^o, {\tau}_1^o+1,T)\\
     =&L_p(\hat{\psi}_1, 1,\hat{\tau}_1)-L_p({\psi}_1^o, 1,\hat{\tau}_1)\\
     +&L_p(\hat{\psi}_2, \hat{\tau}_1+1,\tau_1^o)-L_p(\psi_1^o, \hat{\tau}_1+1,\tau_1^o)\\
     +&L_p(\hat{\psi}_2, \tau_1^o+1,T)-L_p(\psi_2^o, \tau_1^o+1,T)\\
     +&L_m(\hat{\psi}_1,\hat\tau_1)+L_m(\hat{\psi}_2,\hat\tau_1+1)-L_c(\psi_1^o,\hat{\tau}_1)+L_c(\hat\psi_2,{\tau}_1^o)-L_m(\psi_1^o,\tau_1^o)-L_m({\psi}_2^o,\tau_1^o+1)\\
     :=&R_1+R_2+R_3+R_4.
   \end{align*}
   By the uniform convergence result in \Cref{lem_newULLN}(i,iii,iv), we have that
   $R_1=\mathbb E(R_1)+ o_p(1)S\hat\tau_1$, $R_2=\mathbb E(R_2)+o_p(1)S(\tau_1^o-\hat\tau_1)$, $R_3=\mathbb E(R_3) + o_p(1)S(T-\tau_1^o)$ and $R_4=\mathbb E(R_4)+o_p(1)S$, where the expectation is taken with $\hat{\psi}_1$ and $\hat{\psi}_2$ fixed and the $o_p(1)$ term is uniform over $\hat{\tau}_1.$ 

   First, by \Cref{ass.model.id}(i) and the information inequality~(i.e.\ $\psi_1^o,\psi_2^o$ are the true model parameters), $\mathbb E(R_i)\leq 0 $ for all $i=1,\cdots, 4.$ Suppose the model order $\hat\xi_2$ is under-estimated, by \Cref{ass.model.id}(i), we have that $\mathbb E(R_3)\leq -cS(T-\tau_1^o)$ for some constant $c>0.$ Since $T-\tau_1^o\geq \epsilon_\lambda^o T$, we have that $R_1+R_2+R_3+R_4 \leq -cS(T-\tau_1^o) +o_p(ST) <-c/2S(T-\tau_1^o)<0$ as $S,T$ increase, which is a contradiction. Thus, $\hat\xi_2$ cannot be under-estimated. Now, suppose the model order $\hat\xi_1$ is under-estimated. Note that by the implementation of CLMDL~(see \eqref{A} in the main text), we have that $\hat\tau_1\geq \epsilon_\lambda T$ where $\epsilon_\lambda\leq \epsilon_\lambda^o.$ Thus, we have that $R_1+R_2+R_3+R_4 \leq -cS\epsilon_\lambda T +o_p(ST) <-c/2S\epsilon_\lambda T<0$ as $S,T$ increase, which is a contradiction and $\hat{\xi}_1$ cannot be under-estimated either.
   
   \textit{[Localization error rate].} We now focus on the localization error rate. By the above result, both $\hat{\xi}_1$ and $\hat{\xi}_2$ cannot be under-estimated. By definition in \Cref{ass.model.id}(i), if $\xi^*$ over-estimate $\xi^o$, we have that $L((\xi^*,\theta^*_{j+1}),\tau_j^o+1,\tau_{j+1}^o)=L((\xi^o,\theta^o_{j+1}),\tau_j^o+1,\tau_{j+1}^o)$ for all $j=0,\cdots,m_o.$ Thus, in the following, without loss of generality, we assume that $\hat{\xi}_1=\hat{\xi}_2=\xi^o$. The argument goes through for any over-estimated $\hat{\xi}_1$ and $\hat\xi_2.$ For notational simplicity, we replace $\psi$ with $\theta$ in the log-likelihood functions when no confusion arises.
   
   By elementary algebra, we have that
   \begin{align}\label{eq:loglik_comp}
      &\text{CL}(m_o, \hat{\Lambda}, \hat{\Psi})-\text{CL}(m_o, {\Lambda}^o, {\Psi}^o)\nonumber\\
     =&L(\hat{\theta}_1, 1,\hat{\tau}_1)+L(\hat{\theta}_2, \hat{\tau}_1+1,T)-L({\theta}_1^o, 1,{\tau}_1^o)+L({\theta}_2^o, {\tau}_1^o+1,T)\nonumber\\
     =&L_p(\hat{\theta}_1, 1,\hat{\tau}_1)-L_p({\theta}_1^o, 1,\hat{\tau}_1)\nonumber\\
     +&L_p(\hat{\theta}_2, \hat{\tau}_1+1,\tau_1^o)-L_p(\theta_1^o, \hat{\tau}_1+1,\tau_1^o)\nonumber\\
     +&L_p(\hat{\theta}_2, \tau_1^o+1,T)-L_p(\theta_2^o, \tau_1^o+1,T)\nonumber\\
     +&L_m(\hat{\theta}_1,\hat\tau_1)+L_m(\hat{\theta}_2,\hat\tau_1+1)-L_c(\theta_1^o,\hat{\tau}_1)+L_c(\hat\theta_2,{\tau}_1^o)-L_m(\theta_1^o,\tau_1^o)-L_m({\theta}_2^o,\tau_1^o+1)\nonumber\\
     :=&R_1+R_2+R_3+R_4.
   \end{align}
   We now analyze $R_1,R_2,R_3$ and $R_4$ one by one. For $R_1$, by Taylor expansion, we have that
   \begin{align*}
       R_1&=L_p(\hat{\theta}_1, 1,\hat{\tau}_1)-L_p({\theta}_1^o, 1,\hat{\tau}_1)\\
       &=L_p'(\theta_1^o,1,\hat{\tau}_1)(\hat{\theta}_1-\theta_1^o)+\frac{1}{2}(\hat{\theta}_1-\theta_1^o)L_p''(\theta_1^+,1,\hat{\tau}_1)(\hat{\theta}_1-\theta_1^o),
   \end{align*}
   where $\theta_1^+$ is between $\hat{\theta}_1$ and $\theta_1^o$. By \Cref{lil}, we have that $\hat{\theta}_1=O_p(1/\sqrt{ST})$ and by \Cref{lem_BM}, we have that $L_p'(\theta_1^o,1,\hat{\tau}_1)=O_p(\sqrt{ST})$, which implies that $L_p'(\theta_1^o,1,\hat{\tau}_1)(\hat{\theta}_1-\theta_1^o)=O_p(1)$ uniformly over $\hat{\tau}_1.$

   By the uniform convergence result in \Cref{lem_newULLN}(i) and a standard consistency argument, we have that $\|\hat{\theta}_{t_1,t_2}-\theta_{t_1,t_2}^*\|\to_p 0$ uniformly over all $t_2-t_1>\epsilon_\lambda T$. Together with \Cref{lem_theta_star}, we have that
   $\|\hat\theta_1-\theta_1^o\|_2\to_p 0$ and thus $\|\theta_1^+-\theta_1^o\|_2\to_p 0$ uniformly over $\hat\tau_1.$ Combined with the uniform convergence result in \Cref{lem_newULLN}(ii) and the definition of $\theta_1^o$ in \Cref{ass.model.id}(i), this implies that $L_p''(\theta_1^+,1,\hat{\tau}_1)/(S\hat\tau_1)$ converge to a negative definite matrix, making $(\hat{\theta}_1-\theta_1^o)L_p''(\theta_1^+,1,\hat{\tau}_1)(\hat{\theta}_1-\theta_1^o)$ asymptotically negative. Thus, $R_1<O_p(1)$.

   We analyze $R_2$ and $R_3$ together. First, by Taylor expansion, we have that
   \begin{align*}
       R_2&=L_p(\hat{\theta}_2, \hat{\tau}_1+1,\tau_1^o)-L_p(\theta_1^o, \hat{\tau}_1+1,\tau_1^o)\\
       &=L_p'(\theta_1^o,\hat{\tau}_1+1,\tau_1^o)(\hat{\theta}_2-\theta_1^o)+\frac{1}{2}(\hat{\theta}_2-\theta_1^o)L_p''(\theta_{12}^+, \hat{\tau}_1+1,\tau_1^o) (\hat{\theta}_2-\theta_1^o),
   \end{align*}
   where $\theta_{12}^+$ is between $\hat{\theta}_2$ and $\theta_1^o$, and
   \begin{align*}
       R_3&=L_p(\hat{\theta}_2, \tau_1^o+1,T)-L_p(\theta_2^o, \tau_1^o+1,T)\\
       &=L_p'(\theta_2^o, \tau_1^o+1,T)(\hat\theta_2-\theta_2^o)+ \frac{1}{2}(\hat\theta_2-\theta_2^o) L_p''(\theta_2^+, \tau_1^o+1,T)(\hat\theta_2-\theta_2^o).
   \end{align*}
   
   By the same argument as before, we have that $L_p''(\theta_{12}^+, \hat{\tau}_1+1,\tau_1^o)/(S(\tau_1^o-\hat{\tau}_1))$ and $L_p''(\theta_2^+, \tau_1^o+1,T)/(S(T-\tau_1^o))$ converge to negative definite matrices uniformly over $\hat{\tau}_1$. In addition, by triangle inequality, we have that
   $\|\hat\theta_2-\theta_2^o\|_2+\|\hat\theta_2-\theta_1^o\|_2\geq \|\theta_1^o-\theta_2^o\|_2=O(\kappa).$ Thus, we have asymptotically,
   \begin{align}\label{eq:second_order}
       &(\hat{\theta}_2-\theta_1^o)L_p''(\theta_{12}^+, \hat{\tau}_1+1,\tau_1^o) (\hat{\theta}_2-\theta_1^o)+(\hat\theta_2-\theta_2^o) L_p''(\theta_2^+, \tau_1^o+1,T)(\hat\theta_2-\theta_2^o)\nonumber\\
       <& -c\min(\tau_1^o-\hat{\tau}_1,T-\tau_1^o)(1+o_p(1))S\kappa^2,
   \end{align}
   for some $c>0$ uniformly over $\hat{\tau}_1.$ The case where $\tau_1^o-\hat\tau_1\geq T-\tau_1^o$ is easier to handle, we thus focus on $\tau_1^o-\hat\tau_1<T-\tau_1^o$ and have that
   \begin{align}\label{eq:loglik_part1}
       (\hat{\theta}_2-\theta_1^o)L_p''(\theta_{12}^+, \hat{\tau}_1+1,\tau_1^o) (\hat{\theta}_2-\theta_1^o)+(\hat\theta_2-\theta_2^o) L_p''(\theta_2^+, \tau_1^o+1,T)(\hat\theta_2-\theta_2^o)< -c(\tau_1^o-\hat{\tau}_1)(1+o_p(1))S\kappa^2.
   \end{align}

   We now control the first order derivative terms. By Taylor expansion again, we have that
   \begin{align*}       0&=L'(\hat{\theta}_2,\hat{\tau}_1,T)=L'(\theta_2^o,\hat{\tau}_1,T)+L''(\theta_2^+,\hat{\tau}_1,T)(\hat{\theta}_2-\theta_2^o)\\   &=L_p'(\theta_2^o,\tau_1^o+1,T)+L_p'(\theta_1^o,\hat{\tau}_1,\tau_1^o)+L_p''(\theta_1^+,\hat\tau_1,\tau_1^o)(\theta_2^o-\theta_1^o)+O_p(S)+ L''(\theta_2^+,\hat{\tau}_1,T)(\hat{\theta}_2-\theta_2^o),
   \end{align*}
   where the $O_p(S)$ term incorporates the marginal likelihood terms and pairwise likelihood terms at $\hat{\tau}_1$ and $\tau_1^o$, and is uniformly $O_p(S)$ by \Cref{lem_newULLN}(iii-iv). Thus, by similar uniform convergence arguments as above, we have that
   \begin{align}\label{eq:theta2}
       \hat\theta_2-\theta_2^o &= (O_p(\sqrt{ST})+ O_p((\tau_1^o-\hat{\tau}_1)S\kappa)+ O_p(S))/(S(T-\hat{\tau}_1))\nonumber\\&=O_p(\sqrt{1/(ST)})+O_p((\tau_1^o-\hat{\tau}_1)\kappa/T)+O_p(1/T).
   \end{align}
   This implies that
   \begin{align}\label{eq:loglik_part2}
       &L_p'(\theta_1^o,\hat{\tau}_1+1,\tau_1^o)(\hat{\theta}_2-\theta_1^o)+L_p'(\theta_2^o, \tau_1^o+1,T)(\hat\theta_2-\theta_2^o)\nonumber\\
      =&L_p'(\theta_1^o,\hat{\tau}_1+1,\tau_1^o)(\theta_2^o-\theta_1^o)+L_p'(\theta_1^o,\hat{\tau}_1+1,\tau_1^o)(\hat{\theta}_2-\theta_2^o)+L_p'(\theta_2^o, \tau_1^o+1,T)(\hat\theta_2-\theta_2^o)\nonumber\\
      =&L_p'(\theta_1^o,\hat{\tau}_1+1,\tau_1^o)(\theta_2^o-\theta_1^o)+O_p(\sqrt{ST})(\hat\theta_2-\theta_2^o)\nonumber\\
      =&L_p'(\theta_1^o,\hat{\tau}_1+1,\tau_1^o)(\theta_2^o-\theta_1^o)+O_p(1)+O_p(\sqrt{S/T}(\tau_1^o-\hat\tau_1)\kappa)+O_p(\sqrt{S/T}).
   \end{align}

    Thus, combine the above results including \eqref{eq:loglik_part1} and \eqref{eq:loglik_part2}, we have that
   \begin{align}\label{eq:loglik_part12}
       R_2+R_3< 
       &- c(1+o_p(1))S\kappa^2(\tau_1^o-\hat{\tau}_1)+O_p(1)\nonumber\\
       &+L_p'(\theta_1^o,\hat{\tau}_1+1,\tau_1^o)(\theta_2^o-\theta_1^o)+O_p(\sqrt{S/T}(\tau_1^o-\hat\tau_1)\kappa)+O_p(\sqrt{S/T})
   \end{align}
   where the $o_p(1)$ and $O_p(1)$ terms are uniform over $\hat\tau_1$.
      
   We now analyze $R_4=L_m(\hat{\theta}_1,\hat\tau_1)+L_m(\hat{\theta}_2,\hat\tau_1+1)-L_c(\theta_1^o,\hat{\tau}_1)+L_c(\hat\theta_2,{\tau}_1^o)-L_m(\theta_1^o,\tau_1^o)-L_m({\theta}_2^o,\tau_1^o+1).$ It is easy to see that $R_4$ arises due to the compensating mechanism of the composite likelihood for correcting the edge effect at $\hat\tau_1$ and $\tau_1^o$. Since $\hat\tau_1<\tau_1^o$, the part $R_4^1:=L_m(\hat{\theta}_1,\hat\tau_1)+L_m(\hat{\theta}_2,\hat\tau_1+1)-L_c(\theta_1^o,\hat{\tau}_1)$ is the difference between the compensating marginal likelihood~(which incorrectly imposes temporal independence on a stationary segment) and the true pairwise likelihood at the estimated change-point $\hat\tau_1$ (recall $\hat\tau_1<\tau_1^o)$. Thus, by \Cref{lem_newULLN}(iii-iv) and the information inequality~(i.e.\ $\theta_1^o$ is the true model parameter), with probability going to 1, we have that $R_4^1/S<-c_1'$ uniformly over $\hat\tau_1<\tau_1^o.$ Similarly, the part $R_4^2:=L_c(\hat\theta_2,{\tau}_1^o)-L_m(\theta_1^o,\tau_1^o)-L_m({\theta}_2^o,\tau_1^o+1)$ is the difference between a pairwise likelihood and the compensating marginal likelihood~(which is the true likelihood due to the temporal independence across two stationary segments) at the true change-point $\tau_1^o.$ Thus, again by \Cref{lem_newULLN}(iii-iv) and the information inequality, we have $R_4^2/S< -c_1''$ uniformly over $\hat\tau_1<\tau_1^o$ with probability going to 1. Together, we have $R_4=-c_1S(1+o_p(1))$ for some $c_1>0$, which is asymptotically negative.

   Thus, combine all the above results, we have that
   \begin{align}\label{eq:loglik_part3}
       R_1+R_2+R_3+R_4<
       &- c(1+o_p(1))S\kappa^2(\tau_1^o-\hat{\tau}_1)-c_1S(1+o_p(1))+O_p(1)\nonumber\\
       &+L_p'(\theta_1^o,\hat{\tau}_1+1,\tau_1^o)(\theta_2^o-\theta_1^o)+O_p(\sqrt{S/T}(\tau_1^o-\hat\tau_1)\kappa)+O_p(\sqrt{S/T})\nonumber\\
       <&- \frac{1}{2}c(1+o_p(1))S\kappa^2(\tau_1^o-\hat{\tau}_1)-\frac{1}{2}c_1S(1+o_p(1))+L_p'(\theta_1^o,\hat{\tau}_1+1,\tau_1^o)(\theta_2^o-\theta_1^o),
   \end{align}
   where the $o_p(1)$ and $O_p(1)$ terms are uniform over $\hat\tau_1$ and the second inequality follows from the fact that $TS\kappa^2\to\infty$ and both $S,T$ diverge.

   \textit{[Preliminary rate]}: By \Cref{lem_BM}, we have that $L_p'(\theta_1^o,\hat{\tau}_1+1,\tau_1^o)(\theta_2^o-\theta_1^o)=O_p(\sqrt{ST})\kappa$. Thus, we have that
   \begin{align}\label{eq:prelim}
       &\mathbb P(|\tau_1^o-\hat\tau_1|>M\sqrt{T/(S\kappa^2)})   = \mathbb P(|\tau_1^o-\hat\tau_1|>M\sqrt{T/(S\kappa^2)}, R_1+R_2+R_3+R_4\geq 0) \nonumber\\
       \leq &\mathbb P(O_p(\sqrt{ST})\kappa-c'(1+o_p(1))S\kappa^2 M\sqrt{T/(S\kappa^2)}\geq 0)=
       \mathbb P( \sqrt{ST}\kappa (O_p(1)-c'(1+o_p(1))M)\geq 0),
   \end{align}
   which can be made arbitrarily small by increasing $M.$ Thus, we have that
   $|\tau_1^o-\hat\tau_1|=O_p(\sqrt{T/(S\kappa^2)}).$ Importantly, this implies that $|\tau_1^o-\hat\tau_1|/T=O_p(\sqrt{1/(TS\kappa^2})\to 0$ as $TS\kappa^2\to\infty$, i.e.\ we have relative consistency.

   \textit{[Sharper rate]}: We further give a sharper bound of $|\tau_1^o-\hat\tau_1|$ by better bounding $L_p'(\theta_1^o,\hat{\tau}_1+1,\tau_1^o)$. By union bound, we have that for any $\epsilon>0,$
   \begin{align}\label{eq:key}
      &\mathbb P\left(|L_p'(\theta_1^o,\hat{\tau}_1+1,\tau_1^o)(\theta_2^o-\theta_1^o)|>\epsilon S\kappa^2(\tau_1^o-\hat{\tau}_1) \text{ and }
      \tau_1^o-\hat\tau_1\geq M \right)\nonumber\\
      \leq &\sum_{t=1}^{\tau_1^o-M} \mathbb P\left(|L_p'(\theta_1^o,t+1,\tau_1^o)(\theta_2^o-\theta_1^o)|>\epsilon S\kappa^2(\tau_1^o-t)\right)\nonumber\\
      \leq & \sum_{t=1}^{\tau_1^o-M} \frac{\mathbb E|L_p'(\theta_1^o,t+1,\tau_1^o)(\theta_2^o-\theta_1^o)|^r}{(\epsilon S\kappa^2(\tau_1^o-t))^r}\leq \sum_{t=1}^{\tau_1^o-M} \frac{(\tau_1^o-t)^{r/2}S^{r/2} \kappa^r}{(\epsilon S\kappa^2(\tau_1^o-t))^r}\nonumber\\
      =& \frac{1}{\epsilon^r (S\kappa^2)^{r/2}} \sum_{t=1}^{\tau_1^o-M} \frac{1}{(\tau_1^o-t)^{r/2}} \leq \frac{1}{\epsilon^r (S\kappa^2)^{r/2}} M^{-r/2+1},
   \end{align}
   where the second inequality follows from Markov inequality and the third inequality follows from the moment inequality in \Cref{s_converge}.
   
   Thus, (A).\ For the case where $\liminf\limits_{S,T\to\infty}S\kappa^2 =\infty$, i.e.\ $S\kappa^2 \to \infty$, we have that
   \begin{align*}
        &\mathbb P(\tau_1^o-\hat\tau_1\geq 1) = \mathbb P(\tau_1^o-\hat\tau_1\geq 1, R_1+R_2+R_3+R_4\geq 0)\\
        \leq & \mathbb P\left[\tau_1^o-\hat\tau_1\geq 1, |L_p'(\theta_1^o,\hat{\tau}_1+1,\tau_1^o)(\theta_2^o-\theta_1^o)| \geq \frac{1}{2}c(1+o_p(1))S\kappa^2(\tau_1^o-\hat{\tau}_1)
        \right]\leq  2^r/(c^r (S\kappa^2)^{r/2}) \to 0,
   \end{align*}
   where the last inequality follows from \eqref{eq:key}. This implies the exact recovery of the change-point $\tau_1^o$, i.e.\ $\mathbb P(\tau_1^o=\hat\tau_1)\to 1.$

   (B).\ For the case where $S\kappa^2 \to 0$, set $M=M_*(S\kappa^2)^{-r/(r-2)}$, we have that
   \begin{align*}
      &\mathbb P((S\kappa^2)^{r/(r-2)}(\tau_1^o-\hat\tau_1)>M_*)   = \mathbb P(\tau_1^o-\hat\tau_1\geq M)\\
       = &\mathbb P(\tau_1^o-\hat\tau_1\geq M, R_1+R_2+R_3+R_4\geq 0)\\
    \leq &\mathbb P\left[\tau_1^o-\hat\tau_1\geq M,  |L_p'(\theta_1^o,\hat{\tau}_1+1,\tau_1^o)(\theta_2^o-\theta_1^o)| \geq \frac{1}{2}c(1+o_p(1))S\kappa^2(\tau_1^o-\hat{\tau}_1)M
        \right]\\
    \leq & (M^*)^{-r/2+1}/\epsilon^r,
   \end{align*}
   which can be arbitrarily close to 0 as $M^*$ increases. Thus, we have
   $(S\kappa^2)^{r/(r-2)}|\tau_1^o-\hat\tau_1|=O_p(1).$

   (C).\ For the case where $\liminf\limits_{S,T\to\infty}S\kappa^2= c''$ where $c''>0$ is a generic constant, we have for all sufficiently large $S,T$, it holds that $S\kappa^2>c''/2.$ Thus, we have that 
   \begin{align*}
        \mathbb P(\tau_1^o-\hat\tau_1\geq M) &= \mathbb P(\tau_1^o-\hat\tau_1\geq M, R_1+R_2+R_3+R_4\geq 0)\leq  M^{-r/2+1}/(\epsilon^r (S\kappa^2)^{r/2}) <c'''M^{-r/2+1},
   \end{align*}
   for some $c'''>0$, where the first inequality follows from \eqref{eq:key}. Thus, we have $|\tau_1^o-\hat\tau_1|=O_p(1)$. However, we show that this in fact further leads to exact recovery of $\tau_1^o$ as well. In particular, we have that for any $\epsilon>0$,
   \begin{align*}
       &\mathbb P\left(|L_p'(\theta_1^o,\hat{\tau}_1+1,\tau_1^o)(\theta_2^o-\theta_1^o)|>\epsilon S \text{ and }
      1\leq \tau_1^o-\hat\tau_1\leq M \right)\\
    \leq & \sum_{t=\tau_1^o-M}^{\tau_1^o-1} \mathbb P\left(|L_p'(\theta_1^o,t+1,\tau_1^o)(\theta_2^o-\theta_1^o)|>\epsilon S \right)\\
    \leq & \sum_{t=\tau_1^o-M}^{\tau_1^o-1} \frac{\mathbb E|L_p'(\theta_1^o,t+1,\tau_1^o)(\theta_2^o-\theta_1^o)|^r}{(\epsilon S)^r} \leq \sum_{t=\tau_1^o-M}^{\tau_1^o-1} \frac{(\tau_1^o-t)^{r/2}S^{r/2}\kappa^r}{(\epsilon S)^r} \leq c'''' M^{r/2+1} S^{-r/2} \to 0
   \end{align*}
   for any fixed $M.$ 
   
   Recall that $R_4=-c_1 S(1+o_p(1))$ uniformly over $\hat\tau_1<\tau_1^o.$ Thus, together it implies that for any $\delta>0$, we have that for all sufficiently large $S,T$, $\mathbb P(\tau_1^o-\hat\tau_1\geq 1) < \delta.$ This implies that $\mathbb P(\tau_1^o=\hat\tau_1) \to 1$, i.e. the exact recovery of $\tau_1^o.$

   \textbf{[II. $m_o=2$]}. We now extend the above result to the case of two change-points. First, using similar arguments as before, we can easily show that the model order cannot be under-estimated on each segment. We thus focus on the localization error rate. 

   Same as the case for $m_o=1$, the analysis is based on the comparison of log-likelihoods as in \eqref{eq:loglik_comp}, which is again analyzed by Taylor expansion. We first consider the general case of $m_o\geq 2$ and derive a preliminary localization error rate.
   
   By \Cref{lem_theta_hat}, for any $t_2-t_1\geq \epsilon_\lambda^o T$, we have that $\hat\theta_{t_1,t_2}-\theta_l^o=O_p(1/\sqrt{ST})+O_p(1/T)+O_p(\kappa)$, for all $l=1,\cdots,m_o+1.$ Moreover, note that all the first order derivative terms in the Taylor expansion take the form
   \begin{align}\label{eq:first_order}
       L_p'(\theta_j^o, t_1', t_2')(\hat{\theta}_{\hat\tau_{l-1}+1,\hat\tau_{l}}-\theta_j^o),
   \end{align}
   for some $j$ and $l$ in $1,\cdots,m_o+1.$ In particular, depending on the relative locations between $(\hat\tau_{l-1}, \hat\tau_l)$ and $(\tau_{j-1}^o,\tau_{j}^o)$, the index $(t_1',t_2')$ can be
   \begin{itemize}
       \item $(\hat\tau_{l-1}+1,\tau_j^o)$ if $\tau_{j-1}^o\leq \hat\tau_{l-1}<\tau_j^o\leq \hat{\tau}_{l}$,
       \item $(\tau_{j-1}^o+1,\tau_j^o)$ if $\hat\tau_{l-1}\leq \tau_{j-1}^o <\tau_j^o\leq \hat{\tau}_{l}$,
       \item $(\tau_{j-1}^o+1,\hat\tau_l)$ if $\hat\tau_{l-1}\leq \tau_{j-1}^o < \hat{\tau}_{l}\leq \tau_j^o$,
       \item  $(\hat\tau_{l-1}+1,\hat\tau_{l})$ if $\tau_{j-1}^o \leq \hat\tau_{l-1} < \hat{\tau}_{l}\leq \tau_j^o$,
   \end{itemize}
   where recall $\hat\tau_0=\tau_0^o\equiv 0$ and $\hat\tau_{m_o+1}=\tau_{m_o+1}^o\equiv T.$ In other words, all possible $L_p'(\theta_j^o, t_1', t_2')$ are the first order derivatives of the log-likelihood function on a stationary segment evaluated at the true pseudo-parameter. Thus, by \Cref{lem_BM}, we have that
   $$\eqref{eq:first_order}=\sqrt{ST}\cdot (O_p(1/\sqrt{ST})+O_p(1/T)+O_p(\kappa))=O_p(1)+O_p(\sqrt{S/T}) + O_p(\sqrt{ST}\kappa).$$
   
   In addition, similar to the term $R_4$ in \eqref{eq:loglik_comp}, the log-likelihood terms due to the compensating mechanism for correcting the edge effect take the form
   \begin{align}\label{eq:marg}
       &L_m(\hat{\theta}_{\hat\tau_{l-1}+1,\hat\tau_{l}}, \hat\tau_{l})+ L_m(\hat{\theta}_{\hat\tau_{l}+1,\hat\tau_{l+1}}, \hat\tau_{l}+1) - L_c(\theta_j^o,\hat\tau_l) \nonumber\\
       \text{or }&L_c(\hat{\theta}_{\hat\tau_{l-1}+1,\hat\tau_{l}}, \tau_{j}^o)-L_m(\theta_{j}^o,\tau_j^o)-L_m(\theta_{j+1}^o,\tau_{j}^o+1),
   \end{align}
   for some $j$ and $l$ in $1,\cdots,m_o+1.$ Using the same uniform convergence result in analyzing $R_4$ and \Cref{ass.model.id}(i), we have that all terms in \eqref{eq:marg} are $-O_p(S)$ asymptotically. Thus, we have that the summation of terms \eqref{eq:first_order} + \eqref{eq:marg} take the order
   \begin{align*}
       O_p(1)+O_p(\sqrt{S/T}) + O_p(\sqrt{ST}\kappa)-O_p(S)< O_p(\sqrt{ST}\kappa)-O_p(S).
   \end{align*}

   By same arguments as that in analyzing $R_2+R_3$ in the case of $m_o=1$, we can easily show that all second order derivative terms between $\hat\tau_{l-1}$ and $\hat\tau_l$ are asymptotically negative. We claim that for each $\tau_j^o$, there exists a $\hat\tau_l$ such that $|\hat\tau_l-\tau_j^o|/T \to 0$ with probability going to 1. We prove by contradiction, suppose there exists a $j$ such that the above claim does not hold. Thus, we can find an $\epsilon>0$ and $\hat\tau_{l-1}$ and $\hat\tau_l$ such that $\tau_j^o-\hat\tau_{l-1}>\epsilon T$ and $\hat\tau_l-\tau_j^o>\epsilon T$ for some large $T.$ Thus, it can be easily shown that the second order derivatives resulted from $L(\hat{\theta}_{\hat\tau_{l-1}+1,\hat\tau_{l}}, \hat\tau_{l-1}+1,\hat\tau_l)$ is asymptotically smaller than $-c\epsilon(TS\kappa^2)$ for some $c>0$, due to the fact that $[\hat\tau_{l-1}+1,\hat\tau_l]$ covers two stationary segments of length $O(T).$ Note that $O_p(\sqrt{ST}\kappa)=o_p(TS\kappa^2)$ due to the fact that $TS\kappa^2 \to\infty.$ Thus, a contradiction arises and we have that with probability going to 1, the claim holds. Since $m_o$ is known, we indeed have that $|\hat\tau_j-\tau_j^o|/T \to 0$ for $j=1,\cdots, m_o.$

   We now fix $m_o=2.$ Note that if $\hat\tau_1\geq \tau_1^o$, the analysis of $|\hat{\tau}_2-\tau_2^o|$ can be reduced to the analysis of $m_o=1$. Same holds for the case where $\hat\tau_2\leq \tau_2^o.$ Thus, to conserve space, we only analyze the more complicated event where $\hat\tau_1<\tau_1^o$ and $\hat\tau_2>\tau_2^o.$ By Taylor expansion, we have that
   \begin{align}\label{eq:loglik_compare_pts2}
      &\text{CL}(m_o, \hat{\Lambda}, \hat{\Psi})-\text{CL}(m_o, {\Lambda}^o, {\Psi}^o)\nonumber\\
     =&L_p(\hat{\theta}_1, 1,\hat{\tau}_1)-L_p({\theta}_1^o, 1,\hat{\tau}_1) + L_p(\hat\theta_3,\hat\tau_2+1,T)-L_p(\theta_3^o,\hat\tau_2+1,T)\nonumber\\
     +&L_p(\hat{\theta}_2, \hat{\tau}_1+1,\tau_1^o)-L_p(\theta_1^o, \hat{\tau}_1+1,\tau_1^o)\nonumber\\
     +&L_p(\hat{\theta}_2, \tau_1^o+1,\tau_2^o)-L_p(\theta_2^o, \tau_1^o+1,\tau_2^o)\nonumber\\
     +&L_p(\hat{\theta}_2, \tau_2^o+1,\hat\tau_2)-L_p(\theta_3^o, \tau_2^o+1,\hat\tau_2) \nonumber\\
     -&O_p(S):= R_1+R_2+R_3+R_4+R_5,
   \end{align}
   where for notational simplicity, $R_5=-O_p(S)$ denotes the terms due to compensating mechanism of the composite likelihood. By the same arguments as the one for $m_o=1$, we can show that $R_1=O_p(1)$. In addition, by Taylor expansion, we have that
   \begin{align}\label{eq:half_loglik}
       &R_2+\frac{1}{2}R_3\nonumber\\
    =&L_p'(\theta_1^o, \hat{\tau}_1+1,\tau_1^o)(\hat\theta_2-\theta_1^o)+\frac{1}{2}(\hat\theta_2-\theta_1^o)L_p''(\theta_1^+, \hat{\tau}_1+1,\tau_1^o)(\hat\theta_2-\theta_1^o)\nonumber \\
    +& \frac{1}{2}L_p'(\theta_2^o, \tau_1^o+1,\tau_2^o)(\hat\theta_2-\theta_2^o)+\frac{1}{4}(\hat\theta_2-\theta_2^o)L_p''(\theta_2^+, \tau_1^o+1,\tau_2^o)(\hat\theta_2-\theta_2^o)\nonumber\\
    <& L_p'(\theta_1^o, \hat{\tau}_1+1,\tau_1^o)(\hat\theta_2-\theta_1^o) + \frac{1}{2}L_p'(\theta_2^o, \tau_1^o+1,\tau_2^o)(\hat\theta_2-\theta_2^o) - c S\kappa^2 (\tau_1^o-\hat\tau_1),
   \end{align}
   where the inequality follows using exactly the same argument as that in analyzing \eqref{eq:second_order}. Similarly, we have that 
   \begin{align*}
       R_4+\frac{1}{2}R_3< L_p'(\theta_3^o, \tau_2^o+1,\hat\tau_2)(\hat\theta_2-\theta_3^o) + \frac{1}{2}L_p'(\theta_2^o, \tau_1^o+1,\tau_2^o)(\hat\theta_2-\theta_2^o) - c S\kappa^2 (\hat\tau_2-\tau_2^o).
   \end{align*}
   Using the same argument for \eqref{eq:theta2}, we can show that
   \begin{align*}
       \hat\theta_2-\theta_2^o =O_p(\sqrt{1/(ST)})+O_p((\tau_1^o-\hat{\tau}_1)\kappa/T)+O_p((\hat{\tau}_2-\tau_2^o)\kappa/T)+O_p(1/T).
   \end{align*}
   Thus, combine the above results and using the same argument as that for \eqref{eq:loglik_part2}, we have that
   \begin{align*}
       \eqref{eq:loglik_compare_pts2} <& L_p'(\theta_1^o, \hat{\tau}_1+1,\tau_1^o)(\theta_2^o-\theta_1^o)- c (1+o_p(1))S\kappa^2 (\tau_1^o-\hat\tau_1)\\
       +&L_p'(\theta_3^o, \tau_2^o+1,\hat\tau_2)(\theta_2^o-\theta_3^o)-c (1+o_p(1)) S\kappa^2(\hat\tau_2-\tau_2^o)\\
       -&c_1S(1+o_p(1)).
   \end{align*}
   Using the exact same argument as the one to prove the preliminary rate in \eqref{eq:prelim}, we can easily show that
   \begin{align*}
       \max_{l=1,2}|\hat{\tau}_l-\tau_l^o|=O_p(\sqrt{T/(S\kappa^2)}.
   \end{align*}
   Similarly, using the exact same argument as the one to prove the sharper rate (A),(B),(C) (see \eqref{eq:key}), we have that
   \begin{align*}
       \mathbb P(\hat\tau_l=\tau_l^o, l=1,2) \to 1
   \end{align*}
   given that $\liminf\limits_{S,T\to\infty}S\kappa^2>0$, and we have
   \begin{align*}
       (S\kappa^2)^{r/(r-2)}\max_{l=1,2}|\hat{\tau}_l-\tau_l^o|=O_p(1).
   \end{align*}
   
   \textbf{[III. $m_o\geq 3$]}. The proof for $m_o\geq 3$ can be directly built on the proof for $m_o=2$. First, recall that for a general $m_o\geq 3$, we have shown that $|\hat\tau_j-\tau_j^o|/T\to 0$ for all $j=1,\cdots,m_o$. Thus, for any $[\hat\tau_j+1,\hat\tau_{j+1}]$, it can only cover no, one or two change-points. Thus, we can directly use the analysis of the comparison of CL in the case of $m_o=1$ and $m_o=2$, and thus finish the proof.   
\end{proof}

\begin{proof}[Proof of \Cref{lem_rate_m_unknown}]
For notational convenience, in the proof, we use $\hat{\tau}$ instead of $\tilde{\tau}$ to denote the CL-based estimators.

First, note that if $\hat{m}<m_o$, for at least one change-point $j$, there is no $\hat{\tau}_{l}, l=1,2,\cdots,\hat{m}$ that is within $\epsilon_\lambda T/2$ distance from it. Using similar arguments as the proof of \Cref{lem_rate}, we can show that such scenario will not happen with probability one. In other words, the number of change-points cannot be underestimated. Given that $\hat{m}\geq m_o$, using the same arguments as the proof of \Cref{lem_rate} for the preliminary rate~(see \eqref{eq:prelim}), we can show that a true change-point $\tau_j^o$ must be accompanied by a $\hat\tau_{i_j}$ that is within $O_p(\sqrt{T/(S\kappa^2)})$ distance.

In addition, using the same argument as that in the proof of \Cref{lem_rate}, we can show that the model order cannot be under-estimated on each segment. By definition in \Cref{ass.model.id}(i), if $\xi^*$ over-estimates $\xi^o$, we have that $L((\xi^*,\theta^*_{j+1}),\tau_j^o+1,\tau_{j+1}^o)=L((\xi^o,\theta^o_{j+1}),\tau_j^o+1,\tau_{j+1}^o)$ for all $j=0,\cdots,m_o.$ Thus, in the following, without loss of generality, we assume that $\hat{\xi}_j=\xi^o$. The argument goes through for any over-estimated $\hat{\xi}_j$. For notational simplicity, we replace $\psi$ with $\theta$ in the log-likelihood functions when no confusion arises.

In the following, we show that with probability going to 1, CL will not make false positive detections, i.e.\ $\hat{m}=m_o.$ First, since $TS\kappa^2\to\infty$, based on the preliminary localization rate, we know that any false positive must happen $(1-c)\epsilon_\lambda T$ \textit{away} from the true change-point, where $c>0$ is an arbitrarily small constant.

Without loss of generality, consider the stationary segment from $\tau^o_{l-1}+1$ to $\tau^o_l$ and assume that $\hat{\tau}_{i_{l-1}}\geq \tau^o_{l-1}$ and $\hat{\tau}_{i_{l}}> \tau^o_{l}$. The proof for other scenarios follow similar but simpler arguments. We prove by contradiction. Suppose there exists a false positive change-point $\hat{\tau}$ such that $\hat\tau_{i_{l-1}} < \hat\tau < \hat\tau_{i_l}.$ We consider the difference between the log-likelihood functions. In particular, we have 
\begin{align}\label{eq:like_diff}
    &L(\hat{\theta}, \hat{\tau}_{i_{l-1}}+1, \hat{\tau}_{i_{l}})- L(\hat{\theta}_1, \hat{\tau}_{i_{l-1}}+1, \hat\tau) - L(\hat\theta_2, \hat\tau+1, \hat{\tau}_{i_{l}})\nonumber\\
    =&L(\theta_{l}^o, \hat{\tau}_{i_{l-1}}+1, \hat{\tau}_{i_{l}})-\frac{1}{2}(\hat\theta-\theta_l^o)^\top L''(\theta^+, \hat{\tau}_{i_{l-1}}+1, \hat{\tau}_{i_{l}})(\hat\theta-\theta_l^o)\nonumber\\
    -&L(\theta_{l}^o, \hat{\tau}_{i_{l-1}}+1, \hat{\tau})+\frac{1}{2}(\hat\theta_1-\theta_l^o)^\top L''(\theta^+_1, \hat{\tau}_{i_{l-1}}+1, \hat{\tau})(\hat\theta_1-\theta_l^o)\nonumber\\
    -&L(\theta_{l}^o, \hat{\tau}+1, \hat{\tau}_{i_{l}})+\frac{1}{2}(\hat\theta_2-\theta_l^o)^\top L''(\theta^+_2, \hat{\tau}+1, \hat{\tau}_{i_{l-1}})(\hat\theta_2-\theta_l^o),
\end{align}
where the equality follows from a Taylor expansion and the fact that $\hat{\theta}$, $\hat{\theta}_1$ and $\hat{\theta}_2$ are maximizers of $L({\theta}, \hat{\tau}_{i_{l-1}}+1, \hat{\tau}_{i_{l}}), L({\theta}, \hat{\tau}_{i_{l-1}}+1, \hat\tau),$ and $L(\theta, \hat\tau+1, \hat{\tau}_{i_{l}})$ respectively. Here, $\theta^+, \theta_1^+$ and $\theta_2^+$ are between $\theta_l^o$ and $\hat{\theta},\hat\theta_1,\hat\theta_2$ respectively. In the following, we show that the presence of a false positive detection $\hat{\tau}$ will make \eqref{eq:like_diff} positive with probability going to 1, which indicates that such false positive detection will not happen eventually. Note that by \Cref{MPLE}, we have that $\hat{\theta}$, $\hat{\theta}_1$ and $\hat{\theta}_2$ all converge to $\theta_l^o$ in probability. 

We first show that the three second order derivative terms in \eqref{eq:like_diff} are of order $o_p(S).$ We analyze the term $(\hat\theta-\theta_l^o)^\top L''(\theta^+, \hat{\tau}_{i_{l-1}}+1, \hat{\tau}_{i_{l}})(\hat\theta-\theta_l^o)$, the other two terms can be analyzed using the exact same argument. By \Cref{lem_newULLN}, we have that $L''(\theta^+, \hat{\tau}_{i_{l-1}}+1, \hat{\tau}_{i_{l}})/(S(\hat{\tau}_{i_l}-\hat\tau_{i_{l-1}})) = \mathbb E\left\{L''(\theta^+, \hat{\tau}_{i_{l-1}}+1, \hat{\tau}_{i_{l}})/(S(\hat{\tau}_{i_l}-\hat\tau_{i_{l-1}})) \right\}+o_p(1)$, where the expectation is taken only w.r.t.\ $y_{t,s}$ with all other parameters treated as fixed. Note that we have $\hat{\theta}\to_p \theta_l^o$, and thus $\theta^+\to_p \theta_l^o$. By the definition of $\theta^o_l$ and the fact that $|\hat\tau_{i_l}-\tau_l^o|=o({T})$ and $\hat{\tau}_{i_l}-\hat{\tau}_{i_{l-1}}>\epsilon_\lambda T$, we have that $L''(\theta^+, \hat{\tau}_{i_{l-1}}+1, \hat{\tau}_{i_{l}})/(S(\hat{\tau}_{i_l}-\hat\tau_{i_{l-1}}))$ eventually becomes a negative definite matrix.

Thus, the key is to further establish an asymptotic order of $\hat\theta$. Note that due to the localization error of $\hat\tau_{i_l}$, $\hat{\theta}$ is estimated based on two stationary segments, one is from $\hat\tau_{i_{l-1}}+1$ to $\tau_l^o$, and another is from $\tau_l^o+1$ to $\hat\tau_{i_l}$. Note that by the preliminary localization error rate, the second segment length $\hat\tau_{i_l}-\tau_l^o$ is upper bounded by $O_p(\sqrt{T/(S\kappa^2)})=o(T)$. Thus, we further have that $\tau_l^o-\hat{\tau}_{i_{l-1}}$ is lower bounded by $(1-c)\epsilon_\lambda T$ for any arbitrarily small $c>0.$

We have that
\begin{align*}    0=&L'(\hat{\theta}, \hat{\tau}_{i_{l-1}}+1, \hat{\tau}_{i_{l}})\\
=&L'(\theta_{l}^o,\hat{\tau}_{i_{l-1}}+1, \hat{\tau}_{i_{l}}) + L''(\theta^+,\hat{\tau}_{i_{l-1}}+1, \hat{\tau}_{i_{l}})(\hat{\theta}-\theta_{l}^o)\\
=&L_p'(\theta_{l}^o,\hat\tau_{i_{l-1}}+1,\tau_l^o) + L_p'(\theta_{l}^o,\tau_l^o+1, \hat\tau_{i_l}) + L''(\theta^+,\hat\tau_{i_{l-1}}+1,\hat\tau_{i_l})(\hat{\theta}-\theta_{l}^o)+O_p(S)\\
=&L_p'(\theta_{l}^o,\hat\tau_{i_{l-1}}+1,\tau_l^o)  + L_p'(\theta_{l+1}^o,\tau_l^o+1, \hat\tau_{i_l})+ L_p''(\theta_{l+1}^+,\tau_l^o+1, \hat\tau_{i_l})(\theta_{l}^o-\theta_{l+1}^o)+O_p(S)\\
& +  L''(\theta^+,\hat\tau_{i_{l-1}}+1,\hat\tau_{i_l})(\hat{\theta}-\theta_{l}^o)\\
=& O_p(\sqrt{ST}) + O_p(\sqrt{ST}) + O_p( (\hat\tau_{i_l}-\tau_l^o)S\kappa) + O_p(ST)(\hat{\theta}-\theta_{l}^o)+O_p(S),
\end{align*}
where the first equality follows from the definition of $\hat{\theta}$ and the second equality follows from a Taylor expansion where $\theta^+$ is between $\hat{\theta}$ and $\theta^o_{l}$. The third equality follows by decomposing $L'(\theta_{l}^o,\hat\tau_{i_{l-1}}+1,\hat\tau_{i_l})$ into $L_p'(\theta_{l}^o,\hat\tau_{i_{l-1}}+1,\tau_l^o)$ and $L_p'(\theta_{l}^o,\tau_l^o+1, \hat\tau_{i_l})$ plus the additional $O_p(S)$ edge effect term. The fourth equality follows from a Taylor expansion of $L_p'(\theta_{l}^o,\tau_l^o+1, \hat\tau_{i_l})$ where $\theta_{l+1}^+$ is between $\theta_{l}^o$ and $\theta_{l+1}^o.$ For the last equality, by \Cref{lem_BM}, we have $L_p'(\theta_{l}^o,\hat\tau_{i_{l-1}}+1,\tau_l^o)=O_p(\sqrt{ST})$ and $L_p'(\theta_{l}^o,\tau_l^o+1, \hat\tau_{i_l})=O_p(\sqrt{ST})$ as both sequences are mean zero; by the fact that $\|\theta_{l+1}^+-\theta_{l+1}^o\|_2\to 0$ and the uniform convergence result in \Cref{lem_newULLN}, we have $L_p''(\theta_{l+1}^+,\tau_l^o+1, \hat\tau_{i_l})(\theta_{l}^o-\theta_{l+1}^o)=O_p( (\hat\tau_{i_l}-\tau_l^o)S\kappa)$ and $L''(\theta^+,\hat\tau_{i_{l-1}}+1,\hat\tau_{i_l})=O_p(ST).$

Thus, we have that
\begin{align}\label{eq:theta_bias_variance0}
    \hat{\theta}-\theta_{l}^o= O_p(1/\sqrt{ST})+  O_p( (\hat\tau_{i_l}-\tau_l^o)\kappa/T)+O_p(1/T),
\end{align}
where the first term can be viewed as the variance and the second and third term can be viewed as the bias due to the localization error of change-point detection. Plug in the preliminary localization error rate, this implies that $\hat{\theta}-\theta_{l}^o=O_p(1/\sqrt{ST})+O_p(\sqrt{T/(S\kappa^2)}\kappa /T)+O_p(1/T)=O_p(1/\sqrt{ST}+1/T)$. Thus, together, we have that
\begin{align*}
    (\hat\theta-\theta_l^o)^\top L''(\theta^+, \hat{\tau}_{i_{l-1}}+1, \hat{\tau}_{i_{l}})(\hat\theta-\theta_l^o)= &O_p(ST)\times (O_p(1/(ST)) + O_p(1/T^2))\\
    =&O_p(1)+ O_p(S/T)=o_p(S).
\end{align*}
The same holds for the other two second order derivative terms in \eqref{eq:like_diff}.

On the other hand, for the likelihood part, we have that
\begin{align*}
     & L(\theta_{l}^o, \hat{\tau}_{i_{l-1}}+1, \hat{\tau}_{i_{l}})-L(\theta_{l}^o, \hat{\tau}_{i_{l-1}}+1, \hat{\tau})-L(\theta_{l}^o, \hat{\tau}+1, \hat{\tau}_{i_{l}})\\
    =& L_c(\theta_{l}^o, \hat\tau, \hat\tau+1)-L_m(\theta_l^o,\hat{\tau})-L_m(\theta_l^o,\hat{\tau}+1)\\
    >& \mathbb E \left [L_c(\theta_{l}^o, \hat\tau, \hat\tau+1)-L_m(\theta_l^o,\hat{\tau})-L_m(\theta_l^o,\hat{\tau}+1)\right] - \big| L_c(\theta_{l}^o, \hat\tau, \hat\tau+1)-L_m(\theta_l^o,\hat{\tau})-L_m(\theta_l^o,\hat{\tau}+1) \\
    &- \mathbb E\left[ L_c(\theta_{l}^o, \hat\tau, \hat\tau+1)-L_m(\theta_l^o,\hat{\tau})-L_m(\theta_l^o,\hat{\tau}+1)\right] \big| := R_1 + R_2.
\end{align*}
Note that by \Cref{lem_newULLN}, we have that $\mathbb P(R_2>\epsilon S)\to 0$ for any $\epsilon>0.$ In other words, $R_2=o_p(S)$. On the other hand, by the definition of $\theta_l^o$, we know that $R_1>CS$ for some $C>0$. Thus, we know that with probability going to 1, $ L(\theta_{l}^o, \hat{\tau}_{i_{l-1}}+1, \hat{\tau}_{i_{l}})-L(\theta_{l}^o, \hat{\tau}_{i_{l-1}}+1, \hat{\tau})-L(\theta_{l}^o, \hat{\tau}+1, \hat{\tau}_{i_{l}})>CS$ for some $C>0.$

Plug everything in \eqref{eq:like_diff}, we have that 
\begin{align*}
    L(\hat{\theta}, \hat{\tau}_{i_{l-1}}+1, \hat{\tau}_{i_{l}})- L(\hat{\theta}_1, \hat{\tau}_{i_{l-1}}+1, \hat\tau) - L(\hat\theta_2, \hat\tau+1, \hat{\tau}_{i_{l}})>0
\end{align*}
with probability going to 1. Thus, the false positive detection $\hat{\tau}$ appears with probability going to 0 and we have that
\begin{align*}
    \mathbb P(\hat m=m_o) \to 1.
\end{align*}
Given that $\hat m=m_o$, using the same argument as that in \Cref{lem_rate}, we have can further show that \Cref{lem_rate}(i) and (ii) hold, which concludes the proof.
\end{proof}

\begin{proof}[Proof of \Cref{lem_newULLN}]
    The proof follows from a moment inequality, a union bound and the standard argument for establishing uniform law of large numbers. In the following, we give the proof for (i), as the proof for (ii)-(iv) follows the same arguments.

    By symmetry, we show that
    \begin{align*}
        \mathbb P \left(\sup_{\substack{t_1,t_2,\\ t_2-t_1\geq 0}}\sup_{\theta\in\Theta} \frac{1}{(t_2-t_1+1)S}
       \left(L_p(\theta,t_1,t_2)-\mathbb E(L_p(\theta,t_1,t_2))\right)>\epsilon\right)\to 0.
    \end{align*}
    Given $\delta>0$ and a fixed $\theta_0\in\Theta$, define $\Delta_\delta(y_{t,s_1},y_{t+i,s_2},\theta_0)=\sup_{\theta \in B(\theta_0,\delta)} l(\theta;y_{t,s_1},y_{t+i,s_2})-\inf_{\theta \in B(\theta_0,\delta)} l(\theta;y_{t,s_1},y_{t+i,s_2})$, where $B(\theta_0,\delta)$ denotes the $\delta$-ball around $\theta_0.$ By \Cref{ass.mom} and the dominated convergence theorem, we have that $\lim_{\delta\to 0}\mathbb E(\Delta_\delta(y_{t,s_1},y_{t+i,s_2},\theta_0))=0.$ In fact, due to the stationarity of each segment and the finite number of segments, given $\theta_0,$ for any $\epsilon>0$, we can find a $\delta_\epsilon(\theta_0)>0$ such that for any $(t,t+i,s_1,s_2)\in D_{t_1,t_2}$ with $t_2-t_1\geq 0$, we have that $\mathbb E(\Delta_{\delta_\epsilon(\theta_0)}(y_{t,s_1},y_{t+i,s_2},\theta_0))<\epsilon.$ Clearly, $\bigcup_{\theta\in\Theta} B_{\theta,\delta_\epsilon(\theta)}$ covers $\Theta$. In addition, since $\Theta$ is compact, we can find a finite sub-cover, such that $\Theta$ is covered by $\bigcup_{k=1}^K B_{\theta_k,\delta_\epsilon(\theta_k)}$
    
    Furthermore, we have that
    \begin{align*}
       &\sup_{t_1,t_2}\sup_{\theta\in\Theta} \frac{1}{(t_2-t_1+1)S}
       \left(L_p(\theta,t_1,t_2)-\mathbb E(L_p(\theta,t_1,t_2))\right)\\
       =&\sup_{t_1,t_2}\max_{k} \sup_{\theta\in B_{\theta_k,\delta_\epsilon(\theta_k)}}\frac{1}{(t_2-t_1+1)S}
       \left(L_p(\theta,t_1,t_2)-\mathbb E(L_p(\theta,t_1,t_2))\right)\\
       \leq &\sup_{t_1,t_2} \max_{k}\left[ \frac{1}{(t_2-t_1+1)S} \left( \sum_{(t,i,s_1,s_2)\in D_{t_1,t_2}} \sup_{\theta\in B_{\theta_k,\delta_\epsilon(\theta_k)}}l(\theta;y_{t,s_1},y_{t+i,s_2})-\mathbb E(\sup_{\theta\in B_{\theta_k,\delta_\epsilon(\theta_k)}}l(\theta;y_{t,s_1},y_{t+i,s_2})\right)\right.\\
        + & \left. 
        \frac{1}{(t_2-t_1+1)S} \left( \sum_{(t,i,s_1,s_2)\in D_{t_1,t_2}}\mathbb E(\sup_{\theta\in B_{\theta_k,\delta_\epsilon(\theta_k)}}l(\theta;y_{t,s_1},y_{t+i,s_2})-\mathbb E(\inf_{\theta\in B_{\theta_k,\delta_\epsilon(\theta_k)}}l(\theta;y_{t,s_1},y_{t+i,s_2})\right)
       \right]\\
       =&\max_{k}\left[\sup_{t_1,t_2} \frac{1}{(t_2-t_1+1)S} \left( \sum_{(t,i,s_1,s_2)\in D_{t_1,t_2}} \sup_{\theta\in B_{\theta_k,\delta_\epsilon(\theta_k)}}l(\theta;y_{t,s_1},y_{t+i,s_2})-\mathbb E(\sup_{\theta\in B_{\theta_k,\delta_\epsilon(\theta_k)}}l(\theta;y_{t,s_1},y_{t+i,s_2})\right)\right]
       \\+& C\epsilon,
    \end{align*}
    where the last equality follows from $\mathbb E(\Delta_{\delta_\epsilon(\theta_0)}(y_{t,s_1},y_{t+i,s_2},\theta_0))<\epsilon$ and the fact that each $(t,s)$ has a finite number of neighbors.

    We now further bound the first term. In particular, we have
    \begin{align*}
        &\mathbb P \left(\sup_{t_1,t_2} \frac{1}{(t_2-t_1+1)S} \left( \sum_{(t,i,s_1,s_2)\in D_{t_1,t_2}} \sup_{\theta\in B_{\theta_k,\delta_\epsilon(\theta_k)}}l(\theta;y_{t,s_1},y_{t+i,s_2})-\mathbb E(\sup_{\theta\in B_{\theta_k,\delta_\epsilon(\theta_k)}}l(\theta;y_{t,s_1},y_{t+i,s_2})\right)>\epsilon\right)\\
        \leq& \sum_{t_1,t_2}\mathbb P \left( \frac{1}{(t_2-t_1+1)S} \left( \sum_{(t,i,s_1,s_2)\in D_{t_1,t_2}} \sup_{\theta\in B_{\theta_k,\delta_\epsilon(\theta_k)}}l(\theta;y_{t,s_1},y_{t+i,s_2})-\mathbb E(\sup_{\theta\in B_{\theta_k,\delta_\epsilon(\theta_k)}}l(\theta;y_{t,s_1},y_{t+i,s_2})\right)>\epsilon\right)\\
        \leq& C\sum_{t_1,t_2} ((t_2-t_1+1)S)^{-r/2} \leq C T S^{-r/2} /(r/2-1) \to 0.
    \end{align*}
    where the first inequality follows from union bound, the second inequality follows from \Cref{ass.mom}(r) and the moment inequality in \Cref{s_converge}, and the third inequality follows from elementary algebra. Thus, we conclude the proof.    
\end{proof}

\begin{proof}[Proof of \Cref{lem_BM}]
    The proof follows the same arguments as the one for the proof of \Cref{lil}. Specifically, we have that for any $\tau_{l-1}^o+1\leq t_1\leq t_2 \leq \tau_l^o$,
    \begin{align*}
        L'(\theta_l^o, t_1, t_2)=L_p'(\theta_l^o, t_1, t_2) + L_m'(\theta_l^o, t_1) + L_m'(\theta_l^o,t_2),
    \end{align*}
    where the three terms on the right hand side are mean zero by the definition of $\theta_l^o.$
    
    By \Cref{s_converge} and union bound, we can easily show that $\frac{1}{\sqrt{ST}}\sup_{\tau_{l-1}^o+1\leq t \leq \tau_l^o }L_m'(\theta_l^o, t)=o_p(1)$ and $\frac{1}{\sqrt{ST}}\sup_{\tau_{l-1}^o+1\leq t \leq \tau_l^o-1 }L_c'(\theta_l^o, t, t+1)=o_p(1)$. Thus, we can bound $\frac{1}{\sqrt{ST}}L'(\theta_l^o, t_1, t_2)$ by
    \begin{align}\label{BrownianSheet1}
	\frac{1}{\sqrt{ST}}\left(\sup_{\tau_{l-1}^o+1\leq t \leq \tau_l^o}L_p'(\theta_l^o, \tau_{l-1}^o+1, t)-\inf_{\tau_{l-1}^o+1\leq t \leq \tau_l^o}L_p'(\theta_l^o, \tau_{l-1}^o+1, t)\right).	
    \end{align}
	By Assumption \ref{ass.mom}($r$), Assumption \ref{ass.mixtime} and Theorem 2.2 in \cite{Yang2007}, we can show that $\eqref{BrownianSheet1}$ is $O_p(1)$, which concludes the proof.    
\end{proof}

\begin{proof}[Proof of \Cref{lem_theta_star}]
We prove by contradiction. For expositional simplicity, we assume that $\theta_1^o$ is fixed and thus we have that $\max_{2\leq l\leq m_o+1}\|\theta_l^o-\theta_1^o\|_2 \to 0$ due to the vanishing change sizes and the fact that $m_o$ is finite.

Suppose the convergence result in \Cref{lem_theta_star} does not hold, we have that for any $\delta>0$, there always exist $S,T$ large enough and $t_1,t_2$ such that
$\|\theta_{t_1,t_2}^*-\theta_1^o\|_2>\delta$, where $\theta_{t_1,t_2}^*$ is the maximizer of $\mathbb E(L(\theta,t_1,t_2)).$ If $[t_1,t_2]$ is within a single stationary segment, say $l$th stationary segment, we have that $\theta_{t_1,t_2}^*=\theta_l^o$ and thus contradiction arises. Thus, we need to consider the case where $[t_1,t_2]$ covers more than one stationary segment, say from $l_1-1$ to $l_2+1$. We thus decompose the expected log-likelihood as
\begin{align*}
    \mathbb E(L(\theta,t_1,t_2))=\mathbb E(L_p(\theta,t_1,\tau_{l_1}^o)) + \sum_{l=l_1}^{l_2-1} \mathbb E(L_p(\theta,\tau_{l}^o+1,\tau_{l+1}^o)) + \mathbb E(L_p(\theta,\tau_{l_2}^o+1,t_2)) + O(S),
\end{align*}
where for simplicity, we group the $\mathbb E (L_c)$ and $\mathbb E(L_m)$ terms as $O(S)$. This holds due to \Cref{ass.mom}(i) and since there are finite number of change-points. Given that $\|\theta_{t_1,t_2}^*-\theta_1^o\|_2>\delta$, due to the vanishing change sizes, we have that $\min_{1\leq l\leq m_o+1}\|\theta_{t_1,t_2}^*-\theta_l^o\|_2>\delta$ for all sufficiently large $S,T$. Thus, by \Cref{ass.model.id}(i), we have that
\begin{align*}
    \mathbb E(L(\theta_{t_1,t_2}^*,t_1,t_2)) - \Big[ \mathbb E(L_p(\theta_{l_1}^o,t_1,\tau_{l_1}^o)) + \sum_{l=l_1}^{l_2-1} \mathbb E(L_p(\theta_{l+1}^o,\tau_{l}^o+1,\tau_{l+1}^o)) + \mathbb E(L_p(\theta_{l_2+1}^o,\tau_{l_2}^o+1,t_2)) \Big] < - cST,
\end{align*}
with some $c>0$ uniformly for all $t_1,t_2$ such that $t_2-t_1>\epsilon_\lambda T$.

On the other hand, by Taylor expansion, we have that
\begin{align*}
     &\mathbb E(L(\theta_1^o,t_1,t_2)) - \Big[ \mathbb E(L_p(\theta_{l_1}^o,t_1,\tau_{l_1}^o)) + \sum_{l=l_1}^{l_2-1} \mathbb E(L_p(\theta_{l+1}^o,\tau_{l}^o+1,\tau_{l+1}^o)) + \mathbb E(L_p(\theta_{l_2+1}^o,\tau_{l_2}^o+1,t_2)) \Big]\\
    =&O(S) + (\theta_1^o-\theta_{l_1}^o)^\top\mathbb E(L_p''(\theta_{l_1}^+,t_1,\tau_{l_1}^o))(\theta_1^o-\theta_{l_1}^o) + \sum_{l=l_1}^{l_2-1}(\theta_1^o-\theta_{l+1}^o)^\top\mathbb E(L''_p(\theta_{l+1}^+,\tau_{l}^o+1,\tau_{l+1}^o))(\theta_1^o-\theta_{l+1}^o)\\
    +& (\theta_1^o-\theta_{l_2+1}^o)^\top\mathbb E(L''_p(\theta_{l_2+1}^+,\tau_{l_2}^o+1,t_2))(\theta_1^o-\theta_{l_2+1}^o) > -cST\kappa^2,
\end{align*}
where the last inequality follows from \Cref{ass.mom}(i). Thus, we have uniformly for all $t_1,t_2,$ for large enough $S,T$, it holds that $\mathbb E(L(\theta_1^o,t_1,t_2))>\mathbb E(L(\theta_{t_1,t_2}^*,t_1,t_2))$, which contradicts the definition of $\theta_{t_1,t_2}^*.$ Thus, the claim holds, i.e. we have $\sup_{\substack{t_1,t_2,\\t_2-t_1\geq \epsilon_\lambda T}}\|\theta_{t_1,t_2}^*-\theta_1^o\|_2 \to 0.$
\end{proof}

\begin{proof}[Proof of \Cref{lem_theta_hat}]
     First, if $[t_1,t_2]$ does not cover any true change-points, it is clear that $\hat\theta_{t_1,t_2}=O_p(1/\sqrt{ST})$. Thus, in the following, we assume that $[t_1,t_2]$ covers $\tau_{l_1}^o,\cdots,\tau_{l_2}^o$. Denote $\hat\theta=\hat\theta_{t_1,t_2}$, by Taylor expansion, we have that
   \begin{align}\label{eq:loglik_Taylor}
       0=&L'(\hat{\theta}, t_1, t_2)=L'(\theta_1^o,t_1,t_2) + L''(\theta^+,t_1,t_2)(\hat\theta-\theta_1^o)\nonumber\\
       =&L_p'(\theta_1^o,t_1,\tau_{l_1}^o)+\sum_{j=l_1}^{l_2-1}L_p'(\theta_1^o,\tau_{j}^o+1,\tau_{j+1}^o)+L_p'(\theta_1^o,\tau_{l_2}^o+1,t_2)\nonumber\\
       +&\left[L_p''(\theta^+,t_1,\tau_{l_1}^o)+\sum_{j=l_1}^{l_2-1}L_p''(\theta^+,\tau_{j}^o+1,\tau_{j+1}^o)+L_p''(\theta^+,\tau_{l_2}^o+1,t_2)\right](\hat\theta-\theta_1^o)+O_p(S),
   \end{align}
   where $\theta^+$ is between $\hat{\theta}$ and $\theta_1^o$. Here, for simplicity, the $O_p(S)$ incorporates all terms due to compensating mechanism of the composite likelihood for correcting the edge effect, which is $O_p(S)$ uniformly over $t_1,t_2$ by \Cref{lem_newULLN}(iii,iv) and \Cref{ass.mom}(ii) and the fact that $m_o$ is finite. 

   By a further Taylor expansion on the first order derivative, we have that
   \begin{align}\label{eq:loglik_firstorder}
       &L_p'(\theta_1^o,t_1,\tau_{l_1}^o)+\sum_{j=l_1}^{l_2-1}L_p'(\theta_1^o,\tau_{j}^o+1,\tau_{j+1}^o)+L_p'(\theta_1^o,\tau_{l_2}^o+1,t_2)\nonumber\\
      =&L_p'(\theta_{l_1}^o,t_1,\tau_{l_1}^o)+\sum_{j=l_1}^{l_2-1}L_p'(\theta_{j+1}^o,\tau_{j}^o+1,\tau_{j+1}^o)+L_p'(\theta_{l_2+1}^o,\tau_{l_2}^o+1,t_2)\nonumber\\
      +&L_p''(\theta_{l_1}^+,t_1,\tau_{l_1}^o)(\theta_1^o-\theta_{l_1}^o)+\sum_{j=l_1}^{l_2-1}L_p''(\theta_{j+1}^+,\tau_{j}^o+1,\tau_{j+1}^o)(\theta_1^o-\theta_{j+1}^o)+L_p''(\theta_{l_2+1}^+,\tau_{l_2}^o+1,t_2)(\theta_1^o-\theta_{l_2+1}^o)\nonumber\\
      =&O_p(\sqrt{ST})+O_p(ST\kappa),
   \end{align}
   where $\theta_j^+$ is between $\theta_j^o$ and $\theta_1^o$, and the last equality follows from \Cref{lem_BM}, \Cref{ass.model.id}(i), \Cref{lem_newULLN}(ii) and the fact that $\max_{1\leq l\leq m_o+1}\|\theta_1^o-\theta_l^o\| \leq C\kappa \to 0$.

   In addition, by the uniform convergence result in \Cref{lem_newULLN}(i) and a standard consistency argument, we have that $\|\hat{\theta}-\theta_{t_1,t_2}^*\|\to_p 0$ uniformly over all $t_2-t_1>\epsilon_\lambda^o T$ (recall we denote $\hat\theta=\hat\theta_{t_1,t_2}$). Together with \Cref{lem_theta_star}, we have that
   $\|\hat\theta-\theta_1^o\|_2\to_p 0$ and thus $\max_{1\leq l\leq m_o+1}\|\theta^+-\theta_l^o\|_2\to_p 0$ uniformly over $t_1,t_2$. Combined with the uniform convergence result in \Cref{lem_newULLN}(ii) and the definition of $\theta_1^o$ in \Cref{ass.model.id}(i), this implies that $$\left[L_p''(\theta^+,t_1,\tau_{l_1}^o)+\sum_{j=l_1}^{l_2-1}L_p''(\theta^+,\tau_{j}^o+1,\tau_{j+1}^o)+L_p''(\theta^+,\tau_{l_2}^o+1,t_2)\right]=O_p(ST),$$
   and is a negative definite matrix. Together with \eqref{eq:loglik_firstorder} and \eqref{eq:loglik_Taylor}, we have that
   \begin{align*}
       \hat\theta_{t_1,t_2}-\theta_1^o=O_p(1/\sqrt{ST})+O_p(1/T)+O_p(\kappa).
   \end{align*}
   for any $t_2-t_1\geq \epsilon_\lambda^o T$, which finishes the proof.    
\end{proof}
}

\section{Proof for infill asymptotics in Section \ref{sec:main_infill}}\label{sec:proof_infill}
\subsection{Technical lemmas for the main theorems}
This section states some lemmas that will be used in the proof of the main results for the infill setting in Section \ref{sec:main_infill} of the supplement. The proofs of these lemmas are deferred to Section \ref{proof.prop.lem_infill}. The following lemma provides a preliminary result for the convergence of the change-point estimators when the number of change-points is known.

\begin{lemma} 
	\label{known_infill}
	Let $\bfY$ be observations from a piecewise stationary random field specified by the vector $(m_o,\Lambda^o,\Psi^o)$, and Assumptions \ref{ass.infill.consis} and \ref{ass.diff.model} hold. Suppose that the number of change-points $m_o$ is known. 
	We estimate the change-points and model parameters by 
	\begin{align}
	\{\hat{\Lambda}_{ST},\hat{\Psi}_{ST}\}=\arg\min_{\substack{ \Psi\in \mathcal{M}^{m_o+1}\\ \Lambda\in A_{\epsilon_\lambda}^{m_o}}}\mathrm{CLMDL}(m_o,\Lambda,\Psi)\,,
	\end{align}
	where $A_{\epsilon_\lambda}^{m_o}$ is defined in \eqref{A}. Then, $\hat{\Lambda}_{ST}\longrightarrow\Lambda^o$ almost surely, provided that $S,T\longrightarrow \infty$ and $S=O(T^{1/(\nu+1)})$ for any $\nu>-\min_{1\leq j\leq m_o+1}\delta_j$.
\end{lemma}

The following lemma states some properties of the proposed procedure when the number of change-points is unknown.

\begin{lemma} \label{coro1_infill}
	Under the conditions of Lemma \ref{known_infill}, if the number of change-points $m_o$ is unknown and is estimated from the data using \eqref{minMDL} with the modified CLMDL criterion in \eqref{MDLform_infill}, then we have the following.
	
	(a) The number of change-points cannot be underestimated. In other words, $\hat{m}\geq m_o$ for sufficiently large $(S,T)$, almost surely.
	
	(b) When $\hat{m}\geq m_o$, then $\Lambda^o$ must be a subset of the limit points of $\hat{\Lambda}_{ST}$, i.e., for any  $\epsilon>0$ and $\lambda_j^o\in\Lambda^o$, there exists a $\hat{\lambda}_k\in \hat{\Lambda}_{ST}$ such that $|\lambda_j^o-\hat{\lambda}_k| < \epsilon$ for sufficiently large $(S,T)$, almost surely. 
%
\end{lemma}

Lemma \ref{coro1_infill} is refined in the following Lemma \ref{thm2_infill}, where an exact recovery property of the change-point estimator is established for the change-points $\tau_j^o$ for $j \in M_o$.

\begin{lemma}\label{thm2_infill}
	Under the conditions of Lemma \ref{known_infill}, with $(\hat{m},\hat{\Lambda}_{ST})$ estimated from the data using \eqref{minMDL} with the modified CLMDL criterion \eqref{MDLform_infill}, we have that, for each $j \in M_o$, there exists a 
	$\hat{\lambda}_{i_j}\in \hat{\Lambda}_{ST}, 1\leq i_j \leq \hat{m}$, such that
	$$\mathbb P\left(\left|[T\lambda_j^o] - [T\hat{\lambda}_{i_j}]\right| = 0\right) \longrightarrow 1\,,$$
	and for each $j \in {M_o^c}$, there exists a 
	$\hat{\lambda}_{i_j}\in \hat{\Lambda}_{ST}, 1\leq i_j \leq \hat{m}$, such that
	$$[T\lambda_j^o] - [T\hat{\lambda}_{i_j}]  = O_p(1) \,,$$
	as $(S,T)\longrightarrow \infty$.
\end{lemma}

The following lemma shows the consistency of the model parameter estimators.

\begin{lemma}\label{lil_infill}
	Under the conditions of Lemma \ref{known_infill}, for each $j=1,\ldots,m_o+1$, we have 
	\begin{align} \label{suppartcon_infill}
	\sup_{\underline{\lambda_d},\overline\lambda_u} \left|\hat{\theta}_{ST}^{(j)}(\lambda_d,\lambda_u)-\theta_j^o\right|\longrightarrow 0 \,,
	\end{align}
	almost surely. If, in addition, Assumption \ref{ass.infill.mix.time}($p$) for $p \geq 1$ holds, then we have
	\begin{align} \label{liltheta_infill}
	\sup_{\underline\lambda_d,\overline\lambda_u}\left|\hat{\theta}_{ST}^{(j)}(\lambda_d,\lambda_u)-\theta_j^o\right|=O_p(T^{-1/2})\,,
	\end{align}
	where $\theta_j^o$ is defined in Assumption \ref{ass.infill.consis}, and $\hat{\theta}_{ST}^{(j)}(\lambda_d,\lambda_u)$ is defined in 
	\eqref{eq:hat.theta} with $\xi_j=\xi^o$.
\end{lemma}

\subsection{Proofs of main theorems}
\begin{proof}[Proof of Theorem \ref{asym_infill}]	

    We first prove that $\hat m =m_o$ with probability going to 1. We prove by contradiction. First, by \Cref{coro1_infill} we have that $\hat m\geq m_o$ with probability going to 1. Thus, we can assume that $\hat m = m_o+d$ for $d\geq 1.$ We prove the case for $d=1$. The case for $d>1$ follows the same arguments with more tedious algebra.

    Given $\hat m=m_o+1$, denote $\hat\Lambda=(\hat{\lambda}_1,\cdots,\hat\lambda_{m_o+1})$ as the CLMDL based change-point estimator. Without loss of generality, we assume that $\hat\lambda_{m_o+1}$ gives the extra false positive change-point estimation. Thus, by \Cref{coro1_infill}, we have $|T\hat\lambda_j-T\lambda_j^o|=O_p(1)$ for $j=1,\cdots,m_o.$ Denote $\hat\Psi=(\hat\theta_1,\cdots,\hat\theta_{m_o+2})$ as the estimated model parameters.

    Based on $(\hat\Lambda, \hat\Psi)$, we construct $(\tilde\Lambda, \tilde\Psi)$ by removing $\hat\lambda_{m_o+1}.$ In particular, define $\tilde\Lambda=(\tilde{\lambda}_1,\cdots,\tilde\lambda_{m_o})=(\hat{\lambda}_1,\cdots,\hat\lambda_{m_o})$ and $\tilde\Psi=(\hat\theta_1,\cdots,\hat\theta_{m_o},\tilde\theta_{m_o+1})$, where $\tilde\theta_{m_o+1}$ is the parameter estimate based on data from $[T\hat\lambda_{m_o}]+1$ to $T.$ Thus, by definition, we have that
    \begin{align*}
        0\geq\mathrm{CLMDL}(\hat m, \hat\Lambda,\hat\Psi) - \mathrm{CLMDL}(m_o, \tilde\Lambda,\tilde\Psi).
    \end{align*}

    On the other hand, by a Taylor's expansion of the loglikelihood functions, we further have that
    \begin{align}
        &\mathrm{CLMDL}(\hat m, \hat\Lambda,\hat\Psi) - \mathrm{CLMDL}(m_o, \tilde\Lambda,\tilde\Psi)
        \label{argminprob_infill}\\
        =&C_1' - C_2' \nonumber\\
	+ &\left[\sum_{j=1}^{m_o+1}L_{ST}\{\theta_j^o,\tilde{\lambda}_{j-1},\tilde{\lambda}_j; \bfY\}-\sum_{l=1}^{m_o+2}L_{ST}\{\theta_l^o,\hat{\lambda}_{l-1},\hat{\lambda}_l; \bfY\}\right] \label{sum2_infill}\\
	 + &\frac{1}{2}\sum_{l=m_o+1}^{m_o+2}(\hat{\theta}_{l}-\theta_{m_o+1}^o)  L_{ST}^{''}\{\theta_l^+,\hat{\lambda}_{l-1},\hat{\lambda}_l; \bfY\}(\hat{\theta}_l-\theta_{m_o+1}^o) \label{taylor2sum_infill} \\
  - & \frac{1}{2}(\tilde{\theta}_{m_o+1}-\theta_{m_o+1}^o)  L_{ST}^{''}\{\tilde\theta_{m_o+1}^+,\hat{\lambda}_{m_o},1; \bfY\}(\tilde{\theta}_{m_o+1}-\theta_{m_o+1}^o)\label{taylor2sum_infill_2},
    \end{align}
	where $C_1'-C_2'$ is positive and of order $O(S\log ST)$, and $|\theta_l^+-\theta_{m_o+1}^o|<|\hat{\theta}_l-\theta_{m_o+1}^o|$ for $l=m_o+1,m_o+2$, and $|\tilde\theta_{m_o+1}^+-\theta_{m_o+1}^o|<|\tilde{\theta}_{m_o+1}-\theta_{m_o+1}^o|$.	

	
	By Lemma \ref{lil_infill}, we have $\hat{\theta}_{ST}^{(j)}- \theta_j^o = O_p(T^{-1/2})$ and thus the summation in \eqref{taylor2sum_infill} and \eqref{taylor2sum_infill_2} are at most of order $O_p(S)$. Note that the quantity $C_1'-C_2'$, which is of order $O(S\log ST)$, and therefore dominates the expression. 
	Define 
	$D_{cST}^{\lambda}=\{([T\lambda],[T\lambda]+1,\bs_1,\bs_2):, \bs_1 \in \mathcal{S}, \bs_2 \in \bs_1\cup\mathcal{N}(\bs_1)\}$.
	Due to the presence of the one extra pseudo-change point at $\hat\lambda_{m_o+1}$, \eqref{sum2_infill} is
	\begin{align}
	\sum_{D_{cST}^{\hat{\lambda}_l}} \left[l_{pair}(\theta_{l}^o; y_{t,s_1}^{(l)}, y_{t+1,s_2}^{(l)})-  l_{marg}({\theta}_{l}^o;y_{t,s}^{(l)})
	-l_{marg}({\theta}_{l}^o;y_{t,s}^{(l)})\right]\,,\label{cross-overterm_infill}
	\end{align}
 where we set $l=m_o+1$ for notational simplicity. Denote
	\begin{align*}
	B_{ \lambda ST}=
	\sum_{D_{cST}^{\lambda}}\left[l_{pair}(\theta_{l}^o; y_{t,s_1}^{(l)}, y_{t+1,s_2}^{(l)})-  l_{marg}({\theta}_{l}^o;y_{t,s}^{(l)})
	-l_{marg}({\theta}_{l}^o;y_{t,s}^{(l)})\right]\,.
	\end{align*}
	If we can show that
	\begin{align}
	P\left(\inf_{\lambda^o_{l-1}+\epsilon_\lambda<\lambda<\lambda^o_l-\epsilon_\lambda}B_{\lambda ST}< -S(\log ST)\right)\longrightarrow 0,
	\label{exponentialinequality_infill}
	\end{align}
	then we know the quantity in \eqref{argminprob_infill} is indeed positive with probability going to $1$.
	
	By union bound, closure of the exponential type Orlicz spaces and the exponential-type decaying tail probabilities, we have that
	\begin{align*}
	P\left(\inf_{\lambda^o_{l-1}+\epsilon_\lambda<\lambda<\lambda^o_l-\epsilon_\lambda}B_{ \lambda ST}< -S\log ST\right)
	\leq &P\left(\inf_{\lambda^o_{l-1}+\epsilon_\lambda<\lambda<\lambda^o_l-\epsilon_\lambda} \frac{B_{ \lambda ST}}{S}< -\log ST\right)\\
	\leq &T(\lambda_l^o-\lambda_{l-1}^o)P\left(\frac{B_{ \lambda ST}}{S}< -\log ST\right)\\
	\leq &T(\lambda_l^o-\lambda_{l-1}^o)P\left(\left|\frac{B_{ \lambda ST}}{S}\right|> \log ST\right)\\
	\leq &2T(\lambda_l^o-\lambda_{l-1}^o)\exp{(-((\log{ST})/K_{\psi_p})^p)}\longrightarrow 0,
 \end{align*}
where the last term clearly goes to zero if $p>1$ and goes to zero if $T^{K_{\psi_p}-1}/S \longrightarrow 0$ for $p=1$. Thus, we have a contradiction, which in turn proves that $\hat m=m_o$ with probability going to 1.
	
Lastly, the convergence rate follows from Lemma \ref{thm2_infill} and the weak consistency of the estimated number of change-points. Hence the proof of \eqref{weak.consist_infill} is complete. 
\end{proof}

\subsection{Proofs of propositions and lemmas}\label{proof.prop.lem_infill}
\begin{proof}[Proof of Lemma \ref{known_infill}]	
	Let $B$ be the probability one set on which Assumption \ref{ass.infill.consis} and \eqref{suppartcon_infill} in Lemma \ref{lil_infill} hold. We will show that for each $\tilde{\omega} \in B$, we have $\hat{\Lambda}_{ST}\rightarrow \Lambda^o$ as $(S,T)\rightarrow \infty$. To begin, for any $\tilde{\omega} \in B$, suppose on the contrary that $\hat{\Lambda}_{ST}\nrightarrow \Lambda^o$. Since the values of $\hat{\Lambda}_{ST}$ are bounded, there exists a subsequence $\{(S_l,T_l)\}$ such that $\hat{\Lambda}_{ST} \rightarrow \Lambda^*\neq\Lambda^o$ along the subsequence. Note that $\Lambda^* \in A_{\epsilon_\lambda}^{m_o}$ for all $(S,T)$. Recall that $(\xi^o,\hat{\theta}_{ST}^{(j)})$ contains the model order and model parameter estimators for the $j$th segment. 
	Since $\Theta\equiv\Theta(\xi^o)$ is compact, we can assume that $\hat{\theta}_{ST}^{(j)}\rightarrow \theta_j^*$ along $\{(S_l, T_l)\}$ for some $\theta_j^* \in \Theta$. To lighten the notation, we replace $(S_l, T_l)$ by $(S,T)$ and it follows that for all sufficiently large $(S,T)$,
	\begin{equation}\label{clmdl_infill}
		{\mathrm{CLMDL}} (m_o, \hat{\Lambda}_{ST},\hat{\Psi}_{ST})=c_{ST}- \sum_{j=1}^{m_o+1}L_{ST}\{\hat{\theta}_{ST}^{(j)},\hat{\lambda}_{j-1},\hat{\lambda}_j; \bfY\},
	\end{equation}
	where $c_{ST}$ is deterministic, positive and of order $O(S\log ST)$. 
	
	For each limiting estimated segment $I_j^*=(\lambda_{j-1}^*,\lambda_j^*) ,j=1,\ldots,m+1$, 
	there are two possible cases. First, $I_j^*$ is nested in the $i$th segment $(\lambda_{i-1}^o, \lambda_i^o)$. 
	Second, $I_j^*$ fully or partly covers $k+2$ $(k\geq0)$ true intervals: $(\lambda_{i-1}^o,\lambda_i^o),\ldots,(\lambda_{i+k}^o,\lambda_{i+k+1}^o)$. 
	We consider these two cases separately.
 
	Case 1. If $\lambda_{i-1}^o\leq\lambda_{j-1}^* < \lambda_j^* \leq \lambda_i^o$, in particular, 
	if $\lambda_{i-1}^o < \lambda_{j-1}^* < \lambda_j^* <\lambda_i^o$, then for sufficiently large $(S,T)$, 
	the estimated $j$th segment is a proportion of the stationary process from the true $i$th segment. 
	If $\lambda_j^*=\lambda_i^o$ or $\lambda_{j-1}^*=\lambda_{i-1}^o$, then as $\hat{\lambda}_j \rightarrow \lambda_i^o$ or $\hat{\lambda}_{j-1}\rightarrow \lambda_{i-1}^o$,
	the estimated segment includes only a decreasing proportion of observations from an adjacent segment. Hence, Assumption \ref{ass.infill.consis} implies that $\hat{\theta}_{ST}^{(j)}\rightarrow \theta_j^* = \theta_i^o$ almost surely.

	Case 2. If $\lambda_{i-1}^o\leq\lambda_{j-1}^* < \lambda_i^o <\cdots<\lambda_{i+k}^o < \lambda_j^*\leq \lambda_{i+k+1}^o$ for some $k \geq0$, then for sufficiently large $(S,T)$, the estimated $j$th segment contains observations from at least two pieces of different stationary processes. Partitioning the composite likelihood by the true configuration of the series, we have
	\begin{align}\label{partition_infill}
		& L_{ST}\{\hat{\theta}_{ST}^{(j)}, \hat{\lambda}_{j-1},\hat{\lambda}_j;\bfY\}  \nonumber \\
		=&L_{ST}\{\hat{\theta}_{ST}^{(j)}, \hat{\lambda}_{j-1},\lambda_i^o; \bfY\}+\sum_{l=i}^{i+k-1}L_{ST}\{\hat{\theta}_{ST}^{(j)}, \lambda_l^o,\lambda_{l+1}^o; \bfY\} + L_{ST}\{\hat{\theta}_{ST}^{(j)}, \lambda_{i+k}^o, \hat{\lambda}_j; \bfY\} \nonumber \\
		+& \sum_{l=i}^{i+k}\sum_{D_{cST}^{\lambda_l^o}}\left[l_{pair}\{\hat{\theta}_{ST}^{(j)}; y_{[\lambda_l^oT],s_1}, y_{[\lambda_l^oT]+1,s_2})\} - l_{marg}(\hat{\theta}_{ST}^{(j)}; y_{[\lambda_l^oT],s_1}) - l_{marg}(\hat{\theta}_{ST}^{(j)}; y_{[\lambda_l^oT]+1,s_2})\right]\,.
	\end{align}	
	The last terms in \eqref{partition_infill} involves observations from one piece of the stationary process and is of order $O_p(S)$. By Assumptions \ref{ass.infill.consis}, we have for sufficiently large $(S,T)$,
	\begin{eqnarray*}
	\frac{1}{S^{\delta_i}(\lambda_i^o-\hat\lambda_{j-1})T}\left(	L_{ST}\{\theta_i^o, \hat{\lambda}_{j-1},\lambda_i^o;\bfY\}-L_{ST}\{\hat{\theta}_{ST}^{(j)}, \hat{\lambda}_{j-1},\lambda_i^o;\bfY\}\right)\geq 0 \,,\\
	\frac{1}{S^{\delta_{l+1}}(\lambda_{l+1}^o-\lambda_{l}^o)T}\left(	L_{ST}\{\theta_{l+1}^o, \lambda_l^o,\lambda_{l+1}^o;\bfY\} - L_{ST}\{\hat{\theta}_{ST}^{(j)}, \lambda_l^o,\lambda_{l+1}^o;\bfY\}\right)\geq 0  \,,\\
	\frac{1}{S^{\delta_{i+k+1}}(\hat{\lambda}_j-\lambda_{i+k}^o)T}\left(	L_{ST}\{\theta_{i+k+1}^o, \lambda_{i+k}^o, \hat{\lambda}_j;\bfY\}-L_{ST}\{\hat{\theta}_{ST}^{(j)},\lambda_{i+k}^o, \hat{\lambda}_j; \bfY\}\right) \geq 0 \,,
	\end{eqnarray*}
	almost surely. 
	Note that strict inequalities hold for at least one of the preceding equations, since $\hat{\theta}_{ST}^{(j)}$ cannot correctly specify the model for all different segments.
	Thus, for sufficiently large $(S,T)$,
	\begin{align} \label{limineqcase2_infill}
		L_{ST}\{\hat{\theta}_{ST}^{(j)},\hat{\lambda}_{j-1},\hat{\lambda}_j; \bfY\} <&  L_{ST}\{(\xi_i^o,\theta_i^o), \hat{\lambda}_{j-1},\lambda_i^o; \bfY\}+\sum_{l=i}^{i+k-1}L_{ST}\{\theta_{l+1}^o, \lambda_l^o,\lambda_{l+1}^o; \bfY\}   \nonumber \\
		+&L_{ST}\{\theta_{i+k+1}^o, \lambda_{i+k}^o, \hat{\lambda}_j; \bfY\}-O(S^\delta T)\,,
	\end{align}
	almost surely, where $\delta=\min_{1\leq j\leq m_o+1}\delta_j.$ By the assumption that $S=O(T^{1/(\nu+1)})$ for any $\nu>-\delta$, we have that the order of the difference $O(S^\delta T)$ in \eqref{limineqcase2_infill} dominates the order of $c_{ST}=O(S\log ST)$ in the expression \eqref{clmdl_infill}.
	As the number of estimated segments is equal to the true number of segments and $\lambda^*\neq\lambda^o$, there is at least one segment in which Case 2 applies. Thus for sufficiently large $(S,T)$,
	\begin{align} \label{case2con_infill}
		0&\geq {\mathrm{CLMDL}}(m_o,\hat{\Lambda}_{ST},\hat{\Psi}_{ST})- {\mathrm{CLMDL}}(m_o,\Lambda^o,\Psi^o)\text{\hspace{ 10mm} [property of the estimator]}\nonumber \\
		&> 0, \text{\quad [\eqref{limineqcase2_infill} holds for at least one piece]} 
	\end{align}
	which is a contradiction. Hence $\hat{\Lambda}_{ST}\rightarrow\Lambda^o$ for each $\tilde{\omega}\in B$. This completes the proof.
\end{proof}

\begin{proof}[Proof of Lemma \ref{coro1_infill}]	
	In the proof of Lemma \ref{known_infill}, the assumption of a known number of change-points is used only to ensure that Case 2 applies for at least one $j$. 
	The contradiction in the proof of Lemma \ref{known_infill} arises whenever Case 2 applies. 
	By observing this fact, (a) and (b) follow.
\end{proof}

\begin{proof}[Proof of Lemma \ref{thm2_infill}]
	From Lemma \ref{coro1_infill}, we can assume that $\hat{m}\geq m_o$ and for each $\lambda_j^o$, there exists a $\hat{\lambda}_{i_j}$ such that 
	$|\lambda_j^o - \hat{\lambda}_{i_j}| = o_p(1)$. This lemma further refines the rate 
 for the change-point estimators. First, consider all change-points $\lambda_j^o$ such that $j\in M_o.$ Suppose that for some $\lambda_l^o$ with $l\in M_o$, there does not exist an $\hat{\lambda}_{i_l}$ such that 
	$P\left(\left|[T\lambda_l^o] - [T\hat{\lambda}_{i_l}]\right|=0\right) \longrightarrow 1$, then either one of the events
	\begin{eqnarray} \label{thm2case_infill}
		&& \text{i) \quad } [T\lambda_l^o] - [T\hat{\lambda}_{i_l}] \geq 1, \nonumber  \text{\quad or} \\
		&& \text{ii) \quad } [T\hat{\lambda}_{i_l}] - [T\lambda_l^o]\leq 1\,,
	\end{eqnarray}
	holds with nonzero probability, where $\hat{\lambda}_{i_l}$ is the estimated change-point closest to $\lambda_l^o$. From Lemma \ref{coro1_infill}, we know that $\hat{\lambda}_{i_l}\to \lambda_l^o$ in probability. We show that $P\left([T\lambda_l^o]-[T\hat{\lambda}_{i_l}]\geq 1\right) \longrightarrow 0$. The proof for the other direction is the same.

	
	On the event $\{T\lambda_l^o-T\hat{\lambda}_{i_l}\geq 1\}$ and from the fact that $\hat{\lambda}_{i_l}$ is the estimated change point and the consistency of $\hat{\theta}$, we have that
	\begin{align*}
		L_{ST}^{(j)}(\hat{\theta}_{i_l}, \hat{\lambda}_{i_l-1},\hat{\lambda}_{i_l}; \bfY) + 
		L_{ST}^{(j)}(\hat{\theta}_{i_l+1}, \hat{\lambda}_{i_l},\hat{\lambda}_{i_l+1}; \bfY) 
		> L_{ST}^{(j)}(\theta^o_{l}, \hat{\lambda}_{i_l-1},\lambda_l^o; \bfY) + 
		L_{ST}^{(j)}(\theta^o_{l+1}, \lambda_l^o,\hat{\lambda}_{i_l+1}; \bfY)\,.
	\end{align*}

    By Assumption \ref{ass.infill.consis} and \Cref{lil_infill}, we have
    $L_{ST}^{(j)}(\hat{\theta}_{i_l}, \hat{\lambda}_{i_l-1},\hat{\lambda}_{i_l}; \bfY) + 
		L_{ST}^{(j)}(\hat{\theta}_{i_l+1}, \hat{\lambda}_{i_l},\hat{\lambda}_{i_l+1}; \bfY)=L_{ST}^{(j)}({\theta}_l^o, \hat{\lambda}_{i_l-1},\hat{\lambda}_{i_l}; \bfY) + 
		L_{ST}^{(j)}({\theta}_{l+1}^o, \hat{\lambda}_{i_l},\hat{\lambda}_{i_l+1}; \bfY)+O_p(1)$ and thus we have
  	\begin{align*}
		L_{ST}^{(j)}({\theta}_{l}^o, \hat{\lambda}_{i_l-1},\hat{\lambda}_{i_l}; \bfY) + 
		L_{ST}^{(j)}({\theta}_{l+1}^o, \hat{\lambda}_{i_l},\hat{\lambda}_{i_l+1}; \bfY) 
		+O_p(1)> L_{ST}^{(j)}(\theta^o_{l}, \hat{\lambda}_{i_l-1},\lambda_l^o; \bfY) + 
		L_{ST}^{(j)}(\theta^o_{l+1}, \lambda_l^o,\hat{\lambda}_{i_l+1}; \bfY)\,.
	\end{align*}
 
	Note that \Cref{ass.diff.model}, the probability of the above inequality to hold goes to zero for $l\in M_o$ as long as $S \to \infty$, which is the essential ingredient for the exact recovery property of the estimated change-points. Thus we have
	\begin{align*}
		\mathbb P\left( [T\lambda_l^o] - [T\hat{\lambda}_{i_l}] \geq 1\right)=&\mathbb P\Bigg([T\lambda_l^o] - [T\hat{\lambda}_{i_l}] \geq 1, L_{ST}^{(j)}(\hat{\theta}_{i_l}, \hat{\lambda}_{i_l-1},\hat{\lambda}_{i_l}; \bfY) + 
		L_{ST}^{(j)}(\hat{\theta}_{i_l+1}, \hat{\lambda}_{i_l},\hat{\lambda}_{i_l+1}; \bfY) 
		\nonumber\\
		&\quad\quad\quad\quad> L_{ST}^{(j)}(\theta^o_{l}, \hat{\lambda}_{i_l-1},\lambda_l^o; \bfY) + 
		L_{ST}^{(j)}(\theta^o_{l+1}, \lambda_l^o,\hat{\lambda}_{i_l+1}; \bfY)\Bigg) \longrightarrow 0,
	\end{align*}
	On the other hand, for each $\lambda_j^o$ with $j \in {M_o^c}$, it can be shown that there exists a 
	$\hat{\lambda}_{i_j}\in \hat{\Lambda}_{ST}, 1\leq i_j \leq \hat{m}$, such that
	$[T\lambda_j^o] - [T\hat{\lambda}_{i_j}]  = O_p(1) \,,$
	as $(S,T)\longrightarrow \infty$.
	The proof follows a similar argument as above and thus is omitted.
\end{proof}

\begin{proof}[Proof of Lemma \ref{lil_infill}]	
	Let $B$ be the probability one set on which Assumption \ref{ass.infill.consis} holds. We will show that for each $\tilde{\omega} \in B$, $\sup_{\underline{\lambda_d},\overline{\lambda_u}} \left|\hat{\theta}_{ST}^{(j)}(\lambda_d,\lambda_u)-\theta_j^o\right|\longrightarrow 0 \,,$ as $(S,T)\rightarrow \infty$. To begin, for any $\tilde{\omega} \in B$, suppose on the contrary that $\sup_{\underline{\lambda_d},\overline{\lambda_u}} \left|\hat{\theta}_{ST}^{(j)}(\lambda_d,\lambda_u)-\theta_j^o\right|\nrightarrow 0$. Since $\Theta$ is compact, there exists a subsequence $\{(S_l,T_l)\}$ such that $\hat{\theta}_{ST}^{(j)}(\lambda_d^*,\lambda_u^*) \rightarrow \theta_j^*\neq\theta_j^o$ along the subsequence for fixed $\lambda_d^*$ and $\lambda_u^*$.
	By the definition of $\hat{\theta}_{ST}^{(j)}(\lambda_d,\lambda_u)$, we have 	
	\begin{equation}\label{consist_contradict}
	L_{ST}\{\theta_j^*,\lambda_d^*,\lambda_u^*;\bfX_j^o\} \geq L_{ST}\{\theta_j^o,\lambda_d^*,\lambda_u^*;\bfX_j^o\}
	\end{equation}
	for sufficiently large $(S,T)$. It follows that \eqref{consist_contradict} contradicts Assumption \ref{ass.infill.consis}. 
	Thus, it follows that \eqref{suppartcon_infill} holds. Furthermore, by Assumptions \ref{ass.infill.consis} and \ref{ass.infill.mix.time}(ii), \eqref{suppartcon_infill} and Corollary 5.53 in \cite{van1998asymptotic}, we can show that \eqref{liltheta_infill} holds, thus complete the proof.
\end{proof}

\subsection{Illustrative examples}\label{infill_example}
In this subsection, we consider the widely used autoregressive spatial model,
\begin{equation} \label{sim.process.infill}
	x_{t,\bs} = \phi x_{t-1,\bs} + \varepsilon_{t,\bs}\,,
\end{equation}
where $\phi\in(-1,1)$, $t \in \{1,\ldots,T\}$, $\bs \in \mathcal{S}$ and $\{ \varepsilon_{t,\bs}: \bs \in \mathcal{S} \}$ is a zero-mean error process with covariance function ${\rm Cov}(\varepsilon_{t,\textbf{u}}, \varepsilon_{t,\textbf{v}})$. Two illustrative examples of model \eqref{sim.process.infill} with Gaussian noises, which satisfy Assumptions \ref{ass.infill.consis}, \ref{ass.diff.model} and \ref{ass.infill.mix.time}($p$), are given below.

\subsubsection{Gaussian noises with isotropic exponential covariance function}\label{ill_example_exp}
Consider the autoregressive spatial model in \eqref{sim.process.infill} with stationary isotropic Gaussian noises $\{\varepsilon_{t,\bs}: \bs \in  \mathcal{S}\}$, with mean $0$ and an isotropic covariogram ${\rm Cov}(\varepsilon_{t,\bs_1}, \varepsilon_{t,\bs_2})=\sigma^2 \exp (-\varphi \lVert \bs_2-\bs_1\lVert_2)$ with ${\rm Cov}(\varepsilon_{t,\bs_1}, \varepsilon_{t',\bs_2}) =0$, when $t\neq t'$. \cite{Zhang2004} studied the estimability of $\sigma^2\varphi$, $\sigma^2$ and $\varphi$ under spatial context, and also studied the consistency of model parameter estimators. 

Consider that data are collected along a straight line. Although spatial data are usually collected over a spatial region, there are situations when data are collected along lines. One example is the International H2O project, where measurements of meteorological data were collected by surface stations and aircraft along three straight flight paths in the southern Great Plains. Ecological data are sometimes collected along line transects as well. Similar setting can also be found in \cite{du2009fixed}. 

Below we show that Assumption \ref{ass.infill.consis} holds with $\delta=1$.	
For simplicity of presentation, we give the detailed proof for the time lag $k=1$, and the maximum cardinality of the neighborhood set $B_{\mathcal{N}}=2$. Also, the data $X$ is observed on $\mathcal{Z}_n=\mathcal{T}_n\times \mathcal{S}_n$ with $\mathcal{T}_n=[1,T_n]\cap \mathbb{N}$ and $\mathcal{S}_n=\{\frac{i}{n}:i=1,\ldots,n\} \subset \mathbb{R}$ after parametrization along a straight line in $\mathbb{R}^2$. For notational simplicity, in the following we use $(S,T)$ instead of $(|\mathcal{S}_n|,|\mathcal{T}_n|)$ when there is no confusion. Hence we have $S=n$. Extensions to $k>1$, larger $B_{\mathcal{N}}$ and other sampling configuration $\mathcal{S}_n$ are possible but technically involved and we do not pursue them here. We define several index sets that will be used throughout this section, where
\begin{align*}
	E_{1ST}&=\{(t,\bs):t=1, \bs\in \mathcal{S}, \text{ each } \bs  \text{ repeats } 3 \text{ times} \},\\
	E_{2ST}&=\{(t,\bs):t=T, \bs\in \mathcal{S}, \text{ each } \bs  \text{ repeats } 3 \text{ times} \},\\
	D_{ST}&=\{(t,i,\bs_1,\bs_2): 1\leq t, t+i \leq T, 0\leq i\leq 1, \bs_1 \in \mathcal{S}, \bs_2 \in \bs_1\cup \mathcal{N}(\bs_1), \text{if }i=0, \bs_1\neq \bs_2\}.
\end{align*}

Note that we have the neighborhood set $\mathcal{N}(\bs)=\{\bs-\frac{1}{n},\bs+\frac{1}{n}\}$, and hence $|\mathcal{N}(\bs)|=2$. The sets $E_{1ST}$ and $E_{2ST}$ are used to index the marginal loglikelihood for compensating the edge effect. 
The set $D_{ST}$ is used to index the pairwise loglikelihood between $1$ and $T$. Under this setting, for each point $(t,\bs) \in \mathcal{T}_n\times \mathcal{S}_n$, there are four possible directions to pair up: (1): $(t,\bs+\frac{1}{n})$, (2): $(t+1,\bs)$, (3): $(t+1,\bs+\frac{1}{n})$ and (4): $(t+1,\bs-\frac{1}{n})$ and define the pairs in those directions as $D^{(1)}_{ST}$, $D^{(2)}_{ST}$, $D^{(3)}_{ST}$ and $D^{(4)}_{ST}$, respectively. Hence, we have $D_{ST}=\cup_{i=1,2,3,4} D^{(i)}_{ST}$. To study the asymptotic behavior of the new composite likelihood in \eqref{eq:CPL}, we can first study the asymptotic behavior of the pairwise likelihood along each direction and then aggregate them. Note that the covariance function for $x_{t,\bs}$ is ${\rm Cov}(x_{t,\bs},x_{t+i,\bs+\frac{j}{n}})=\frac{\sigma^2}{1-\phi^2} \phi^{|i|} e^{-\varphi\frac{|j|}{n}}=\sigma_x^2 \phi^{|i|} e^{-\varphi\frac{|j|}{n}}$ where $\sigma_x^2=\frac{\sigma^2}{1-\phi^2}$ is the variance of $x_{t,\bs}$. Denote the true parameter values as $\theta_0=(\sigma_0^2,\phi_0,\varphi_0)$ and $\sigma_{x0}^2=\frac{\sigma_0^2}{1-\phi_0^2}$. Define $\theta=(\sigma^2,\phi,\varphi)$ and $\sigma_{x}^2=\frac{\sigma^2}{1-\phi^2}$, consider the following
\begin{eqnarray*}
	L_{ST}(\theta; X)&=&\sum_{(t,i,\bs_1,\bs_2)\in D_{ST}} l_{pair}\left(\theta; x_{t,\bs_1},x_{t+i,\bs_2}\right) + \sum_{(t,\bs)\in E_{1ST}} l_{marg}\left(\theta; x_{t,\bs}\right) + \sum_{(t,\bs)\in E_{2ST}} l_{marg}\left(\theta; x_{t,\bs}\right) \nonumber\\
	&=&\sum_{(t,i,\bs_1,\bs_2)\in D^{(1)}_{ST}} \log f\left(x_{t,\bs_1},x_{t+i,\bs_2};\theta\right) + \sum_{(t,i,\bs_1,\bs_2)\in D^{(2)}_{ST}}\log f\left(x_{t,\bs_1},x_{t+i,\bs_2};\theta\right) \\
	& & + \sum_{(t,i,\bs_1,\bs_2)\in D^{(3)}_{ST}}\log f\left(x_{t,\bs_1},x_{t+i,\bs_2};\theta\right) + \sum_{(t,i,\bs_1,\bs_2)\in D^{(4)}_{ST}} \log f\left(x_{t,\bs_1},x_{t+i,\bs_2};\theta\right) \\
	& & + \sum_{(t,\bs)\in E_{1ST}} \log f\left(x_{t,\bs};\theta\right) + \sum_{(t,\bs)\in E_{2ST}}\log f\left(x_{t,\bs};\theta\right)\,.
\end{eqnarray*}
First, multiply on both sides by $-2$ and consider pairs $(t,\bs)$ and  $(t,\bs+\frac{1}{n})$ in $D^{(1)}_{ST}$, we have $\sum_{(t,i,\bs_1,\bs_2)\in D^{(1)}_{ST}} -2 \log f\left(x_{t,\bs_1},x_{t+i,\bs_2};\theta\right) = \sum_{t=1}^{T} \sum_{i=1}^{n-1} -2 \log f\left(x_{t,\frac{i}{n}},x_{t,\frac{i+1}{n}};\theta\right)$. We have the following equation
\begin{eqnarray*}
-2 \log f\left(x_{t,\frac{i}{n}},x_{t,\frac{i+1}{n}};\theta\right) & = & -2 \log f\left(x_{t,\frac{i+1}{n}}\Big|x_{t,\frac{i}{n}};\theta\right) -2  \log f\left(x_{t,\frac{i}{n}};\theta\right)\,,
\end{eqnarray*}
with 
$-2 \log f\left(x_{t,\frac{i+1}{n}}\Big|x_{t,\frac{i}{n}};\theta\right)= \frac{\left(x_{t,\frac{i+1}{n}}-e^{-\frac{\varphi}{n}}x_{t,\frac{i}{n}}\right)^2}{\sigma_{x}^2 \left(1-e^{-\frac{2\varphi}{n}}\right)} + \log \sigma_{x}^2 (1-e^{-\frac{2\varphi}{n}}) +\log 2\pi$ and marginal log-likelihood $-2 \log f\left(x_{t,\frac{i}{n}};\theta\right)=\frac{x_{t,\frac{i}{n}}^2}{\sigma_{x}^2} + \log \sigma_{x}^2 +\log 2\pi$. Applying the same techniques used in \cite{Ying1991} and \cite{Ying1993}, we have
\begin{eqnarray}\label{microergodic_div}
& & \sum_{t=1}^{T} \sum_{i=1}^{n-1} -2 \log f\left(x_{t,\frac{i+1}{n}}\Big|x_{t,\frac{i}{n}};\theta\right) - \sum_{t=1}^{T} \sum_{i=1}^{n-1} -2 \log f\left(x_{t,\frac{i+1}{n}}\Big|x_{t,\frac{i}{n}};\theta_0\right)\nonumber\\
& = & \left(\frac{\sigma_{x0}^2 \varphi_0}{\sigma_{x}^2 \varphi}-1\right)\sum_{t=1}^{T} \sum_{i=1}^{n-1} W_{t,k,n}^2 -(n-1)T \log \frac{\sigma_{x0}^2 \varphi_0}{\sigma_{x}^2 \varphi}\\
& & + \left(\frac{\sigma_{x0}^2 \varphi_0}{\sigma_{x}^2 \varphi}-1\right)(\varphi-\varphi_0) T + \frac{\sigma_{x0}^2 (\varphi-\varphi_0)^2}{2\sigma_{x}^2 \varphi}T+o(T)\nonumber\,,
\end{eqnarray}
where $W_{t,k,n}$ are standard normal random variables, and 
\begin{eqnarray*}
	& & \sum_{t=1}^{T} \sum_{i=1}^{n-1} -2  \log f\left(x_{t,\frac{i}{n}};\theta\right) - \sum_{t=1}^{T} \sum_{i=1}^{n-1} -2  \log f\left(x_{t,\frac{i}{n}};\theta_0 \right)\\
	& = & \left(\frac{1}{\sigma_{x}^2}-\frac{1}{\sigma_{x0}^2} \right) \sum_{t=1}^{T} \sum_{i=1}^{n-1} x_{t,\frac{i}{n}}^2 - (n-1)T  \log \frac{\sigma_{x0}^2}{\sigma_{x}^2}\,.
\end{eqnarray*}

Similarly, multiply on both sides by $-2$ and consider pairs $(t,\bs)$ and  $(t+1,\bs+\frac{1}{n})$ in $D^{(3)}_{ST}$, we have $\sum_{(t,i,\bs_1,\bs_2)\in D^{(3)}_{ST}} -2 \log f\left(x_{t,\bs_1},x_{t+i,\bs_2};\theta\right) = \sum_{t=1}^{T-1} \sum_{i=1}^{n-1} -2 \log f\left(x_{t,\frac{i}{n}},x_{t+1,\frac{i+1}{n}};\theta\right)$. We have the following equation
\begin{eqnarray*}
	-2 \log f\left(x_{t,\frac{i}{n}},x_{t+1,\frac{i+1}{n}};\theta\right) & = & -2 \log f\left(x_{t+1,\frac{i+1}{n}}\Big|x_{t,\frac{i}{n}};\theta\right) -2  \log f\left(x_{t,\frac{i}{n}};\theta\right)\,,
\end{eqnarray*}
with 
$-2 \log f\left(x_{t+1,\frac{i+1}{n}}\Big|x_{t,\frac{i}{n}};\theta\right)= \frac{\left(x_{t+1,\frac{i+1}{n}}-\phi e^{-\frac{\varphi}{n}}x_{t,\frac{i}{n}}\right)^2}{\sigma_{x}^2 \left(1-\phi^2e^{-\frac{2\varphi}{n}}\right)} + \log \sigma_{x}^2 (1-\phi^2e^{-\frac{2\varphi}{n}}) +\log 2\pi$ and marginal log-likelihood $-2 \log f\left(x_{t,\frac{i}{n}};\theta\right)=\frac{x_{t,\frac{i}{n}}^2}{\sigma_{x}^2} + \log \sigma_{x}^2 +\log 2\pi$. Applying similar techniques, we have
\begin{eqnarray}\label{sigma_phi_div}
	& & \sum_{t=1}^{T-1} \sum_{i=1}^{n-1} -2 \log f\left(x_{t+1,\frac{i+1}{n}}\Big|x_{t,\frac{i}{n}};\theta\right) - \sum_{t=1}^{T-1} \sum_{i=1}^{n-1} -2 \log f\left(x_{t+1,\frac{i+1}{n}}\Big|x_{t,\frac{i}{n}};\theta_0\right)\nonumber\\
	& = & \left[\frac{\sigma_{x0}^2 (1-\phi_0^2)}{\sigma_{x}^2 (1-\phi^2)}-1\right]\sum_{t=1}^{T-1} \sum_{i=1}^{n-1} \widetilde{W}_{t,k,n}^2 -(n-1)(T-1) \log \frac{\sigma_{x0}^2 (1-\phi_0^2)}{\sigma_{x}^2 (1-\phi^2)} + \frac{\sigma_{x0}^2 (\phi_0-\phi)^2}{\sigma_{x}^2 (1-\phi^2)} \sum_{t=1}^{T-1} \sum_{i=1}^{n-1} W_{t,k,n}^{*2}\nonumber\\
	& &  + \left[\frac{2\sigma_{x0}^2 \phi_0^2 \varphi_0}{\sigma_{x}^2 (1-\phi^2)}-\frac{2 \phi_0^2 \varphi_0}{1-\phi_0^2}+\frac{2 \phi^2 \varphi}{1-\phi^2}-\frac{2\sigma_{x0}^2 (\phi_0-\phi) (\phi_0 \varphi_0-\phi \varphi)}{\sigma_{x}^2 (1-\phi^2)}\right] T +o(T)\nonumber\\
	& = & \left(\frac{\sigma_{0}^2}{\sigma^2}-1\right)\sum_{t=1}^{T-1} \sum_{i=1}^{n-1} \widetilde{W}_{t,k,n}^2 -(n-1)(T-1) \log \frac{\sigma_{0}^2}{\sigma^2} + \frac{\sigma_{x0}^2 (\phi_0-\phi)^2}{\sigma^2} \sum_{t=1}^{T-1} \sum_{i=1}^{n-1} W_{t,k,n}^{*2}\\
	& &  + \left[\frac{2\sigma_{x0}^2 \phi_0^2 \varphi_0}{\sigma^2}-\frac{2 \sigma_{x0}^2 \phi_0^2 \varphi_0}{\sigma_{0}^2}+\frac{2 \sigma_x^2 \phi^2 \varphi}{\sigma^2}-\frac{2\sigma_{x0}^2 (\phi_0-\phi) (\phi_0 \varphi_0-\phi \varphi)}{\sigma^2}\right] T +o(T)\nonumber\,,
\end{eqnarray}
where $\widetilde{W}_{t,k,n}$ and $W_{t,k,n}^{*}$ are standard normal random variables, and 
\begin{eqnarray*}
	& & \sum_{t=1}^{T-1} \sum_{i=1}^{n-1} -2  \log f\left(x_{t,\frac{i}{n}};\theta\right) - \sum_{t=1}^{T-1} \sum_{i=1}^{n-1} -2  \log f\left(x_{t,\frac{i}{n}};\theta_0 \right)\\
	& = & \left(\frac{1}{\sigma_{x}^2}-\frac{1}{\sigma_{x0}^2} \right) \sum_{t=1}^{T-1} \sum_{i=1}^{n-1} x_{t,\frac{i}{n}}^2 - (n-1)(T-1)  \log \frac{\sigma_{x0}^2}{\sigma_{x}^2}\,.
\end{eqnarray*}
Similar arguments can be applied for pairs $(t,\bs)$ and  $(t+1,\bs)$ in $D^{(2)}_{ST}$, and also pairs $(t,\bs)$ and $(t+1,\bs-\frac{1}{n})$ in $D^{(4)}_{ST}$. 
By apply similar arguments in \cite{bachoc2019composite}, we have from \eqref{microergodic_div} and \eqref{sigma_phi_div}, for any $\epsilon>0$,
$$L_{ST}(\theta_0;X)-\max_{\theta \in \Theta, |\theta-\theta_0|>\epsilon}L_{ST}(\theta;X)=O(ST)>0,\text{ almost surely}\,.$$
Hence, Assumption \ref{ass.infill.consis} holds with $\delta=1$. 

Also, for each consecutive stationary segments $\bfX_{j}^o$ and $\bfX_{j+1}^o$ of the random field, where $j=1,2,\ldots,m_o$, recall that the parameter in $j$-th segment is $\theta_{j}^o=(\sigma_{j}^{o2},\phi_{j}^o,\varphi_{j}^o)$. Partition the parameter space $\Theta_j \subset\mathbb{R}^+ \times (-1,1) \times \mathbb{R}^+$ into two subspaces as follows: $\Theta_{1j}=\{(\sigma^2,\phi,\varphi)\in\Theta: \sigma_{x}^2\varphi=\sigma_{xj}^{o2}\varphi_{j}^o\}$ and $\Theta_{2j}=\Theta	\setminus \Theta_{1j}$.
We can easily show that if the parameter in $(j+1)$-th segment $\theta_{j+1}^o=(\sigma_{j+1}^{o2},\phi_{j+1}^o,\varphi_{j+1}^o) \in \Theta_{1j}$, then $W_{ST}^{(j)}(q; \theta_{j}^o, \theta_{j+1}^o)=O_p(1)$. On the other hand, if the parameter in $(j+1)$-th segment $\theta_{j+1}^o=(\sigma_{j+1}^{o2},\phi_{j+1}^o,\varphi_{j+1}^o) \in \Theta_{2j}$, then $W_{ST}^{(j)}(q; \theta_{j}^o, \theta_{j+1}^o)=O(S)<0$, almost surely, see \cite{Ying1991} and \cite{bachoc2019composite}.
Hence, Assumption \ref{ass.diff.model} holds. Also, it can be easily shown that Assumption \ref{ass.infill.mix.time}($p$) holds with $p=1$.

\subsubsection{Gaussian noises with multiplicative exponential covariance function}\label{ill_example_exp2}
Consider the autoregressive spatial model in \eqref{sim.process.infill} with stationary Gaussian noises $Z=\{z_{\bs}: \bs \in  \mathcal{S}\subset \mathbb{R}^2\}$, with mean $0$ and a covariogram
${\rm Cov}(z_{\bs_1}, z_{\bs_2})=\sigma^2 \exp (-\varphi_1 |\bs_{2}^1-\bs_{1}^1|-\varphi_2|\bs_{2}^2-\bs_{1}^2|)$, with ${\rm Cov}(\varepsilon_{t,\textbf{u}}, \varepsilon_{t',\textbf{v}}) =0$, when $t\neq t'$. \cite{Ying1993} studied the estimability of $\sigma^2\varphi_1\varphi_2$, $\sigma^2$, $\varphi_1$ and $\varphi_2$ under spatial context, and also studied the consistency of model parameter estimators. We show that Assumption \ref{ass.infill.consis} holds with $\delta=1$.

For simplicity of presentation, we give the proof for the time lag $k=1$, and the maximum cardinality of the neighborhood set $B_{\mathcal{N}}=4$. Also, the data $\bfY$ is observed on $\mathcal{Z}_n=\mathcal{T}_n\times \mathcal{S}_n$ with $\mathcal{T}_n=[1,T_n]\cap \mathbb{N}$ and $\mathcal{S}_n=\{(\frac{i}{n},\frac{j}{n}):i,j =1,\ldots,n\}$. Extensions to $k>1$, larger $B_{\mathcal{N}}$ and other sampling configuration $\mathcal{S}_n$ are possible but technically involved and we do not pursue them here. We define several index sets that will be used throughout this section, where
\begin{align*}
	E_{1ST}&=\{(t,\bs):t=1, \bs\in \mathcal{S}, \text{ each } \bs  \text{ repeats } 5 \text{ times} \},\\
	E_{2ST}&=\{(t,\bs):t=T, \bs\in \mathcal{S}, \text{ each } \bs  \text{ repeats } 5 \text{ times} \},\\
	D_{ST}&=\{(t,i,\bs_1,\bs_2): 1\leq t, t+i \leq T, 0\leq i\leq 1, \bs_1 \in \mathcal{S}, \bs_2 \in \bs_1\cup \mathcal{N}(\bs_1), \text{if }i=0, \bs_1\neq \bs_2\}.
\end{align*}

Note that we have the neighborhood set $\mathcal{N}(\bs)=\{\bs+\textbf{e}_1,\bs-\textbf{e}_1,\bs+\textbf{e}_2,\bs-\textbf{e}_2\}$ where $\textbf{e}_1=(\frac{1}{n},0)$ and $\textbf{e}_2=(0,\frac{1}{n})$, and hence $|\mathcal{N}(\bs)|=4$. The sets $E_{1ST}$ and $E_{2ST}$ are used to index the marginal loglikelihood for compensating the edge effect. 
The set $D_{ST}$ is used to index the pairwise loglikelihood between $1$ and $T$. 
By using similar argument as in Section \ref{ill_example_exp} and taking summation with an additional dimension, we can show that by apply similar arguments in \cite{bachoc2019composite}, we have for any $\epsilon>0$,
$$L_{ST}(\theta_0;X)-\max_{\theta \in \Theta, |\theta-\theta_0|>\epsilon}L_{ST}(\theta;X)=O(ST)>0,\text{ almost surely}\,.$$
Hence, Assumption \ref{ass.infill.consis} holds with $\delta=1$. 

Also, for each consecutive stationary segments $\bfX_{j}^o$ and $\bfX_{j+1}^o$ of the random field, where $j=1,2,\ldots,m_o$, recall that the parameter in $j$-th segment is $\theta_{j}^o=(\sigma_{j}^{o2},\phi_{j}^o,\varphi_{1j}^o,\varphi_{2j}^o)$. Partition the parameter space $\Theta_j \subset\mathbb{R}^+ \times (-1,1) \times \mathbb{R}^+ \times \mathbb{R}^+$ into three subspaces as follows:
$\Theta_1=\{(\sigma^2,\phi,\varphi_1,\varphi_2)\in\Theta: \sigma_{x}^2\varphi_1=\sigma_{xj}^{o2}\varphi_{1j}^o\}$, $\Theta_2=\{(\sigma^2,\phi,\varphi_1,\varphi_2)\in\Theta: \sigma_{x}^2\varphi_2=\sigma_{xj}^{o2}\varphi_{2j}^o\}$ and $\Theta_3=\Theta \setminus (\Theta_1 \cup \Theta_2)$. By similar argument in Section \ref{ill_example_exp} and the results of \cite{bachoc2019composite}, we can show that Assumption \ref{ass.diff.model} holds with $r_j=1$ when the value of parameter $\sigma_x^2\varphi_1$,  $\sigma_x^2\varphi_2$ or $\phi$ changes, and with $r_j=1/2$ when only the value of parameter $\varphi_1$, $\varphi_2$ or $\sigma^2$ changes. Also, it can be easily shown that Assumption \ref{ass.infill.mix.time}($p$) holds with $p=1$.

\section{Proof for solving CLMDL via PELT} \label{PELT.issue}
With Optimal Partitioning in \cite{Jackson2005}, the computational cost to minimize CLMDL is $O(ST^2)$. To operationalize PELT in \cite{Killick2012}, we need to calculate a $K$ term, such that 
$$\mathbb{C}(Y_{[(t_1+1):t_2]})+\mathbb{C}(Y_{[(t_2+1):T]})+K \leq \mathbb{C}(Y_{[(t_1+1):T]}),$$
for all $0\leq t_1<t_2<T$ where $\mathbb{C}(Y_{[(t_1+1):t_2]})$ is the cost function for segment of $Y_{[(t_1+1),t_2]}$ as defined in the main text.

A typical/standard assumption used to derive $K$ for PELT, e.g. see \cite{Killick2012} and \cite{Ma2016}, is that the likelihood function satisfies that
\begin{equation} \label{PELT.lik}
L_{ST}(\hat{\psi}_1;Y_{[(t_1+1),t_2]})+L_{ST}(\hat{\psi}_2;Y_{[(t_2+1),T]}) \geq L_{ST}(\hat{\psi};Y_{[(t_1+1):T]}).
\end{equation}
Note that \eqref{PELT.lik} is true if we have
$$L_{ST}({\psi};Y_{[(t_1+1),t_2]})+L_{ST}({\psi};Y_{[(t_2+1),T]}) = L_{ST}({\psi};Y_{[(t_1+1):T]}),$$
simply due to the definition of maximum likelihood. 
However, due to the temporal dependence and the edge effect, we instead have
$$L_{ST}({\psi};Y_{[(t_1+1),t_2]})+L_{ST}({\psi};Y_{[(t_2+1),T]}) + O_p(S) = L_{ST}({\psi};Y_{[(t_1+1):T]}).$$


When $S$ is fixed, as in the multivariate time series setting, \eqref{PELT.lik} can be shown to hold asymptotically since $O_p(S)=O_p(1)$. However, this argument does not work for $S \longrightarrow \infty$ as the $O_p(S)$ term is non-ignorable. In Lemma \ref{PELTK}, we first make the wrong assumption that \eqref{PELT.lik} holds and thus derive a operational $K$, and then prove that PELT with the derived $K$ is still asymptotically valid.

\begin{proof}[Proof of Lemma \ref{PELTK}]
	We first derive the formula of $K$ as stated in Lemma \ref{PELTK}. To fit into the framework of \eqref{costmin}, let 
	\begin{align*}
		\mathbb{C}(Y_{[(\tau_{j-1}+1):\tau_{j}]})&= C_{k,\mathcal{N}} \left\{ \sum_{i=1}^{\hat{c}_j}\log \hat{\xi}_{i,j} 
		+ \left(\frac{\hat{d}_j}{2}+1\right)\log (\tau_j-\tau_{j-1})+\frac{\hat{d}_j}{2}\log S\right\} \\
		&- L_{ST}^{(j)}(\hat{\psi}_j;Y_{[(\tau_{j-1}+1):\tau_{j}]})\,,
	\end{align*}
	$\beta=1$ and $f(m)=\log m$, so that 
	$\mathrm{CLMDL}(m,\Lambda,\Psi)=\sum_{j=1}^{m+1}\mathbb{C}( Y_{[(\tau_{j-1}+1):\tau_{j}]})+\beta f(m)$.
	We seek a constant $K$ such that for all $t<s<T$,
	\begin{equation} \label{costPELT}
	\mathbb{C}( Y_{[(t+1):s]})+\mathbb{C}(Y_{[(s+1):T]}) + K \leq \mathbb{C}( Y_{[(t+1):T]})\,.
	\end{equation}
	Let $\hat{\psi}_{1}=(\hat{\xi}_{1},\hat{\theta}_{1}), 
	\hat{\psi}_{2}=(\hat{\xi}_{2},\hat{\theta}_{2})$ and $\hat{\psi}_{\cdot}=(\hat{\xi}_{\cdot},\hat{\theta}_{\cdot})$ be 
	the model parameter estimators based on the segments $ Y_{[(t+1):s]},  Y_{[(s+1):T]}$ and $ Y_{[(t+1):T]}$ respectively. 
	Also, let $\hat{d}_{1}$, $\hat{d}_{2}$ and $\hat{d}_{\cdot}$ be the dimensions of $\hat{\theta}_{1}$, $\hat{\theta}_{2}$
	and $\hat{\theta}_{\cdot}$ respectively.
	
	
	Assume that \eqref{PELT.lik} holds\footnote{Note that \eqref{PELT.lik} obviously does not hold for the spatio-temporal setting, we make the assumption here just for the derivation of $K$.}, \eqref{costPELT} reduces to
	\begin{align*}\label{costPELT.a}
	&\left[\hat{\xi}_1+\left(\frac{\hat{d}_1}{2}+1\right)\log \left(\frac{s-t}{2}\right) +\frac{\hat{d}_1}{2}\log S\right] + 
	\left[\hat{\xi}_2+\left(\frac{\hat{d}_2}{2}+1\right)\log \left(\frac{T-s}{2}\right) +\frac{\hat{d}_2}{2}\log S\right] + \frac{K}{C_{k,\mathcal{N}}} \\
	&\leq \hat{\xi}_{\cdot}+\left(\frac{\hat{d}_{\cdot}}{2}+1\right)\log (T-t) +\frac{\hat{d}_\cdot}{2}\log S \,.
	\end{align*}
	To ensure that $K$ fulfills \eqref{costPELT} for all $t<s <T$, we determine the $K$ such that 
	the first two terms on the left of \eqref{costPELT} attain their maximum values and 
	that the term on the right of \eqref{costPELT} attains its minimum value. 
	Therefore, using the fact that $\arg\max_{0<x<T} \log \{x(T-x)\}=T/2$ and putting 
	$\hat{d}_1=\hat{d}_2=d_{max}, \hat{d}_{\cdot}=d_{min},\hat{\xi}_1=\hat{\xi}_2=\xi_{max}^*$ and $\hat{\xi}_{\cdot}=\xi_{min}^*$, 
	we can solve for the least possible $K$ that satisfies
	\begin{align*}
		 & \xi_{max}^*+\left(\frac{d_{max}}{2}+1\right)\log \left(\frac{T}{2}\right) + \frac{d_{max}}{2}\log S + \xi_{max}^*+\left(\frac{d_{max}}{2}+1\right)\log \left(\frac{T}{2}\right) + \frac{d_{max}}{2}\log S +
		\frac{K}{C_{k, \mathcal{N}}}  \\
		&=\xi_{min}^*+\left(\frac{d_{min}}{2}+1\right)\log T+\frac{d_{min}}{2}\log S\,.
	\end{align*}
	After some algebra, we obtain
	\begin{equation*}
		K=C_{k,\mathcal{N}}\left \{ \left(\frac{d_{min}}{2}-d_{max}\right)\log ST+(2+d_{max})\log 2 + \xi_{min} - 2\xi_{max}-\log T\right\}.
	\end{equation*}
	
	Note that $K=-O(\log ST) <0$. We now prove Lemma \ref{PELTK} that the true change points $T\Lambda^o$ will not be pruned by \eqref{prunning} asymptotically. We prove the result by induction. We first prove that asymptotically $T\lambda_1^o$ will not be pruned by any $t$ satisfying $T(\lambda_1^o+\epsilon_\lambda)\leq t<T\lambda_2^o.$
	\begin{align*}
	&\lim\limits_{S,T\to\infty}\mathbb P(T\lambda_1^o \text{ is pruned})\\
	=&\lim\limits_{S,T\to\infty}\mathbb P(\exists~T(\lambda_1^o+\epsilon_\lambda)\leq t<T\lambda_2^o, \text{ s.t. } F(T\lambda_1^o) + \mathbb{C}(Y_{[(T\lambda_1^o+1):t]})+K\geq F(t))\\
	\leq & \lim\limits_{S,T\to\infty}\mathbb P(\exists~T(\lambda_1^o+\epsilon_\lambda)\leq t<T\lambda_2^o, \text{ s.t. } F(T\lambda_1^o) + \mathbb{C}(Y_{[(T\lambda_1^o+1):t]})> F(t))\\
	\leq & \sum_{t=T(\lambda_1^o+\epsilon_\lambda)}^{T\lambda_2^o-1}\mathbb P(F(T\lambda_1^o) + \mathbb{C}(Y_{[(T\lambda_1^o+1):t]})> F(t)).
	\end{align*}
	Under the conditions of Theorem \ref{unknownprob}, the minimizer of $F(T\lambda_1^o)$ is $(0, T\lambda_1^o)$ asymptotically since there is no change point between $t=1$ and $t=T\lambda_1^o$. Note that $F(T\lambda_1^o) + \mathbb{C}(Y_{[(T\lambda_1^o+1):t]})> F(t)$ means that $(0,T\lambda_1^o,t)$ is not the minimizer of $F(t)$. By the same argument as in the proof of Lemma \ref{thm2} and Theorem \ref{unknownprob}, we can show that the probability $(0,T\lambda_1^o,t)$ is not the minimizer of $F(t)$ converges to 0 for any $T(\lambda_1^o+\epsilon_\lambda)\leq t<T\lambda_2^o$, where
    $$\mathbb P(F(T\lambda_1^o) + \mathbb{C}(Y_{[(T\lambda_1^o+1):t]})> F(t))= O(S^{-r/2}) + O(T\cdot S^{-r/2}),$$
    which then implies that $\lim\limits_{S,T\to\infty}\mathbb P(T\lambda_1^o \text{ is pruned})\longrightarrow 0$ as $T^2\cdot S^{-r/2} \longrightarrow 0$. Thus, for the evaluation of $F(T\lambda_2^o)$, $T\lambda_1^o$ is not pruned. By Theorem \ref{unknownprob}, the minimizer of $F(T\lambda_2^o)$ is $(0, T\lambda_1^o, T\lambda_2^o)$ asymptotically and thus can be found by PELT. The same argument can be used again to prove that asymptotically $T\lambda_2^o$ will not be pruned by any $t$ such that $T(\lambda_2^o+\epsilon_\lambda)\leq t<T\lambda_3^o$ and so on. Since the number of change point is finite, we finish the proof.
    
    Similarly, we can prove the result under the conditions of Theorem \ref{finerate} and $T\cdot S^{-r/2} \longrightarrow 0$, thus Lemma \ref{PELTK} is proved.
\end{proof}

\pagebreak
\setlength{\bibsep}{4pt plus 1ex}
\begin{spacing}{1.2}
	\bibliographystyle{apalike}
	\bibliography{Reference}
\end{spacing}